\documentclass[a4paper,11pt,twoside]{report}

\usepackage[english]{babel}
\usepackage[utf8]{inputenc}
\usepackage[T1]{fontenc}
\usepackage{listings}
\usepackage{lscape}
\usepackage{amsmath}
\usepackage{subcaption}
\usepackage{graphicx}
\usepackage{newpxtext,newpxmath}
\usepackage{comment}
\usepackage{microtype}
\usepackage{slashed}
\usepackage[colorinlistoftodos]{todonotes}
\usepackage{setspace}
\usepackage{tabularx}
\usepackage{array}
\usepackage{appendix}
\usepackage[
  a4paper,
  top=3cm,
  bottom=3cm,
  left=2.5cm,
  right=2.5cm,
  marginparwidth=1.75cm
]{geometry}

\usepackage[
  backend=bibtex,
  style=numeric-comp,
  sorting=none,
  giveninits=true,
  doi=false,
  url=false,
  isbn=false,
  eprint=true
]{biblatex}

\usepackage[
  colorlinks=true,
  linkcolor=blue,
  citecolor=purple,
  urlcolor=blue
]{hyperref}

\renewbibmacro{in:}{}

\DeclareFieldFormat[article]{title}{\mkbibquote{#1}}

\DeclareFieldFormat{pages}{#1}

\DeclareFieldFormat{eprint:arxiv}{%
  \href{https://arxiv.org/abs/#1}{%
    arXiv\addcolon\space#1%
    \iffieldundef{eprintclass}
      {}
      {\space[\printfield{eprintclass}]}%
  }%
}

\renewbibmacro*{journal+issuetitle}{%
  \iffieldundef{doi}
    {%
      \printfield{journaltitle}%
      \setunit{\addspace}%
      \printfield{volume}%
      \setunit{\addspace}%
      \printtext[parens]{\printfield{year}}%
      \iffieldundef{number}
        {}
        {\setunit{\addspace}\printtext{no.~\printfield{number}}}%
      \iffieldundef{pages}
        {}
        {\setunit{\addcomma\space}\printfield{pages}}%
    }
    {%
      \href{https://doi.org/\thefield{doi}}{%
        \printfield{journaltitle}%
        \setunit{\addspace}%
        \printfield{volume}%
        \setunit{\addspace}%
        \printtext[parens]{\printfield{year}}%
        \iffieldundef{number}
          {}
          {\setunit{\addspace}\printtext{no.~\printfield{number}}}%
        \iffieldundef{pages}
          {}
          {\setunit{\addcomma\space}\printfield{pages}}%
      }%
    }%
}

\renewbibmacro*{note+pages}{%
  \printfield{note}%
}

\newcommand{\ket}[1]{| #1 \rangle}
\newcommand{\bra}[1]{ \langle #1 |}

\title{From Instantons to AdS Boundaries}
\author{Leonardo Sanhueza}

\begin{document}

\begin{titlepage}

\newcommand{\HRule}{\rule{\linewidth}{0.5mm}} 


\begin{center}
    \includegraphics[width=2cm]{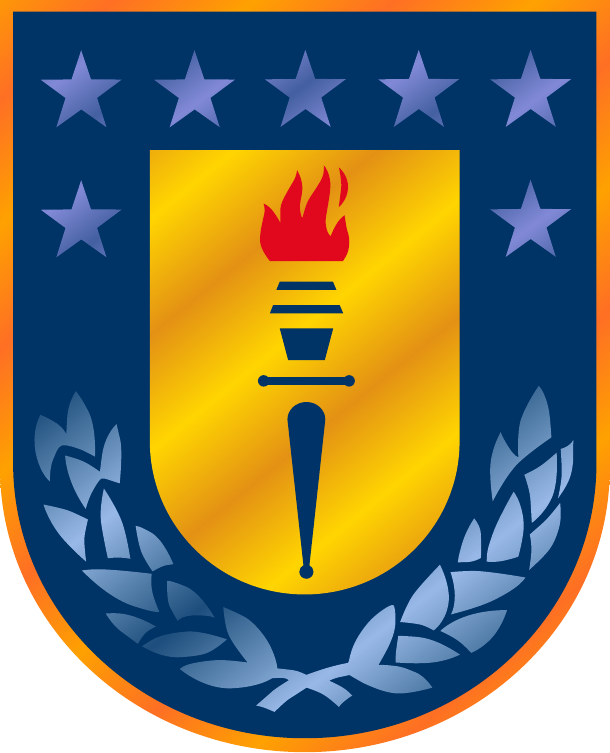}
\end{center} 
 

\center 

\textsc{\Large Universidad de Concepción}\\[0.5cm] 
\textsc{\large Facultad de Ciencias Físicas y Matemáticas - Departamento de Física}\\[0.5cm] 

\makeatletter
\HRule \\[0.4cm]
{ \huge \bfseries \@title}\\[0.4cm] 
{\Large \textit{Tracing the Geometric Path from Quantum Anomalies to Massive Gravity}}
\HRule \\[1.5cm]
 

\begin{minipage}{0.4\textwidth}
\begin{flushleft}
\large
\emph{Author:}\\
\@author
\end{flushleft}
\end{minipage}
~
\begin{minipage}{0.4\textwidth}
\begin{flushleft}
\large
\emph{Supervisor:}\\
Dr. Cristóbal Corral Badiola
\end{flushleft}
\end{minipage}\\[1.5cm]

\noindent
\begin{minipage}[t]{0.54\textwidth}
\vspace{0pt}
\begin{flushleft}

{\normalsize\textbf{External Committee}}\\[0.4cm]

\textbf{Dr. Luca Ciambelli}\\
{\small Université Libre de Bruxelles}\\[0.3cm]

\textbf{Dr. Gastón Giribet}\\
{\small New York University}\\[0.3cm]

\textbf{Dr. Rodrigo Olea}\\
{\small Pontificia Universidad Católica de Valparaíso}\\[0.3cm]

\end{flushleft}
\end{minipage}
\hfill
\begin{minipage}[t]{0.40\textwidth}
\vspace{0pt}
\begin{flushright}

{\normalsize\textbf{Internal Committee (UdeC)}}\\[0.4cm]

\textbf{Dr. Andrés Anabalón}\\
{\small Universidad de Concepción}\\[0.3cm]

\textbf{Dr. Julio Oliva}\\
{\small\emph{Co-supervisor}}\\
{\small Universidad de Concepción}

\end{flushright}
\end{minipage}

\vspace{1.5cm}
\makeatother

{\large A thesis submitted for the degree of}\\[0.5cm]
{\large \emph{PhD in Physics}}\\[0.5cm]
{\large \today}\\[0.5cm] 
\vfill 

\end{titlepage}

\renewcommand{\abstractname}{\LARGE Publications}
\begin{abstract}
The research presented in this thesis is based on the following works: 

\begin{enumerate}
    \item C. Corral, R. Olea, and \textbf{L. Sanhueza}, \textit{Extrinsic Renormalization in New Massive Gravity}, 	arXiv:2609.33690 (2026)~\cite{Corral:2026eyw}
    \item F. Colipí-Marchant, C. Corral, D. Flores-Alfonso, and
    \textbf{L. Sanhueza}, \textit{Axial anomaly in nonlinear conformal electrodynamics}, \textit{Physical Review D} \textbf{107}, 104042 (2023).~\cite{Colipi-Marchant:2023awk} 
\end{enumerate}

During the course of this thesis, I also participated in the
following publications and research collaborations:

\begin{enumerate}
    \item M. Adamo, L. Ciambelli, F. Diaz, M. Imseis,
    \textbf{L. Sanhueza}, and S. Speziale,
    \textit{Null hypersurfaces approaching null infinity}, in preparation.
    \item F. Diaz, S. Hüsnügil, O. Labrin, and
    \textbf{L. Sanhueza},
    \textit{Gravitational memory beyond null infinity through
    finite-distance Carrollian screens}, arXiv:2607.18675, (2026).~\cite{Diaz:2026igh} 
    \item G. Barriga, C. Henríquez-Baez,
    \textbf{L. Sanhueza}, and A. Vera,
    \textit{Self-gravitating baryonic tubes supported by
    $\pi$- and $\omega$-mesons and their flat limit}, arXiv:2511.11382, (2025).~\cite{Barriga:2025mky}
    \item C. Corral, B. Diez, D. Flores-Alfonso, N. Merino, and
    \textbf{L. Sanhueza}, \textit{Inhomogeneous metrics on complex bundles in Lovelock gravity},  \textit{Physical Review D} \textbf{111}, 124016 (2025).~\cite{Corral:2025yvr}
\end{enumerate}

\end{abstract}

\renewcommand{\abstractname}{\LARGE Abstract}
\begin{abstract}
This thesis presents two independent investigations of the role of symmetry, geometry, and boundary data in gravitational field theories. The first one concerns the axial anomaly of charged Dirac fermions on gravitational instantons coupled to nonlinear conformal electrodynamics. The second one, on the other hand, deals with extrinsic holographic renormalization of three-dimensional New Massive Gravity.
 
In the first part, we consider Einstein gravity coupled to ModMax electrodynamics: a nonlinear deformation of Maxwell theory that preserves classical conformal invariance and electromagnetic duality. We analyze nonlinearly charged Taub--NUT and Eguchi--Hanson configurations and evaluate the Dirac index using the Atiyah--Patodi--Singer theorem, including the spectral asymmetry of the boundary Dirac operator. The comparison identifies how the nonlinear gauge background fields modify the bulk and boundary contributions to the axial anomaly. In the Taub--NUT case, the result agrees with its Maxwell counterpart, whereas the Eguchi--Hanson configuration exhibits an explicit nonlinear dependence in the gauge-field contribution when expressed in terms of the solution parameters.

In the second part, we construct extrinsic boundary counterterms for asymptotically anti-de Sitter solutions of New Massive Gravity. For the Brown-Henneaux boundary conditions, a term linear in the extrinsic curvature reproduces the finite boundary response obtained from the auxiliary-field formulation. At the degenerate point, however, the field equations allow for relaxed AdS boundary conditions an additional Fefferman--Graham coefficient is allowed, and a quadratic extrinsic boundary functional accommodates the weakened fall-offs. The variational principle, boundary Ward identities, and Noether--Wald charges are obtained, and their equivalence with previous methods is shown.

Together, these studies demonstrate complementary ways in which global and asymptotic information enters the definition of quantum and gravitational observables.
\end{abstract}

\cleardoublepage
\thispagestyle{empty}

\vspace*{0.3\textheight}

\begin{flushleft}
\begin{minipage}{1\textwidth}

\itshape
``In its passing into another, something only comes together
with itself, and this relation to itself in the passing and in
the other is genuine infinity.''

\vspace{1cm}

\begin{flushright}
--- \textsc{G.\,W.\,F. Hegel}\\
\textit{Encyclopaedia Logic}
\end{flushright}

\end{minipage}
\end{flushleft}

\cleardoublepage

\renewcommand{\abstractname}{\LARGE Acknowledgements}
\begin{abstract}
Durante el proceso de mi doctorado, uno de los pilares más importantes ha sido mi familia. En especial, quiero agradecer a la persona más importante de mi vida: mi mamá. Muchas gracias por tu apoyo emocional y económico durante todos estos años; sin ti, nada de esto habría sido posible.

También quiero agradecer a mis hermanos, Dar, Rorro y Gustavo; a mis sobrinos, Matilda, Mati y Pablo; y a Edgardo. Gracias por escucharme durante estos años, por acompañarme y por darme sus consejos.

Durante mi estadía en Viña del Mar, tuve la fortuna de contar con tío Lito, Nano, Kika, tía Pato, Mati, Gilo, Michi y Juanito. Muchas gracias por acompañarme y compartir conmigo cuando me mudé a esa ciudad. Fue maravilloso encontrarlos allá. 

En la última etapa de mi doctorado, quiero agradecer especialmente a mi tío por darme la oportunidad de quedarme en Santiago. Gracias a usted y al Nano por escucharme y acompañarme durante el período más estresante de este proceso. No tengo palabras para expresar cuánto agradezco todo lo que han hecho por mí.

I do not have enough words to thank my advisor, Cristóbal Corral. Thank you for everything: for guiding me over so many years, for everything you taught me, and for the patience and dedication you brought to my education. We also shared many memorable moments along the way. Thank you for supporting me both emotionally and professionally. I am sorry if I was difficult at times. I am especially grateful for your support during the final stage of this thesis, for helping me through such a stressful period, and for giving me the strength to continue. My sincere thanks also go to Tania.

Another person who marked a turning point in my career and in the way I think about physics is Luca Ciambelli. Thank you for giving me the opportunity to visit Perimeter, for introducing me to new ideas, for sharing your way of thinking, and for giving me so much professional and personal advice. I acquired many new skills during that research stay. Thank you also for being patient with my English! My warmest thanks to you, Stefi and your daughter.

I would also like to thank Javier Peraza, who supported me greatly during my time in Canada. Thank you for patiently helping me with many applications, for answering all my questions, and for sharing your advice about both academic life and life in general. Thanks to Karen and los perritos. 

I am also deeply grateful to Julio Oliva. Much of what I learned during my studies is due to the dedication you brought to your courses. Many of the things I know today began in your classroom. Beyond teaching me physics, you gave me valuable advice and helped me with numerous documents and administrative matters. Thank you very much for your time and generosity, prof.

I cannot miss the opportunity to thank Professor Pablo Solano, who taught an excellent course on statistical mechanics and also helped me with several administrative processes.

I would like to thank Hernán González, who supported me during the final stage of my PhD by providing me with a working space at USS and helping me with my applications. I am equally grateful to Rodrigo Olea, who taught me how to write a scientific paper during the final stage of my PhD. Thank you for your patience, prof, and for all the advice you gave me during this period.

I would also like to acknowledge Giorgos Anastasiou, Ignacio Araya, Daniel Flores, Olivera Miskovic, Luis Avilés, Omar Valdivia, Pepe Díaz, Nelson Merino, Fabrizio Canfora, Patricio Salgado-Rebolledo, Diego Molina, Adolfo Cisterna, Ricardo Troncoso, Marcela Lagos, Aldo Vera, Carla Henríquez, Francisco Rojas, Alfredo Pérez, Fernando Izaurieta, Marcelo Loewe, Marco Bianchi, Jorge Zanelli, Felipe Díaz, Óscar Fuentealba, Céline Zwikel, Laurent Freidel, Bianca Dittrich, Luis Lehner, Simone Speziale, Jorge Cuadra, Tom Maedler, and Astor Sandoval. I am probably forgetting many people, and for that I sincerely apologize.

During my PhD, I had the opportunity to meet many wonderful people whom I would also like to mention here. Donah, thank you for supporting me during my time in Canada and for everything you did for me. I am also grateful to Mike, Sercan, Sonja, Rodrigo, Raquel, Nico, Simon, Jackie, Federico, Sruthi, José, Dawit, Thanos, Adrian, Anto, Kiana, Giulio, Ilaria, Gonzalo Barriga, Ángelo Yáñez, Marcelo Oyarzo, Jean Cortés, Lucas Acito, Facu, Mati Sempé, Mai, Camilo, Keanu, Yoel, Briceño, Oriana, Colipí, Pablo, Chinaski, and Marciano, as well as to my friends and colleagues at UNAP, UAI, USS, PI, PUCV, UFSM, and PUC. 

A special mention goes to Nico and Borja: thank you for the strange adventures, the discussions, and all the unforgettable moments!

I would also like to thank Nilsa and Alicia for their help with the many administrative procedures behind this process and for assisting me with all the required documents.

Finally, I am grateful to the members of my thesis committee for taking the time to review my thesis so carefully within such a short period.

The author declares the use of ChatGPT to assist in the preparation of figures and to improve the grammar and clarity of the manuscript. The scientific content, mathematical derivations, and conclusions remain the sole responsibility of the author.
\end{abstract}

\tableofcontents

\setcounter{tocdepth}{2}

\listoffigures
\listoftables

\chapter{Introduction}

Symmetry provides one of the principal organizing ideas of the modern theoretical physics. It restricts the form of interactions, relates apparently different configurations, and determines which quantities can be conserved. Its physical implications, however, depend on more than the local form of the equations of motion. Quantization may break a classical symmetry, global topology may distinguish configurations with similar local properties, and boundary conditions may determine whether a transformation represents a gauge redundancy or a physical symmetry. Understanding a field theory therefore requires a particular attention not only to its dynamical equations, but also to the structures needed to define its observables.

This thesis investigates these issues through two independent approaches. The first is related to the axial anomaly of charged Dirac fermions on gravitational instantons coupled to nonlinear conformal electrodynamics. The second concerns the holographic renormalization of New Massive Gravity through boundary terms constructed from the extrinsic curvature. These investigations address different physical questions and employ different mathematical techniques. Their common perspective is that local dynamics must be considered together with global geometry and boundary data in order to obtain a complete description of the quantities being calculated.

The axial anomaly provides a particularly direct illustration of this principle. A massless fermionic action may possess a classical axial symmetry even though the corresponding current fails to be conserved after quantization. In the path-integral formulation, this failure can be understood through the transformation of the regularized fermionic measure~\cite{Fujikawa:1979ay}. The anomaly is therefore not obtained by examining the classical equations alone. Its computation requires a quantum prescription that preserves the appropriate gauge symmetries and identifies the breaking of axial current conservation.

On nontrivial Euclidean backgrounds, this quantum statement has a geometric interpretation. The index of the chiral Dirac operator measures the difference between the numbers of zero modes of opposite chirality. On manifolds with boundary, its relation to bulk characteristic forms includes a contribution from the spectral asymmetry of the boundary Dirac operator. The Atiyah--Patodi--Singer theorem provides the corresponding framework and introduces the $\eta$-invariant as an essential part of the calculation~\cite{APS-eta}. In applications to gravitational instantons, the boundary contribution must consequently be evaluated together with the bulk gauge and gravitational terms.

In the first part of this thesis, we ask how this structure is realized when the electromagnetic background is governed by a nonlinear theory. The model considered is ModMax electrodynamics, a deformation of Maxwell theory that preserves classical conformal invariance and electric--magnetic duality of the source-free field equations~\cite{Bandos:2020jsw}. These properties make it possible to introduce nonlinear constitutive relations while retaining substantial control over the symmetry structure. They do not, however, imply that all solutions or quantum observables coincide with those of Maxwell theory. The nonlinear parameter can modify the gauge configurations supported by a given geometry and thereby affect the backgrounds entering the fermionic spectral problem.

We study this question for nonlinearly charged Taub- NUT and Eguchi--Hanson configurations in Einstein--ModMax theory. The analysis combines the construction of the classical fields, the identification of regular instanton sectors, and the computation of the Dirac index including the boundary spectral contribution. A central objective is to distinguish effects that arise from the nonlinear background solution from those fixed by the anomaly formula for a minimally coupled Dirac field. The resulting comparison shows how different geometries can retain or suppress explicit dependence on the nonlinear parameter in the quantities entering the index calculation~\cite{Colipi-Marchant:2023awk}.

The second part of this thesis explores the role of boundary data in gravitational theories. Three-dimensional gravity provides a particularly controlled setting for this purpose. Although vacuum Einstein gravity has no local propagating degrees of freedom, it possesses nontrivial global geometries, boundary excitations, and asymptotic symmetries. The BTZ black hole gives the clearest example: it is locally equivalent to \(\mathrm{AdS}_3\), but its global identifications produce horizons and nonvanishing conserved charges~\cite{Banados:1992wn,Banados:1992gq}. Moreover, the Brown--Henneaux analysis showed that suitable AdS boundary conditions lead to two copies of the Virasoro algebra with a nonvanishing classical central charge~\cite{Brown:1986nw}. These features make three-dimensional gravity a useful framework for studying how physical information can be encoded at a boundary even when the local bulk dynamics is simple.

Higher-curvature interactions modify this picture by introducing local gravitational degrees of freedom. New Massive Gravity is a parity-preserving extension of three-dimensional general relativity whose generic linearized spectrum contains the two helicities of a massive spin-two field~\cite{Bergshoeff:2009hq,Bergshoeff:2009aq}. The theory retains locally AdS solutions such as the BTZ black hole but also admits non-Einstein AdS waves and black holes with weakened asymptotic behavior~\cite{Ayon-Beato:2009cgh,Oliva:2009ip,Giribet:2009qz}. Consequently, the asymptotic expansion appropriate to Einstein gravity cannot be imposed on every NMG solution without first examining the higher-curvature field equations.

The gravitational action must also be compatible with the data prescribed at the boundary. In Einstein gravity, the Gibbons--Hawking--York term removes the normal derivatives of the metric variation at a finite boundary~\cite{York:1972sj,Gibbons:1976ue}. For asymptotically AdS spacetimes, additional boundary terms are required to obtain finite observables when the regulator is removed. A geometrical alternative to the standard intrinsic counterterm expansion is provided by extrinsic boundary terms, or Kounterterms, constructed using the extrinsic curvature of radial hypersurfaces~\cite{Aros:1999id,Olea:2005gb,Olea:2006vd,Miskovic:2009bm}. Their relation to standard holographic renormalization, as well as their limitations for general boundary geometries and conformal anomalies, has been studied in Refs.~\cite{Anastasiou:2020zwc,Anastasiou:2020mik}.

The parameter space of NMG contains two different special loci. At the critical point, the massive and massless linearized spin-two operators become degenerate, the Brown--Henneaux central charges vanish, and logarithmic modes may appear. At the degenerate point, by contrast, the two maximally symmetric vacua coincide. The radial field equations then allow a term \(g_{(1)ij}\) linear in the Fefferman--Graham coordinate, producing a weakened AdS fall-off and an additional boundary source~\cite{Alishahiha:2010bw,Giribet:2010ed,Kwon:2011jz,Cunliff:2013en}. These two conditions have different origins and do not coincide for the NMG conventions adopted in this thesis.

The second original investigation constructs extrinsic boundary terms for the power-law asymptotic sectors of NMG, including the relaxed sector arising at vacuum degeneracy. For configurations with \(g_{(1)ij}=0\), a boundary term linear in the extrinsic curvature renders the on-shell action and its variation finite under the boundary conditions considered. The resulting holographic stress tensor agrees with that obtained from the auxiliary-field formulation of Hohm and Tonni~\cite{Hohm:2010jc}. This agreement holds for locally AdS solutions and for the nonconstant-curvature configurations within the same asymptotic sector.

At the degenerate point, the additional coefficient \(g_{(1)ij}\) changes the boundary analysis. Although it does not generate a new \(g_{(1)ij}\)-dependent power-law divergence in the bulk action at the orders examined, it produces a subleading divergent contribution when the linear extrinsic boundary term is evaluated on the relaxed expansion. A quadratic extrinsic term removes this contribution and leads to a finite variation involving responses conjugate to both \(g_{(0)ij}\) and \(g_{(1)ij}\). The construction is tested independently through Noether--Wald charges. It reproduces the standard NMG charges of the BTZ black hole, gives a vanishing charge for the null Killing vector of the AdS waves considered, and agrees with previous results for the rotating hairy black hole at the degenerate point. Logarithmic branches associated with the critical point are not included in this analysis.

The thesis is organized to separate the necessary background from the original investigations. Chapter~\ref{Chapter2} introduces symmetries, fermionic fields, and the axial anomaly. Chapter~\ref{chapter3} develops the Einstein--ModMax instanton analysis and the associated Dirac-index calculation. Chapter~\ref{Chapter4} introduces three-dimensional gravity, the geometry of the AdS boundary, radial foliations, and the variational foundations of holographic renormalization. Chapter~\ref{chapter5} develops the extrinsic renormalization of NMG, derives the corresponding boundary responses and conserved charges, and tests the construction on representative solutions. Chapter~\ref{chapter6} summarizes the principal conclusions and identifies directions for further research.

\chapter{Symmetries and Anomalies} \label{Chapter2}
Symmetries lie at the core of modern theoretical physics. At the classical level, Noether's theorem~\cite{Noether01011971} establishes an elegant and precise relation between continuous symmetries of the action of a theory and conserved quantities, such as energy, momentum, and electric charge. This correspondence has shaped our understanding of the fundamental interactions and provides an essential tool in both classical and quantum field theory. 

It is important, however, to distinguish between global and gauge symmetries. A global symmetry is characterized by transformations parameters that are constant throughout spacetime and acts non-trivially on physically distinct states. According to Noether's first theorem, every continuous global symmetry gives rise to a conserved current and an associated conserved charge. By contrast, a gauge symmetry depends on arbitrary functions of spacetime and represents a redundancy in the description of the physical degrees of freedom, rather than an ordinary transformation between distinct physical states. Such symmetries are governed by Noether's second theorem. Because their transformation parameters are arbitrary functions of spacetime, gauge invariance implies that some of the equations of motion are related to one another and therefore are not independent. This reflects the fact that gauge related field configurations describe the same physical state and that the gauge formulation contains redundant variables. 

The distinction between global and gauge symmetries becomes particularly important upon quantization, since a symmetry of the classical theory need not survive in the quantum theory. This failure is known as a quantum anomaly and, for a continuous global symmetry, may manifest as a violation of the corresponding classical conservation law. Its physical significance depends on the symmetry involved. An anomalous global symmetry need not compromise the consistency of the theory and can reveal exact information about its structure and infrared behavior. By contrast, an non vanishing anomaly in a gauge symmetry obstructs the consistency of the quantum theory. Gauge-anomaly cancellation therefore imposes stringent constraints on the allowed matter content and couplings.

A paradigmatic example is provided by the Standard Model, in which gauge anomaly cancellation requires a precise balance among the fermionic degrees of freedom. The contributions of quarks and leptons cancel within each generation, ensuring the consistency of the electroweak gauge theory. The bottom quark belongs to a left-handed weak isospin doublet and therefore requires an up-type partner, namely the top quark. The importance of the introduction of the top quark is therefore essential for restoring anomaly cancellation and ensuring gauge invariance and renormalizability. 

\begin{figure}[ht]
    \centering
    \includegraphics[width=0.65\linewidth]{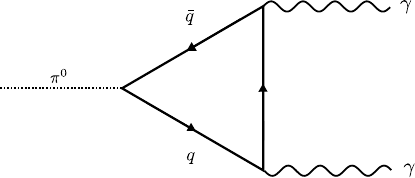}
    \caption{Triangle diagram representing the neutral pion decay process $\pi^0 \to \gamma \gamma$}
    \label{fig:decay}
\end{figure}

The first examples of anomalies were discovered in the context of quantum electrodynamics~\cite{Tomonaga1946,Schwinger1948,Feynman1949a,Feynman1949b,Dyson1949}. It was known that the axial, or chiral, current associated with massless fermions is classically conserved. This conservation law follows directly from Noether's theorem and reflects an underlying chiral symmetry of the classical action. Nevertheless, experimental observations revealed a puzzling inconsistency with this statement: the neutral pion was observed to decay into two photons $\pi^{0}\to\gamma\gamma$  with a well-defined decay rate (See figure \ref{fig:decay}).

From the classical viewpoint, this process appears to be forbidden by the conservation of the axial current. 

During the 1950s and 1960s, considerable effort was devoted to understanding this decay process. In 1949, Steinberger proposed the Pion-Nucleon $\pi-N$ model~\cite{Steinberger} to describe the $\pi^{0}\to\gamma\gamma$ decay (see also Fukuda and Miyamoto~\cite{FukudaMiyamoto1949}). This model, which contains a $\gamma_{5}$ vertex, correctly reproduced the experimental decay rate when evaluated using the triangle diagram. On the other hand, since electrodynamics possesses a conserved vector current, isospin symmetry is associated with a conserved isospin current. It is therefore natural to attempt to construct a conserved axial current corresponding to chiral symmetry, leading to an axial Ward identity.

To develop this idea, an axial Ward identity was introduced within the $\pi-N$ model. An attempt involved replacing the pseudoscalar $\gamma_{5}$ vertex with an axial-vector $\gamma_{\mu}\gamma_{5}$ vertex. Although this approach was well motivated from the symmetry perspective, there was a tension between axial current conservation, gauge invariance, and the experimental observations, predicting the presence of the axial anomaly.

The analysis of the conservation of the axial current was considered by Julian Schwinger in 1951~\cite{Schwinger:1951nm}. He pointed out that, in quantum electrodynamics, the axial symmetry leads to a violation of the corresponding conservation law at the quantum level. This observation suggested that quantum effects could obstruct the preservation of classical symmetries.

Several years later, in 1963, Ken Johnson analyzed the axial-vector-vector triangle diagram~\cite{JOHNSON1963} while addressing a fundamental question: is it possible to define the axial current in such a way that both the conservation of the vector current and the conservation of the axial current hold simultaneously? Or is the incorrect axial current under consideration? His work clarified that this requirement leads to conflict once quantum corrections are taken into account. 

During the 1960s, the framework of current algebra, developed by Murray Gell-Mann~\cite{Gell-MannCurrent}, was extensively used to address the problem of the axial current. Within this context, and relying on the hypothesis of a partially conserved axial current (PCAC), the neutral pion was interpreted as a pseudo-Goldstone boson associated with the spontaneous breaking of chiral symmetry. This interpretation was motivated by the experimental data indicating that the axial current is almost conserved. The PCAC relation is given by 
\begin{equation} \label{PCAC relation}
    \partial_{\mu}\left(J_{5}^{\mu}\right)_{A}=f_{\pi}m_{\pi}^{2}\pi_{A}\,,
\end{equation}
where $f_{\pi}$ is the pion decay constant, $m_{\pi}$ is the pion mass, $\pi_{A}$ denotes the pion field, and the capital index A labels the generators of the internal flavor symmetry. The current $J_{5}^{\mu}$ is the axial current defined in Eq.~\eqref{AxialCurrent}.

Within this approach, Sutherland and Veltman derived a result that is in conflict with experimental observations!~\cite{Veltman,Sutherland:1967vf}. Assuming PCAC, electromagnetic gauge invariance, and the absence of additional contributions to the divergence of the axial current, they found that the amplitude for the decay $\pi^{0}\to\gamma\gamma$ vanishes in the soft-pion limit and is therefore strongly suppressed by the pion mass. Indeed, considering the limit of exact axial current conservation $m_{\pi}\to0$, the PCAC relation gives the conservation of the axial current. This result evidently exposes a tension between symmetry principles and experimental data, indicating the necessity of an anomalous breaking of the axial symmetry.

Despite these difficulties, the puzzle was finally resolved in 1969 independently by Bell and Jackiw on one hand, and by Adler on the other, using different approaches~\cite{Bell:1969ts,Adler:1969gk}. Bell and Jackiw analyzed the problem within the framework of the $\sigma$  model, while Adler considered spinor electrodynamics. Although their methods were different, both analyzes led to the same conclusion: it is not possible to simultaneously maintain exact axial current conservation and gauge invariance in quantum theory. 

The resolution of the puzzle came from the analysis of the AVV triangle Feynman diagram. The quantum corrections associated with this diagram revealed that, while the conservation of the vector current can be preserved, the conservation law for the axial current is violated. One finds that the divergence of the axial current acquires an additional quantum contribution given by 
\begin{equation} \label{axial current}
    \partial_{\mu}J_{5}^{\mu}=\frac{e^{2}}{16\pi^{2}}F_{\mu\nu}\tilde{F}^{\mu\nu}\:,
\end{equation}
where $e$ denotes the electric charge, $F_{\mu\nu}$ is the electromagnetic field strength tensor, and 
\begin{equation} \label{dual F}
    \tilde{F}^{\mu\nu}=\frac{1}{2}\epsilon^{\mu\nu\lambda\rho}F_{\lambda\rho}\,,
\end{equation}
its dual tensor. This is the well known Adler-Bell-Jackiw (ABJ) anomaly, or simply the axial anomaly.

The discovery of this result represented a turning point in quantum field theory. It demonstrated that symmetries valid at the classical level may fail after quantization, and the conservation laws can be modified by quantum effects. In the years that followed, the axial anomaly was profoundly studied and was eventually understood to possess a close relation to geometry and topology~\cite{Jackiw:1976dw,Jackiw:1977pu,Nielsen:1977aw,Nielsen:1977qm}. In particular, it was shown that the singlet axial anomaly is related to the index of the Dirac operator through the Atiyah-Singer index theorem~\cite{Atiyah:1963zz,Atiyah:1968mp,Atiyah:1967ih,Atiyah:1970ws,Atiyah:1971rm} (See also~\cite{freed2021atiyahsingerindextheorem} for a review). This is because the anomaly can be expressed as a sum over the eigenfunctions of the Dirac operator $\varphi_n(x)$, where only the zero-modes of definite chirality contribute,
\begin{equation} \label{IndexD}
    \text{index }D_{+}=n_{+}-n_{-}=\int dx\sum_{n}\varphi_{n}^{\dagger}(x)\gamma_{5}\varphi_{n}(x)\,,
\end{equation}
where $n_{+}$ and $n_{-}$ denote the positive and negative chiral zero modes, respectively, and $D_{+}$ is the Weyl operator. 

The Atiyah-Singer theorem further allows one to express this index in terms of topological invariants. In four dimensions, and in the absence of gravitational contributions, the index of the chiral Dirac operator is related to the Pontryagin index according to 
\begin{equation} \label{IndexD2}
    \text{index }D_{+}=-\frac{1}{8\pi^{2}}\int\text{tr}\,\left(F\wedge F\right)\,,
\end{equation}
where 
\begin{equation}
    F= \frac{1}{2} F_{\mu \nu}^A t_A dx^\mu dx^\nu \,,
\end{equation}
is the gauge-field strength two-form. Here, $t_{A}$ are the generators of the gauge Lie algebra in the representation carried by the fermions, while the $tr$ denotes the trace over the corresponding internal representation indices. 
This relation reveals that the anomaly admits a natural topological interpretation. This result establishes a profound connection between quantum field theory and differential geometry, showing that the non-trivial topology of fiber bundles can have direct physical consequences. In particular, topologically non-trivial gauge configurations, such as instantons, play a fundamental role in this context, where the axial anomaly can be tracked back to the imbalance of spinors with different chirality, produced by their interaction with gauge fields possessing nontrivial topology. 

It is important to emphasize that the original derivation of the axial anomaly was performed within the framework of perturbation theory. Although this approach resolved the $\pi^{0}\to\gamma\gamma$  puzzle, it left open several conceptual questions concerning the fundamental origin of the anomaly. 

This issue was addressed by Fujikawa, who, using the path integral formulation introduced by Feynman, developed an alternative derivation of the axial anomaly based on the transformation properties of the fermionic path integral measure under axial transformations~\cite{Fujikawa:1979ay,Fujikawa:1980eg}. 

Within this framework, it became clear that anomalies arise from the non-invariance of the functional measure under symmetry transformations. In particular, the Fujikawa method allowed us to show that the axial anomaly admits a geometric interpretation through the properties of the fermionic measure and its regularization. 

Additionally, together with the Adler-Bardeen theorem~\cite{Adler:1969er}, it is well established that the axial anomaly is one-loop exact, meaning that the anomaly receives no contributions from radiative corrections. Therefore, Fujikawa's method derives the anomaly without relying on a diagrammatic perturbative expansion.

The concepts described above reveal that anomalies are not merely artifacts of perturbation theory, but rather fundamental features of quantum field theories with profound physical and mathematical consequences. Motivated by these ideas, in this chapter, we review the symmetries of a system of scalar fields as a simple setting in which to introduce internal and spacetime symmetries and their associated Noether currents. We then turn to the Dirac action and its symmetries, focusing in particular on the vector $U(1)$ symmetry and the axial symmetry. At the classical level, both symmetries give rise to conserved currents. However, after quantization, the conservation of the axial current may fail, leading to the axial anomaly. To understand this phenomenon, we introduce the path integral formalism and then present the Fujikawa method, which provides a geometrically motivated framework for deriving anomalies from the non-invariance of the functional measure under axial transformation. 

\section{Internal and Spacetime Symmetries}\label{SectionInternalSpacetime}

In classical mechanics, the dynamics of a system are described by a finite number of generalized coordinates $q^{i}(t)$. In field theory, by contrast, the fundamental objects are fields $\phi^{i}(\vec{x},t)$, which are defined at every point in space $\vec{x}$ and time $t$. A field is therefore a mathematical object that assigns to each point in spacetime a scalar, vector, spinor, or tensor, depending on the physical theory under consideration. The configuration space of a field theory is the set of all possible field configurations, where each point in this infinite-dimensional space represents a particular configuration of the fields throughout spacetime. 

Since the value of the fields can vary independently at each point in space, a field theory possesses infinitely many degrees of freedom, in contrast to the finite-dimensional systems of classical mechanics. Nevertheless, the dynamics of classical field theories are still determined by an action principle, where the Lagrangian density is integrated over spacetime, as we will see. 

As discussed in the introduction of this chapter, Noether's theorem establishes a profound connection between the continuous symmetries of the action and conserved quantities. In field theory, these symmetries give rise to locally conserved currents, whose associated charges are constant in time under appropriate boundary conditions.

\begin{equation}
    \eta_{\mu\nu}=\eta^{\mu\nu}=\text{diag }\left(-,+,...,+\right)\,.,
\end{equation}
with $\eta^{\mu\nu}$ the inverse metric. The fields satisfy the Klein-Gordon equation 
\begin{equation}
    \Box\phi^{i}(x)=m^{2}\phi^{i}(x)\,,
\end{equation}
where $\Box=\eta^{\mu\nu}\partial_{\mu}\partial_{\nu}$ is the Lorentz-invariant d'Alambertian and m is the mass. This equation follows from the variational principle applied to the action 
\begin{equation} \label{FreeScalarAction}
    I\left[\phi\right]=-\frac{1}{2}\int d^{D}x\left[\eta^{\mu\nu}\partial_{\mu}\phi^{i}\partial_{\nu}\phi^{i}+m^{2}\phi^{i}\phi^{i}\right]\,, 
\end{equation}
where the repeated internal index $i$ is summed from $1$ to $n$. 

We are interested in symmetries. A symmetry of the system is a transformation of the fields
\begin{equation}
    \phi^{i}(x)\to\phi'{}^{i}(x)\,,
\end{equation}
 with the property that if $\phi^{i}(x)$ is a solution of the equations of motion, then $\phi^{'i}(x)$ is also a solution. For the systems considered in this thesis, it is sufficient to restrict attention to transformations that leave the action invariant up to a boundary term.\footnote{More precisely, the Lagrangian may be quasi-invariant: its infinitesimal variation can take the form $\delta_\epsilon\mathcal L=\partial_\mu K^\mu_\epsilon$, where $\mathcal L$ denotes the Lagrangian density. The action then changes by a boundary integral, $\delta_\epsilon I=\int_{\partial M}K^\mu_\epsilon \,\mathrm d\Sigma_\mu$. Such a total derivative does not affect the bulk Euler--Lagrange equations. In the presence of boundaries, however, this does not by itself establish a symmetry of the full variational problem: the transformation must also preserve the allowed boundary conditions, and its boundary contribution must be treated consistently with the boundary terms included in the action.} 

We distinguish two classes of symmetries. A transformation is called an \textbf{internal symmetry} if it acts only on the field indices, mixing the $n$ fields among themselves at each fixed spacetime point $x^{\mu}$, while leaving the spacetime coordinates invariant. A transformation is called a \textbf{spacetime symmetry} if it acts on the coordinates of Minkowski spacetime, such as a translation or a Lorentz transformation, and therefore changes the argument $x^{\mu}$ of the fields. 

In this section, we study the internal and spacetime symmetries of a system of scalar fields and derive the corresponding conserved currents. 

\subsection{Internal Symmetries}

To illustrate these ideas, we consider the simplest internal symmetry of this system, namely the $SO(n)$ symmetry. The $n$ fields $\phi^{i}$ may be regarded as the components of a vector in an abstract internal space $\mathbb{R}^{n}$. The action is invariant under rotations in this internal space, and the transformation of the fields is given by 
\begin{equation}
    \phi^{i}(x)\to\phi^{i}{}'(x)=R^i{}_{j}\phi^{j}(x)\,,
\end{equation}
where $R^{i}{}_{j}$ is a matrix of the orthogonal group $SO(n)$, satisfying
\begin{equation}
    R^{T}R=\mathbb{I}\,,\qquad\det R=1\,.
\end{equation}
This transformation mixes the fields at every spacetime point in the same way, independently of the coordinate $x^{\mu}$. Therefore, it is called a \textbf{global internal symmetry}. The invariance of the action~\eqref{FreeScalarAction} follows from the fact that $SO(n)$ rotations preserve both the internal metric $\delta_{ij}$, which define the Killing-Cartan metric of $SO(n)$. In particular, 
\begin{equation}
    \delta_{ij}\phi^{i}{}'\phi^{j}{}'=\delta_{ij}R^{i}{}_{k}R^{j}{}_{l}\phi^{k}\phi^{l}=\delta_{kl}\phi^{k}\phi^{l}\,.
\end{equation}
Since $R$ is constant, the derivatives transform in the same way, 
\begin{equation}
    \partial_{\mu}\phi^{i}{}'=R^{i}{}_{j}\partial_{\mu}\phi^{j}\,,
\end{equation}
and hence the kinetic term is also invariant. 

Since we are interested in Noether's theorem, it is sufficient to work with infinitesimal transformations. To first order in a small parameter $\epsilon\ll1$, an $SO(n)$ transformation may be written as 
\begin{equation}
    R^{i}{}_{j}=\delta^{i}{}_{j}+\epsilon r^{i}{}_{j}+\mathcal{O}(\epsilon^{2})\,,
\end{equation}
where $r^{i}{}_{j}$ is a generator of $SO(n)$ and is antisymmetric $r_{ij}=-r_{ji}$. This antisymmetric condition is precisely the condition that ensures $R^{T}R=\mathbb{I}$ to first order in $\epsilon$. The corresponding infinitesimal field variation is 
\begin{equation}
    \delta\phi^{i}=\phi^{i}{}'-\phi^{i}=\epsilon r^{i}{}_{j}\phi^{j}\,.
\end{equation}
The previous discussion of the orthogonal group $SO(n)$ can be generalized to an arbitrary connected Lie group $G$. This generalization is important because the internal symmetry groups that appear in physics, such as $U(1)$, $SU(2)$, and $SU(3)$, are not always orthogonal groups. 

Let $G$ be a connected Lie group of dimension $\dim G$. We consider a representation of $G$ in which the generators of its Lie algebra are matrices $(t_{A})^{i}{}_{j}$, with $A=1,...,{\rm dim}G$. These generators satisfy the commutation relation
\begin{equation} \label{StructureConstant}
    [t_{A},t_{B}]=f_{AB}{}^{C}t_{C}\,,
\end{equation}
where $f_{AB}{}^{C}$ are the structure constants of the Lie algebra. They encode the algebraic structure of the group and are antisymmetric in the lower indices, $f_{AB}{}^{C}=-f_{BA}{}^{C}$. In what follows, we take the generators to be anti-Hermitian matrices, so that no explicit factors of i appear in the commutation relations. 

A general element of the Lie algebra is a linear combination of the generators with real parameters $\theta^{A}$, that is, 
\begin{equation}
    \Theta=\theta^{A}t_{A}\,.
\end{equation}
The corresponding group element is obtained by exponentiation,
\begin{equation}
    U(\theta)=e^{\Theta}=e^{\theta^{A}t_{A}}\,.
\end{equation}
The scalar fields transform in the representation just described as follows 
\begin{equation}
    \phi^{i}(x)\to\phi'{}^{i}(x)=U(\Theta)^{i}{}_{j}\phi^{j}(x)\,.
\end{equation}
At the infinitesimal level, expanding $U(\Theta)$ to first order in $\theta^{A}$, one finds 
\begin{equation}
    U(\Theta)^{i}{}_{j}=\delta_{j}^{i}+\theta^{A}\left(t_{A}\right)^{i}{}_{j}+\mathcal{O}(\theta^{2})\,,
\end{equation}
so that, the field variation is given by 
\begin{equation} \label{InternalSym}
    \delta\phi^{i}=\theta^{A}\left(t_{A}\right)^{i}{}_{j}\phi^{j}=\Theta^{i}{}_{j}\phi^{j}\,.
\end{equation}
In compact notation, one can write 
\begin{equation}
    \delta\phi=\Theta\phi\,.
\end{equation}
This is the general form of an infinitesimal internal symmetry transformation. The $SO(n)$ case, is recovered by taking $\Theta=\epsilon r$, with $r$ an antisymmetric matrix. 

\subsection{Spacetime Symmetries}

We now study the spacetime symmetries of a system of scalar fields. Unlike internal symmetries, which act on the internal degrees of freedom of the fields while leaving the spacetime coordinates unchanged, spacetime symmetries act directly on the coordinates of Minkowski spacetime. The most important examples are the Lorentz transformations and spacetime translations, which together form the Poincaré group.

The Lorentz group is defined as the group of homogeneous linear transformations in $D-$ dimensional Minkowski spacetime that preserves the spacetime interval. Thus, 
\begin{equation} \label{IntervalInvariance}
    x^{\mu}\eta_{\mu\nu}x^{\nu}=x'{}^{\mu}\eta_{\mu\nu}x'{}^{\nu}\,.
\end{equation}
The coordinates transform as 
\begin{equation} \label{CoordinateTrans}
    x^{\mu}{}'=\Lambda^{\mu}{}_{\nu}x^{\nu}\,,\qquad x^{\mu}=\left(\Lambda^{-1}\right)^{\mu}{}_{\nu}x^{\nu}\,.
\end{equation}
Substituting into Eq.~\eqref{IntervalInvariance} gives, 
\begin{equation}
    \Lambda^{\mu}{}_{\lambda}\eta_{\mu\nu}\Lambda^{\nu}{}_{\rho}=\eta_{\lambda\rho}\,,
\end{equation}
or, in matrix notation 
\begin{equation} \label{LorentzCond}
    \Lambda^{T}\eta\Lambda=\eta\,.
\end{equation}
This is the defining property of the Lorentz group. Its connected component containing the identity is the proper orthochronous Lorentz group\footnote{This means that the Lorentz transformations satisfy $\text{det }\Lambda = 1$ and $\Lambda^0{}_{0}\geq1$.}, generated by spatial rotations and Lorentz boosts, which is denoted by $SO(1,D-1)$. 

By adjoining constant spacetime translations, one obtains the Poincaré group, as we mentioned before. Its finite transformations are 
\begin{equation}
    x'{}^{\mu}=\Lambda^{\mu}{}_{\nu}x^{\nu}+a^{\mu}\,,
\end{equation}
where $a^{\mu}$ is a constant vector. 

Similarly to the case of internal symmetries, continuous Lorentz transformations can be studied by expanding around the identity. Since the Lorentz group is a continuous Lie group, any transformation sufficiently close to the identity is controlled by the generators of its Lie algebra. It is therefore convenient to write a Lorentz transformation as 
\begin{equation} \label{ExponentialRepresentation}
    \Lambda=\exp\left[\frac{1}{2}\lambda^{\mu\nu}m_{\mu\nu}\right]\,,
\end{equation}
where $\lambda_{\mu\nu}=-\lambda_{\nu\mu}$ are infinitesimal parameters and $m_{\mu\nu}$ are the generators of the Lorentz algebra. This exponential form makes explicit the relation between finite Lorentz transformations and their infinitesimal generators, which are the objects that enter the Noether formalism. 

Expanding~\eqref{ExponentialRepresentation} to first order gives 
\begin{equation}
    \Lambda^{\mu}{}_{\nu}=\delta^{\mu}{}_{\nu}+\frac{1}{2}\lambda^{\lambda\rho}\left(m_{\lambda\rho}\right)^{\mu}{}_{\nu}+\mathcal{O}(\lambda^{2})\,.
\end{equation}
In the vector representation, the generators are chosen so that 
\begin{equation}
    \left(m_{\lambda\rho}\right)^{\mu}{}_{\nu}=\delta_{\lambda}^{\mu}\eta_{\rho\nu}-\delta_{\rho}^{\mu}\eta_{\lambda\nu}\,.
\end{equation}
Thus, in this representation, the infinitesimal Lorentz transformations read
\begin{equation}
    \Lambda^{\mu}{}_{\nu}=\delta^{\mu}_{\nu}+\lambda^{\mu}{}_{\nu}+\mathcal{O}(\lambda^{2})\,.
\end{equation}
Therefore, the coordinates transform infinitesimally as 
\begin{equation}
    x^{\mu}{}'=x^{\mu}+\lambda^{\mu}{}_{\nu}x^{\nu}\,.
\end{equation}
Notice that the antisymmetry of $\lambda_{\mu\nu}$ follows from the Lorentz condition~\eqref{LorentzCond}. Since $\lambda_{\mu\nu}$ is antisymmetric, the Lorentz group $D(D-1)/2$ has independent generators. In four dimensions, these correspond to three rotations and three boosts. 

It is also worth mentioning that the generators $m_{\mu\nu}$ satisfy the Lorentz algebra,\footnote{The generators $m_{\mu\nu}$ are said to be in the vector representation because they act on objects carrying one Lorentz index, such as spacetime vectors $x^{\mu}$. Other representations are obtained by finding different matrices that satisfy the same commutation relations. In particular, the spinor representation is generated by matrices $\Sigma_{\mu\nu}$, constructed from the gamma matrices, and is essential for introducing Dirac spinors and the Dirac equation.}
\begin{equation} \label{AlgebraGen}
    \left[m_{\mu\nu},m_{\lambda\rho}\right]=\eta_{\nu\lambda}m_{\mu\rho}-\eta_{\mu\lambda}m_{\nu\rho}-\eta_{\nu\rho}m_{\mu\lambda}+\eta_{\mu\rho}m_{\nu\lambda}\,.
\end{equation}

We now introduce the infinitesimal transformation associated with the Lorentz transformations acting on scalar fields. Using $\phi'(x)=\phi(\Lambda x)$, and expanding to first order in $\lambda$, one obtains
\begin{equation} \label{LorentzInf}
    \delta\phi(x)=-\frac{1}{2}\lambda^{\mu\nu}L_{\mu\nu}\phi(x)\,,
\end{equation}
 where $L_{\mu\nu}=x_{\mu}\partial_{\nu}-x_{\nu}\partial_{\mu}$ are the differential Lorentz generators acting on scalar fields. In the case of the translations, the scalar field is 
\begin{equation} \label{Scalarfieldtranslation}
    \phi'(x)=\phi(x+a)\,,
\end{equation}
where $a$ is a constant vector. Thus, the infinitesimal transformation corresponds 
\begin{equation} \label{TranslationInf}
    \delta\phi(x)=a^{\mu}\partial_{\mu}\phi(x)\,.
\end{equation}
Thus, defining $P_{\mu}=\partial_{\mu}$, the infinitesimal Poincaré transformation of a scalar field is 
\begin{equation}
    \delta\phi(x)=a^{\mu}P_{\mu}\phi(x)+\frac{1}{2}\lambda^{\mu\nu}L_{\mu\nu}\phi(x)\,.
\end{equation}
 The first term in this equation generates spacetime translations, whereas the second term generates Lorentz rotations and boosts. Together, they provide the infinitesimal realization of the Poincaré group on scalar fields. With the conventions
 \begin{equation}
    P_\mu=-i\partial_\mu\,, \qquad M_{\mu\nu} = x_\mu P_\nu-x_\nu P_\mu = -i\left(x_\mu\partial_\nu-x_\nu\partial_\mu\right)\,,
\end{equation}
the generators satisfy the Poincar\'e algebra,
\begin{align}
    [P_\mu,P_\nu] &=0\,, \label{eq:poincare-PP} \\
    [M_{\mu\nu},P_\rho]&= i\left(\eta_{\mu\rho}P_\nu -\eta_{\nu\rho}P_\mu\right)\,, \label{eq:poincare-MP}\\
    [M_{\mu\nu},M_{\rho\sigma}] &=i\left(\eta_{\mu\rho}M_{\nu\sigma} -\eta_{\mu\sigma}M_{\nu\rho} -\eta_{\nu\rho}M_{\mu\sigma}  +\eta_{\nu\sigma}M_{\mu\rho} \right)\,,
    \label{eq:poincare-MM}
\end{align}
where $\eta_{\mu\nu}$ is the Minkowski metric. Translations form an Abelian subalgebra, while the mixed commutator expresses the transformation of $P_\mu$ as a Lorentz vector. The final commutator closes the Lorentz subalgebra. These relations exhibit the semidirect-sum structure
\begin{equation}
    \mathfrak{iso}(1,D-1) = \mathfrak{so}(1,D-1)\ltimes\mathbb{R}^{1,D-1}\,.
\end{equation}
For scalar fields, $M_{\mu\nu}$ contains only the orbital contribution. Thus, fields carrying Lorentz indices also require generators acting on those indices.

The transformations introduced in this section define how scalar fields behave under spacetime symmetries. A physically relevant symmetry, however, is not merely a transformation of the fields, but one that leaves the dynamics of the theory invariant. In the Lagrangian formulation, this condition is expressed by the invariance of the action. Continuous symmetries satisfying this condition have profound physical consequences; Noether's theorem establishes that every continuous symmetry of the action gives rise to a locally conserved current and an associated conserved charge. In the case of internal symmetries, they lead to conserved internal charges, whereas spacetime translations and Lorentz transformations give rise to the conserved energy-momentum tensor and angular momentum tensor, respectively. Since these conserved quantities play a fundamental role throughout classical and quantum field theory, we now review the Noether formalism to analyze them. 

\subsection{Noether's Theorem and Conserved Currents}

Noether's theorem is one of the most fundamental results in classical and quantum field theory. It establishes a direct connection between continuous symmetries of the action and conserved quantities. This theorem provides the mathematical origin of many conservation laws encountered in physics, including the conservation of electric charge, energy, linear momentum, and angular momentum. Since symmetries play a fundamental role in quantum field theory and, in particular, in the study of anomalies, we now introduce the Noether formalism. 

Let us continue considering the system of scalar fields that we introduced in Sec.~\ref{SectionInternalSpacetime}, whose dynamics are described by the action 
\begin{equation}
    I[\phi]=\int d^{D}x\,\mathcal{L}\left(\phi^{i},\partial_{\mu}\phi^{i}\right)\,,
\end{equation}
where the Lagrangian density $\mathcal{L}$ depends on the fields and their first derivatives. The arbitrary variation of the Lagrangian density is 
\begin{equation}
    \delta\mathcal{L}=\left[\frac{\partial\mathcal{L}}{\partial\phi^{i}}-\partial_{\mu}\frac{\partial\mathcal{L}}{\partial(\partial_{\mu}\phi^{i})}\right]\delta\phi^{i}+\partial_{\mu}\left(\frac{\partial\mathcal{L}}{\partial(\partial_{\mu}\phi^{i})}\delta\phi^{i}\right)\,.
\end{equation}
From this expression, we can read off the Euler-Lagrange equations for the field $\phi^{i}$, corresponding to 
\begin{equation}
    \frac{\partial\mathcal{L}}{\partial\phi^{i}}-\partial_{\mu}\frac{\partial\mathcal{L}}{\partial(\partial_{\mu}\phi^{i})}=0\,.
\end{equation}
Therefore, on-shell, the variation of the Lagrangian density reduces to a total derivative, 
\begin{equation} \label{ArbitraryVar}
    \delta\mathcal{L}\,\hat{=}\,\partial_{\mu}\left(\frac{\partial\mathcal{L}}{\partial(\partial_{\mu}\phi^{i})}\delta\phi^{i}\right)\,,
\end{equation}
where $\hat{=}$ means equality upon using the equations of motion. 

Suppose now that the theory admits a continuous infinitesimal symmetry. We denote by $\delta_{A}\phi^{i}$ the infinitesimal variation of the field generated by a symmetry $A$. If the transformation is a symmetry of the action, the Lagrangian density may change at most by a total derivative, 
\begin{equation} \label{VariationUnderSym}
    \delta_{A}\mathcal{L}=\partial_{\mu}K_{A}^{\mu}\,.
\end{equation}
where $K_{A}^{\mu}$ is a local function of the fields and their derivatives determined by the symmetry transformation. Note that this notation includes both the internal and spacetime symmetries discussed in the previous sections. For example, recall that internal symmetries are described by~\eqref{InternalSym}, while spacetime symmetries are generated by~\eqref{LorentzInf}. 

If the arbitrary variation~\eqref{ArbitraryVar} is promoted to a symmetry variation, i.e. $\delta\phi^{i}=\delta_{A}\phi^{i}$, then on-shell one has 
\begin{equation}
    \delta_{A}\mathcal{L}\,\hat{=}\,\partial_{\mu}\left(\frac{\partial\mathcal{L}}{\partial(\partial_{\mu}\phi^{i})}\delta_{A}\phi^{i}\right)\,.
\end{equation}
Comparing this expression with~\eqref{VariationUnderSym} after imposing the equations of motion, we obtain 
\begin{equation}
    \partial_{\mu}\left[\frac{\partial\mathcal{L}}{\partial(\partial_{\mu}\phi^{i})}\delta_{A}\phi^{i}-K_{A}^{\mu}\right]\hat{=}0
\end{equation}
This equation motivated the definition of the Noether current, which is given by 
\begin{equation} \label{Current}
    J_{A}^{\mu}=\frac{\partial\mathcal{L}}{\partial(\partial_{\mu}\phi^{i})}\delta_{A}\phi^{i}-K_{A}^{\mu}\,.
\end{equation}
Therefore, whenever the equations of motion hold, the current satisfies the local conservation law 
\begin{equation} \label{CurrentConserv}
    \partial_{\mu}J_{A}^{\mu}=0\,.
\end{equation}
From this local conservation law, one can define an integrated Noether charge. Let $\Sigma$  be a spacelike hypersurface, with $d\Sigma_{\mu}$ its oriented surface element, which is 
\begin{equation}
    d\Sigma_{\mu}=\sqrt{|\gamma|}\,n_{\mu} d^{D-1}x\,,
\end{equation}
where $n_{\mu}$ is a timelike normal vector to $\Sigma$. The charge associated with the current~\eqref{Current} is defined by 
\begin{equation}
    Q_{A}=\int_{\Sigma}d\Sigma_{\mu}J_{A}^{\mu}\,.
\end{equation}
If the current has no flux through the boundary of the integration region, this charge is independent of the choice of hypersurface. In particular, on $t=cte$ hypersurface, this definition reduces to
\begin{equation}
    Q_{A}(t)=\int d^{D-1}xJ_{A}^{0}(\vec{x},t)\,.
\end{equation}
Taking the time derivative of this expression and using~\eqref{CurrentConserv}, one finds 
\begin{equation}
    \frac{dQ_{A}}{dt}=\int d^{D-1}x \partial_0 J_{A}^{0}=-\int d^{D-1}x\partial_{i}J_{A}^{i}\,.
\end{equation}
Using Gauss' theorem, this becomes
\begin{equation}
    \frac{dQ_{A}}{dt}=-\int_{\partial\Sigma}dS_{i}J_{A}^{i}\,,
\end{equation}
where $dS_{i}$ is the surface element. 

Therefore, if the spatial components of the current vanish sufficiently rapidly at infinity, or if there is no flux through the boundary $\partial\Sigma$, the boundary term vanishes, and the charge is conserved over time, 
\begin{equation}
    \frac{dQ_{A}}{dt}=0\,.
\end{equation}
This shows that the local conservation law~\eqref{CurrentConserv} is stronger that the conservation of the integrated charge. It states not only that the total charge is constant in time, but also that any change of charge inside a finite spatial region must be accounted for by a flux of current through the boundary of the region under consideration. 

We want to study the Noether currents for internal and spacetime transformations. To illustrate these ideas, let us consider the canonical scalar-field Lagrangian
\begin{equation}
    \mathcal{L}=-\frac{1}{2}\partial_{\mu}\phi^{i}\partial^{\mu}\phi^{i}-V(\phi^{i})\,.
\end{equation}
For an internal symmetry, the infinitesimal transformation was obtained in Eq.~\eqref{InternalSym}. Since an internal global symmetry does not act on the spacetime coordinates, the Lagrangian is invariant without a spacetime total derivative, provided that the potential satisfies 
\begin{equation}
    V(\phi')=V(\phi)\,.
\end{equation}
Therefore, using the Noether current~\eqref{Current}, with the fact that $K_{A}^{\mu}=0$, one obtains 
\begin{equation}
    J_{A}^{\mu}=-\partial^{\mu}\phi^{i}(t_{A})_{\,j}^{i}\phi^{j}\,.
\end{equation}
This is the conserved current associated with the internal global symmetry generated by $t_{A}$.

Let us now consider spacetime symmetries. Under an infinitesimal translation, a scalar field transforms as in Eq.~\eqref{Scalarfieldtranslation}. With this, the infinitesimal variation of the field is given by Eq.~\eqref{TranslationInf}. Thus, the variation associated with the translation parameter $a^{\mu}$ is 
\begin{equation}
    \delta_{a}\phi^{i}=a^\mu\partial_{\mu}\phi^{i}\,.
\end{equation}
The Lagrangian density transforms in the same way, 
\begin{equation}
    \delta\mathcal{L}=a^{\mu}\partial_{\mu}\mathcal{L}=a^{\nu}\partial_{\mu}\left(\delta_{\nu}^{\mu}\mathcal{L}\right)\,.
\end{equation}
Therefore, for spacetime translations, one identifies 
\begin{equation}
    K_{\nu}^{\mu}=\delta_{\nu}^{\mu}\mathcal{L}\,.
\end{equation}
Substituting these expression into the Noether current~\eqref{Current} gives, 
\begin{equation}
    J_{\nu}^{\mu}=\frac{\partial\mathcal{L}}{\partial(\partial_{\mu}\phi^{i})}\partial_{\mu}\phi^{i}-\delta_{\nu}^{\mu}\mathcal{L}\,.
\end{equation}
This quantity is denoted by $T_{\,\nu}^{\mu}$ and is the canonical energy-momentum tensor associated with spacetime translations. Its conservation law is 
\begin{equation}
    \partial_{\mu}T_{\,\nu}^{\mu}=0\,,
\end{equation}
which is the Noether conservation law following from the invariance under translations. The corresponding conserved charges are 
\begin{equation}
    E=\int d^{D-1}xT^{00}\,,\qquad\text{and}\qquad P^{i}=\int d^{D-1}T^{0i}\,.
\end{equation}
These charges correspond to the energy and spatial momentum of the field theory, respectively. 

Similarly, Lorentz transformations lead to the conserved angular momentum current. For scalar fields, the infinitesimal Lorentz transformation is given by Eq.~\eqref{LorentzInf}. Since the Lagrangian density is a Lorentz scalar, it transforms in the same way. Integrating by parts, it can be written as a total derivative, that is
\begin{equation}
    \delta\mathcal{L}=\frac{1}{2}\lambda^{\mu\nu}\partial_{\lambda}(M^\lambda)_{\mu\nu}\,,
\end{equation}
where we have defined 
\begin{equation}
    (M^\lambda)_{\mu\nu}\equiv\left(x_{\mu}\delta_{\nu}^{\lambda}-x_{\nu}\delta_{\mu}^{\lambda}\right)\mathcal{L}\,.
\end{equation}
Using the Noether current~\eqref{Current}, and extracting the coefficient of $\lambda^{\mu\nu}$, one obtains the current associated with Lorentz transformations,
\begin{equation}
    M^{\mu}{}_{\lambda\rho}=x_{\lambda}T_{\,\rho}^{\mu}-x_{\rho}T_{\,\lambda}^{\mu}\,.
\end{equation}

For scalar fields, there is no intrinsic spin contribution, so this is the full Lorentz current. Its conservation follows from the conservation and symmetry of the stress-energy tensor, satisfying 
\begin{equation}
    \partial_{\mu}M^{\mu}{}_{\lambda\rho}=0\,.
\end{equation}
The associated conserved charges are obtained by integrating the time component of the current over a $t=cte$ hypersurface, leading to
\begin{equation}
    M_{\lambda\rho}=\int d^{D-1}xM^{0}{}_{\lambda\rho}\,.
\end{equation}
In four dimensions, the spatial components $M_{ij}$ generate rotations, while the mixed components $M_{0i}$ generate Lorentz boosts. Thus, Lorentz invariance gives rise to the conservation of angular and boost charges. Since scalar fields carry no intrinsic spin, the current above is purely orbital. For fields with spin, such as Dirac fields, the Lorentz current receives an additional spin contribution.

\section{Dirac Spinors}

So far, we have considered scalar fields for the sake of simplicity. Scalar fields transform trivially under Lorentz transformations, in the sense that their components do not mix. However, recall that from an active transformation viewpoint, a scalar field transforms as~\eqref{LorentzInf}. This means that the Lorentz transformation acts only on the spacetime argument of the field and not on any additional field components. Upon quantization, scalar fields give rise to spin-zero particles. However, many particles in nature, such as electrons, carry intrinsic angular momentum, or spin. In field theory, such particles are described by fields whose components transform non-trivially under the Lorentz group. 

To illustrate this idea, let us consider a general multi-component\footnote{This notation includes scalar fields as a particular case.} field $\Phi^{a}(x)$, whose transformation law takes the form 
\begin{equation}
    \Phi^{a}\to\Phi^{a}{}'(x)=R(\Lambda)^{a}{}_{b}\Phi^{b}(\Lambda^{-1}x)\,,
\end{equation}
where $R(\Lambda)$ is a representation of the Lorentz group acting on the field components.\footnote{A representation of a Lie group $G$ on a vector space $V$ is a map $R:G\to GL(V)$, assigning to each group element $g\in G$ a linear transformation $R(g)$ acting on $V$. The associated Lie algebra representation assigns to each generator of $\mathfrak{g}$ a matrix $t_{A}$ satisfying~\eqref{StructureConstant}. Thus, for a Lie algebra element $\Theta=\theta^{A}t_{A}$, the corresponding group element in the representation is $R(g)=e^{\Theta}$. More generally, when a group acts on a space that is not necessarily a vector space, we are talking about a realization of the symmetry.}

Different choices of the representation $R(\Lambda)$ describe different types of fields. For instance, a vector field $A^{\mu}(x)$ carries a Lorentz vector index $\mu$ , and therefore transforms with the vector representation, 
\begin{equation}
    A^{\mu}\to A^{\mu}{}'(x)=\Lambda^{\mu}{}_{\nu}A^{\nu}(\Lambda^{-1}x)\,.
\end{equation}
Here, the index $\mu$  labels spacetime components, and the Lorentz matrix $\Lambda^{\mu}{}_{\nu}$ mixes those components. By contrast, a spinor field also has many components, but its index is not a spacetime vector index. A Dirac spinor, denoted by $\psi^{\alpha}(x)$, carries a spinor index $\alpha$, which transforms with a different representation of the Lorentz group, denoted by $S(\Lambda)$. Thus, 
\begin{equation}
    \psi^{a}(x)\to\psi^{\alpha}{}'(x)=S(\Lambda)^{\alpha}{}_{\beta}\psi^{\beta}(\Lambda^{-1}x)\,.
\end{equation}
Therefore, although both $A^{\mu}(x)$ and $\psi^{\alpha}(x)$ are multi-component fields, they transform in different representations of the Lorentz group. The field $A^{\mu}$ is a Lorentz vector, while $\psi^{\alpha}$ is a spinor. 

To construct the spinor representation, one introduces the gamma matrices $\gamma^{\mu}$, which satisfy the Clifford algebra 
\begin{equation} \label{CliffordAlgebra}
    \{\gamma^{\mu},\gamma^{\nu}\}=2\eta^{\mu\nu}\mathbb{I}\,.
\end{equation}
From these matrices, one defines the spinor generators 
\begin{equation} \label{SpinorsGenerators}
    \Sigma_{\mu\nu}=\frac{1}{4}[\gamma_{\mu},\gamma_{\nu}]\,.
\end{equation}
These generators satisfy the same Lorentz algebra as the vector generators $m_{\mu\nu}$ (c.f Eq.~\eqref{AlgebraGen}), but they act on spinor indices rather than Lorentz vector indices. Following the notation used for internal symmetries, we define the Lie algebra element 
\begin{equation}
    \Theta=\frac{1}{2}\lambda^{\mu\nu}\Sigma_{\mu\nu}\,.
\end{equation}
The corresponding Lorentz transformation in the spinor representation is obtained by exponentation, 
\begin{equation}
    S(\Lambda)=e^{\Theta}=\exp\left(\frac{1}{2}\lambda^{\mu\nu}\Sigma_{\mu\nu}\right)\,.
\end{equation}
This gives the appropriate transformation law for Dirac spinors. The same Lorentz transformation is therefore represented by $\Lambda_{\,\nu}^{\mu}$ on vector indices and by $(\Lambda)^{\alpha}{}_{\beta}$ on spinor indices. 

The Dirac action is then constructed by requiring a Lorentz invariant action for a spinor field whose equations of motion are first order in spacetime derivatives.\footnote{This is in contrast to the Klein--Gordon equation, which is second order in derivatives.}

The basic Lorentz covariant first order differential operator acting on spinors is
\begin{equation}
    \slashed{\partial}-m\,,\qquad\slashed{\partial}\equiv\gamma^{\mu}\partial_{\mu}.
\end{equation}
where $m$ denotes the mass of the field. The Clifford algebra ensures that the square of this operator reproduces the Klein-Gordon operator, so each component of the Dirac spinor satisfies the Klein-Gordon equation on shell.

To build Lorentz invariant scalars from spinors, one introduces the Dirac adjoint spinor 
\begin{equation}
    \bar{\psi}=i\psi{}^{\dagger}\gamma^{0}\,.
\end{equation}
The reason for the use of $\bar{\psi}$, rather than the ordinary Hermitian conjugate $\psi^{\dagger}$, is that finite-dimensional representations of the Lorentz group are not unitary. The Dirac adjoint allows one to construct Lorentz covariant bilinears such as $\bar{\psi}\psi$ or $\bar{\psi}\gamma^{\mu}\psi$, that transforms as a scalar and vector under $SO(1,D-1)$, respectively. Therefore, in the conventions used in this thesis, the Dirac action is 
\begin{equation}
    I[\psi,\bar{\psi}]=-\int d^{D}x\bar{\psi}\left[\slashed{\partial}-m\right]\psi\,.
\end{equation}
Varying this action with respect to $\bar{\psi}$ gives the Dirac equation, 
\begin{equation}
    \left(\slashed{\partial}-m\right)\psi=0\,,
\end{equation}
while the variation with respect to $\psi$  gives the adjoint Dirac equation. These structures are fundamental in the construction of vector and axial currents. 

Let us now continue with the symmetries of the Dirac action. The gamma matrices satisfy the Clifford algebra (c.f. Eq.~\eqref{CliffordAlgebra}). In our conventions, their Hermiticity properties are
\begin{equation} \label{Hermiticity}
    (\gamma^{\mu})^{\dagger}=\gamma^{0}\gamma^{\mu}\gamma^{0}\,,
\end{equation}
which implies
\begin{equation}
    (\gamma^{0})^{\dagger}=-\gamma^{0}\,,\qquad(\gamma^{i})^{\dagger}=\gamma^{i}\,,\qquad i=1,2,3\,.
\end{equation}
Thus, $\gamma^{0}$ is anti-Hermitian, while the spatial gamma matrices $\gamma^{i}$ are Hermitian. Additionally, in four dimensions, one can define the chiral matrix as
\begin{equation}
    \gamma^{5}\equiv i\gamma^{0}\gamma^{1}\gamma^{2}\gamma^{3}\,,
\end{equation}
which satisfies
\begin{equation} \label{AntiCommutation}
    \left(\gamma^{5}\right)^{2}=\mathbb{I}\,,\qquad\{\gamma^{5},\gamma^{\mu}\}=0\,.
\end{equation}

\section{Symmetries of the Dirac Action}

The Dirac action is invariant under a global $U(1)$ transformation\footnote{This transformation is considered because it is the simplest internal symmetry of the Dirac field; it changes the phase of the spinor without changing spacetime coordinates. Moreover, the opposite phase of $\bar{\psi}$ ensures that the bilinear $\bar{\psi}\psi$  is invariant.} of the fermionic fields, 
\begin{equation}
    \psi\to\psi'=e^{i\alpha}\psi\:,\qquad\bar{\psi}\to\bar{\psi}'=e^{-i\alpha}\bar{\psi}\,,
\end{equation}
where $\alpha$  is a constant parameter. Indeed, under this transformation of the action behaves as
\begin{equation}
    I'[\psi',\bar{\psi}']	=-\int d^{D}x\bar{\psi}'\left[\slashed{\partial}-m\right]\psi'=-\int d^{D}xe^{-i\alpha}\bar{\psi}\left[\slashed{\partial}-m\right]e^{i\alpha}\psi=I[\psi,\bar{\psi}]\,.
\end{equation}
Therefore, the global $U(1)$ transformation is a symmetry of the Dirac action. By Noether's theorem, it gives rise to a conserved current. 

To find the corresponding Noether current, we consider the infinitesimal form of the transformation, 
\begin{equation}
    \delta\psi=i\alpha\psi\,,\qquad\delta\bar{\psi}=-i\alpha\bar{\psi}\,.
\end{equation}
This allows one to identify, using the Noether current~\eqref{Current}, that $K^{\mu}=0$. Thus, the conserved current associated with this symmetry is 
\begin{equation}
    J^{\mu}=i\bar{\psi}\gamma^{\mu}\psi\,.
\end{equation}
This current is conserved on-shell, satisfying the conservation law 
\begin{equation}
    \partial_{\mu}J^{\mu}=0\,.
\end{equation}

In addition to this global $U(1)$ symmetry, the massless Dirac action admits a global axial $U(1)$ symmetry.\footnote{The axial transformation is generated by $\gamma_{5}$ because this matrix distinguishes the two chiral components of a Dirac spinor. Indeed, using the projectors 
\begin{equation}
    P_{\pm}=\frac{1}{2}\left(\mathbb{I}\pm\gamma^{5}\right)\,,
\end{equation}
one can decompose the spinor as $\psi=\psi_{+}+\psi_{-}$ such that $\psi_{+}=P_{+}\psi$  and $\psi_{-}=P_{-}\psi\,$. These components are eigenstates of $\gamma^{5}$, satisfying $\gamma^{5}\psi_{+}=+\psi_{+}$ and $\gamma^{5}\psi_{-}=-\psi_{-}$. Therefore, under the axial transformation~\eqref{AxialTransfor}, the exponential acts differently on each chiral component. On $\psi_{+}$, $\gamma^{5}$ is replaced by its eigenvalues $+1$, while on $\psi_{-}$, it is replaced by its eigenvalue $-1$. Hence $\psi_{+}\to e^{i\beta}\psi_{+}$ and $\psi_{-}\to e^{-i\beta}\psi_{-}$. Thus, the axial transformation rotates the two chiral components with opposite phases. For this reason, it is also called a chiral transformation. }
This transformation is defined by
\begin{equation} \label{AxialTransfor}
    \psi\to\psi'=e^{i\beta\gamma_{5}}\psi\:,\qquad\bar{\psi}\to\bar{\psi}'=\bar{\psi}e^{i\beta\gamma_{5}}\,,
\end{equation}
where $\beta$ is a constant parameter. Notice that the transformation of $\bar{\psi}$ has the same sign as the transformation of $\psi$. This differs from the global $U(1)$ symmetry and follows from the anticommutation relation given in Eq.~\eqref{AntiCommutation}. Indeed, this relation implies $e^{-\beta\gamma^{5}}\gamma^{0}=\gamma^{0}e^{i\beta\gamma^{5}}$, which gives
\begin{equation}
    \bar{\psi}'=i(\psi')^{\dagger}\gamma^{0}=i\psi^{\dagger}e^{-\beta\gamma^{5}}\gamma^{0}=i\psi^{\dagger}\gamma^{0}e^{i\beta\gamma^{5}}=\bar{\psi}e^{i\beta\gamma^{5}}\,.
\end{equation}

At the infinitesimal level, the axial transformation is 
\begin{equation}
    \delta\psi=i\beta\gamma^{5}\psi\,,\qquad\delta\bar{\psi}=i\beta\bar{\psi}\gamma^{5}\,.
\end{equation}
Under this transformation, the kinetic term of the Dirac Lagrangian is invariant for constant $\beta$. However, the mass term is not invariant, since 
\begin{equation} 
    \delta(\bar{\psi}\psi)=2i\beta\bar{\psi}\gamma^{5}\psi\,.
\end{equation}
Therefore, the axial $U(1)$ symmetry is a classical symmetry only in the massless case. In this case, the Noether current associated with the axial transformation is the axial current, 
\begin{equation} \label{AxialCurrent}
    J_{5}^{\mu}=i\bar{\psi}\gamma^{\mu}\gamma^{5}\psi\,.
\end{equation}
Classically, in the massless case, this current is conserved,
\begin{equation}
    \partial_{\mu}J_{5}^{\mu}=0\,.
\end{equation}
This classical conservation law is the starting point for the study of the axial anomaly. After quantization, the conservation of the axial current may fail even in the massless theory. In the path integral formulation, Fujikawa's method relates this quantum anomaly to the non-invariance of the functional measure under axial transformations. For this reason, the next section is dedicated to reviewing the path integral formalism, which will allow us to introduce Fujikawa's computation of the axial anomaly. 

\section{Axial Anomaly in QED: The Fujikawa Method}

In this section, we review the Fujikawa method through the path integral formulation of quantum field theory (See appendix~\ref{pathintegralformalism})~\cite{Fujikawa:1979ay,Fujikawa:1980eg}. The main idea of this method is that the axial anomaly can be understood as arising from the non-invariance of the fermionic functional measure under local chiral transformations. More precisely, although a change of variables in the path integral does not change the value of the functional integral itself, the fermionic measure may acquire a nontrivial Jacobian. Fujikawa showed that, after suitable regularization, this Jacobian gives rise to an additional quantum contribution to the divergence of the axial current. 

To illustrate this idea, let us consider the massless Euclidean path integral of QED,
\begin{equation} \label{PathIntegralQED}
    \mathcal{Z}_{\text{QED}}=\int\mathcal{D}\bar{\psi}\mathcal{D}\psi\mathcal{D}A_{\mu}\exp\left(-\int d^{4}\left[x\bar{\psi}\slashed{D}\psi+\frac{1}{4}F_{\mu\nu}F^{\mu\nu}\right]\right)\,,
\end{equation}
where the Dirac operator is defined by $\slashed{D}:=\gamma^{\mu}\left(\partial_{\mu} +iA_{\mu}\right)$, and the $\gamma$  matrices satisfy \eqref{CliffordAlgebra}, \eqref{Hermiticity}, and \eqref{AntiCommutation}. We choose conventions in which $\slashed{D}$ is Hermitian. Therefore, it admits a complete orthonormal set of eigenstates $\{\psi_{n}(x)\}$, such that 
\begin{equation} \label{Eigenvalueproblem}
    \slashed{D}\psi_{n}(x)=\lambda_{n}\psi_{n}(x)\,\qquad\lambda_{n}\in\mathbb{R}\,,
\end{equation}
These eigenfunctions obey the orthonormality and completeness relations 
\begin{equation} \label{OrthogonalityCond2}
    \int d^{4}x\:\psi_{n}^{\dagger}(x)\psi_{\ell}(x)=\delta_{\ell n}\,,\qquad{\rm and}\,\qquad\sum_{n}\psi_{n}(x)\psi_{n}^{\dagger}(y)=\delta^{(4)}(x-y)\,\mathbb{I}\,,
\end{equation}
where $\mathbb{I}$ denotes the identity matrix in spinor space. The Dirac spinors can be expanded in this basis as 
\begin{equation} \label{FieldDecomposition}
    \psi(x)=\sum_{n}a_{n}\psi_{n}(x)=\sum_{n}a_{n}\langle x|n\rangle\,,\qquad\bar{\psi}(x)=\sum_{n}\psi_{n}^{\dagger}(x)\bar{b}_{n}=\sum_{n}\langle n|x\rangle\bar{b}_{n}\,,
\end{equation}
where $a_{n}$ and $\bar{b}_{n}$ are independent Grassmann variables, while $\psi_{n}(x)$ is a complex valued spinor function, with $\psi_{n}^{\dagger}(x)$ its transpose conjugate. Here we have used Dirac notation 
\begin{equation} \label{DiracNotation}
    \langle x|n\rangle=\psi_{n}(x)\,,\qquad\langle n|x\rangle=\psi_{n}^{\dagger}(x)\,.
\end{equation}
In terms of these Grassmann coefficients, the fermionic measure becomes 
\begin{equation} \label{IntegralMeasure}
    \mathcal{D}\bar{\psi}\mathcal{D}\psi=\left(\det\langle n|x\rangle\det\langle x|\ell\rangle\right)^{-1}\prod_{n}d\bar{b}_{n}\prod_{\ell}da_{\ell}=\prod_{n}d\bar{b}_{n}da_{n}\,.
\end{equation}
where the Jacobian appears to a negative power because the field transformation is performed over Grassmann variables \footnote{For details, see chapter 9 in~\cite{Weinberg:1995mt}, and chapter 5 in~\cite{Peskin:1995ev}.}. The Jacobian associated with the change from the fields to the expansion coefficients is unity because the eigenfunctions form a complete orthonormal basis. 

We now consider the chiral transformation given in Eq.~\eqref{AxialTransfor}. Its infinitesimal version corresponds to 
\begin{subequations} \label{ExpandedFields}
\begin{align}
    \psi'(x)	&=\psi(x)+i\alpha(x)\gamma_{5}\psi(x)\,, \\
\bar{\psi}'(x)	&=\bar{\psi}(x)+\bar{\psi}(x)i\alpha(x)\gamma_{5}\,.
\end{align}
\end{subequations}
The transformed fields~\eqref{ExpandedFields} can again be expanded in the same eigenstates 
\begin{equation} \label{StatesinGrassmann}
    \psi'(x)=\sum_{n}a_{n}^{'}\psi_{n}\,,\qquad\bar{\psi}'(x)=\sum\psi_{n}^{\dagger}b_{n}^{'}\,,
\end{equation}
where, the transformed Grassmann coefficients are related to the original ones by 
\begin{equation} \label{GrassmanTransfor}
    a_{n}^{'}=\sum_{\ell}B_{n\ell}a_{\ell}\,,\qquad\bar{b}_{n}'=\sum_{\ell}B_{n\ell}\bar{b_{\ell}}\,.
\end{equation}
Here the coefficients $B_{n\ell}$ are define as follows  
\begin{equation} \label{MatrixExpansion}
    B_{n\ell}=\int d^{4}x\psi_{n}^{\dagger}(x)e^{i\alpha(x)\gamma_{5}}\psi_{\ell}(x) \simeq \delta_{n\ell}+i\int d^{4}x\:\alpha(x)\psi_{n}^{\dagger}(x)\gamma_{5}\psi_{\ell}(x)+\mathcal{O}(\alpha^{2})\,.
\end{equation}
Since the variables are Grassmann, a linear transformation given by~\eqref{GrassmanTransfor} changes the Fermionic measure accordingly
\begin{equation} \label{TransformationGrassmannDet}
    \prod_{n}da_{n}^{'}=(\det B)^{-1}\prod_{n}da_{n}\,.
\end{equation}
The same argument applies to the $\bar{b}_{n}$ variables. Therefore, under axial transformations the fermionic measure transforms as 
\begin{equation} \label{FermionicMeasureTrans}
    \mathcal{D}\bar{\psi}\mathcal{D}\psi\to\mathcal{D}\bar{\psi}'\mathcal{D}\psi'=\left(\det\sum_{p}B_{\ell p}B_{pm}\right)^{-1}\prod_{m}d\bar{b}_{n}\prod_{\ell}da_{\ell}=\Big(\det B_{n\ell}\Big)^{-2}\prod_{m}d\bar{b}_{n}\prod_{\ell}da_{\ell}\,.
\end{equation}
This is the \textbf{axial Jacobian}. However, it is worth mentioning that this determinant is formally divergent and it must be regularized. To do so, we use the following formula 
\begin{equation} \label{FormulaDet}
    \det B=\exp\left(\rm{tr}\ln B\right)\,.
\end{equation}
Expanding to first order in $\alpha(x)$ and using that $\ln(1+\epsilon)\approx\epsilon+\mathcal{O}(\epsilon^{2})$, when $\epsilon\ll1$, one obtains 
\begin{align} \label{DeterminantExpansion}
    \det B	&=\exp\left(\rm{tr}\ln B\right)=\exp\Big(\rm{tr}\,\ln\Big[\mathbb{I}+i\int d^{4}x\,\alpha(x)\psi_{n}^{\dagger}(x)\gamma_{5}\psi_{\ell}(x)\Big]\Big)\,, \nonumber \\
	&\simeq \exp\Big(i\int d^{4}x\alpha(x)\sum_{n}\psi_{n}^{\dagger}(x)\gamma_{5}\psi_{n}(x)\Big) +\mathcal{O}(\alpha^2)\,.
\end{align}
Therefore, one can obtain the inverse of the determinant of the quantity $B$, which is given by 
\begin{equation} \label{detexp2}
    \left(\det B\right)^{-2}=\exp\Big(-2i\int d^{4}x\alpha(x)\mathcal{F}(x)\Big)\,, \qquad \mathcal{F}(x)=\sum_{n}\psi_{n}^{\dagger}(x)\gamma_{5}\psi_{n}(x)\,.
\end{equation}
The quantity $\mathcal{F}(x)$ is ultraviolet divergent because the sum runs over infinitely many modes. This divergence can be seen formally from the completeness relation, which gives
\begin{equation}
    \mathcal{F}(x)=\text{tr }\left(\gamma_{5}\,\delta^{(4)}(0)\right)\,.
\end{equation}
To regularize this divergence, Fujikawa introduced a gauge-invariant Gaussian regulator as follows 
\begin{equation}
    \mathcal{F}(x)=\lim_{M\to\infty}\sum_{n}\psi_{n}^{\dagger}(x)\gamma_{5}\,\exp\left[-\frac{\lambda_{n}^{2}}{M^{2}}\right]\psi_{n}(x)\,.
\end{equation}
Using the eigenvalue equation (c.f. Eq.~\eqref{Eigenvalueproblem}), this can be written as
\begin{equation}
    \mathcal{F}(x)=\lim_{M\to\infty}\sum_{n}\psi_{n}^{\dagger}(x)\gamma_{5}\exp\left[-\frac{\slashed{D}^{2}}{M^{2}}\right]\psi_{n}(x)\,.
\end{equation}
By inserting the completeness relation, this expression can be written as the coincident point limit of the heat kernel\footnote{The coincident point limit is an operation commonly used in quantum field theory in which two spacetime points appearing in a propagator or heat kernel are brought to the same point.}
\begin{equation}
    \mathcal{F}(x)=\lim_{M\to\infty}\lim_{y\to x}\text{tr }\Big(\gamma_{5}\exp\left[-\frac{\slashed{D}^{2}}{M^{2}}\right]\delta^{(4)}(x-y)\Big)\,,
\end{equation}
where the trace is over spinor indices and $\slashed{D}$ acts on the $x-$dependence of the delta function. In Dirac notation, 
\begin{equation}
    \langle x|y\rangle=\delta^{(4)}(x-y)\,.
\end{equation}
Together with the completeness relation for momentum eigenstates,
\begin{equation}
    \mathbb{I}=\int\frac{d^{4}k}{(2\pi)^{4}}|k\rangle\langle k|\,,
\end{equation}
this allows us to write the regulated expression as
\begin{equation} \label{RegulatedF}
    \mathcal{F}(x)=\lim_{M\to\infty}\int\frac{d^{4}k}{(2\pi)^{4}}e^{-ik\cdot x}\text{tr }\Big(\gamma_{5}\exp\left[-\frac{\slashed{D}^{2}}{M^{2}}\right]\Big)e^{ik\cdot x}\,.
\end{equation}
On the other hand, the square of the Dirac operator can be decomposed into its symmetric and antisymmetric parts as follows 
\begin{align}
    \slashed{D}^{2}	&=\gamma^{\mu}\gamma^{\nu}D_{\mu}D_{\nu}=\frac{1}{2}\left(\{\gamma^{\mu},\gamma^{\nu}\}+\left[\gamma^{\mu},\gamma^{\nu}\right]\right)D_{\mu}D_{\nu} \,.
\end{align}
Using the Clifford algebra (c.f. Eq.~\eqref{CliffordAlgebra}) and the definition $\gamma_{\mu_{!}...\mu_{p}}=\gamma_{[\mu_{1}}...\gamma_{\mu_{p}]}$, together with $\left[D_{\mu},D_{\nu}\right]=F_{\mu\nu}$, one obtains, with our Euclidean conventions 
\begin{align} \label{Lichnerowicz}
    \slashed{D}^{2}	&=-D_{\mu}D^{\mu}+\gamma^{\mu\nu}D_{\mu}D_{\nu}\,, \nonumber \\
	&=-D_{\mu}D^{\mu}+\frac{1}{2}\gamma^{\mu\nu}F_{\mu\nu}\,.
\end{align}
This is the gauge-covariant version of the \textbf{Lichnerowicz-Weitzenböck identity}. It is worth mentioning that in curved space, the corresponding identity also contains the gravitational curvature contribution, which is necessary in our case. 

Using the identity~\eqref{Lichnerowicz}, together with the fact that $e^{ik\cdot x}$ shifts the covariant derivative accordingly,
\begin{equation}
    e^{-ik\cdot x}D_{\mu}e^{ik\cdot x}=D_{\mu}+ik_{\mu}\,,
\end{equation}
and rescaling the integration variable as $k^{\mu}\to Mk^{\mu}$ , the regulated expression~\eqref{RegulatedF} can be written as 
\begin{equation}
    \mathcal{F}(x)=\lim_{M\to\infty}M^{4}\int\frac{d^{4}k}{(2\pi)^{4}}e^{-k_{\mu}k^{\mu}}\text{tr }\Big(\gamma_{5}\exp\left[\frac{D_{\mu}D^{\mu}}{M^{2}}+\frac{2ik^{\mu}D_{\mu}}{M}-\frac{\gamma_{\mu}\gamma_{\nu}}{2M^{2}}F_{\mu\nu}\right]\Big)\,.
\end{equation}
In the limit $M\to\infty$, only the terms containing four $\gamma$  matrices contribute to the trace with $\gamma_{5}$. Indeed, 
\begin{equation}
    \text{tr }(\gamma_{5}\gamma_{\mu})=\text{tr }(\gamma_{5}\gamma_{\mu}\gamma_{\nu})=\text{tr }(\gamma_{5}\gamma_{\mu}\gamma_{\nu}\gamma_{\lambda})=0\,,
\end{equation}
whereas
\begin{equation}
    \text{tr }\left(\gamma_{5}\gamma_{\mu}\gamma_{\nu}\gamma_{\lambda}\gamma_{\rho}\right)=-4\epsilon_{\mu\nu\lambda\rho}\,.
\end{equation}
Therefore, the leading nonvanishing contribution comes from the second-order expansion of the field-strength in the exponential, one finds 
\begin{equation}
 \mathcal{F}(x)	=-\frac{1}{2}\int\frac{d^{4}k}{(2\pi)^{4}}e^{-k_{\mu}k^{\mu}}\epsilon_{\mu\nu\lambda\rho}F^{\mu\nu}F^{\lambda\rho}\,.   
\end{equation}
Using the Euclidean Gaussian integral 
\begin{equation}
    \int\frac{d^{4}k}{(2\pi)^{4}}e^{-k_{\mu}k^{\mu}}=\frac{1}{16\pi^{2}}\,,
\end{equation}
this equation becomes 
\begin{equation} \label{FxPontryagin}
    \mathcal{F}(x)	=-\frac{1}{32\pi^{2}}\epsilon_{\mu\nu\lambda\rho}F^{\mu\nu}F^{\lambda\rho} =-\frac{1}{16\pi^{2}}\tilde{F}_{\mu\nu}F^{\mu\nu}\,,
\end{equation}
which is written in terms of the dual field strength. 
Therefore, the regularized Jacobian becomes 
\begin{equation}
    \left(\det B\right)^{-2}=\exp\Big(\frac{i}{8\pi^{2}}\int d^{4}x\alpha(x)\tilde{F}_{\mu\nu}F^{\mu\nu}\Big)\,.
\end{equation}
Considering this result, the path integral measure transforms as 
\begin{equation} \label{MeasurePathIntegral}
    \mathcal{D}\bar{\psi}'\mathcal{D}\psi'=\exp\Big(-2i\int d^{4}x\alpha(x)\mathcal{F}(x)\Big)\mathcal{D}\bar{\psi}\mathcal{D}\psi\,.
\end{equation}

It is important to emphasize that the path integral is invariant under field redefinitions. In particular, under a local $U(1)$ axial transformation, the classical Euclidean Dirac Lagrangian in~\eqref{PathIntegralQED} transforms as\footnote{This result is obtained after performing an integration by parts and neglecting the boundary term.}
\begin{equation} \label{LagTransformation}
    \mathcal{L}_{D}\to\mathcal{L}_{D}^{'}=\mathcal{L}_{D}-i\alpha(x)\partial_{\mu}j_{5}^{\mu}\,,
\end{equation}
where the sign depends on the convention used for the Euclidean action and the definition of the axial current $j_{5}^{\mu}$.
Combining, \eqref{FxPontryagin}, \eqref{MeasurePathIntegral}, and~\eqref{LagTransformation}, the path integral~\eqref{PathIntegralQED} becomes 
\begin{equation}
    \mathcal{Z}[U(x)]=\int\mathcal{D}A\mathcal{D}\bar{\psi}\mathcal{D}\psi\exp\left[-I_{QED}^{(E)}+i\int d^{4}x\alpha(x)\left(\partial_{\mu}j_{5}^{\mu}+\frac{1}{8\pi^{2}}\tilde{F}_{\mu\nu}F^{\mu\nu}\right)\right]\,.
\end{equation}
Since the change of variables must leave the path integral invariant, and since $\alpha(x)$ is arbitrary, one obtains that the divergence of the axial current is proportional to the Pontryagin density of $U(1)$, namely
\begin{equation}
   \langle\partial_{\mu}j_{5}^{\mu}\rangle=-\frac{1}{8\pi^{2}}\tilde{F}_{\mu\nu}F^{\mu\nu}\,. 
\end{equation}
This is the axial anomaly equation, also known as the Adler-Bell-Jackiw anomaly~\cite{Bell:1969ts,Adler:1969gk}. Adler, Bell, and Jackiw originally obtained this result from the one-loop triangle diagram in perturbation theory, whereas Fujikawa's method identifies the anomaly with the regularized Jacobian of the fermionic measure. 

One of the most important features of this anomaly is its close relation to topology. Indeed, the Chern-Pontryagin density is sensitive to the topological structure of the gauge field configurations. In particular, instantons backgrounds can give rise to nontrivial contributions to the anomaly. 

Moreover, this anomaly is directly related to the Atiyah-Patodi-Singer index theorem~\cite{APS-eta} important in the presence of boundaries, as we will discuss in Sec.~\ref{Sec.IndexTheorem}. The Fujikawa method, therefore, provides a physical derivation of a result with deep mathematical significance, bridging the non-invariance of the quantum fermionic measure to topological invariants of gauge fields. 
\chapter{Axial Anomaly in Nonlinear Conformal Electrodynamics}  \label{chapter3}

As discussed previously, the axial anomaly is a remarkable quantum phenomenon arising from the breakdown of axial symmetry by quantum effects. Beyond representing merely a violation of a classical conservation law, it provides a clear example in which the nontrivial topology of fiber bundles has direct and measurable physical consequences. Historically, the axial anomaly played a fundamental role in resolving the puzzle associated with the neutral pion decay process $\pi^{0}\to\gamma\gamma$  , answering one of the most challenging problems in quantum field theory during the 60's~\cite{Steinberger:1949wx,10.1143/ptp/4.3.347,Schwinger:1951nm,Adler:1969gk,Bell:1969ts,Bardeen:1969md}. 

The importance of the axial anomaly goes far beyond high-energy physics. In the context of condensed matter systems, it has been associated with spectral asymmetry in three-dimensional paramagnetic topological insulators such as (Hg, Mn, Te)~\cite{Wang_2024}. Furthermore, the anomalous half-quantized Hall conductance arising from occupied electronic states below the Fermi level has been shown to be directly related to parity and axial anomalies~\cite{Mogi_2022,Wang_20242}. Similarly, topological transport phenomena and the chiral magnetic effect in Weyl semimetals can be understood as a consequence of the axial anomaly~\cite{Goswami:2015uxa, Zyuzin:2012tv}. 

Experimental signatures of these effects have been reported in zirconium pentatelluride~\cite{Li:2014bha}, and more recently, indirect evidence has also appeared in relativistic heavy-ion collisions through the analysis of quark-gluon plasma~\cite{Belmont:2014lta,STAR:2015wza}. 

The axial anomaly also plays an important role in topological insulators~\cite{1985ZhPmR..42..145V,PANKRATOV198793,doi:10.1126/science.1148047,PhysRevLett.95.226801,PhysRevLett.96.106802}, where it governs electromagnetic response properties~\cite{Ryu_2012}. These materials exhibit remarkable phenomena such as spin-momentum locking~\cite{Hsieh:2009nature}, Majorana states in superconducting phases~\cite{PhysRevLett.100.096407}, and the quantum spin Hall effect~\cite{PhysRevLett.95.226801}. Such features make them candidates for spintronic devices and dissipationless transistors in quantum computing. 

The study of axial anomaly has also been extended to gauge theories coupled to gravity~\cite{Kimura:1969axial,Delbourgo:1972xb,AlvarezGaume:1983ig}, where the nontrivial topology of spacetime may significantly influence matter fields~\cite{Charap:1977ww}. In this framework, different classes of instanton solutions arise on manifolds with inequivalent topologies. In particular, there exist solutions to the Yang-Mills equations that either become singular in flat space or cannot exist there~\cite{Charap:1977re}. 

Such configurations are frequently studied through (anti-) self-dual gauge fields on Euclidean spacetimes. In these cases, the stress-energy tensor vanishes as a consequence of conformal symmetry together with the (anti-)self-duality condition. As a result, the absence of backreaction renders their gravitational effects subtle. Nevertheless, as Pope pointed out, there are situations in which nontrivial interactions between the gravitational and gauge sectors lead to physically relevant consequences~\cite{Pope:1978zx,Pope:1981jx}. 

Motivated by these developments, in this work, we study the axial anomaly in the context of nonlinear electrodynamics, focusing on the recently proposed ModMax electrodynamics~\cite{Bandos:2020jsw}. ModMax constitutes a nonlinear extension of Maxwell theory that preserves two of its most important properties: conformal invariance and electromagnetic duality symmetry. 

There are many contexts which nonlinear electrodynamics arises. For instance, integrating out fermionic degrees of freedom in quantum electrodynamics generates effective nonlinear interactions that capture one-loop quantum effects due to virtual fermions~\cite{Heisenberg:1936nmg}. Moreover, nonlinear constitutive relations have been introduced to address divergences associated with the self-energy of charged point ~\cite{Born:1933qff,Born:1934ji,Born:1934gh}. 

In the presence of gravity, nonlinear electrodynamics has found applications in black holes ~\cite{Ayon-Beato:1998hmi,Bronnikov:2000vy,Ayon-Beato:2000mjt,Cataldo:2000we,Burinskii:2002pz,Hassaine:2007py,Miskovic:2010ey,Cisterna_2020,Bravo-Gaete:2022mnr,DiazGarcia:2022jpc,Ayon-Beato:2022dwg,Bravo-Gaete:2022mnr,Alvarez:2022upr}, conserved charges~\cite{Miskovic:2010ui}, holographic superconductors~\cite{Jing:2011vz,Aranguiz:2013cwa}, and black strings solutions~\cite{Cisterna:2021ckn}, among others~\cite{Babaei-Aghbolagh:2022uij,Babaei-Aghbolagh:2022itg,Ferko:2022iru,Arenas-Henriquez:2022ntz,Ferko:2023ruw}. 

Our analysis focuses on (anti-)self-dual configurations in conformal nonlinear electrodynamics, which automatically satisfy the electrodynamic equations of motion. However, electromagnetic duality transformations does not necessarily preserve the constitutive relations of a nonlinear theory. Consequently, establishing duality invariance becomes a nontrivial problem. 

Notably, imposing invariance under duality rotations on general nonlinear electrodynamics does not uniquely determine the theory~\cite{Gaillard:1981rj,Bialynicki-Birula:1983}. Instead, one finds plenty of possible theories parameterized by real functions of a single variable~\cite{Salazar:1987ap,Gibbons:1995cv}. Within this family of theories, the one-parameter family of Lagrangians known as ModMax electrodynamics is included. Besides ModMax, another theory possessing similar properties is the nonlinear Bialynicki-Birula electrodynamics~\cite{Bialynicki-Birula:1992rcm}, whose Lagrangian vanishes as a consequence of its defining constraints. Furthermore, this theory possesses an even larger $SL\left(2,\mathbb{R}\right)$ duality symmetry. 

Explicit constructions of duality-invariant nonlinear electrodynamics have been studied using formulations where duality symmetry becomes manifest~\cite{Avetisyan:2021heg,Avetisyan:2022zza,Mkrtchyan:2019opf,Pasti:1995ii}. Outside these approaches, ModMax has attracted considerable attention and has been studied in different contexts, including black holes, birefringence, supersymmetry, gravitational instantons, wormholes, and the 't Hooft anomalies~\cite{Flores-Alfonso:2020euz,Amirabi:2020mzv,Bandos:2021rqy,Chatzistavrakidis:2021dqg,Barrientos:2022bzm,BallonBordo:2020jtw,Flores-Alfonso:2020nnd}. 

In this work, we show that gravitational instantons coupled to (anti-)self-dual fields in nonlinear electrodynamics induce an axial anomaly for charged Dirac spinors. To the best of our knowledge, this represents the first explicit realization of the axial anomaly in the context of ModMax electrodynamics. To this end, we study the field equations of the Einstein-ModMax system in backgrounds possessing a nontrivial circle fibration over a Kähler manifold, as occurs in the Taub-NUT and Eguchi-Hanson spaces. We begin by reviewing the Taub-NUT case, which has been discussed in the literature~\cite{BallonBordo:2020jtw,Flores-Alfonso:2020nnd}. We then introduce an Eguchi-Hanson type ansatz and obtain a new self-gravitating ModMax configuration. In both cases, the electromagnetic and gravitational curvatures satisfy (nonlinear) self-duality relations.

The imbalance between positive and negative chiral charged fermions is computed using the Atiyah-Patodi-Singer index theorem, taking into account the boundary contribution. Particular attention is devoted to the computation of the $\eta_{D}$-invariant, which measures the spectral asymmetry of the Dirac operator evaluated on the tangential components at the boundary. 

We find that the nonlinearly charged Eguchi-Hanson and Taub-NUT instantons generate an imbalance between Dirac spinors of opposite chirality, thereby triggering the axial anomaly. Finally, we comment on possible extensions of these results and their implications for other types of anomalies.

\section{Nonlinear Electrodynamics}

The Maxwell theory of electrodynamics represents one of the most successful theories in physics. It unifies electric and magnetic phenomena within a single framework and provides an accurate description of a vast range of electromagnetic processes. Nevertheless, despite its success, Maxwell electrodynamics presents certain conceptual difficulties. 

One of its most notable problems arises when considering the self-energy of a point-like charged particle. An explicit computation of the electromagnetic energy associated with a static electric field shows that the electromagnetic self-energy diverges at the origin. This suggests that classical electrodynamics may become inadequate at short distances. Additionally, in the article \textit{Foundations of the New Field Theory} by Born and Infeld~\cite{Born:1934gh}, it was argued that the appearance of the classical radius $r_{e}=e^{2}/mc^{2}$ without involving Planck's constant $\hbar$ indicated that the electromagnetic laws themselves should be modified rather than just quantized. 

Motivated by these facts, the authors proposed a nonlinear generalization of Maxwell electrodynamics in order to provide a new perspective on the relation between field theory and quantum mechanics. Their proposal was inspired by the earlier work of Gustav Mie, who attempted to construct a purely electromagnetic theory of matter in which charged particles would emerge entirely from the electromagnetic field~\cite{Mie:1912sea,Mie:1912usi,Mie:1913ice}. In particular, he wanted to construct a theory in which the electromagnetic field constituted the sole fundamental ingredient. However, his formulation was found unacceptable because it lacked gauge invariance, a symmetry that later became a cornerstone of modern field theories. 

Considering Maxwell electrodynamics, which is described by the Lagrangian 
\begin{equation}
    \mathcal{L}=-\frac{1}{4}F_{\mu\nu}F^{\mu\nu}\,.
\end{equation}
Born and Infeld proposed the following Lagrangian density 
\begin{align} \label{BornInfeldLag}
    \mathcal{L}_{{\rm BI}}	&=T-T\sqrt{-{\rm det}\left(\eta_{\mu\nu}+T^{- 1/2}F_{\mu\nu}\right)}\,, \nonumber \\
	&=-\frac{1}{4}F_{\mu\nu}F^{\mu\nu}+\frac{1}{32T}\left((F_{\mu\nu}F^{\mu\nu})^{2}+(F_{\mu\nu}\tilde{F}^{\mu\nu})^{2}\right)+\mathcal{O}((F_{\mu\nu}F^{\mu\nu})^{3})\,.
\end{align}
Here $T$ is a parameter with dimensions of energy. In the context of string theory, $T$ is naturally interpreted as the tension of a $D$-brane, and the Born-Infeld theory arises as the low-energy effective action governing the dynamics of open strings~\cite{FRADKIN1985123}. 

Although the theory did not achieve its original objective of providing a complete unification between quantum mechanics and field theory, it possesses many remarkable properties. In particular, it is the only known nonlinear electrodynamics that does not exhibit vacuum birefringence and possesses physically consistent shock-wave propagation properties~\cite{Boillat:1970gw,BOILLAT19729}. Furthermore, during the development of string theory in the 1980's, it was discovered that in bosonic open string theories, the Born-Infeld theory naturally emerges as the effective low energy action~\cite{Fradkin:1985qd,Leigh:1989jq}.

In parallel, Heisenberg became interested in the implications of the new Dirac's positron theory~\cite{Dirac:1928electron,Dirac:1931monopole} and the role of the quantum fluctuations associated with the Dirac sea picture of the quantum electrodynamics vacuum. He noticed that the quantum fluctuations would lead to quantum nonlinear effects in electromagnetic interactions~\cite{Heisenberg:1934pza}. 

Subsequently, Heisenberg's students, Euler and Kockel, and independently, Akhiezer, Landau, and Pomeranchuk studied light by light scattering and derived quantum corrections to Maxwell electrodynamics~\cite{Heisenberg:1936nmg,Landau:1953um,Akhiezer:1936vzu}. The results led to the following effective Lagrangian 
\begin{equation} \label{EulerKockelLag}
    \mathcal{L}=-\frac{1}{4}F_{\mu\nu}F^{\mu\nu}+\frac{\hbar^{3}\alpha^{2}}{90\mu_{0}^{2}m_{e}^{4}c^{5}}\left[\left(F_{\mu\nu}F^{\mu\nu}\right)^{2}+\frac{7}{4}\left(F_{\mu\nu}\tilde{F}^{\mu\nu}\right)^2\right]+\mathcal{O}\left((FF)^3\right)\,,
\end{equation}
where $\alpha=\tfrac{e^{2}}{4\pi\epsilon_{0}\hbar c}$ is the fine structure constant and $\mu_{0}$ denotes the vacuum magnetic permeability. Notice that these nonlinear corrections, which appear as higher-order powers of the electromagnetic field strengths $F_{\mu\nu}$, arise from one-loop effects generated by virtual electrons and positrons circulating in closed loops. 

Shortly after the work by Euler and Kockel, Heisenberg and Euler extended these results and obtained a non-perturbative expression for the effective electromagnetic action~\cite{Heisenberg:1936nmg}. This achievement allowed them to identify a remarkable prediction of Dirac's positron theory, previously noted by Sauter: the instability of the quantum electrodynamics vacuum when subjected to a classical electric field. As a consequence, the vacuum becomes unstable against the production of electron-positron pairs, a phenomenon now known as the Sauter-Schwinger effect~\cite{Sauter:1931zz,Schwinger:1951nm}. 

Despite these remarkable features, the Euler-Heisenberg theory does not preserve all the symmetries of Maxwell electrodynamics. In particular, the nonlinear contributions break both conformal and electric-magnetic duality invariance. It is important to emphasize that, although Maxwell's equations are invariant under duality rotations, the Maxwell action is not. 

Nevertheless, it was later shown that the effective Euler-Heisenberg-type Lagrangian describing a supersymmetric system consisting of one charged spin-$\tfrac{1}{2}$ field and two charged spin-$0$ fields is duality invariant~\cite{Duff:1979bk}. The search for nonlinear electrodynamics theories sharing the same symmetry structure as Maxwell electrodynamics eventually led to the work of Iwo Bialynicki-Birula in 1983~\cite{Bialynicki-Birula:1983,Bialynicki-Birula:1992rcm}. By imposing electric-magnetic duality invariance on the equations of motion of a general nonlinear electrodynamics, he showed that the corresponding Lagrangian density must satisfy restrictive conditions, leading to a family of duality-invariant theories. 

In particular, these conditions imply that the Born-Infeld Lagrangian is duality invariant, a property that had already been predicted by Schrödinger in 1935~\cite{Schrodinger:1935oqa}. Bialynicki-Birula also investigated two regimes of the Born-Infeld theory, namely the weak-field limit $T\to\infty$  and the strong-field limit $T\to0$. 

In the weak field regime, it is straightforward to show that the Born-Infeld theory reduces to Maxwell electrodynamics. The strong-field limit, however, is more subtle. To analyze it, Bialynicki-Birula considered the Hamiltonian density form of the Born-Infeld theory, which corresponds to 
\begin{equation} \label{HamiltonianBI}
    \mathcal{H}_{{\rm BI}}=\sqrt{T^{2}+T(\vec{D}^{2}+\vec{B}^{2})+|\vec{D}\times\vec{B}|^{2}}-T\,,
\end{equation}
where $\vec{D}=\tfrac{\partial\mathcal{L}}{\partial\vec{E}}$ is the electric displacement vector field. Taking the limit $T\to0$, one obtains
\begin{equation} \label{BBHamiltonian}
    \mathcal{H}_{{\rm BB}}=|\vec{D}\times\vec{B}|\,.
\end{equation}

Remarkably, the energy density of this ultra-Born-Infeld theory is equal to the length of the Poynting vector. Moreover, this limit exhibits an important family of symmetries: it is invariant under conformal transformations and possesses an $SL(2,\mathbb{R})$-duality symmetry. Although these symmetries exist, the Bialynicki-Birula theory in  Hamiltonian form does not include Maxwell's theory. Furthermore, it can be shown that this theory describes all possible configurations of null electromagnetic fields. 

Similarly to the Born-Infeld theory, the Bialynicki-Birula theory also appears in the context of string theory. In particular, it can be understood as a limiting case of the dynamics of a $D3-$brane~\cite{Bandos:2020jsw,Bandos_2021} and is closely related to the strong-field regime of Born-Infeld electrodynamics~\cite{Perry:1996mk,Pasti_1997}. 

\section{ModMax Electrodynamics}

Motivated by the search for nonlinear electrodynamics theories sharing the same symmetry structure as Maxwell theory, namely conformal and duality invariance, a natural question arises: are there other nonlinear theories belonging to this family? 

The theory was provided by Bandos, Lechner, Sorokin,and Townsend, who constructed a nonlinear electrodynamics theory known as Modified Maxwell (ModMax). As we shall see, this theory provides the natural framework for the analysis developed in the present work. For a complete review of nonlinear electrodynamics, see~\cite{Sorokin:2021tge,DUNNE_2012}.

Departing from the superposition principle of electromagnetic fields leads to a generalization of the source-free Maxwell equations, 
\begin{equation} \label{MaxwelEOM}
    \nabla_{\mu}P^{\mu\nu}=0\,,\qquad\text{and}\qquad\nabla_{\mu}\tilde{F}^{\mu\nu}=0\,,
\end{equation}
where $P^{\mu\nu}=P^{\mu\nu}\left(F^{\lambda\rho}\right)$ depends nonlinearly on the electromagnetic field strength, and $\tilde{F}^{\mu\nu}=\tfrac{1}{2}\epsilon_{\mu\nu\lambda\rho}F^{\lambda\rho}$ is the dual field strength. The second equation corresponds to the Bianchi identity. 

One of the most remarkable features of ModMax electrodynamics that it constitutes the unique nonlinear extension of Maxwell theory that preserves both conformal symmetry and $SO(2)$ electric-magnetic duality invariance. Originally, the theory was formulated in a Legendre-dual representation, while an explicit Lagrangian formulation was obtained shortly thereafter by Kosyakov~\cite{Kosyakov:2020wxv}. Another important property of ModMax is that it is Legendre self-dual~\cite{Flores-Alfonso:2020euz}, in close analogy with Born-Infeld electrodynamics~\cite{Gibbons:2000xe}. 

Despite these similarities, the two theories exhibit important differences. Born-Infeld electrodynamics is well known for the absence of vacuum birefringence, whereas ModMax exhibits birefringence when the field equations are linearized around nontrivial electromagnetic backgrounds~\cite{Bandos:2020jsw}. Nevertheless, the theory remains continuously connected to Maxwell electrodynamics and reduces to it in the linear limit. 

In this work, we use ModMax electrodynamics as the matter sector coupled to Einstein gravity. For this reason, it is convenient to formulate the theory through the action 
\begin{equation} \label{ModMaxAction}
    I_{{\rm M}}=-\int_{\mathcal{M}}d^{4}x\sqrt{|g|}\mathcal{L}_{{\rm M}}\,,
\end{equation}
where $g=\det(g_{\mu\nu})$ and the Lagrangian is given by 
\begin{equation} \label{ModMaxLagden}
    \mathcal{L}_{{\rm M}}=X\cosh\gamma-\sqrt{X^{2}-Y^{2}}\sinh\gamma\,.
\end{equation}
Here, 
\begin{equation} \label{LorentzLagMod}
   X=\frac{1}{4}F_{\mu\nu}F^{\mu\nu}\,,\qquad\text{and}\qquad Y=\tfrac{1}{4}F_{\mu\nu}\tilde{F}^{\mu\nu}\,, 
\end{equation}
are the two independent electromagnetic invariants that can be constructed from the field strength and its dual. In this work, we focus on the Euclidean sector of the theory \footnote{In the Lorentzian theory, one should replace $Y^{2}\to-Y^{2}$.}.

Under parity, the electromagnetic invariants transform as $X\to X$ and $Y\to - Y$. Although $Y$ is parity odd, the ModMax Lagrangian depends on it only through $Y^2$, so that
\begin{equation}
    \mathcal L_M(X,-Y)=\mathcal L_M(X,Y)\,.
\end{equation} 
Consequently, ModMax provides a parity-preserving nonlinear extension of Maxwell electrodynamics, a property shared with Born--Infeld theory.

The parameter $\gamma$  controls the degree of nonlinearity of the electromagnetic fields. In the limit $\gamma\to 0$, one recovers standard Maxwell electrodynamics. Since $\gamma$  is dimensionless, conformal invariance is preserved \footnote{This is the opposite of Born-Infeld. The theory is not conformally invariant because the tension $T$ parameter is dimensional.}. Furthermore, causality considerations in the Lorentzian theory require $\gamma\geq0$. 

Additionally, the Bianchi identity implies the local existence of a gauge potential $A=A_{\mu}dx^{\mu}$ such that $F_{\mu\nu}=2\partial_{[\mu}A_{\nu]}$. 

On the other hand, varying the action~\eqref{ModMaxAction} with respect to $A_{\mu}$ yields the generalized Maxwell equations~\eqref{MaxwelEOM} together with the constitutive relation 
\begin{equation} \label{ConstitutiveRelation}
    P_{\mu\nu}=\left(\cosh\gamma-\frac{X\sinh\gamma}{\sqrt{X^{2}-Y^{2}}}\right)F_{\mu\nu}+\frac{Y\sinh\gamma}{\sqrt{X^{2}-Y^{2}}}\tilde{F}_{\mu\nu}\,.
\end{equation}
A straightforward calculation reveals that 
\begin{equation} \label{DualityInvariance}
   \tilde{F}_{\mu\nu}F^{\mu\nu}=\tilde{P}_{\mu\nu}P^{\mu\nu}\,, 
\end{equation}
which is a sufficient condition for duality~\cite{Gibbons:1995cv}. 

The stress-energy tensor is obtained by varying the ModMax action with respect to the metric. In particular, using Eqs.~\eqref{ModMaxLagden} and~\eqref{ConstitutiveRelation}, one finds 
\begin{equation}\label{StressTensor}
    T_{\mu\nu}=P_{(\mu}{}^{\lambda}F_{\nu)\lambda}-g_{\mu\nu}\mathcal{L}_{{\rm M}}\,.
\end{equation}
Considering the limit $\gamma\to 0$, this expression reduces to the standard Maxwell stress-energy tensor. Moreover, since ModMax is conformally invariant, the stress tensor~\eqref{StressTensor} satisfies the Bessel-Hagen criterion~\cite{Bessel1921} 
\begin{equation} \label{Bessel-Hagen}
    T_{\mu}^{\mu}=0\,.
\end{equation}
Whenever this condition is broken at the quantum level, it is known as the trace anomaly~\cite{Capper:1974ic,Deser:1976yx,Duff:1977ay}. To couple ModMax electrodynamics to gravity, we consider the Einstein-Hilbert action with cosmological constant, 
\begin{equation} \label{EH-M action}
I[g_{\mu\nu},A_{\mu}]=\kappa\int_{\mathcal{M}}d^{4}x\sqrt{|g|}\left(R-2\Lambda\right)+I_{{\rm M}}\,,
\end{equation}
where $\kappa=\left(16\pi G\right)^{-1}$ with $G$ Newton's constant, $\Lambda$ is the cosmological constant, $R=g^{\mu\nu}R^{\lambda}{}_{\mu\lambda\nu}$ is the Ricci scalar, and $I_{\rm M}$ is given in Eq.~\eqref{ModMaxAction}. The field equations are obtained by varying Eq.~\eqref{EH-M action} with respect to the metric 
\begin{equation} \label{EinsteinEOM}
    R_{\mu\nu}-\frac{1}{2}g_{\mu\nu}R+\Lambda g_{\mu\nu}=\frac{1}{2\kappa}T_{\mu\nu}\,.
\end{equation}
Taking the trace and using the conformal condition, i.e. $T_{\mu}^{\mu}=0$, one obtains $R=4\Lambda$. Therefore, every solution of Einstein-ModMax theory possesses a constant Ricci scalar curvature, and the Einstein equations reduce to 
\begin{equation} \label{EinsteinEOMcond}
    R_{\mu\nu}=\Lambda g_{\mu\nu}+8\pi GT_{\mu\nu}\,.
\end{equation}
Equation~\eqref{EinsteinEOMcond}, together with the generalized Maxwell equations~\eqref{MaxwelEOM}, forms a system of second-order nonlinear partial differential equations governing the dynamics of the Einstein-ModMax system. 

\section{(Anti-) Self-dual Instantons in Yang-Mills Theory} \label{InstantonsSection}

Gauge theory is a field theory whose dynamics are invariant under local symmetry transformations, known as gauge transformations. The origin of this idea can be traced back to the work of Hermann Weyl in 1918, who attempted to unify electromagnetism and gravity through a local symmetry principle~\cite{Weyl:1918ib}. Although this novel proposal of gauge invariance was not physically successful, this work introduced the concept of gauge invariance, which later became one of the most important ingredients in modern theoretical physics. 

Motivated by these ideas, Yang and Mills generalized the abelian gauge symmetry of Maxwell electrodynamics to non-abelian Lie groups~\cite{Yang-Mills}. The construction of this theory led to a new class of interacting spin-1 field theories, now known as Yang-Mills theories, which form the basis of the Standard Model. 

One of the most important features of gauge theories is their close relation with geometry~\cite{Lubkin:1963zz,Hermann1970,Trautman:1970cy}. In modern language, gauge fields $A_{\mu}$ are called connections on principal fiber bundles, while the field strength $F_{\mu\nu}$ is referred to as the curvature of these bundles. This geometrical perspective has profound physical consequences. An example of this corresponds to the reformulation of the Dirac monopole in terms of gauge potentials provided by Wu and Yang~\cite{Wu:1975es,Wu:1976ge}. They introduced overlapping coordinate patches related by gauge transformations, which revealed the topological nature of magnetic charge and established a direct connection between gauge theories and the topology of fiber bundles. 

In this framework, another important quantity is the Pontryagin density, which enters the Yang--Mills action through the
$\theta$-term. In Lorentzian signature and units $\hbar=1$, we write
\begin{equation}
    I_{\theta}
    =
    -\frac{\theta}{16\pi^{2}}
    \int_{\mathcal M}d^{4}x\,\sqrt{|g|}\,
    \operatorname{tr}
    \left(F_{\mu\nu}\widetilde F^{\mu\nu}\right),
    \label{thetaterm}
\end{equation}
where $F_{\mu\nu}=F_{\mu\nu}^{a}t_{a}$ is the non-Abelian field
strength, the gauge coupling is absorbed into the connection,
and the anti-Hermitian generators satisfy
$\operatorname{tr}(t_{a}t_{b})=-\tfrac12\delta_{ab}$.
The Pontryagin density is locally a total derivative, so a constant
$\theta$ does not modify the bulk classical field equations,
although boundary contributions must be treated separately.
Nevertheless, it affects the quantum theory through the relative
phases of configurations with different topological charges.
This is particularly relevant in Yang--Mills theory, whose
Euclidean formulation admits smooth finite-action instantons
with nonzero integer Pontryagin number
\cite{tHooft:1976rip,tHooft:1976snw}.

The $\theta-$term has important physical consequences. For instance, it plays a fundamental role in the description of topological insulators and in axion electrodynamics (see Ref.~\cite{Sekine:2021axionreview} for a review). 

A remarkable development occurred in 1975 with the discovery of instantons in non-abelian gauge theories by Belavin, Polyakov, Schwarz, and Tyupkin~\cite{Belavin:1975fg}. They found finite-action solutions of the Euclidean $SU(2)$ Yang-Mills equations, now known as BPST instantons. These solutions revealed that gauge theories possess a variety of topological structures that cannot be captured within perturbation theory. 

One of the most important properties of instantons is that their field strength satisfies an (anti-) self duality condition, 
\begin{equation} \label{anti-self-dualcondition}
    F_{\mu\nu}=\pm\tilde{F}_{\mu\nu}\,.
\end{equation}
This condition automatically solves the Yang-Mills equations, and at the same time minimizes
the Euclidean action. Additionally, instantons are characterized by the SU(2) Chern-Pontryagin index in four dimensions, corresponding to
\begin{equation}
    C_2[A] = -\frac{1}{8\pi^2} \int_{\mathcal{M}} d^4x\, \sqrt{|g|} \, \rm{tr} (\widetilde{F}_{\mu \nu} F^{\mu \nu})\,,
\end{equation}
which is an integer topological invariant. This quantity classifies inequivalent $SU(2)$ principal fiber bundles. 

Instantons also provide a physical interpretation for the existence of homotopically inequivalent Yang--Mills vacua~\cite{Jackiw:1976pf,Callan:1976je}. Different vacuum sectors are labeled by distinct winding numbers, and instantons describe quantum tunneling processes between them. In the saddle point approximation, the Euclidean instanton action determines the leading contribution to the tunneling amplitude. Moreover, instantons are asymptotically pure gauge configurations and may be viewed as topologically nontrivial maps between the boundary of spacetime and the gauge group manifold (see Ref.~\cite{CCgauge} for a comprehensive review of instantons and gauge theories, and the references therein). 

The BPST instanton carries topological index $C_{2}=\pm1$ for self-dual and anti-self-dual configurations, respectively. From this perspective, the community tried to extend these results leading to solutions with arbitrary integer topological charge $k\in\mathcal{\mathbb{Z}}$, corresponding to multi-instanton solutions~\cite{Corrigan:1976vj,Witten1977}, among others. The existence of these solutions revealed a profound connection between topology and quantum field theory. 

An even deeper relation emerged through the Atiyah-Singer index theorem, which relates the topological charge of the gauge field to the index of the chiral operator 
\begin{equation} \label{indexD}
    \text{index }D=N_{+}-N_{-}\,,
\end{equation}
where $N_{+}$ and $N_{-}$ denote the numbers of positive and negative chiral fermionic zero modes, respectively. This result establishes a bridge between geometry, topology, and quantum field theory, showing that the topological properties of gauge fields have direct physical consequences for those systems.

\section{Gravitational Instantons}

The discovery of the Yang-Mills instantons and their role in the path integral formulation of quantum field theory opened the possibility of applying similar ideas to quantum gravity~\cite{DeWitt1,DeWitt2,DeWitt3}. The main question was whether Euclidean solutions of Einstein's equations could contribute nontrivially to the gravitational path integral and play a role analogous to that of instantons in Yang-Mills theory.

During the 1970's, several authors investigated this interesting topic, and in 1977, Hawking introduced the notion of gravitational instantons ~\cite{Hawking:1976jb}. These objects are regular solutions of the Euclidean Einstein equations with non-trivial topology and finite Euclidean action. Some gravitational instantons can be obtained through analytic continuations of Lorentzian black hole solutions of Einstein or Einstein-Maxwell theory. For instance, in this category, one can find the Euclidean Schwarzschild and Kerr metrics, which, although they are gravitational instantons, do not satisfy the (anti)-self duality condition on the Riemann curvature
\begin{equation} \label{AntiSelfDualRiemann}
    R_{\lambda\rho}^{\mu\nu}=\pm\frac{1}{2}\epsilon^{\mu\nu\alpha\beta}R_{\alpha\beta\lambda\rho}\,.
\end{equation}
There also exists an important class of gravitational instantons whose Riemann tensor is (anti)-self dual. For instance, in this family, it is possible to find the Taub-NUT~\cite{Taub:1950ez,Newman:1963yy}, Eguchi-Hanson~\cite{Eguchi:1978xp,Eguchi:1978gw}, multi-center instantons~\cite{Hawking:1976jb,Gibbons:1978tef}, and Chen-Teo~\cite{Chen:2011tc,Chen:2015vva} geometries. These solutions closely resemble Yang-Mills instantons, since they are characterized by topological invariants, satisfy the (anti)-self duality condition, and play a fundamental role in the semiclassical description of gravity. 

As in the Yang-Mills case, in four dimensions, gravitational instantons are characterized by the Pontryagin index, 
\begin{equation} \label{Pontryaginindex}
    P	=\frac{1}{32\pi^{2}}\int_{\mathcal{M}}d^{4}x\sqrt{|g|}\epsilon_{\mu\nu\lambda\rho}R_{\sigma\tau}^{\lambda\rho}R^{\sigma\tau\mu\nu}\,,
\end{equation}
and the Euler-Poincaré characteristic, 
\begin{equation} \label{EulerCharacteristic}
    \chi=\frac{1}{128\pi^{2}}\int_{\mathcal{M}}d^{4}x\,\sqrt{|g|}\,\epsilon_{\mu\nu\lambda\rho}\epsilon^{\sigma\tau\gamma\delta}R_{\gamma\delta}^{\lambda\rho}R_{\sigma\tau}^{\mu\nu}\,.
\end{equation}
Here $g$ is the determinant of the metric, and $\epsilon_{\mu\nu\lambda\rho}$ is the Levi-Civita tensor. 

The gravitational Pontryagin number~\eqref{Pontryaginindex} characterizes a global topological property of the tangent bundle through an integral of the Riemann curvature. On a closed, oriented four-manifold, it is unchanged by smooth deformations of the metric and contributes to the index governing the imbalance between positive- and negative-chirality fermionic zero modes. For gravitational instantons with an asymptotic boundary, the corresponding index also requires boundary contributions~\cite{Hirzebruch:1995,APS-eta}.
 
The Euler--Poincaré characteristic~\eqref{EulerCharacteristic} can also be related to the fixed-point sets of a circle action generated by a Killing vector field.\footnote{The fixed-point set consists of the points where the generating Killing vector vanishes.} For a compact, oriented four-manifold without boundary, it equals the sum of the Euler characteristics of the connected fixed-point components. Each isolated fixed point, or nut, contributes one, whereas each two dimensional fixed surface, or bolt, contributes its own Euler characteristic; a spherical bolt therefore contributes two~\cite{Gibbons:1979xm}. For noncompact geometries, the applicability of this relation additionally depends on the behavior at infinity.

The relevance of these invariants goes beyond a topological perspective. Through index theorems, they can be related to properties of fermionic operators defined over manifolds. In particular, in the Atiyah-Patodi-Singer  theorem, the Pontryagin term contributes directly to the index of the Dirac operator and therefore plays a fundamental role in the gravitational contribution to the axial anomaly, as we will see. 

In the present work, the central ingredient in the computation of the Dirac index is the existence of nontrivial Pontryagin invariants. Indeed, the gravitational Pontryagin density can be expressed in terms of the electric and magnetic components of the Weyl tensor and vanishes whenever these tensors are orthogonal. Consequently, gravitational backgrounds with (anti-) self-dual curvature are of particular interest to us, since they give rise to nontrivial topological invariants. 

In this section, we study nonlinearly charged self-gravitating configurations with a nontrivial Pontryagin number. These solutions satisfy the Einstein-ModMax field equations and can be described geometrically as inhomogeneous metrics on complex line bundles over Kähler manifolds~\cite{Page:1985bq}. The gauge potential is aligned along the fiber direction, allowing the electromagnetic field to simultaneously satisfy both the ModMax constitutive relation and the equations of motion. 

In particular, we focus on solutions belonging to the Taub-NUT and Eguchi-Hanson families. We begin our discussion with the construction of the Taub-NUT within the Einstein-ModMax theory and then develop an analogous construction for the Eguchi-Hanson geometry.

\subsection{Nonlinearly Charged Taub-NUT-AdS instanton}

The first case corresponds to the Taub-NUT spacetime. The Lorentzian Taub-NUT spacetime is a one-parameter generalization of the Schwarzschild black hole that is often interpreted as a gravitational dyon, since it carries both (electric) mass and NUT charge (magnetic mass) ~\cite{Lynden-Bell:1996dpw,Araneda:2016iiy}. These quantities source the electric and magnetic components of the Weyl tensor, respectively. 

In Euclidean signature, metrics of this type can be written as 
\begin{equation} \label{TNmetric}
    ds^{2}=f(r)\left(d\psi+2n\mathcal{B}_{(k)}\right)^{2}+\frac{dr^{2}}{f(r)}+(r^{2}-n^{2})d\Sigma_{(k)}^{2}\,,
\end{equation}
where the metric function $f(r)$ is determined by the field equations, and $n$ denotes the NUT charge. Here, $\mathcal{B}_{(k)}$ is the Kähler potential $1$-form whose exterior derivative is,
\begin{equation} \label{KhalerPotential}
    \Omega_{(k)}=d\mathcal{B}_{(k)}\,,
\end{equation}
defines the Kähler form associated with the Einstein-Kähler base manifold of constant curvature $d\Sigma_{(k)}^{2}$. For $k=1,0,-1$, the base manifold is topologically $\mathbb{S}^{2}$, $\mathbb{T}^{2}$, and $\mathbb{H}^{2}$, respectively. The corresponding Kähler potentials and line elements are summarized in Table~\ref{TableKhaler}. 
\begin{table}[htbp]
	\centering
	\begin{tabular}{cccc}
    \cline{1-4}
		& $k=1$ \,  &  $k=0$ \, & $k=-1$ \, \\ 
 \hline
 $\mathcal{B}_{(k)}$ \,  & $\cos\vartheta d\varphi$ \, & $\vartheta d\varphi$ \, & $\cosh\vartheta d\varphi$  \\ 
 \hline
  $d\Sigma_{(k)}^2$ \, & $d\vartheta^2 + \sin^2\vartheta d\varphi^2$ \, & $d\vartheta^2+d\varphi^2$ \,
  & $d\vartheta^2 + \sinh^2\vartheta d\varphi^2$ \\
 \hline
	\end{tabular}
\caption{Line element of the transverse sections and Kähler potential $1$-form for $\mathbb{S}^2$, $\mathbb{T}^2$, and $\mathbb{H}^2$ topologies, respectively.}
\label{TableKhaler}
\end{table}

The first charged Taub-NUT solution was obtained by Brill in the context of Einstein-Maxwell theory ~\cite{Brill:1964tn}. More recently, higher-dimensional generalizations have been constructed in ~\cite{Awad:2005ff}. In these configurations, the gauge potential is assumed to be aligned along the fiber direction and takes the form 
\begin{equation} \label{potentialansatz}
    A=A_{\mu}dx^{\mu}=\alpha(r)\left(d\psi+2n\mathcal{B}_{(k)}\right)\,,
\end{equation}
where $\alpha(r)$ is an arbitrary function of the radial coordinate $r$. Henceforth, we shall assume the metric and gauge field ansätze given in~\eqref{TNmetric} and~\eqref{potentialansatz}, respectively.

Substituting these ansätze into the Einstein-ModMax field equations~\eqref{EinsteinEOM} and~\eqref{MaxwelEOM} yields the following linearly independent ordinary differential equations: 
\begin{align}
    0	&=\alpha''+\frac{2r\alpha'}{\left(r^{2}-n^{2}\right)}-\frac{4n^{2}e^{-2\gamma}\alpha}{\left(r^{2}-n^{2}\right)^{2}}\,, \label{EMMeom1} \\
0	&=4\kappa\left[rf'+\left(\frac{r^{2}+n^{2}}{r^{2}-n^{2}}\right)f+\Lambda\left(r^{2}-n^{2}\right)-k\right] -\left(r^{2}-n^{2}\right)e^{\gamma}\alpha'{}^{2}+\frac{4n^{2}e^{-\gamma}\alpha^{2}}{\left(r^{2}-n^{2}\right)}\,,\label{EMMeom2}
\end{align}
where we define $\alpha=\alpha(r)$ and $f=f(r)$, and the prime denotes differentiation with respect to the radial coordinate. 

The system can be solved by first integrating the equation for $\alpha(r)$ [cf. Eq.~\eqref{EMMeom1}] and then substituting the resulting expression into the gravitational equation~\eqref{EMMeom2}, which yields 
\begin{subequations}\label{TNFunction}
    \begin{align}
    f(r)	&=k\left(\frac{r^{2}+n^{2}}{r^{2}-n^{2}}\right)-\frac{2mGr}{r^{2}-n^{2}}-\frac{e^{-\gamma}}{4\kappa}\left(\frac{q^{2}-p^{2}}{r^{2}-n^{2}}\right)
	-\frac{\Lambda}{3}\left(\frac{r^{4}-6n^{2}r^{2}-3n^{4}}{r^{2}-n^{2}}\right)\,, \\
\alpha(r)	&=\frac{1}{2n}\left[q\sinh\Phi(r)+p\cosh\Phi(r)\right]\,,
\end{align}
\end{subequations}
where $m$, $q$, and $p$ are integration constants, and we have defined 
\begin{equation} \label{PotentialFunction}
    \Phi(r)=e^{- \gamma}\ln\left(\frac{r-n}{r+n}\right)\,.
\end{equation}
This solution was first obtained in Refs.~\cite{BallonBordo:2020jtw,Flores-Alfonso:2020nnd}. A simple zero $r=r_{+}$ of $f(r)$, bounding the positive definite region of the Euclidean metric, generally produces a conical defect unless the Euclidean time coordinate is identified as $\tau\sim\tau+\beta_{\tau}$, with $\beta_{\tau}=4\pi/|f'(r_{+})|$. The fixed-point set of $\partial_{\tau}$ is called a nut when it is an isolated point, and a bolt when it is a two dimensional surface~\cite{Gibbons:1979xm}.

Furthermore, for $k=1$, the fiber coordinates exhibit an additional apparent singularity at $\vartheta=0$ and $\vartheta=\pi$, known as the Misner string~\cite{Misner:1963fr}.
For the standard spherical fibration, removing this singularity requires $\beta_{\tau}=8\pi|n|$, which must be compatible with the period imposed by regularity at $r=r_{+}$.

In Lorentzian signature, the Misner string may give rise to closed timelike curves. For this reason, Taub-NUT geometries have been studied in Euclidean signature, where they bear a close resemblance to Yang-Mills instantons. Remarkably, it has been shown that the presence of the Misner string contributes nontrivally to the entropy of the spacetime~\cite{Mann:1999pc, Mann:1999bt, Hawking:1998ct, Mann:2004mi, Ciambelli:2020qny, Garfinkle:2000ms}. From the geometrical viewpoint, this occurs because the strings represent an obstruction to the existence of a globally defined time function generating the Hamiltonian evolution of the system~\cite{Hawking:1998jf}. 

More recently, the Lorentzian Taub-NUT solution has attracted renewed interest after it was shown that the Misner string is transparent to geodesic observers~\cite{Clement:2015cxa}. This observation led to a new thermodynamic interpretation of spacetime, in which the NUT charge is treated as a genuine thermodynamic variable with an associated conjugate potential, enlarging the thermodynamic phase space of the system~\cite{Durka:2019ajz, Hennigar:2019ive, Bordo:2019slw, BallonBordo:2019vrn, Frodden:2021ces}. 

Let us focus on the case $k=1$. First, notice that by performing the coordinate transformation $\psi\to\tfrac{\psi}{2n}$, the line element~\eqref{TNmetric} can be written as
\begin{equation} \label{GralMetric}
    ds^{2}=\frac{dr^{2}}{f(r)}+\left(r^{2}-n^{2}\right)\left[\sigma_{1}^{2}+\sigma_{2}^{2}+\frac{4n^{2}f(r)\sigma_{3}^{2}}{\left(r^{2}-n^{2}\right)}\right]\,,
\end{equation}
where $\sigma_{i}$ are the left-invariant Maurer-Cartan one-forms of $SU(2)$, satisfying 
\begin{equation} \label{Maurer-Cartan}
    d\sigma_{i}+\tfrac{1}{2}\epsilon_{ijk}\sigma_{j}\wedge\sigma_{k}=0.
\end{equation}
Using Euler angles with $0\leq\vartheta\leq\pi$ , $0\leq\varphi\leq2\pi$ , and $0\leq\psi\leq\beta_{\psi}$, they can be explicitly represented as
\begin{subequations}\label{LeftInvariant}
    \begin{align}  
    \sigma_{1}	&=\cos\psi\,d\vartheta+\sin\vartheta\,\sin\psi\,d\varphi\,,\\
\sigma_{2}	&=-\sin\psi\,d\vartheta+\sin\vartheta\,\cos\psi\,d\varphi\,,\\
\sigma_{3}	&=d\psi+\cos\vartheta\,d\varphi\,.
\end{align}
\end{subequations}

In the presence of a negative cosmological constant, say $\Lambda=-\tfrac{3}{\ell^{2}}$, we find that if the conditions
\begin{equation} \label{ConditionsSelf}
    p=\pm q \qquad \text{and } \qquad mG= 1-\frac{4n^2}{\ell^2}
\end{equation}
are satisfied, the metric~\eqref{GralMetric} possesses an (anti-)self-dual Weyl tensor \footnote{Notice that the (anti-)self-duality condition is imposed on the Weyl tensor rather than on the Riemann tensor. The reason is that, for Einstein spaces with nonvanishing cosmological constant, the Riemann tensor contains a background contribution proportional to $\Lambda$. In the literature, this observation is often encoded through the definition of the displaced Riemann curvature, $F_{\lambda\rho}^{\mu\nu}=R_{\lambda\rho}^{\mu\nu}+\frac{1}{\ell^{2}}\delta_{\lambda\rho}^{\mu\nu}\,.$ For Einstein spaces satisfying $ R_{\nu}^{\mu}=-\frac{3}{\ell^{2}}\delta_{\nu}^{\mu}\,$, the displaced curvature coincides with the Weyl tensor, $ W_{\lambda\rho}^{\mu\nu}|_{{\rm E}}=F_{\lambda\rho}^{\mu\nu}$. Therefore, one can impose the (anti-)self-duality condition on the displaced curvature, $F_{\lambda\rho}^{\mu\nu}=\pm\tilde{F}_{\lambda\rho}^{\mu\nu}\,,$ which is equivalent to requiring the Weyl tensor to be (anti-)self-dual. It is worth mentioning that in the Ricci-flat limit $\Lambda=0$, the notion of (anti-)self duality is equivalent for the Weyl and Riemann tensor.}
\begin{equation} \label{ASDWeyl}
    W^{\mu\nu}_{\lambda\rho}=\pm\tilde{W}^{\mu\nu}_{\lambda\rho}\,,
\end{equation}
and the Killing vector field $\partial_{\psi}$ acquires a zero-dimensional set of fixed points at $r=n$, corresponding to a nut. 
Then, the solution becomes
\begin{subequations}\label{Fnut}
    \begin{align} 
    f_{{\rm nut}}(r)	&=\frac{r-n}{r+n}+\frac{\left(r-n\right)^{2}\left(r+3n\right)}{\ell^{2}(r+n)}\,,  \\
\alpha_{{\rm nut}}(r)	&=\frac{q}{2n}\left(\frac{r-n}{r+n}\right)^{e^{-\gamma}}\,. 
\end{align} 
\end{subequations}
This geometry corresponds to the (anti-)self-dual Taub-NUT-AdS solution coupled to a ModMax field. Notice that the electromagnetic field does not backreact on the metric, despite the fact that the field strength is nontrivial. Indeed, the stress energy tensor vanishes identically, 
\begin{equation} \label{STvanishes}
    T_{\mu\nu}=0\,.
\end{equation}
This remarkable property follows from the (anti-)self duality condition Eq.~\eqref{anti-self-dualcondition}. \footnote{In Maxwell electrodynamics, the stress-energy tensor is given by $ T_{\mu\nu}=F_{\mu\lambda}F_{\nu}^{\,\lambda}-\frac{1}{4}g_{\mu\nu}F_{\lambda\rho}F^{\lambda\rho}\,,$
which can be written as $T_{\mu\nu}=\frac{1}{2}g^{\lambda\rho}\left(F_{\mu\lambda}+\tilde{F}_{\mu\lambda}\right)\left(F_{\nu\rho}-\tilde{F}_{\nu\rho}\right).$
From this expression, it follows immediately that $T_{\mu\nu}=0$ whenever the field strength is (anti-)self-dual, i.e. $F_{\mu\nu}=\pm\tilde{F}_{\mu\nu}$.}

In nonlinear electrodynamics, the appropriate (anti-) self-duality condition is described by, 
\begin{equation} \label{ASDnonlinear}
    F_{\mu\nu}=\pm\tilde{P}_{\mu\nu}\,.
\end{equation}

Consequently, the electromagnetic field given by the ModMax theory (cf. Eq.~\eqref{ModMaxLagden}) contributes non-trivially to the topology of the configuration while leaving the spacetime geometry unchanged.

For this solution, the coordinate ranges are \begin{equation} \label{rangeofcoord}
    n\leq r<\infty \,,\qquad0\leq\vartheta\leq\pi\,,\qquad0\leq\varphi\leq2\pi\,.
\end{equation}
The absence of conical singularities further requires the following periodic identification 
\begin{equation} \label{ConicalSing}
    0\leq\psi\leq4\pi\,.
\end{equation}
As a result, the geometry is locally equivalent to $\mathbb{R}^{4}$ in the neighbourhood of the zero-dimensional fixed-point set located at $r=n$. 
In the limit $\gamma\to0$, one recovers the charged Euclidean (anti-)self-dual Taub-NUT solution of Ref.~\cite{Brill:1964tn}. Furthermore, the ModMax field approaches a Maxwell configuration asymptotically, 
\begin{equation} \label{AsymptoticAlpha}
    \alpha(r)\to\frac{q}{2n}\,,\qquad\text{as}\qquad r\to\infty 
\end{equation}
Therefore, although the non-linearly charged ModMax Taub-NUT solution is locally inequivalent to its Einstein-Maxwell counterpart, both solutions share the same asymptotic behavior.

Another gravitational instanton belonging to the same class discussed above is the Taub-Bolt solution~\cite{Page:1978hdy}. It is obtained when the Killing vector field $\partial_{\psi}$ possesses a two-dimensional set of fixed points located at $r=r_{b}$, with $r_{b}>n$, where $r_{b}$ is defined as the largest root of the equation for $f(r_{b})=0$. Such fixed point sets are known as bolts, in contrast to the zero-dimensional fixed point sets known as nuts. The Taub-Bolt solution in ModMax electrodynamics has been studied previously, although its contribution to the axial anomaly has not yet been explored. Here, however, we shall focus on the globally (anti-)self-dual Taub-NUT solution presented in Eq.~\eqref{Fnut}, since it provides the simplest setting for our analysis. Nevertheless, outside of the Taub-NUT family there exists another gravitational instanton with a bolt structure that is of particular interest to us, and we devote the following section to its study.

\subsection{Nonlinearly Charged Eguchi-Hanson Instanton} \label{SecEH}

Having reviewed a previously known nonlinearly charged gravitational instanton, we are now in a position to introduce a new one. Our goal in this section is to construct a ModMax generalization of the Eguchi-Hanson solution coupled to Maxwell electrodynamics found in Ref.~\cite{Eguchi:1978gw}. The latter is a purely Euclidean configuration and therefore admits no Lorentzian counterpart. Geometrically, it can be understood as a $U(1)$ fibration over a $2$-dimensional Kähler manifold and possesses the topology of the cotangent bundle $T^{*}S^{2}$, which is diffeomorphic to the tangent bundle $TS^{2}$, and consequently its asymptotic boundary is a lens space \footnote{The lens space is a $3$-dimensional manifold obtained by taking the $3$-sphere $S^{3}$ and identifying points under the action of a discrete cyclic symmetry group $\mathcal{\mathbb{Z}}_{2}$. In other words, the result quotient space is $\mathbb{RP}^{3}=S^{3}/\mathbb{Z}_{2}$, which is known as the three-dimensional real projective space. }.

Motivated by these geometrical properties, we consider the following Eguchi-Hanson type line element  
\begin{equation} \label{EHmetric}
    ds^{2}=\frac{r^{2}f(r)}{4}\left(d\psi+\mathcal{B}_{(k)}\right)^{2}+\frac{dr^{2}}{f(r)}+\frac{r^{2}}{4}d\Sigma_{(k)}^{2}\,,
\end{equation}
where $\mathcal{B}_{(k)}$ and $d\Sigma_{(k)}^{2}$ are defined in Table~\ref{TableKhaler}. As in the Taub-NUT case, it is natural to assume that the gauge field is aligned along the fiber direction, 
\begin{equation} \label{AnsatzPotential}
    A\equiv A_{\mu}dx^{\mu}=\alpha(r)\left(d\psi+\mathcal{B}_{(k)}\right)\,.
\end{equation}
Substituting these ansätze into the Einstein-ModMax field equations~\eqref{EinsteinEOM} and ~\eqref{MaxwelEOM} yields the following linearly independent ordinary differential equations 
\begin{align} 
    0	&=\alpha''+\frac{\alpha'}{r}-\frac{4\alpha e^{-2\gamma}}{r^{2}}\,, \label{ODESystem1}\\
0	&=\kappa\left[rf'+4\left(f-k\right)+\Lambda r^{2}\right]-\alpha'{}^{2}e^{\gamma}+\frac{4\alpha^{2}e^{-\gamma}}{r^{2}}\,, \label{ODESystem2}
\end{align}
respectively. 

This system of equations can be solved using the same strategy employed in the Taub-NUT case. One first integrates the ModMax equation~\eqref{ODESystem1} and then substitutes the result into the gravitational equation~\eqref{ODESystem2}. This procedure yields 
\begin{subequations} \label{EHfunctions}
    \begin{align}
    f(r)	&=k-\frac{a^{4}}{r^{4}}-\frac{8pqe^{-\gamma}}{\kappa r^{2}}-\frac{\Lambda r^{2}}{6}\,, \\
\alpha(r)	&=q\,r^{-2e^{-\gamma}}+p\,r^{2e^{-\gamma}}\,,
\end{align}
\end{subequations}
where $a$, $p$, and $q$ are integration constants. This geometry possesses a bolt located at the largest positive root $r_{b}$ satisfying $f(r_{b})=0$. 

Similar to the Taub-NUT case, for $k=1$ the metric~\eqref{EHmetric} can be written in terms of the left-invariant Maurer-Cartan one-forms of $SU(2)$ [cf. Eq.~\eqref{LeftInvariant}]
\begin{equation} \label{metricMC}
    ds^{2}=\frac{r^{2}}{f(r)}dr^2+\frac{r^{2}}{4}\left(\sigma_{1}^{2}+\sigma_{2}^{2}+f(r)\sigma_{3}^{2}\right)\,.
\end{equation}
Notice that the hypersurfaces of constant radial coordinate are conformally equivalent to the Berger spheres~\cite{Berger}. 

Although this solution possesses a bolt, the Weyl tensor for this space is globally (anti-)self dual only if $\Lambda=0$, independently of the values of $p$ and $q$. This contrasts with the Taub-Bolt solution coupled to ModMax electrodynamics, which remains asymptotically (anti-)self dual even in presence of a nonvanishing cosmological constant~\cite{Flores-Alfonso:2020nnd}. Nevertheless, the electromagnetic field satisfies the nonlinear (anti-)self dual condition of Eq.~\eqref{ASDnonlinear} if and only if $p=0$. Therefore, from this point onward we shall restrict our attention to the case $p=\Lambda=0$ in~\eqref{EHfunctions}, for which the bolt radius reduces to $r_{b}=a$. 

In this case, the coordinate ranges are 
\begin{equation} \label{EHrangecoordinates}
    a\leq r<\infty\,,\qquad0\leq\vartheta\leq\pi\,,\qquad0\leq\varphi\leq2\pi.
\end{equation}
On the other hand, the absence of conical singularities at the bolt implies that $0\leq\psi\leq2\pi$. Therefore, the degenerate hypersurface $r=a$ is topologically $\mathbb{S}^{3}/\mathbb{Z}^{2}=\mathbb{RP}^{3}$, while the asymptotic behavior is $\mathbb{R}^{4}/\mathbb{Z}^{2}$ being asymptotically locally Euclidean.

Before closing this section, let us emphasize that the Eguchi-Hanson space has found numerous applications in mathematical physics. For instance, it played an important role in the resolution of orbifold singularities of Calabi-Yau manifolds in string theory~\cite{Polchinski:1998rr}. It has also served as a seed geometry to study gravitational solitons with nontrivial topology on spatial hypersurfaces~\cite{Clarkson:2005qx,Clarkson:2006zk,Chng:2006gh,Wong:2011aa,Hendi:2012zg, Durgut_2023}. More recently, higher-curvature corrections to the Eguchi-Hanson metric have been investigated in Refs.~\cite{Corral:2021xsu,Corral:2022udb}, while the backreaction of conformally coupled scalar fields was analyzed in Ref.~\cite{Barrientos:2022yoz}. These developments suggest that the nonlinearly charged instanton presented here may admit a variety of future applications. 

In the next section, we study how the nonlinearly charged gravitational instantons presented above contribute to the axial anomaly of Dirac spinors. In particular, we compute the Dirac index explicitly, carefully taking into account the boundary contributions prescribed by the Atiyah-Patodi-Singer index theorem. 

\section{Index Theorem and Axial Anomaly} \label{Sec.IndexTheorem}

The Atiyah-Singer index theorem (See~\cite{Atiyah:1968mp, Atiyah:1967ih, Atiyah:1970ws, Atiyah:1971rm} and~\cite{freed2021atiyahsingerindextheorem} for a review) provides a powerful bridge between analysis, geometry, and topology. Broadly speaking, it establishes a relationship between the analytical properties of elliptic differential operators acting on sections of vector bundles and the topological invariants of the underlying manifold and bundles. In particular, it allows one to compute the index of an elliptic operator 
\begin{equation} \label{IndexDef}
    \text{index}\,D=\text{dim ker}\,D-\text{dim coker}\,D\,,
\end{equation}
in terms of characteristic classes and global topological data. Here, $D$ denotes the elliptic operator under consideration, while ${\rm coker}\,D$ is its cokernel.\footnote{Let $V$ and $W$ be vector spaces and $D:V\to W$ a linear operator. The kernel is defined as ${\rm ker }\,D=\{v\in V|Dv=0\}\,,$ and consists of all vectors annihilated by $D$. The image of $D$ is ${\rm Im}\,D=\{w\in W|w=Dv\,,v\in V\}\,,$ which contains all vectors that can be obtained from the action of $D$ on elements of $V$. Therefore, the cokernel is defined as the quotient space ${\rm coker}\,D=W/{\rm Im}\,D\,,$ that is, the codomain modulo the image. In other words, it measures the part of $W$ that is not reached by the action of $D$.} Consequently, quantities associated with the spectrum of a differential operator can be determined without explicitly solving the corresponding differential equations. 

The main idea behind index theory already appears in several classical results of differential geometry. A notable example is the Gauss-Bonnet theorem, which relates the Euler characteristic of a compact manifold to the integral of the Euler form over the manifold. Through cohomology arguments, the Euler characteristic can also be expressed as an alternating sum of Betti numbers, which count linearly independent harmonic forms~\cite{Warner:1983}. Thus, the theorem establishes a connection between the topology of the manifold and the kernel of the Laplace operator acting on differential forms. 

There exist several important index theorems arising from different elliptic complexes. For the  Rham complex, the corresponding result is the Gauss-Bonnet theorem. For the signature complex, one obtains the Hirzebruch signature theorem. The Dolbeault complex leads to the Hirzebruch-Riemann-Roch theorem~\cite{Hirzebruch:1995}, and finally the spin complex gives rise to the Atiyah-Singer index theorem for the Dirac operator, whose index is expressed in terms of the $\widehat{A}$-genus and characteristic classes of the gauge bundle~\cite{Atiyah:1968mp,Atiyah:1967ih}. 

In the applications relevant to this work, the elliptic operator of interest is the Dirac operator acting on spinor fields. To clarify ideas, it is well known that the kernel is defined as 
\begin{equation} \label{Kernel}
    \text{ker}\,D=\{\psi,D\psi=0\}\,.
\end{equation}
This gives the fermionic zero modes of the theory. In even spacetime dimensions, the spinor bundle decomposes into positive and negative chirality sectors. The massless Dirac operator anticommutes with the chirality operator and therefore maps spinors of one chirality to the opposite chirality. We denote these components by $D_{+}$ and $D_{-}$, where $D_{+}$ acts on positive chirality spinors and $D_{-}$ on negative chirality spinors. With the appropriate boundary or normalizability conditions, the index is
\begin{equation} \label{IndexDdiff}
    \text{index}\,D_{+}=N_{+}-N_{-}\,,
\end{equation}
where $N_{\pm}=\text{dim ker}\,D_{\pm}$ counts the positive and negative chirality fermionic zero modes, respectively. Therefore, the index~\eqref{IndexDdiff} measures the chiral imbalance induced by the background geometry and gauge fields. As we will see, this quantity plays a fundamental role in the analysis of the axial anomaly, since the nonconservation of the axial current is related to the difference in the number of positive and negative chirality fermionic zero modes. 

Most of the classical formulations of the index theory apply to compact manifolds without boundaries. Nevertheless, the relevant situations in physics involve manifolds with nontrivial boundaries or noncompact geometries. Examples include gravitational instantons, asymptotically locally Euclidean spaces, black holes, and manifolds arising in gauge theory. In a series of seminal works, Atiyah, Patodi,  and Singer (APS) extended the previous result of Atiyah and Singer to manifolds with boundary~\cite{APS-eta}. Their formulation revealed that, in addition to the bulk topological contribution, the index receives boundary corrections. These corrections are determined by the spectral asymmetry of an induced boundary operator, encoded in the celebrated $\eta$-invariant, together with boundary contributions that may be expressed through Chern-Simons forms. The APS theorem therefore establishes a precise relation between geometrical and topological data in the presence of boundaries, and plays a fundamental role in the study of anomalies and instantons. 

Since the gravitational instantons considered in this work possess nontrivial asymptotic boundaries, the Atiyah-Patodi-Singer index theorem provides a natural framework for computing the index of the Dirac operator. In this section, we apply this formalism to evaluate the chiral asymmetry induced by gravitational instantons in ModMax electrodynamics and study how these backgrounds contribute to the axial anomaly.

In four dimensions, the APS index theorem for the Dirac operator coupled to a $U(1)$ gauge field, 
\begin{equation} \label{DiracOperator}
    \slashed{D}=\gamma^{c}E_{c}^{\mu}\left(\partial_{\mu}+\frac{1}{4}\omega^{ab}{}_{\mu}\gamma_{ab}+iA_{\mu}\right)\,,
\end{equation}
takes the form 
\begin{align} \label{IndexBdy}
    N_{+}-N_{-}	&=-\frac{1}{24}\left[\frac{1}{32\pi^{2}}\int_{\mathcal{M}}d{}^{4}x\,\sqrt{|g|}\,\varepsilon_{\mu\nu\lambda\rho}R_{\sigma\tau}^{\lambda\rho}R^{\sigma\tau\mu\nu}-\frac{1}{4\pi^{2}}\int_{\partial\mathcal{M}}d{}^{3}x\sqrt{|h|}\;n^{\mu}\varepsilon_{\mu\nu\lambda\rho}K^{\sigma\nu}\nabla^{\lambda}K_{\sigma}^{\rho}\right] \nonumber \\
	&+\frac{1}{32\pi^{2}}\int_{\mathcal{M}}d{}^{4}x\sqrt{|g|}\,\varepsilon_{\mu\nu\lambda\rho}F^{\mu\nu}F^{\lambda\rho}-\frac{1}{2}\left[\eta_{D}\left(\partial\mathcal{M}\right)+h_{D}\left(\partial\mathcal{M}\right)\right]\,.
\end{align}
Here, $\omega^{ab}=\omega^{ab}{}_{\mu}\,dx^{\mu}$ denotes the spin connection $1$-form, $\gamma_{ab}\equiv\gamma_{[a}\gamma_{b]}$ are the generators of the Lorentz group in the spinorial representation, $h_{\mu\nu}=g_{\mu\nu}-n_{\mu}n_{\nu}$ is the induced metric on the boundary $\partial\mathcal{M}$, whose space-like unit normal vector is denoted by $n^{\mu}$. The extrinsic curvature is defined as 
\begin{equation} \label{ExtrinsicK}
    K_{\mu\nu}=h_{\mu}^{\lambda}h_{\nu}^{\rho}\nabla_{\lambda}n_{\rho}\,.
\end{equation}
The first line of~\eqref{IndexBdy} corresponds to the gravitational contribution to the index, consisting of the bulk Pontryagin density together with its associated Chern-Simons boundary term. The second line contains the gauge contribution together with the APS boundary correction. 

The quantity $\eta_{D}(\partial\mathcal{M})$ is the Atiyah-Patodi-Singer $\eta-$invariant, which measures the spectral asymmetry of the Dirac operator induced on the boundary $\partial\mathcal{M}$. More precisely, it provides a measure of the imbalance between positive and negative eigenvalues of the boundary Dirac operator. The term $h_{D}(\partial\mathcal{M})$ denotes the dimension of the kernel of the boundary Dirac operator and therefore counts its zero modes. The $\eta-$invariant has recently been interpreted as the axial charge associated with physical states~\cite{Kobayashi:2021jbn}. It is obtained by performing the analytic continuation of the meromorphic function $\eta_{D}(s)$ to $s=0$, namely 
\begin{equation} \label{Etaszero}
    \eta_{D}(\partial\mathcal{M})=\eta_{D}(s)|_{s=0}\,,
\end{equation}
where
\begin{equation} \label{etadef}
    \eta_{D}(s)=\sum_{\lambda\neq0}\text{sign}(\lambda)|\lambda|^{-s}\,,
\end{equation}
with $\lambda$  denoting the eigenvalues of the Dirac operator $\slashed{D}$ evaluated on the tangential components of $\partial\mathcal{M}$. An equivalent integral representation of Eq.~\eqref{etadef} is 
\begin{equation} \label{etarepresentation}
    \eta_{D}(s)=\frac{1}{\Gamma\left(\frac{s+1}{2}\right)}\int_{0}^{\infty}dx\;x^{\frac{s-1}{2}}\,\text{tr}\left(\slashed{D}\,e^{-x\slashed{D}^{2}}\right)\,.
\end{equation}
Whenever $\text{Index}(\slashed{D})=0$, the integrated anomaly vanishes. This is the case, for instance, for Dirac fermions propagating on the (anti-)self-dual Taub-NUT background of general relativity~\cite{Eguchi:1977iu}. However, as first shown by Pope~\cite{Pope:1978zx,Pope:1981jx}, the situation changes in the presence of (anti-)self-dual Maxwell fields. In that case, the index becomes nonvanishing, implying an imbalance between positive and negative chiral fermionic zero modes. Consequently, the axial symmetry is broken at the quantum level for charged Dirac spinors on the (anti-)self-dual Taub-NUT space. As we shall see, a similar phenomenon occurs in the presence of nonlinear electrodynamics. 

Alternatively, the Atiyah-Patodi-Singer $\eta_{D}$-invariant can be computed geometrically through the Hitchin formula~\cite{HITCHIN19741}. To this end, one writes the induced metric on a hypersurface of constant radial coordinate $r=r_{0}$ in the conformal form of a Berger sphere, represented by 
\begin{equation} \label{BergerSphere}
    ds^{2}=\Omega(r_{0})\left[\sigma_{1}^{2}+\sigma_{2}^{2}+\mu^{2}(r_{0})\sigma_{3}^{2}\right]\,,
\end{equation}
where $\Omega(r_{0})$ and $\mu(r_{0})$ are positive functions of $r_{0}$.

It is worth noting that $\eta_{D}(s)$ is invariant under a constant rescaling of the eigenvalues $\lambda$ , as follows directly from its definition in Eq.~\eqref{etadef}. Since the Dirac operator scales uniformly under conformal transformations of the boundary metric~\eqref{BergerSphere}, one may set $\Omega(r_{0})=1$ without loss of generality~\cite{Bakas:2011nq}. In vacuum, Hitchin's formula establishes that
\begin{equation} \label{Hitchin}
    \eta_{D}\left(\partial\mathcal{M}\right)+h_{D}\left(\partial\mathcal{M}\right)=\frac{1}{6}\left(1-\mu_{0}^{2}\right)^{2}\,,
\end{equation}
where $\mu_{0}=\mu(r_{0})$. This relation provides a particularly convenient geometrical prescription for computing the contribution of the boundary to the Dirac index and, consequently, to the axial anomaly.

To compute the $\eta_{D}-$invariant in presence of ModMax fields, we consider the induced boundary metric~\eqref{BergerSphere} on a hypersurface of constant radius. Since the $\eta-$invariant is insensitive to an overall constant rescaling of the spectrum, one may, without loss of generality, choose the dreibein $1$-form $e^{a}=e_{\mu}^{a}dx^{\mu}$ as
\begin{equation} \label{Vielbein}
    e^{1}=\sigma_{1}\,,\qquad e^{2}=\sigma_{2}\,,\qquad e^{3}=\mu_{0}\,\sigma_{3}\,.
\end{equation}
The corresponding dual vector basis $E_{a}=E_{a}^{\mu}\partial_{\mu}$ is determined by the orthogonality condition $e^a (E_b)=\delta^a_b$. This yields 
\begin{equation}
    E_{1}=\Sigma_{1}\,,\qquad E_{2}=\Sigma_{2}\,,\qquad E_{3}=\mu_{0}^{-1}\Sigma_{3}\,,
\end{equation}
where $\Sigma_{i}=\Sigma_{i}^{\mu}\partial_{\mu}$ , with $i=1,2,3$, denote the vector fields dual to the left-invariant Maurer-Cartan one-forms of $SU(2)$ introduced in~\eqref{LeftInvariant}. Explicitly, 
\begin{subequations}\label{LeftInvariants}
    \begin{align} 
    \Sigma_{1}	&=- \cot\vartheta\sin\psi\,\partial_{\psi}+\cos\psi\,\partial_{\vartheta}+\frac{\sin\psi}{\sin\vartheta}\,\partial_{\varphi}\,, \\
\Sigma_{2}	&=-\cot\vartheta\cos\psi\,\partial_{\psi}-\sin\psi\,\partial_{\vartheta}+\frac{\cos\psi}{\sin\vartheta}\,\partial_{\varphi}\,, \\
\Sigma_{3}	&=\partial_{\psi}\,.
\end{align}
\end{subequations}
The nonvanishing components of the spin connection are obtained from the torsion-free condition,
\begin{equation} \label{MaurerCartanEq}
    de^{a}+\omega^{a}{}_{b}\wedge e^{b}=0,
\end{equation}
which gives 
\begin{equation} \label{SpinConnections}
    \omega^{12}=\left(\frac{\mu_{0}^{2}-2}{2\mu_{0}}\right)e^{3}\,,\qquad\omega^{13}=\frac{\mu_{0}\,e^{2}}{2}\,,\qquad\omega^{23}=-\frac{\mu_{0}\,e^{1}}{2}\,.
\end{equation}
For the Dirac matrices, we choose the standard Pauli representation adapted to the three-dimensional boundary hypersurface
\begin{equation}
    \gamma^{1}=\left(\begin{array}{cc}
0 & 1\\
1 & 0
\end{array}\right)\,,\qquad\gamma^{2}=\left(\begin{array}{cc}
0 & -i\\
i & 0
\end{array}\right)\,,\qquad\gamma^{3}=\left(\begin{array}{cc}
1 & 0\\
0 & -1
\end{array}\right)\,.
\end{equation}
Finally, we denote the $U(1)$ gauge connection evaluated on the hypersurface $r=r_{0}$ by \footnote{For the Taub-NUT geometry, the coordinate redefinition $\psi\to\psi/(2n)$ implies that the gauge function appearing in the boundary connection is $b(r)=2n\alpha(r)$. For the Eguchi-Hanson case, no such rescaling is needed, and therefore, the function is simply $b(r)=\alpha(r)$.}
\begin{equation} \label{AnsPotentialBerger}
    A=b(r_{0})\sigma_{3}\equiv b_{0}\sigma_{3}\,.
\end{equation}
Then, defining the self-adjoint operator $K_{i}=i\Sigma_{i}$ and the ladder operator $K_{\pm}=K_{1}\pm iK_{2}$, one can show that they satisfy the angular momentum algebra, 
\begin{equation} \label{momentumalgebra}
    \left[K_{i},K_{j}\right]=i\epsilon_{ijk}K^{k}\,,\qquad\left[K_{3},K_{\pm}\right]=\pm K_{\pm},\qquad\left[K_{+},K_{-}\right]=2K_{3}\,.
\end{equation}
Hence, the Dirac operator acting on a two-component spinor can be expressed in matrix form as
\begin{equation} \label{DiracOperatorModify}
    \slashed{D} \psi=\left(\begin{array}{cc}
\mu_{0}^{-1}\left(K_{3}+b_{0}\right) & K_{-}\\
K_{+} & -\mu_{0}^{-1}\left(K_{3}+b_{0}\right)
\end{array}\right)\left(\begin{array}{c}
\psi_{1}\\
\psi_{2}
\end{array}\right)+\frac{(\mu_{0}^{2}+2)}{4\mu_{0}}\left(\begin{array}{c}
\psi_{1}\\
\psi_{2}
\end{array}\right)=0\,.
\end{equation}
To determine the spectrum of the Dirac operator, we follow Ref.~\cite{Pope:1981jx} and introduce the notation
\begin{equation} \label{SphericalHarmonics}
    |s\rangle\equiv\,_{s}Y_{lm}(\vartheta,\varphi)\,,
\end{equation}
where $_{s}Y_{lm}(\vartheta,\varphi)$ denotes the spin-weighted spherical harmonics~\cite{Goldberg:1966uu}. The action of the operators $K_{\pm}$ and $K_{3}$ on these states is given by
\begin{subequations} \label{States}
    \begin{align} 
    K_{\pm}\,|s\rangle	&=\sqrt{(l\mp s)(l\pm s+1)}\;|s\pm1\rangle\,, \\
K_{3}\,|s\rangle	&=s\,|s\rangle\,,
\end{align}
\end{subequations}
for $|s|\leq l$ and $|m|\leq l$. Notice that $K_{3}$ acts diagonally on the basis states $|s\rangle$ , while the ladder operators $K_{\pm}$ map a state with spin weight s into one with spin weight $s\pm1$. Consequently, the Dirac operator preserves the two-dimensional subspace spanned by $\{|s\rangle,|s+1\rangle\}$. This observation motivates the following ansatz for the eingenvectors
\begin{equation} \label{StateDecomposition}
    \psi=\left(\begin{array}{c}
c_{1}\,|s\rangle\\
c_{2}\,|s+1\rangle
\end{array}\right)\,,
\end{equation}
where $c_{1}$ and $c_{2}$ are constants. Substituting the ansatz~\eqref{StateDecomposition} into~\eqref{DiracOperatorModify} reduces the spectral problem to a $2\times2$ matrix eigenvalue problem. Thus, for $-l\leq s\leq l-1$, the eigenvalues $\lambda$  of the Dirac operator $\slashed{D}$ are~\cite{Pope:1981jx},
\begin{equation} \label{Eigenvalues}
    \lambda=\frac{\mu_{0}}{2}\pm\mu_{0}^{-1}\sqrt{(2s+1+2b_{0})^{2}+4\mu_{0}^{2}(l-s)(l+s+1)}\,,
\end{equation}
where, for each value of $s$, these eigenvalues have degeneracy $d=2l+1$. Notice that there are two special cases corresponding to the endpoints of the $SU(2)$ representation. Indeed, the previous construction assumes that both states $|s\rangle$  and $|s+1\rangle$  belong to the same representation. However, (i) for $s=l$, the state $|l+1\rangle$  does not exist, and the eigenvector becomes 
\begin{equation} \label{State1}
    \psi=\left(\begin{array}{c}
c_{1}\,|s\rangle\\
0
\end{array}\right)\,,
\end{equation}
so that only the first component survives. Similarly, (ii) when $s+1=-l$, or equivalently $s=-\left(l+1\right)$, the state $|-\left(l+1\right)\rangle$  does not belong to the representation, and the eigenvector becomes 
\begin{equation} \label{State2}
    \psi=\left(\begin{array}{c}
0\\
c_{2}\,|s+1\rangle
\end{array}\right)\,,
\end{equation}
leaving only the second component. The corresponding eigenvalues are therefore 
\begin{subequations}\label{speciallambda}
\begin{align} 
    \mbox{(i)}\;\;\;\;\; \lambda &= \frac{\mu_0}{2} \pm \mu_0^{-1}\left|2l+1+2b_0\right|\,,\\
    \mbox{(ii)}\;\;\;\;\; \lambda &= \frac{\mu_0}{2} \pm \mu_0^{-1}\left|2l+1-2b_0\right|\,.
\end{align}
\end{subequations}
We are interested in evaluating the $\eta_{D}-$invariant at the asymptotic boundary, namely in the limit $r_{0}\to\infty$. For simplicity, from now on we restrict ourselves to the case of vanishing cosmological constant, or equivalently $\ell\to\infty$ . For the (anti-)self-dual Taub-NUT (TN) and Eguchi-Hanson (EH) spaces, one then finds 
\begin{align} \label{AsymptoticConditions}
    \mbox{TN:}\qquad\lim_{r_{0}\to\infty}\mu(r_{0})	=0\qquad\wedge\qquad\lim_{r_{0}\to\infty}b(r_{0})=q\,, \\
\mbox{EH:}\qquad\lim_{r_{0}\to\infty}\mu(r_{0})	=1\qquad\wedge\qquad\lim_{r_{0}\to\infty}b(r_{0})=0\,,
\end{align}
respectively. Notice, that in the charged Euclidean Taub-NUT case, the ModMax field approaches its Maxwell counterpart at the asymptotic boundary, as discussed in Sec.~\ref{InstantonsSection}. By contrast, for the Eguchi-Hanson instanton, the ModMax field decays sufficiently fast near the asymptotic boundary. Therefore, the $\eta_{D}-$ invariant of the latter coincides with the one computed in Ref.~\cite{Eguchi:1978gw}, namely 
\begin{equation} \label{Etaplush EH}
    \eta_{D}^{{\rm (EH)}}\left(\partial\mathcal{M}\right)+h_{D}^{{\rm (EH)}}\left(\partial\mathcal{M}\right)=\frac{1}{4}\,.
\end{equation}
Although the boundary contribution to the Dirac index in Eq.~\eqref{IndexBdy} vanishes for the charged Eguchi-Hanson solution of Sec.~\ref{SecEH}, the bulk contribution is nontrivial. Explicitly, one finds 
\begin{align}  
    \frac{1}{32\pi^{2}}\int_{\mathcal{M}}d{}^{4}x\,\sqrt{|g|}\,\varepsilon_{\mu\nu\lambda\rho}R_{\sigma\tau}^{\lambda\rho}R^{\sigma\tau\mu\nu}	&=-3\,,  \label{Topological1} \\
    \frac{1}{32\pi^{2}}\int_{\mathcal{M}}d{}^{4}x\sqrt{|g|}\,\varepsilon_{\mu\nu\lambda\rho}F^{\mu\nu}F^{\lambda\rho}	&=-\frac{q^{2}}{a^{4e^{-\gamma}}}\,.  \label{Topological2}
\end{align}
In this case, the topology of the bolt induces a screening factor in electromagnetic topological charge~\eqref{Topological2}, which is controlled by the nonlinear parameter $\gamma$. In the limit $\gamma\to\infty$, this contribution approaches $-q^{2}$. This region of parameter space is particularly interesting because, in the manifestly duality-symmetric formulation of Ref.~\cite{Avetisyan:2021heg}, ModMax and Bialynicki-Birula electrodynamics belong to the same one-parameter family. Moreover, this formulation suggests that Bialynicki-Birula electrodynamics can be understood, in a suitable sense, as the limiting case of ModMax electrodynamics when we take the limit $\gamma\to\infty$.

In summary, for finite $\gamma$, the Dirac index for the charged Eguchi--Hanson solution of Einstein--ModMax theory is
\begin{equation}
    N_{+}-N_{-}=-\frac{q^{2}}{a^{4e^{-\gamma}}}\,.\label{DiffEH}
\end{equation}
Since the index is integer valued, this expression requires $q^{2}/a^{4e^{-\gamma}}\in\mathbb Z_{\geq0}$, imposing a necessary topological quantization condition on the parameters for the chosen gauge bundle and boundary conditions. For admissible configurations with $q\neq0$, the index is nonzero and describes an excess of negative-chirality zero modes, yielding a nontrivial contribution to the integrated axial anomaly.

The computation of the $\eta_{D}$-invariant for the (anti-)self-dual Taub-NUT instanton is more subtle since the eigenvalues~\eqref{Eigenvalues} diverge in the limit $r_{0}\to\infty$. Nevertheless, one can exploit the fact that $\eta_{D}(s)|_{s=0}$ is invariant under the global rescaling of the eigenvalues, $\lambda\to\mu_{0}\lambda$, to extract a finite asymptotic spectrum. More precisely, one first rescales the eigenvalues and only afterwards takes the limit $r_{0}\to\infty$. Following this prescription, one obtains~\cite{Pope:1981jx}
\begin{equation} \label{EigenValues2}
    \lambda=\pm|2s+1+2b_{0}|\,.
\end{equation}
For the generic modes, the spectrum is symmetric under $\lambda\to-\lambda$, and therefore, their contribution to the $\eta_{D}$-invariant cancels pairwise. The only nontrivial contribution comes from the endpoint modes, namely the special cases $s=l$ and $s=-(l+1)$, whose eigenvalues are given by~\eqref{speciallambda}. After performing the rescaling $\lambda\to\mu_{0}\lambda$ and taking the asymptotic limit $r_{0}\to\infty$, one obtains
\begin{align}
    (i)\qquad\lambda	&=2l+1+2b_{0}\,, \label{Eigenvaluesagain1}\\
(ii)\qquad\lambda	&=2l+1-2b_{0}\,. \label{Eigenvaluesagain2}
\end{align}
Both branches have multiplicity $d=2l+1$. For the angular-momentum labels $l=0,\tfrac12,1,\ldots$, it is convenient to introduce $p=2l+1$, which runs over the positive integers, and $\alpha=2b_{0}$. The exceptional eigenvalues then become $p+\alpha$ and $p-\alpha$. To keep track of their signs, we initially take $\alpha>0$ to be noninteger and define $J=\lfloor\alpha\rfloor$. Their contribution to the eta function is
\begin{align}
    \eta_{\mathrm{D}}(s;\alpha)
    &=
    \sum_{p=1}^{\infty}p(p+\alpha)^{-s}
    -
    \sum_{p=1}^{J}p(\alpha-p)^{-s}
    +
    \sum_{p=J+1}^{\infty}p(p-\alpha)^{-s}.
    \label{EtaInvariantTN}
\end{align}
The middle sum contains the negative eigenvalues, whose absolute values enter the regulator. The remaining sums contain positive eigenvalues.

The analytic continuation of these sums is conveniently expressed through the Hurwitz zeta function,
\begin{equation}
    \zeta(s,u) = \sum_{n=0}^{\infty}(n+u)^{-s} \,,\qquad u>0\,,\quad \operatorname{Re}s>1\,.
\end{equation}
Separating the multiplicity according to $p=(p+\alpha)-\alpha$ in the first sum and $p=(p-\alpha)+\alpha$ in the last gives
\begin{align} \label{etasalpha}
    \eta_{\mathrm{D}}(s;\alpha) &= \zeta(s-1,1+\alpha) -\alpha\zeta(s,1+\alpha) + \zeta(s-1,J+1-\alpha) \nonumber\\
    & +\alpha\zeta(s,J+1-\alpha)- \sum_{p=1}^{J}p(\alpha-p)^{-s}\,.
\end{align}
We can now continue to $s=0$ using
\begin{equation}
    \zeta(0,u)=\frac12-u\,, \qquad \zeta(-1,u) = -\frac12\left(u^{2}-u+\frac16\right)\,.
    \label{ValuesZeta}
\end{equation}
Together with $\sum_{p=1}^{J}p=J(J+1)/2$, these identities yield
\begin{equation}
    \eta_{\mathrm{exc}}(0;\alpha) = \alpha^{2}-J(J+1)-\frac16.
\end{equation}
The Riemann zeta function is the special case $\zeta(s)=\zeta(s,1)$, for which $\zeta(0)=-1/2$ and $\zeta(-1)=-1/12$.

For integer $q>0$, approaching $\alpha=2q$ from above fixes $J=2q$ and gives
\begin{equation}
    \lim_{\alpha\to(2q)^{+}} \eta_{\mathrm{D}}(0;\alpha) = 4q^{2}-2q(2q+1)-\frac16 = -\frac16-2q\,. \label{EtaTN}
\end{equation}
Taking the limit from above gives the Taub--NUT result in Eq.~\eqref{EtaTN}. To identify this result with the complete boundary eta invariant, one must also verify that the other eigenvalues do not contribute. Special care is needed for the mode $p=2q$, because its eigenvalue approaches zero when the boundary is taken to infinity.

Then, inserting~\eqref{EtaTN} into the Dirac index formula~\eqref{IndexBdy}, and using the bulk integrals
\begin{align} 
    \frac{1}{32\pi^{2}}\int_{\mathcal{M}}d{}^{4}x\,\sqrt{|g|}\,\varepsilon_{\mu\nu\lambda\rho}R_{\sigma\tau}^{\lambda\rho}R^{\sigma\tau\mu\nu}	&=2\,,\label{TopologicalInvariantsTN1} \\
\frac{1}{32\pi^{2}}\int_{\mathcal{M}}d{}^{4}x\sqrt{|g|}\,\varepsilon_{\mu\nu\lambda\rho}F^{\mu\nu}F^{\lambda\rho}	&=2q^{2}\,, \label{TopologicalInvariantsTN2}
\end{align}
together with the fact that the boundary integral of the Chern-Simons form vanishes for the (anti-)self-dual Taub-NUT instanton, we obtain
\begin{equation} \label{DiffTN}
    N_{+}-N_{-}=2q\left(q+\frac{1}{2}\right)\,.
\end{equation}
Therefore, for the (anti-)self-dual Taub-NUT solution of Einstein-ModMax theory, the axial anomaly completely determined by the electric charge $q$ of the ModMax field. In particular, the result is independent of the nonlinear parameter $\gamma$  and coincides with the Einstein-Maxwell case~\cite{Pope:1981jx}. This can be traced back to the fact that the asymptotic behavior of the ModMax field does not modify the $\eta_{D}-$ invariant relative to the Maxwell case. Moreover, the $U(1)$ Chern-Pontryagin index~\eqref{TopologicalInvariantsTN2} is also independent of $\gamma$, in contrast to the Eguchi-Hanson case. This difference is likely related to the presence of a bolt in the latter geometry. In view of this, it would be interesting to investigate whether a similar dependence on the nonlinear parameter arises for Taub-Bolt instantons, which we leave for future work.

\chapter{Three-Dimensional Einstein Gravity} \label{Chapter4}

General relativity is the standard classical theory of gravitation and has been successfully tested over a wide range of physical scales. Its fundamental dynamical variable is the spacetime metric $g_{\mu\nu}$, whose dynamics and coupling to matter are governed by Einstein's field equations. Despite its remarkable experimental success, four-dimensional Einstein gravity is perturbatively non-renormalizable by power counting, and the construction of a complete quantum theory of gravity remains an open problem. 

This does not prevent General Relativity from providing a predictive quantum description at sufficiently low energies: interpreted as an effective field theory, it admits a systematic expansion in energy and curvature relative to the relevant ultraviolet scales~\cite{Donoghue:1994dn}. A fundamental ultraviolet description must, however, go beyond the ordinary perturbative Einstein–Hilbert framework. 


Three-dimensional gravity provides a particularly useful theoretical laboratory in which some of the conceptual and mathematical aspects of this problem can be examined in a simplified setting. The reduction of dimensionality eliminates the local propagating degrees of freedom of vacuum Einstein gravity while preserving nontrivial global geometries, boundary excitations, and, for a negative cosmological constant, black-hole solutions and a rich asymptotic structure. Consequently, three dimensional gravity makes it possible to study the interplay between geometry, topology, gauge symmetries, and holography in a setting that is considerably simpler than its four-dimensional counterpart~\cite{Carlip:1995jn,Carlip:1995qv,Carlip:1995zj}.

The modern development of three-dimensional gravity emerged from several complementary directions. In 1986, Achúcarro and Townsend formulated three-dimensional AdS supergravity as a Chern–Simons theory, with Einstein gravity appearing in its bosonic sector~\cite{Achucarro:1986uwr}. In the same year, Brown and Henneaux demonstrated that the canonical algebra of asymptotic charges in AdS$_{3}$, subject to appropriate boundary conditions, consists of two copies of the Virasoro algebra~\cite{Brown:1986nw}, with equal classical central charges 
\begin{equation}
    c_{L}=c_{R}=\frac{3\ell}{2G}\,, \label{centralcharge}
\end{equation}
where $\ell$  is the AdS radius and $G$ is the three-dimensional Newton constant. This result distinguishes diffeomorphisms that are pure gauge from those that generate physical asymptotic symmetries. \footnote{ The asymptotic symmetry group is obtained by quotienting the transformations preserving the boundary conditions by the trivial gauge transformations, whose associated surface charges vanish throughout the allowed phase space. A charge vanishing on a particular background does not, by itself, make its generating transformation trivial.} Acting on the AdS$_{3}$ vacuum with nontrivial asymptotic diffeomorphisms generates boundary gravitons: excitations that remain locally AdS$_{3}$ but are physically distinguishable through their boundary data and charges. 

Witten subsequently developed the relation between three-dimensional Einstein gravity and Chern–Simons theory in a systematic manner and emphasized its importance as a model for quantum gravity~\cite{Witten:1988hc}. This formulation makes the topological character of the bulk theory
manifest: the Einstein equations together with the vanishing-torsion
condition are equivalent to flatness conditions for the corresponding
gauge connections. This also explains the absence of local propagating
degrees of freedom, since a flat connection is locally pure gauge.
Nevertheless, nontrivial physical information can remain encoded in
global holonomies and boundary degrees of freedom.

The subject acquired further importance after the discovery of the Bañados–Teitelboim–Zanelli black hole~\cite{Banados:1992wn,Banados:1992gq}. The BTZ solution established that black hole geometries can exist in three-dimensional Einstein gravity despite the absence of local gravitational waves. It is obtained by a discrete identification of an appropriate region of the covering space of AdS$_{3}$. Consequently, BTZ and AdS$_{3}$ have the same local curvature invariants, but differ in their global identifications, causal structure, and conserved charges.  

These developments anticipated several central ideas of holography and acquired a new interpretation after the formulation of the AdS/CFT correspondence by Maldacena~\cite{Maldacena:1997re,Gubser:1998bc,Witten:1998qj}. In particular, the Brown--Henneaux analysis provides a direct relation between the asymptotic gravitational symmetry algebra and a two-dimensional conformal field theory at the AdS boundary. The appearance of a classical central charge shows that diffeomorphisms acting nontrivially on the asymptotic data can carry physical charges, even though the bulk theory has no local propagating gravitational degrees of freedom.  The absence of local gravitational waves does not make three-dimensional Einstein gravity trivial. Solutions with the same local curvature can differ by their topology, global identifications, holonomies, causal structure, and boundary data. Moreover, the presence of a timelike conformal boundary requires boundary conditions and a well-defined variational principle. These features make Einstein--AdS$_3$ gravity a particularly useful setting in which to study the relation between bulk geometry, asymptotic symmetries, and gravitational boundary terms.  This chapter begins with the algebraic structure of the curvature tensor and the local degree-of-freedom count in three-dimensional Einstein gravity. It then examines global AdS$_3$, its conformal boundary, and the Brown--Henneaux asymptotic symmetries. Radial foliations and the Fefferman--Graham expansion are subsequently introduced to describe asymptotically locally AdS geometries. Finally, the Einstein--Hilbert action is supplemented by the boundary terms required to obtain a finite action and a well-defined asymptotic variational principle. The resulting holographic stress tensor, Ward identity, and conserved boundary charges provide the Einstein-gravity reference for the analysis developed in the following chapter.

\section{Local Geometry and Degrees of Freedom }

The local simplicity of three dimensional Einstein gravity follows from the algebraic properties of the curvature tensor. In a $d$-dimensional spacetime, the Riemann tensor possesses
\begin{equation}
    N_{R}(d)=\frac{d^{2}\left(d^{2}-1\right)}{12}\,,
\end{equation}
algebraically independent components. For $d=3$, this number is equal to six, precisely equal to the number of independent components of the symmetric Ricci tensor. Indeed, the Weyl tensor vanishes identically in three dimensions, and the complete Riemann tensor is determined algebraically by the Ricci tensor and the Ricci scalar \footnote{Here, antisymmetrization is normalized separately over each bracketed pair, and $\delta_{\mu\nu}^{\lambda\rho}:=\delta_{\mu}^{\lambda}\delta_{\nu}^{\rho}-\delta_{\nu}^{\lambda}\delta_{\mu}^{\rho}$, is the generalized Kronecker delta. A derivation of Eq.~\eqref{eq:riemann-three-dimensions} from the general curvature decomposition is presented in Appendix~\ref{AlgebraicDecomposition}.} 
\begin{equation}
    R_{\lambda\rho}^{\mu\nu}=4\delta_{[\lambda}^{[\mu}R_{\rho]}^{\nu]}-\frac{R}{2}\delta_{\lambda\rho}^{\mu\nu}\,.     \label{eq:riemann-three-dimensions}
\end{equation}
The Einstein equations in the presence of a cosmological constant and matter are 
\begin{equation} 
    G_{\mu\nu}+\Lambda g_{\mu\nu}=\frac{1}{2\kappa}T_{\mu\nu}\,,\qquad\kappa=\frac{1}{16\pi G}\,, \label{eq:einstein-equations-three-dimensions}
\end{equation} 
where $G_{\mu\nu}=R_{\mu\nu}-\tfrac{1}{2}Rg_{\mu\nu}$ is the Einstein tensor, and $T_{\mu\nu}$ is the stress-energy tensor. Taking the trace of Eq.~\eqref{eq:einstein-equations-three-dimensions} gives 
\begin{equation}
    R=6\Lambda-\frac{1}{\kappa}T\,,\qquad T:=g^{\mu\nu}T_{\mu\nu}\,.
\end{equation}
Substituting this expression back into Eq.~\eqref{eq:einstein-equations-three-dimensions} yields
\begin{equation}
    R_{\mu\nu}=2\Lambda g_{\mu\nu}+\frac{1}{2\kappa}\left(T_{\mu\nu}-Tg_{\mu\nu}\right)\,.
\end{equation}
In vacuum, the stress energy tensor vanishes, i.e. $T_{\mu\nu}=0$, and therefore one finds
\begin{equation}
    R_{\mu\nu}=2\Lambda g_{\mu\nu}\,,\qquad R=6\Lambda\,.
\end{equation}
Using Eq.~\eqref{eq:riemann-three-dimensions}, the Riemann tensor becomes 
\begin{equation}
    R_{\lambda\rho}^{\mu\nu}=\Lambda\delta_{\lambda\rho}^{\mu\nu}\,.
\end{equation}
Thus, every smooth vacuum solution of three-dimensional Einstein gravity has constant curvature. Depending on the sign of the cosmological constant, its local geometry is that of Minkowski spacetime for $\Lambda=0\,,$ de Sitter spacetime for $\Lambda>0$, or Anti-de Sitter spacetime for $\Lambda<0$. 

For a negative cosmological constant, the AdS radius $\ell$  is defined by $\Lambda=-\tfrac{1}{\ell^{2}}$. Considering this, the corresponding curvature tensors satisfy 
\begin{equation}
    R_{\lambda\rho}^{\mu\nu}=-\frac{1}{\ell^{2}}\delta_{\lambda\rho}^{\mu\nu}\,,\qquad R_{\nu}^{\mu}=-\frac{2}{\ell^{2}}\delta_{\nu}^{\mu}\,,\qquad R=-\frac{6}{\ell^{2}}\,.
\end{equation}
Although the vanishing of the Weyl tensor explains an important aspect of this local simplicity, it does not, by itself, establish the absence of propagating gravitational degrees of freedom. This conclusion also relies on the field equations and their constraint structure. In the canonical formulation of Einstein gravity $d\geq3$ spacetime dimensions, the spatial metric has $\ensuremath{d(d-1)/2}$ independent components. The Hamiltonian constraint and $d-1$ momentum constraints are first class; each removes one canonical pair after accounting for the corresponding gauge freedom. The number of local physical configuration space degrees of freedom is therefore~\cite{Henneaux:1992ig,Zanelli:2000te,  Carlip:2005zn}
\begin{equation}
    N_{\mathrm{dof}}=\frac{1}{2}\left[d(d-1)-2d\right]=\frac{d(d-3)}{2}\,.
\end{equation}
Notice that, for $d=3$ this number vanishes. Vacuum Einstein gravity in three dimensions consequently has no local propagating gravitational waves. This is a bulk statement: it does not exclude global degrees of freedom or physical boundary excitations. 



\section{Global and Asymptotic Properties of \texorpdfstring{$\mathrm{AdS}_{3}$}{AdS3}}

The local curvature identities established in the previous section do not determine the global geometry of a spacetime or the physical information carried by its boundary. In three-dimensional Einstein gravity with a negative cosmological constant, this distinction is particularly important; although every smooth vacuum solution is locally $\mathrm{AdS}_{3}$, different global identifications and asymptotic data can describe physically inequivalent configurations.

This section develops the geometric ingredients needed to understand that distinction. We first discuss global $\mathrm{AdS}_{3}$ and its conformal boundary, before turning to the Brown--Henneaux asymptotic symmetries. Throughout this section, we consider $\Lambda=-\ell^{-2}$ and natural units, $\hbar=c=1$. 

\subsection{Conformal Boundary and Causal Structure }

A convenient starting point is the realization of $\mathrm{AdS}_{3}$ as a hyperboloid embedded in the flat space $\mathbb{R}^{2,2}$
\begin{equation} \label{Embbeding}
    -(X^{-1})^{2}-(X^{0})^{2}+(X^{1})^{2}+(X^{2})^{2}=-\ell^{2}\,,
\end{equation}
with ambient metric
\begin{equation}
    ds^{2}=-(dX^{-1})^{2}-(dX^{0})^{2}+(dX^{1})^{2}+(dX^{2})^{2}\,.
\end{equation}
The parametrization
\begin{equation}
    X^{-1}	=\sqrt{\ell^{2}+r^{2}}\cos(t/\ell)\,,\quad X^{0}=\sqrt{\ell^{2}+r^{2}}\sin(t/\ell)\,,\quad X^{1}=r\cos\phi\,,\quad X^{2}=r\sin\phi\,,
\end{equation}
induces the metric
\begin{equation}   \label{eq:global-ads3-metric}
    ds^{2}=-\left(1+\frac{r^{2}}{\ell^{2}}\right)dt^{2}+\frac{dr^{2}}{1+r^{2}/\ell^{2}}+r^{2}d\phi^{2}\,,
\end{equation}
where $r\geq0$ and $\phi\sim\phi+2\pi$. 

On the embedded hyperboloid, the coordinate $t$ is periodic with period $2\pi\ell$. Since curves at fixed $r$ and $\phi$  are timelike, this identification produces closed timelike curves. The spacetime ordinarily used in gravitational and holographic applications is therefore the universal covering space, obtained by unwrapping the time coordinate so that $t\in\mathbb{R}$. In what follows $\mathrm{AdS}_{3}$ denotes this universal cover unless stated otherwise.

The location and causal character of the AdS boundary become explicit through conformal compactification, following the approach introduced by Penrose~\cite{Penrose:1964ge}. Define 
\begin{equation}
    r=\ell\tan\chi,\qquad\tau=\frac{t}{\ell},\qquad0\leq\chi<\frac{\pi}{2}\,.
\end{equation}
Considering this, the Eq.~\eqref{eq:global-ads3-metric} then becomes
\begin{equation}
    ds^{2}=\frac{\ell^{2}}{\cos^{2}\chi}\left(-d\tau^{2}+d\chi^{2}+\sin^{2}\chi\,d\phi^{2}\right)\,.
\end{equation}
Although the physical metric diverges as $\chi\to\tfrac{\pi}{2}$, the conformally related metric
\begin{equation}
    d\hat{s}^{2}:=\frac{\cos^{2}\chi}{\ell^{2}}\,ds^{2}=-d\tau^{2}+d\chi^{2}+\sin^{2}\chi\,d\phi^{2}\,
\end{equation}
extends regularly to $\chi=\tfrac{\pi}{2}$. The corresponding defining function $\Omega=\tfrac{\cos\chi}{\ell}$ vanishes at the boundary and satisfies $d\Omega\neq0$ there. A representative of the boundary conformal metric is therefore
\begin{equation}
    ds_{(0)}^{2} = g_{(0)ij}\,dx^{i}dx^{j} = -d\tau^{2}+d\phi^{2}\,. \label{eq:boundary-cylinder}
\end{equation}
Here $x^{i}=(\tau,\phi)$ are boundary coordinates, with $\tau\in\mathbb{R}$ and $\phi\sim\phi+2\pi$. Latin indices $i,j$ label the two directions tangent to the boundary. Thus, the conformal boundary is a timelike cylinder, $\partial\mathcal{M}\simeq\mathbb{R}\times S^{1}$. Its geometry is defined only up to a Weyl rescaling: replacing the defining function by $\Omega'=e^{\omega(x)}\Omega$, where $\omega(x)$ is a smooth function changes the boundary representative according to
\begin{equation}
    g_{(0)ij}\longrightarrow e^{2\omega(x)}g_{(0)ij}\,.
\end{equation}
Thus, the bulk geometry determines a conformal class $[g_{(0)}]$, rather than a unique boundary metric. The timelike character of the boundary has direct causal consequences. Radial null curves satisfy
\begin{equation}
    d\tau=\pm d\chi\,.
\end{equation}
A radial light ray emitted from the center reaches the conformal boundary after a finite global coordinate time,
\begin{equation}
    \Delta t=\frac{\pi\ell}{2}\,,
\end{equation}
even though the boundary lies at infinite radial proper distance in the physical metric. Consequently, $\mathrm{AdS}_{3}$, regarded as a spacetime without its conformal boundary, is not globally hyperbolic. Initial data in the bulk must be supplemented by appropriate boundary conditions to formulate a well-posed evolution problem.

\begin{figure}[ht]
  \centering
    \includegraphics[width=0.5\textwidth]{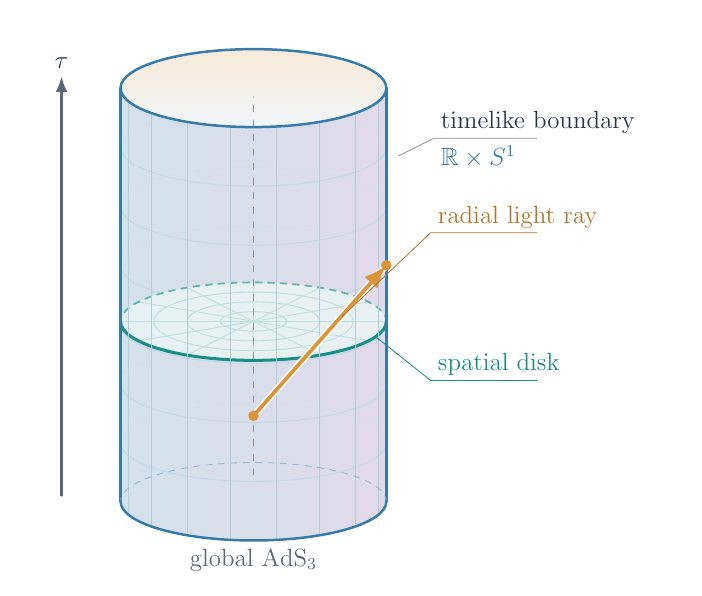}
  \caption{Conformal representation of the universal covering space of
  global $\mathrm{AdS}_3$. The lateral cylinder is the timelike conformal
  boundary $\mathbb R\times S^1$. A constant time spatial disk is in green, and a radial light ray reaches the boundary in finite global
  coordinate time. The top and bottom are limits of the displayed time
  interval, not additional boundaries of the spacetime.}
  \label{fig:ads3-refined}
\end{figure}

The global and causal structure described above is summarized in Fig.~\ref{fig:ads3-refined}. The conformal boundary is timelike and has the topology $\mathbb{R}\times S^{1}$, while radial null curves reach it in a finite interval of global coordinate time.

The timelike conformal boundary can also be approached through a family of timelike radial hypersurfaces. This description will be developed systematically in Section~\ref{sec:radial-foliations-fg}, where we introduce the induced metric, extrinsic curvature, and Fefferman--Graham coordinates. For the moment, we turn to the distinction between the exact isometries of $\mathrm{AdS}_{3}$ and the transformations preserving its asymptotic boundary conditions.

\subsection{Exact Isometries, Brown–Henneaux Boundary Conditions, and Asymptotic Symmetries}

The embedding space description in Eq.~\eqref{Embbeding} makes the exact isometries of $\mathrm{AdS}_{3}$ explicit. The six generators of $SO(2,2)$ given by
\begin{equation}
    J_{AB}=X_{A}\frac{\partial}{\partial X^{B}}-X_{B}\frac{\partial}{\partial X^{A}}\,,
\end{equation}
where ambient indices are lowered with the $\ensuremath{\mathbb{R}^{2,2}}$ metric, restrict to Killing vector fields on the hyperboloid. Their associated algebra is given by 
\begin{equation}
    \mathfrak{so}(2,2)\simeq\mathfrak{sl}(2,\mathbb{R})_{L}\oplus\mathfrak{sl}(2,\mathbb{R})_{R}\,.
\end{equation}
Passing to the universal cover changes the global identifications of the isometry group, but not this Lie algebra.

An exact Killing vector preserves a particular metric, meaning that
\begin{equation}
    \mathcal{L}_{\xi}g_{\mu\nu}=0\,.
\end{equation}
An asymptotic symmetry is less restrictive: it preserves a specified class of boundary conditions and it can map one allowed solution into another. Its definition therefore depends on the gravitational phase space under consideration. Indeed, Brown and Henneaux demonstrated that a suitable asymptotically $\mathrm{AdS}_{3}$ phase space has an infinite-dimensional symmetry algebra~\cite{Brown:1986nw}. In coordinates $(t,r,\phi)$, their boundary conditions can be written as
\begin{align}
    g_{tt}	&=-\frac{r^{2}}{\ell^{2}}+\mathcal{O}(1)\,,\qquad g_{t\phi}=\mathcal{O}(1)\,,\qquad g_{tr}=\mathcal{O}(r^{-3})\,, \\
g_{\phi\phi}	&=r^{2}+\mathcal{O}(1)\,,\qquad g_{rr}=\frac{\ell^{2}}{r^{2}}+\mathcal{O}(r^{-4})\,,\qquad g_{r\phi}=\mathcal{O}(r^{-3})\,.
\end{align}
These conditions include global $\mathrm{AdS}_{3}$ and locally AdS quotients. At this point, it is useful to introduce the dimensionless boundary coordinates $x^{\pm}=\tfrac{t}{\ell}\pm\phi\,$. Thus, the vector fields preserving these fall-offs have the asymptotic form
\begin{align}
    \xi^{+}	&=\epsilon^{+}(x^{+})+\frac{\ell^{2}}{2r^{2}}\partial_{-}^{2}\epsilon^{-}(x^{-})+\mathcal{O}(r^{-4})\,, \\
\xi^{-}	&=\epsilon^{-}(x^{-})+\frac{\ell^{2}}{2r^{2}}\partial_{+}^{2}\epsilon^{+}(x^{+})+\mathcal{O}(r^{-4})\,, \\
\xi^{r}	&=-\frac{r}{2}\left(\partial_{+}\epsilon^{+}+\partial_{-}\epsilon^{-}\right)+\mathcal{O}(r^{-1})\,,
\end{align}
where $\epsilon^{\pm}$ are arbitrary smooth periodic functions. The radial component compensates for the boundary conformal transformation so that the leading asymptotic metric retains its prescribed form. At the level of asymptotic vector fields, the transformations preserving the Brown--Henneaux boundary conditions form two independent families. After quotienting by transformations that are pure gauge, their algebra is given by two copies of the Witt algebra. More precisely, the asymptotic symmetry group (ASG) is defined as
\begin{equation}
    \mathrm{ASG}
    =
    \frac{\mathrm{Diff}_{\mathrm{allowed}}}
         {\mathrm{Diff}_{\mathrm{proper}}},
    \label{eq:ASG-quotient}
\end{equation}
where $\mathrm{Diff}_{\mathrm{allowed}}$ denotes the diffeomorphisms
that preserve the boundary conditions, while
$\mathrm{Diff}_{\mathrm{proper}}$ denotes the subgroup of proper
gauge transformations. The quotient means that two allowed
diffeomorphisms are regarded as the same asymptotic symmetry if they
differ only by a proper gauge transformation.

Proper transformations have vanishing surface charges throughout the
allowed phase space, once the normalization of the charges has been
fixed, and therefore represent redundancies of the gravitational
description. By contrast, an allowed diffeomorphism with a nonzero
surface-charge generator is called an improper diffeomorphism and
acts nontrivially on the physical boundary data.\footnote{Improper
diffeomorphisms are also frequently called large diffeomorphisms. In
the present context, this terminology refers to their nontrivial
action on the asymptotic data and does not necessarily mean that they
are disconnected from the identity.} Notice that the vanishing of a
charge on one particular configuration, such as the AdS$_3$ vacuum,
does not by itself make the transformation proper. To represent a
gauge redundancy, its charge must vanish throughout the complete
allowed phase space.

Acting on the AdS$_3$ vacuum with improper diffeomorphisms outside
its exact isometry group produces physically distinct configurations
known as boundary gravitons~\cite{Brown:1986nw}. These configurations
remain locally AdS$_3$, since they are related locally to the vacuum
by diffeomorphisms, but they differ from it through their asymptotic
data and conserved charges. Their existence therefore does not
contradict the absence of local propagating gravitons in
three-dimensional Einstein gravity: they are physical boundary
excitations rather than local bulk degrees of freedom.

At the level of the asymptotic vector fields, the two independent
families of transformations generate two copies of the Witt algebra.
The canonical generators associated with these transformations
contain boundary contributions. Their Poisson brackets consequently
acquire a central extension, converting the two Witt algebras into
two copies of the Virasoro algebra. Choosing the zero-mode
normalization for which global $\mathrm{AdS}_{3}$ has
$L_{0}=\bar L_{0}=0$, the charge algebra is
\begin{align}
    i\{L_{m},L_{n}\}
    &=
    (m-n)L_{m+n}
    +
    \frac{c_{L}}{12}
    m(m^{2}-1)\delta_{m+n,0},
    \\
    i\{\bar L_{m},\bar L_{n}\}
    &=
    (m-n)\bar L_{m+n}
    +
    \frac{c_{R}}{12}
    m(m^{2}-1)\delta_{m+n,0},
    \\
    \{L_{m},\bar L_{n}\}
    &=
    0.
\end{align}
The coefficients $c_{L}$ and $c_{R}$ are the left- and right-moving
central charges given in Eq.~\eqref{centralcharge}. The central
extension is already present in the classical gravitational charge
algebra.\footnote{Mathematically, the central term is determined by
the Gelfand--Fuchs two-cocycle of the Witt algebra.} The modes
$m=(-1,0,1)$ in each sector form the finite-dimensional
$\mathfrak{sl}(2,\mathbb R)$ subalgebra associated with the exact
AdS$_3$ isometries. The factor $m(m^{2}-1)$ follows from the chosen
zero-mode normalization. An alternative shift of the zero modes
gives the equivalent convention in which the central term is
proportional to $m^{3}$.

A complementary description of the asymptotic data is provided by
the holographic stress tensor. Its definition requires a finite
on-shell action and a well-defined variational principle with respect
to the boundary metric. The required boundary terms and the resulting
stress tensor are derived in
Section~\ref{sec:3D-Einstein}.


\section{Radial foliations and the Fefferman--Graham expansion}
\label{sec:radial-foliations-fg}

The conformal description of the AdS boundary can be complemented by
a foliation of the asymptotic region by timelike hypersurfaces. Their
induced metric describes the intrinsic geometry of each radial
surface, while the extrinsic curvature characterizes its embedding
in the bulk. Coordinates adapted to this foliation provide a natural
framework for expressing the Einstein equations, the variation of
the gravitational action, and the boundary terms required by the
variational principle.

The Fefferman--Graham construction supplies a near-boundary normal
form associated with a chosen representative of the boundary
conformal class~\cite{Fefferamn85,Fefferman:2007rka}. In Einstein gravity, the
field equations then determine the radial expansion of the metric
and separate the prescribed boundary geometry from the
state-dependent asymptotic data~\cite{deHaro:2000vlm,Skenderis:2002wp}.
It is important to distinguish these two steps: the
Fefferman--Graham gauge is a geometric coordinate choice, whereas
the allowed radial powers and the relations among their coefficients
follow from the field equations and the boundary conditions.

This section introduces the intrinsic and extrinsic geometry of the
radial hypersurfaces, the relation between Gaussian normal and
Fefferman--Graham coordinates, and the Einstein asymptotic
expansion. These ingredients are subsequently used to construct the
renormalized Einstein--AdS$_3$ action and its boundary stress tensor.

\subsection{Gaussian Normal Coordinates}

Consider a three-dimensional Lorentzian spacetime with signature $(-,+,+)$ and a timelike conformal boundary. In a neighborhood of this boundary, introduce coordinates $x^{\mu}=(z,x^{i})$ adapted to a family of timelike hypersurfaces $\Sigma_{z}=\{z=\mathrm{constant}\}$. We use a radial coordinate that decreases towards the conformal boundary, located at $z=0$, and write the metric in the zero-shift form, that is 
\begin{equation}     \label{eq:radial-metric-decomposition}
    ds^{2}=N^{2}(z)\,dz^{2}+h_{ij}(z,x)\,dx^{i}dx^{j}\,, 
\end{equation}
where $N(z)>0$ is the radial lapse and $h_{ij}$ is the induced Lorentzian metric on $\Sigma_{z}$. The absence of mixed components $g_{zi}$ means that the radial coordinate lines are orthogonal to these hypersurfaces.

The expression \textbf{Gaussian normal coordinates} will be used here for the radial form~\eqref{eq:radial-metric-decomposition}, allowing a lapse that depends only on $z$. Strict Gaussian normal coordinates use proper distance along the normal geodesics and have unit lapse. The two descriptions are locally related by a reparametrization of the radial coordinate, since the physical normal distance element is 
\begin{equation}
    ds_{\perp}=N(z)\,|dz|\,.
\end{equation}
Consequently, the coordinate $z$ need not itself measure physical proper distance. 

We choose the spacelike unit normal pointing towards increasing $z$,
\begin{equation} \label{eq:radial-unit-normal}
    n^{\mu}=\frac{1}{N}\delta_{z}^{\mu}\,,\qquad n_{\mu}=N\delta_{\mu}^{z}\,,\qquad n^{\mu}n_{\mu}=1\,.
\end{equation}
The radial foliation should not be confused with a foliation by spacelike Cauchy surfaces. Each $\Sigma_{z}$ contains a timelike direction, while its normal is spacelike. Thus, radial evolution describes the dependence of the geometry on distance from the boundary; it is not physical time evolution.

\begin{figure}[ht] 
    \centering
    \includegraphics[width=0.7\textwidth]
    {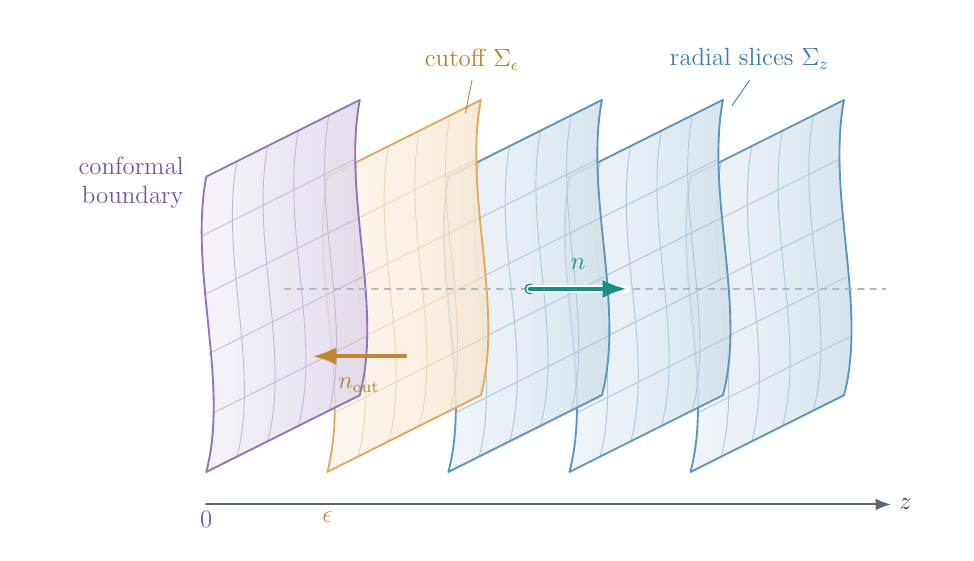}
    \caption[Radial foliation near the conformal boundary]
    {Radial foliation near the conformal boundary. The timelike
    hypersurfaces $\Sigma_z$ are represented by the blue surfaces,
    while $\Sigma_\epsilon$ denotes the radial cutoff. The unit normal
    $n^\mu$ points toward increasing $z$.}
    \label{fig:radial-foliation-refined}
\end{figure}

The resulting radial foliation is represented schematically in
Fig.~\ref{fig:radial-foliation-refined}. The hypersurfaces
$\Sigma_z$ approach the conformal boundary as $z\rightarrow0$, while
$\Sigma_\epsilon$ provides a finite radial cutoff. The unit normal
$n^\mu$ is chosen to point toward increasing $z$. The sign convention
for the extrinsic curvature is specified below and will be used
throughout the remainder of the thesis.

\subsection{Intrinsic and Extrinsic Geometry}

The induced metric determines the intrinsic geometry of each hypersurface. We denote its Levi--Civita derivative by $D_i$ and its curvature tensors by $\mathcal{R}^{ij}_{kl}$, $\mathcal{R}_{ij}$, and $\mathcal{R}$. These quantities describe measurements made within a fixed hypersurface and must be distinguished from the curvature tensors of the bulk metric $g_{\mu\nu}$.

The embedding of $\Sigma_z$ in the bulk is characterized by its
extrinsic curvature. We adopt the convention
\begin{equation}
    K_{ij}
    :=
    -\frac{1}{2}\mathcal L_n h_{ij}
    =
    -\frac{1}{2N}\partial_z h_{ij},
    \label{eq:extrinsic-curvature-convention}
\end{equation}
Its mixed form and trace are
\begin{equation}
    K^i_j=h^{ik}K_{kj},
    \qquad
    K=K^i{}_i.
\end{equation}

In particular, the trace determines the normal variation of the hypersurface volume element,
\begin{equation}
    \frac{1}{N}\partial_{z}\sqrt{|h|}=-K\sqrt{|h|}\,.
\end{equation}
The Gauss--Codazzi relations connect the intrinsic and extrinsic geometry to the bulk curvature. With the Riemann convention adopted in the previous sections, the tangential projection of the bulk Riemann tensor satisfies
\begin{equation}
    R_{kl}^{ij}=\mathcal{R}_{kl}^{ij}-K_{k}^{i}K_{l}^{j}+K_{l}^{i}K_{k}^{j}\,.
\end{equation}
The contracted Codazzi relation is
\begin{equation}
    R_{ni}:=n^{\mu}R_{\mu i}=D_{i}K-D_{j}K_{i}^{j}\,.
\end{equation}
For the radial lapse $N=N(z)$ used here, the bulk Ricci scalar can also be written as
\begin{equation}     \label{eq:ricci-scalar-radial}
    R=\mathcal{R}-K^{2}-K_{j}^{i}K_{i}^{j}+\frac{2}{N}\partial_{z}K\,.
\end{equation}
These are geometric identities and do not require the field equations. They provide the geometric relations needed to decompose the
Einstein equations and the Einstein--Hilbert action along the radial
foliation. In particular, Eq.~\eqref{eq:ricci-scalar-radial} exhibits the normal derivative of the extrinsic curvature that enters the Einstein--Hilbert action, anticipating the importance of boundary terms in its variational principle.

\subsection{The Fefferman--Graham Expansion}

An asymptotically locally anti-de Sitter (AlAdS) spacetime admits a conformal completion with a timelike boundary and has curvature approaching that of AdS,
\begin{equation}
    R^{\mu\nu}_{\lambda\rho}
  =
    -\frac{1}{\ell^{2}}
    \left(
        \delta^{\mu}_{\lambda}\delta^{\nu}_{\rho}
        -
        \delta^{\mu}_{\rho}\delta^{\nu}_{\lambda}
    \right)+\mathcal{O}(z)\,.  \label{eq:AlAdS-curvature}
\end{equation}
More precisely, there exists a defining function $\Omega$ that is positive in the bulk and satisfies
\begin{equation}
    \Omega\big|_{\partial\mathcal{M}}=0,
    \qquad
    d\Omega\big|_{\partial\mathcal{M}}\neq0,
\end{equation}
such that the conformally rescaled metric $\widetilde g_{\mu\nu}=\Omega^{2}g_{\mu\nu}$ extends to the boundary as a regular, nondegenerate metric. The physical metric determines only a conformal class of boundary metrics rather than a preferred representative.

\begin{figure}[ht]
    \centering
    \includegraphics[width=0.7\textwidth]
    {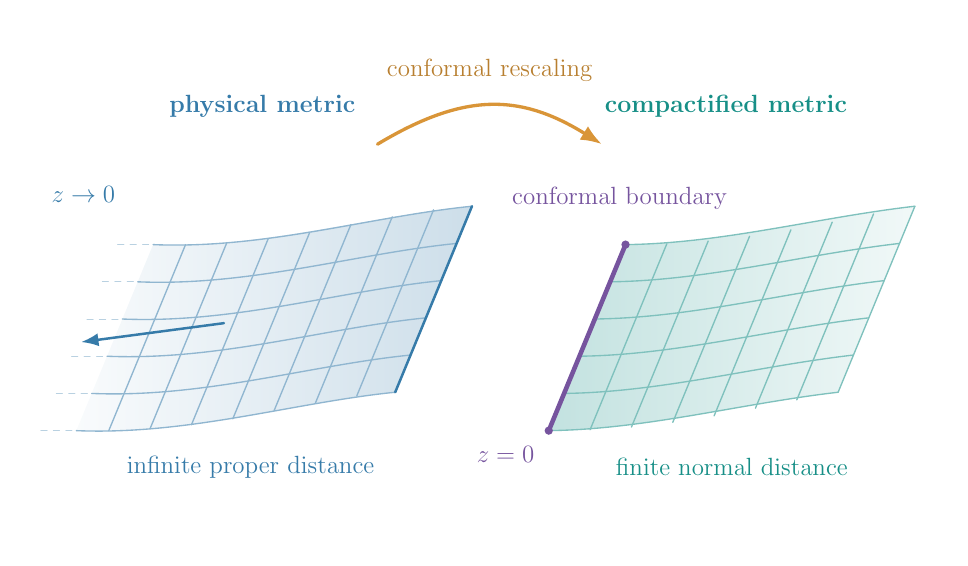}
    \caption[Physical and conformally compactified geometries]
    {Comparison between the physical and conformally compactified
    geometries. The conformal boundary lies at infinite proper radial
    distance in the physical metric, but is represented by the finite
    hypersurface $z=0$ in the rescaled geometry.}
    \label{fig:fg-refined}
\end{figure}

After choosing a representative $g_{(0)ij}$ of this conformal class, one may select the geodesic defining function $z=\ell\Omega$. For a sufficiently regular AlAdS metric, the Fefferman--Graham construction then puts the metric locally near the conformal boundary in the form~\cite{Fefferman:2007rka}
\begin{equation}
    ds^{2}
    =
    \frac{\ell^{2}}{z^{2}}
    \left[
        dz^{2}
        +
        \bar g_{ij}(z,x)\,dx^{i}dx^{j}
    \right],
    \qquad
    \bar g_{ij}(0,x)=g_{(0)ij}(x)\,. \label{eq:fg-metric-background}
\end{equation}
Here $x^{i}$ are coordinates tangent to the boundary and extended
into the bulk along the normal coordinate lines. This
Fefferman--Graham chart is defined locally near the conformal
boundary and need not cover the complete spacetime. This is a special case of Eq.~\eqref{eq:radial-metric-decomposition}, with
\begin{equation} \label{eq:fg-lapse-induced}
    N(z)=\frac{\ell}{z}\,,\qquad h_{ij}(z,x)=\frac{\ell^{2}}{z^{2}}\bar{g}_{ij}(z,x)\,.
\end{equation}
The geometric meaning of this choice becomes explicit after introducing the conformally rescaled metric
\begin{equation}
    \widetilde{g}_{\mu\nu}=\frac{z^{2}}{\ell^{2}}g_{\mu\nu}\,,
\end{equation}
whose line element is
\begin{equation}
    d\widetilde{s}^{2}=dz^{2}+\bar{g}_{ij}(z,x)\,dx^{i}dx^{j}\,.
\end{equation}
Therefore, $z$ is a Gaussian normal coordinate for the rescaled metric: it measures normal distance from the conformal boundary in this auxiliary geometry~\cite{Fefferman:2007rka}. It does not measure proper distance in the physical metric. Indeed, the physical radial distance between $z=\epsilon$ and a fixed $z=z_{0}>\epsilon$ is
\begin{equation}
    L(\epsilon,z_{0})=\int_{\epsilon}^{z_{0}}\frac{\ell}{z}\,dz=\ell\log\left(\frac{z_{0}}{\epsilon}\right)\,,
\end{equation}
which diverges as $\epsilon\rightarrow0$. The boundary metric is the pullback of the rescaled metric,
\begin{equation}
    g_{(0)ij}(x)=\lim_{z\rightarrow0}\bar{g}_{ij}(z,x)=\lim_{z\rightarrow0}\frac{z^{2}}{\ell^{2}}h_{ij}(z,x)\,. \label{limitg0}
\end{equation}
The distinction between the physical and conformally rescaled
geometries is illustrated in Fig.~\ref{fig:fg-refined}. The surface
$z=0$ lies at infinite proper radial distance with respect to the
physical metric, whereas it becomes a regular finite boundary of the
rescaled geometry.

Changing the defining function according to $z'=e^{\omega(x)}z+\cdots$ changes the boundary representation by
\begin{equation}
    g'_{(0)ij}=e^{2\omega(x)}g_{(0)ij}\,.
\end{equation}
The bulk geometry, therefore, determines a conformal class $[g_{(0)}]$, rather than a preferred boundary metric. In particular, an AlAdS spacetime need not have the flat cylindrical boundary representative associated with global AdS$_3$. Substituting Eq.~\eqref{eq:fg-lapse-induced} into
Eq.~\eqref{eq:extrinsic-curvature-convention} gives the exact identity
\begin{equation}
    K^i_j
    =
    \frac{1}{\ell}\delta^i_j
    -
    \frac{z}{2\ell}
    \bar g^{ik}\partial_z\bar g_{kj}.
    \label{eq:FG-extrinsic-curvature}
\end{equation}
For the regular asymptotic behavior considered here,
$z\partial_z\bar g_{ij}\rightarrow0$, and therefore
\begin{equation}
    K^i_j
    \longrightarrow
    \frac{1}{\ell}\delta^i_j\,,
    \qquad
    K
    \longrightarrow
    \frac{2}{\ell}\,,
    \qquad
    z\longrightarrow0\,.
    \label{eq:FG-extrinsic-curvature-limit}
\end{equation}

\subsection{Asymptotic Expansion and Boundary Data}

Once the metric has been written in Fefferman--Graham form, its
remaining radial dependence is determined by the Einstein equations.
The Fefferman--Graham gauge fixes the components $g_{zz}$ and
$g_{zi}$, but it does not by itself determine the coefficients of
$\bar g_{ij}(z,x)$. For vacuum Einstein gravity in three bulk dimensions, the field
equations select the finite expansion
\begin{equation}
    \bar g_{ij}(z,x)
    =
    g_{(0)ij}(x)
    +
    \frac{z^{2}}{\ell^{2}}g_{(2)ij}(x)
    +
    \frac{z^{4}}{\ell^{4}}g_{(4)ij}(x)\,.
    \label{eq:Einstein-FG-expansion}
\end{equation}
This expansion is called even because it contains only even powers of
$z$. Its truncation and the relations among its coefficients follow
from the different components of the Einstein equations. At the first
nontrivial radial order, the radial constraint fixes the trace of
$g_{(2)ij}$ in terms of the Ricci scalar of the boundary
representative:
\begin{equation}
    \text{tr }g_{(2)}
    =
    -\frac{\ell^{2}}{2}\mathcal R_{(0)}\,.
    \label{trg2condition}
\end{equation}
This condition determines only the trace of $g_{(2)ij}$ and not the
complete tensor. Its divergence is constrained independently by the
mixed radial components of the Einstein equations:
\begin{equation}
    \mathcal D_{(0)}^{i}g_{(2)ij}
    =
    \mathcal D_{(0)j}\text{tr }g_{(2)}\,.
\end{equation}
The remaining radial equations determine the next coefficient
algebraically and terminate the expansion:
\begin{align}
    g_{(4)ij}
    &=
    \frac{1}{4}
    g_{(2)ik}g^{k}_{(2)j}\,,
    \label{eq:Einstein-g4}
    \\
    g_{(n)ij}
    &=
    0\,,
    \qquad n\geq5\,.
\end{align}
The finite truncation is therefore a consequence of the
three-dimensional Einstein equations rather than an assumption about
the radial dependence of the metric. The components of $g_{(2)ij}$
not fixed by its trace and divergence contain state-dependent
information and determine the holographic stress tensor.
Consequently, fixing the boundary representative $g_{(0)ij}$ does
not select a unique bulk solution~\cite{deHaro:2000vlm}.

An alternative dimensionless radial coordinate commonly used in the literature is
\begin{equation}
    \rho=\frac{z^{2}}{\ell^{2}}\,.
    \label{eq:rho-coordinate}
\end{equation}
In terms of $\rho$, the Fefferman--Graham metric~\eqref{eq:fg-metric-background} becomes
\begin{equation}
    ds^{2}
    =
    \frac{\ell^{2}}{4\rho^{2}}\,d\rho^{2}
    +
    \frac{1}{\rho}
    \bar g_{ij}(\rho,x)\,dx^{i}dx^{j}\,,
    \label{eq:FG-rho-form}
\end{equation}
where $\bar g_{ij}(\rho,x)
 :=\bar g_{ij}(\ell\sqrt{\rho},x)$.
This change of radial coordinate is convenient when the expansion contains only even powers of $z$, since $(z/\ell)^{2n}=\rho^{n}$. It does not impose evenness: a term $(z/\ell)^{\alpha}$ becomes $\rho^{\alpha/2}$.

Global AdS$_3$ provides a direct example. Its metric is
\begin{equation}
    ds^{2}
    =
    -\left(1+\frac{r^{2}}{\ell^{2}}\right)dt^{2}
    +
    \frac{dr^{2}}{1+r^{2}/\ell^{2}}
    +
    r^{2}d\varphi^{2}\,,
    \qquad
    \varphi\sim\varphi+2\pi \,.
    \label{eq:global-AdS-FG-example}
\end{equation}
Requiring the radial part to take the Fefferman--Graham form gives
\begin{equation}
    \frac{dr}{\sqrt{1+r^{2}/\ell^{2}}}
    =
    -\frac{\ell}{z}\,dz\,,
\end{equation}
where the minus sign ensures that $z$ decreases as $r$ increases.
Choosing the integration constant such that
$r\sim\ell^{2}/z$ near the boundary yields
\begin{equation}
    r=\frac{\ell^{2}}{z}-\frac{z}{4}\,.
\end{equation}
The metric then becomes
\begin{equation}
    ds^{2}
    =
    \frac{\ell^{2}}{z^{2}}
    \left[
        dz^{2}
        -
        \left(1+\frac{z^{2}}{4\ell^{2}}\right)^{2}dt^{2}
        +
        \ell^{2}
        \left(1-\frac{z^{2}}{4\ell^{2}}\right)^{2}d\varphi^{2}
    \right]\,.
    \label{eq:global-AdS-FG}
\end{equation}
This expression is an explicit realization of the even expansion
\eqref{eq:Einstein-FG-expansion}. Its boundary representative is
\begin{equation}
    g_{(0)ij}\,dx^{i}dx^{j}
    =
    -dt^{2}+\ell^{2}d\varphi^{2}\,.
\end{equation}
It belongs to the same conformal class as
$-d\tau^{2}+d\phi^{2}$ after setting $t=\ell\tau$ and
$\varphi=\phi$. On the universal covering space,
$t\in\mathbb{R}$, and the boundary is a timelike cylinder.

The Fefferman--Graham coordinates are adapted to the asymptotic
region rather than to the complete bulk geometry. In the global
AdS$_3$ example, the range $0<z<2\ell$ corresponds to $r>0$,
while $z=2\ell$ reaches the center, where the angular coordinate
degenerates.

To regulate the gravitational action, introduce the cutoff surface
$\Sigma_{\epsilon}$ at $z=\epsilon>0$ and retain the bulk region
$\mathcal{M}_{\epsilon}=\{z\geq\epsilon\}$. The induced metric on
this surface is
\begin{equation}
    h_{ij}(\epsilon,x)
    =
    \frac{\ell^{2}}{\epsilon^{2}}
    \bar g_{ij}(\epsilon,x)\,.
    \label{eq:cutoff-induced-metric}
\end{equation}
Unlike $g_{(0)ij}$, this is a metric on a hypersurface at finite
radial position. The corresponding volume elements are
\begin{equation}
    \sqrt{|h|}
    =
    \frac{\ell^{2}}{\epsilon^{2}}
    \sqrt{|\bar g|}\,,
    \qquad
    \sqrt{|g|}
    =
    \frac{\ell^{3}}{z^{3}}
    \sqrt{|\bar g|}\,.
    \label{eq:FG-volume-elements}
\end{equation}
Bulk and boundary integrals can therefore diverge when
$\epsilon\rightarrow0$, even if the local curvature invariants
remain finite. Removing these divergences and defining the
boundary data held fixed under variations are distinct
requirements. In particular, fixing $h_{ij}(\epsilon,x)$ at a
finite cutoff is not equivalent to fixing the conformal boundary
representative $g_{(0)ij}$ in the limit
$\epsilon\rightarrow0$.

\section{Boundary Terms in Einstein--AdS\texorpdfstring{$_3$}{3} Gravity}
\label{sec:3D-Einstein}

The radial decomposition and Fefferman--Graham expansion developed
above provide the geometric ingredients required to formulate the
Einstein variational problem in an asymptotically locally AdS
spacetime. Specifying the bulk field equations is not sufficient:
one must also identify the data held fixed at the boundary and
supplement the Einstein--Hilbert action with suitable boundary terms.

These boundary terms perform two logically distinct roles. The
Gibbons--Hawking--York term removes the normal derivatives of the
metric variation and establishes a Dirichlet variational principle
at a finite radial cutoff. Additional local counterterms remove the
divergences that arise when the cutoff surface approaches the
conformal boundary. Finiteness of the on-shell action and
well-posedness of its variation must therefore be checked
separately.

Throughout this section, we use the extrinsic-curvature convention
defined in Eq.~\eqref{eq:extrinsic-curvature-convention}. Its exact
Fefferman--Graham expression and asymptotic value are given in
Eqs.~\eqref{eq:FG-extrinsic-curvature} and
\eqref{eq:FG-extrinsic-curvature-limit}.

The Einstein--AdS$_3$ action supplemented by a general boundary Lagrangian is
\begin{equation}
    I[g_{\mu\nu}]
    =
    \kappa
    \int_{\mathcal M_\epsilon}
    d^3x\,\sqrt{|g|}
    \left(
        R-2\Lambda
    \right)
    +
    \int_{\Sigma_\epsilon}
    d^2x\,\sqrt{|h|}\,
    \mathcal B,
    \qquad
    \Lambda=-\frac{1}{\ell^2},
    \label{EAdSaction}
\end{equation}
where \(\kappa=(16\pi G)^{-1}\) and
\(\Sigma_\epsilon=\partial\mathcal M_\epsilon\) is the regulating
surface at \(z=\epsilon\).

The origin of the Gibbons--Hawking--York term can be seen directly
from the radial decomposition~\eqref{eq:ricci-scalar-radial}. The
normal derivative of $K$ contained in the Einstein--Hilbert action
produces a boundary variation involving normal derivatives of
$\delta h_{ij}$. Such contributions are not removed simply by
imposing $\delta h_{ij}=0$ at the cutoff surface. With the
extrinsic-curvature convention~\eqref{eq:extrinsic-curvature-convention},
they are cancelled by
\begin{equation}
    I_{\mathrm{GHY}}
    =
    2\kappa
    \int_{\Sigma_\epsilon}
    d^2x\,\sqrt{|h|}\,K.
    \label{eq:Einstein-GHY-action}
\end{equation}

This term establishes the usual Dirichlet variational principle at a
finite non-null boundary~\cite{York:1972sj,Gibbons:1976ue}.\footnote{
A Dirichlet condition at finite cutoff means that the induced metric
\(h_{ij}\) is fixed on \(\Sigma_\epsilon\). This must be distinguished
from the asymptotic Dirichlet condition used in holography, in which
a representative \(g_{(0)ij}\) of the boundary conformal class is
fixed as the cutoff is removed.}

The Gibbons--Hawking--York term does not, by itself, remove the
divergences generated by the infinite asymptotic volume. For the
boundary conditions considered here, the leading intrinsic
counterterm is the boundary volume term
\begin{equation}
    I_{\mathrm{ct}}
    =
    -\frac{2\kappa}{\ell}
    \int_{\Sigma_\epsilon}
    d^2x\,\sqrt{|h|}.
    \label{eq:Einstein-volume-counterterm}
\end{equation}
The action including the renormalizing boundary terms is therefore
\begin{equation}
    I_{\mathrm{reg}}
    =
    I_{\mathrm{bulk}}
    +
    2\kappa
    \int_{\Sigma_\epsilon}
    d^2x\,\sqrt{|h|}\,K
    -
    \frac{2\kappa}{\ell}
    \int_{\Sigma_\epsilon}
    d^2x\,\sqrt{|h|}.
\end{equation}
The renormalized on-shell action is obtained by evaluating this
expression on AlAdS spacetimes and removing the cutoff,
\begin{equation}
    I_{\mathrm{ren}}[g_{(0)}]
    =
    \lim_{\epsilon\rightarrow0}
    \left.
    I_{\mathrm{reg}}[\mathcal M_\epsilon]
    \right|_{\mathrm{EOM}}.
    \label{eq:Einstein-renormalized-action}
\end{equation}
From the bulk perspective, the divergences removed in this limit are
infrared divergences caused by the infinite asymptotic volume. In the
holographic description, the radial cutoff acts instead as an
ultraviolet regulator for the boundary generating functional
\cite{deHaro:2000vlm,Skenderis:2002wp}.\footnote{For the flat
cylindrical representative used in the Brown--Henneaux phase space,
the boundary volume term removes the relevant divergence. For a
general curved boundary representative, the action at finite cutoff also contains a logarithmic divergence proportional to the
two-dimensional Euler density. Its subtraction gives rise to the
conformal anomaly discussed below.}

An arbitrary variation of $I_{\mathrm{reg}}$ takes the form
\begin{align}
    \delta I_{\mathrm{reg}}
    ={}&
    \kappa
    \int_{\mathcal M_\epsilon}
    d^3x\,\sqrt{|g|}
    \left(
        G_{\mu\nu}
        +
        \Lambda g_{\mu\nu}
    \right)
    \delta g^{\mu\nu}-
    \kappa
    \int_{\Sigma_\epsilon}
    d^2x\,\sqrt{|h|}
    \left(
        K^i_j
        -
        K\delta^i_j
        +
        \frac{1}{\ell}\delta^i_j
    \right)
    \left(
        h^{-1}\delta h
    \right)^j_i\,,
    \label{eq:Einstein-renormalized-variation}
\end{align}
where
\begin{equation}
    \left(
        h^{-1}\delta h
    \right)^j_i
    :=
    h^{jk}\delta h_{ki}.
    \label{eq:Einstein-mixed-metric-variation}
\end{equation}
Up to the factor \(\kappa\sqrt{|h|}\), the tensor multiplying
\((h^{-1}\delta h)^j_i\) is the renormalized radial canonical
momentum.\footnote{The term radial canonical momentum refers to the
momentum conjugate to \(h_{ij}\) when the radial coordinate is
treated as an evolution parameter. This radial evolution is not
physical time evolution because the hypersurfaces of constant \(z\)
are timelike.}

On shell, the bulk term in
Eq.~\eqref{eq:Einstein-renormalized-variation} vanishes. It remains
to verify that the boundary contribution has a finite limit. The
vacuum Einstein equations select the finite even expansion
\eqref{eq:Einstein-FG-expansion}. Using the exact relation between
the Fefferman--Graham metric and the extrinsic curvature gives
\begin{align}
    K^i_j
    &=
    \frac{1}{\ell}\delta^i_j
    -
    \frac{z^2}{\ell^3}g^i_{(2)j}
    +
    \mathcal O(z^4)\,,
    \label{eq:Einstein-K-expansion}
    \\
    K
    &=
    \frac{2}{\ell}
    -
    \frac{z^2}{\ell^3}\text{tr }g_{(2)}
    +
    \mathcal O(z^4)\,.
    \label{eq:Einstein-K-trace-expansion}
\end{align}
Using the FG expansion for Einstein-AdS$_3$ spacetimes, the codimension-1 volume element reads
\begin{equation}
    \sqrt{|h|}
    =
    \frac{\ell^2}{z^2}
    \sqrt{|g_{(0)}|}
    \left[
        1
        +
        \frac{z^2}{2\ell^2}
        \text{tr }g_{(2)}
        +
        \mathcal O(z^4)
    \right]\,.
    \label{eq:Einstein-determinant-expansion}
\end{equation}
We now return to the boundary term in the on-shell variation. Define
\begin{equation}
    \mathcal P^i{}_j
    :=
    -K^i{}_j
    +
    K\delta^i{}_j
    -
    \frac{1}{\ell}\delta^i{}_j.
    \label{eq:Einstein-renormalized-momentum}
\end{equation}
The first two terms arise from the Einstein--Hilbert action together
with the Gibbons--Hawking--York term, while the final term is
produced by the boundary volume counterterm.

Before the counterterm is included, the asymptotic expansion gives
\begin{equation}
   -K^i_j+K\delta^i_j
    =
    \frac{1}{\ell}\delta^i_j
    +
    \frac{z^2}{\ell^3}
    \left(
        g^i_{(2)j}
        -
        \delta^i_j
        \text{tr }g_{(2)}
    \right)
    +
    \mathcal O(z^4).
    \label{eq:Einstein-unrenormalized-momentum}
\end{equation}
The leading term, when multiplied by
\(\sqrt{|h|}\sim z^{-2}\), would produce a divergent on-shell
variation. The counterterm contributes
\(-\ell^{-1}\delta^i_j\) and cancels this leading term.
Consequently,
\begin{equation}
    \mathcal P^i_j
    =
    \frac{z^2}{\ell^3}
    \left(
        g^i_{(2)j}
        -
        \delta^i_j
        \text{tr }g_{(2)}
    \right)
    +
    \mathcal O(z^4).
    \label{eq:Einstein-renormalized-momentum-expansion}
\end{equation}

The remaining factor in the variation also has a regular boundary
limit. Since the conformal factor \(\ell^2/z^2\) is held fixed,
\begin{equation}
    \left(
        h^{-1}\delta h
    \right)^j_i
    =
    \left(
        \bar g^{-1}\delta\bar g
    \right)^j_i
    =
    g_{(0)}^{jk}
    \delta g_{(0)ki}
    +
    \mathcal O(z^2).
    \label{eq:Einstein-metric-variation-expansion}
\end{equation}
The radial orders entering the boundary variation are therefore
\begin{equation}
    \sqrt{|h|}
    =
    \mathcal O(z^{-2}),
    \qquad
    \mathcal P^i_j
    =
    \mathcal O(z^2),
    \qquad
    \left(
        h^{-1}\delta h
    \right)^j_i
    =
    \mathcal O(1).
\end{equation}
Their product is of order \(z^0\), showing explicitly that the
renormalized on-shell variation has a finite limit. Taking \(z\rightarrow0\), one obtains
\begin{align}
    \left.
    \delta I_{\mathrm{ren}}
    \right|_{\mathrm{EOM}}
    &=
    \frac{\kappa}{\ell}
    \int_{\partial\mathcal M}
    d^2x\,
    \sqrt{|g_{(0)}|}
    \left(
        g_{(2)}^{ij}
        -
        g_{(0)}^{ij}
        \text{tr }g_{(2)}
    \right)
    \delta g_{(0)ij}
    \nonumber\\
    &:=
    \frac{1}{2}
    \int_{\partial\mathcal M}
    d^2x\,
    \sqrt{|g_{(0)}|}
    \tau_{\mathrm{(E)}}^{ij}
    \delta g_{(0)ij}.
    \label{FG-GHY}
\end{align}
The holographic stress tensor in Einstein-AdS$_3$ is thus
\begin{equation}
    \tau_{\mathrm{(E)}}^{ij}
    =
    \frac{2\kappa}{\ell}
    \left(
        g_{(2)}^{ij}
        -
        g_{(0)}^{ij}
        \text{tr }g_{(2)}
    \right)\,.
    \label{eq:Einstein-holographic-stress}
\end{equation}
Its trace provides a consistency check. Using
Eq.~\eqref{trg2condition},
\begin{equation}
    \tau^i_{\mathrm{(E)}i}
    =
    -\frac{2\kappa}{\ell}
    \text{tr }g_{(2)}
    =
    \frac{\ell}{16\pi G}
    \mathcal R_{(0)}\,.
\end{equation}
Since the Brown--Henneaux central charge is
\(c=3\ell/(2G)\), this relation becomes
\begin{equation}
    \tau^i_{\mathrm{(E)}i}
    =
    \frac{c}{24\pi}
    \mathcal R_{(0)}\,,
    \label{eq:Einstein-trace-anomaly}
\end{equation}
which is the expected two-dimensional conformal anomaly.

\subsection{The One-Half Gibbons--Hawking Prescription}

The coefficient of the Gibbons--Hawking--York term in
Eq.~\eqref{eq:Einstein-GHY-action} is fixed by the finite-cutoff
Dirichlet problem for \(h_{ij}\). A different prescription is
nevertheless possible when the fixed datum is the conformal boundary
metric \(g_{(0)ij}\), rather than the induced metric at finite cutoff.
In three bulk dimensions, consider
\begin{equation}
    \widetilde I_{\mathrm{reg}}
    =
     \kappa \int_{\mathcal{M}} d^3x \sqrt{|g|}\left( R- 2\lambda \right)
    +
    \kappa
    \int_{\Sigma_\epsilon}
    d^2x\,\sqrt{|h|}\,K\,.
    \label{GHY/2}
\end{equation}
Its boundary contribution is one half of the standard
Gibbons--Hawking--York term in the convention used above. It also
contains no separate boundary volume counterterm.

This relation can be compared directly with the standard
renormalized prescription. At finite cutoff,
\begin{equation}
    \widetilde I_{\mathrm{reg}}-I_{\mathrm{reg}}
    =
    \kappa
    \int_{\Sigma_\epsilon}
    d^2x\,\sqrt{|h|}
    \left(
        \frac{2}{\ell}-K
    \right)\,.
    \label{eq:half-GHY-difference}
\end{equation}
Using Eqs.~\eqref{eq:Einstein-determinant-expansion} and
\eqref{eq:Einstein-K-trace-expansion}, its finite boundary limit is
\begin{align}
    \widetilde I_{\mathrm{ren}}
    -
    I_{\mathrm{ren}}
    &:=
    \lim_{\epsilon\rightarrow0}
    \left(
        \widetilde I_{\mathrm{reg}}
        -
        I_{\mathrm{reg}}
    \right)
    \nonumber\\
    &=
    \frac{\kappa}{\ell}
    \int_{\partial\mathcal M}
    d^2x\,
    \sqrt{|g_{(0)}|}
    \text{tr }g_{(2)}
    \nonumber\\
    &=
    -\frac{\kappa\ell}{2}
    \int_{\partial\mathcal M}
    d^2x\,
    \sqrt{|g_{(0)}|}
    \mathcal R_{(0)}\,.
    \label{eq:half-GHY-Euler-difference}
\end{align}
where the radial Einstein equation
\eqref{trg2condition} was used in the second line. The difference
between the two prescriptions is therefore proportional to the
two-dimensional Euler density.

For a closed orientable two-dimensional manifold, the
Gauss--Bonnet theorem gives
\begin{equation}
    \int_{\partial\mathcal M}
    d^2x\,
    \sqrt{|g_{(0)}|}
    \mathcal R_{(0)}
    =
    4\pi\chi(\partial\mathcal M)\,,
\end{equation}
where \(\chi\) is the Euler characteristic. Its metric variation is
a total derivative and does not modify the local equations or the
holographic stress tensor. If the two-dimensional manifold itself
has a boundary, the corresponding boundary-of-boundary or corner
term must also be included.

The variational principle can now be verified explicitly. Using
Eq.~\eqref{FG-GHY} together with
Eq.~\eqref{eq:half-GHY-Euler-difference}, the on-shell variation of
the one-half Gibbons--Hawking action becomes
\begin{align}
    \delta\widetilde I_{\mathrm{ren}}
    ={}&
    \frac{\kappa}{\ell}
    \int_{\partial\mathcal M}
    d^2x\,\sqrt{|g_{(0)}|}
    \left(
        g_{(2)}^{ij}
        -
        g_{(0)}^{ij}\text{tr }g_{(2)}
    \right)
    \delta g_{(0)ij}
    \nonumber\\
    &-
    \frac{\kappa\ell}{2}
    \delta
    \int_{\partial\mathcal M}
    d^2x\,\sqrt{|g_{(0)}|}\,
    \mathcal R_{(0)}.
    \label{eq:half-GHY-variation}
\end{align}
The first line coincides with the variation obtained from the standard
renormalized Einstein action. The second line is the variation of the
two-dimensional Euler invariant and therefore vanishes for a closed
boundary, or after the corresponding corner data are fixed. The
boundary condition considered here fixes the conformal representative
$g_{(0)ij}$ and should be distinguished from the finite-cutoff
Dirichlet condition on $h_{ij}$, which requires the complete
Gibbons--Hawking--York term. Under the asymptotic Dirichlet condition
$\delta g_{(0)ij}=0$, Eq.~\eqref{eq:half-GHY-variation} vanishes.
Therefore, the one-half Gibbons--Hawking--York prescription defines a
well-posed variational principle at the conformal boundary and
reproduces the standard holographic stress tensor and conserved
charges~\cite{Miskovic:2006tm}.

\subsection{Ward Identity and Conserved Boundary Charges}

Boundary diffeomorphism invariance determines the Ward identity
satisfied by the holographic stress tensor. Under an infinitesimal
boundary diffeomorphism generated by \(\xi^i\), the source transforms
as
\begin{equation}
    \delta_\xi g_{(0)ij}
    =
    2\mathcal D^{(0)}_{(i}\xi_{j)}\,.
\end{equation}
Substitution into Eq.~\eqref{FG-GHY}, followed by an integration by
parts, gives
\begin{align}
    0
    =
    \delta_\xi I_{\mathrm{ren}}
    &=
    \int_{\partial\mathcal M}
    d^2x\,
    \sqrt{|g_{(0)}|}
    \tau_{\mathrm{(E)}}^{ij}
    \mathcal D^{(0)}_i\xi_j \,,
    \nonumber\\
    &=
    -
    \int_{\partial\mathcal M}
    d^2x\,
    \sqrt{|g_{(0)}|}
    \xi_j
    \mathcal D^{(0)}_i
    \tau_{\mathrm{(E)}}^{ij}\,,
\end{align}
where boundary-of-boundary contributions have been omitted. Since
\(\xi^i\) is arbitrary, the diffeomorphism Ward identity is
\begin{equation}
    \mathcal D^{(0)}_i
    \tau_{\mathrm{(E)}}^{ij}
    =
    0\,.
    \label{eq:Einstein-Ward-identity}
\end{equation}

For a boundary Killing vector \(\xi^i\), the corresponding conserved
charge is
\begin{equation}
    Q_{\mathrm{(E)}}[\xi]
    =
    \int_{\Sigma}
    dx\,\sqrt{|\gamma|}\,
    u^i
    \tau^{\mathrm{(E)}}_{ij}
    \xi^j\,,
    \label{eq:Einstein-boundary-charge}
\end{equation}
where \(\Sigma\) is a spacelike section of the two-dimensional
conformal boundary, \(\gamma\) is its induced one-dimensional metric,
and \(u^i\) is its future-directed timelike unit normal. In an
asymptotically AdS$_3$ spacetime, \(\Sigma\) is typically a circle at
fixed boundary time. Conservation of the stress tensor, together
with the Killing equation for \(\xi^i\), ensures that the value of
the charge is independent of the chosen time section, provided that
no flux crosses the boundary.

The standard intrinsic renormalization and the one-half
Gibbons--Hawking prescription therefore produce the same local
boundary response and the same conserved charges under the
asymptotic conditions considered here. This result provides the
Einstein-gravity reference for the extrinsic boundary prescription
developed in the following chapter.

\chapter{Extrinsic Renormalization of New Massive Gravity} \label{chapter5}

The combination of local simplicity and nontrivial global structure
makes three-dimensional gravity a natural setting in which to study
modifications that introduce propagating bulk degrees of freedom.
The linear theory of a massive spin-two field was formulated by
Fierz and Pauli~\cite{Fierz:1939ix}. In three dimensions, a
parity-invariant Fierz--Pauli field carries two local degrees of
freedom, conventionally described as helicities \(+2\) and \(-2\).

An early nonlinear realization of massive gravity in three
dimensions is Topologically Massive Gravity, introduced by Deser,
Jackiw, and Templeton~\cite{Deser:1982vy}. This theory supplements
the Einstein--Hilbert action with a gravitational
Lorentz--Chern--Simons term. Its metric equations are third order
and, at generic couplings, propagate a single massive helicity. The
Chern--Simons interaction is parity odd: a reflection of one spatial
coordinate reverses the orientation and exchanges the helicities
\(+2\) and \(-2\). A theory propagating only one member of this pair
is therefore not parity invariant.

A parity-preserving alternative was introduced by Bergshoeff, Hohm,
and Townsend under the name New Massive Gravity
\cite{Bergshoeff:2009hq}. NMG supplements the Einstein--Hilbert
action with a particular curvature-squared interaction for which the
independent scalar excitation is absent from the generic linearized
spectrum. Around an appropriate maximally symmetric vacuum, the
theory propagates the two helicities of a massive spin-two field.
It consequently provides a setting in which local bulk dynamics
coexists with the global and asymptotic structures characteristic
of three-dimensional gravity.

Related theories combine or modify these mechanisms of mass
generation. General Massive Gravity contains both the
curvature-squared interaction of NMG and the gravitational
Chern--Simons term of Topologically Massive Gravity
\cite{Bergshoeff:2009aq}. It generically propagates two massive
helicities with different masses and reduces to either theory when
the corresponding interaction is removed. Minimal Massive Gravity,
on the other hand, preserves the single local degree of freedom of
Topologically Massive Gravity while modifying its nonlinear field
equations~\cite{Bergshoeff:2014pca}. Its consistency relies on the
mechanism of third-way consistency: the divergence of the additional
curvature tensor vanishes after the complete field equations are
imposed, rather than as an off-shell geometric identity
\cite{Bergshoeff:2015zga}.

The present analysis focuses on NMG. Its parity-preserving
curvature-squared interaction introduces local spin-two dynamics
while admitting maximally symmetric AdS vacua and the corresponding
locally AdS geometries. The theory also possesses non-Einstein
solutions whose curvature and radial behavior differ from those of
the constant-curvature sector. The asymptotic boundary data must
therefore be determined from the NMG field equations rather than
inherited automatically from the Einstein expansion.

This distinction is especially important for the boundary
variational problem. A gravitational action on an asymptotically AdS
spacetime must remain finite when the radial regulator is removed,
but finiteness of its on-shell value is not sufficient. Its on-shell
variation must also be finite and must be expressed in terms of the
variations of the prescribed boundary sources. A boundary term can
cancel divergences in the action while leaving divergent or
unwanted contributions in its variation. Action finiteness and
well-posedness of the variational principle are therefore related
but logically distinct requirements.

For Einstein--AdS$_3$ gravity, Section~\ref{sec:3D-Einstein}
established that the standard Gibbons--Hawking--York term together
with the intrinsic boundary volume counterterm gives a finite
asymptotic Dirichlet problem. It also showed that, under the
asymptotic conditions considered there, an alternative prescription
containing one half of the Gibbons--Hawking--York term differs from
the standard renormalized action only by the two-dimensional Euler
term. This result provides the geometric starting point for an
extrinsic boundary prescription in NMG.

The use of boundary terms depending explicitly on the extrinsic
curvature is closely related to the Kounterterm method developed for
asymptotically AdS gravity
\cite{Miskovic:2006tm,Olea:2006vd,Anastasiou:2020zwc}. Such terms use
the embedding of the regulating hypersurface as part of the
regularization procedure. Their extension to a higher-curvature
theory is natural because the variation of the bulk action already
contains nontrivial contractions between the curvature tensor and
the extrinsic geometry. Nevertheless, the appropriate coefficients
and the sufficiency of the boundary terms must be established from
the field equations and the boundary conditions of the theory under
consideration.

The variational problem and holographic renormalization of NMG have
previously been studied using an auxiliary symmetric tensor and
intrinsic boundary counterterms
\cite{Hohm:2010jc,Kwon:2011jz,Cunliff:2013en}. These constructions
provide the holographic stress tensor for several asymptotic sectors
and show that the theory can admit radial behavior beyond the
Einstein expansion. The formulation adopted in this chapter is
instead written directly in terms of the metric, the Schouten
tensor, and the intrinsic and extrinsic geometry of the radial
hypersurfaces. Its purpose is to determine whether an extrinsic
boundary completion can simultaneously render the on-shell action
and its variation finite.

Three asymptotic sectors are treated separately. Constant-curvature
configurations provide the reference case and permit a direct
comparison with Einstein gravity. The second sector contains
nonconstant-curvature solutions satisfying \(g_{(1)ij}=0\), for
which the standard Brown--Henneaux power-law fall-offs are preserved.
The third sector occurs at the degenerate point, where the
asymptotic field equations no longer require \(g_{(1)ij}\) to
vanish. In that case, \(g_{(1)ij}\) constitutes additional
asymptotic data, subject to the constraints imposed by the field
equations, and its conjugate response appears in the finite
on-shell variation.

The chapter begins with the NMG action, its metric field equations,
and the coupling combinations that distinguish generic,
constant-curvature, degenerate, and critical configurations. The
radial field equations then determine the admissible
Fefferman--Graham coefficients in the asymptotic sectors described
above. The expansion of the on-shell action identifies the divergent
contributions, while the variation of the completed action determines
the finite holographic responses. The result is compared with the
auxiliary-field stress tensor of Hohm and Tonni
\cite{Hohm:2010jc}.

The same boundary completion is subsequently used in the
Noether--Wald construction of conserved charges. The prescription is
tested on the BTZ black hole, nonconstant-curvature AdS waves, and
the rotating hairy black hole. These examples examine,
respectively, the constant-curvature, nonconstant-curvature, and
degenerate sectors. The logarithmic branch associated with critical
boundary conditions lies outside the scope of the present analysis.

\section{New Massive Gravity}
\label{sec:NMG}

This section introduces the dynamics of NMG and fixes the conventions
used in the boundary analysis. The bulk action is
\begin{equation}
    I_{\rm NMG}
    =
    \kappa
    \int_{\mathcal M}
    d^3x\,\sqrt{|g|}
    \left[
        \sigma_1R
        -
        2\lambda
        -
        \frac{\sigma_2}{\mu^2}
        \left(
            R_{\mu\nu}R^{\mu\nu}
            -
            \frac{3}{8}R^2
        \right)
    \right]
    :=
    \int_{\mathcal M}
    d^3x\,\sqrt{|g|}\,
    \mathcal L_{\rm NMG}\,.
    \label{INMG}
\end{equation}
Here, $\lambda$ is the bare cosmological constant and the positive
parameter $\mu^2$ sets the scale of the curvature-squared interaction.
The parameters $\sigma_1,\sigma_2=\pm1$ specify the relative signs of
the Einstein--Hilbert and quadratic-curvature contributions. They are
kept explicit because different sign conventions are commonly used
in the NMG literature.

Varying the action~\eqref{INMG} with respect to the metric gives the
NMG field equations
\begin{equation}
    \sigma_1G_{\mu\nu}
    +
    \lambda g_{\mu\nu}
    -
    \frac{\sigma_2}{2\mu^2}
    \mathcal K_{\mu\nu}
    =
    0\,.
    \label{eq:NMG-conventional-field-equation}
\end{equation}
The tensor $\mathcal K_{\mu\nu}$ contains the complete contribution
of the quadratic-curvature interaction. Its expression in terms of
the Ricci curvature is
\begin{align}
    \mathcal K^\mu_\nu
    ={}&
    2\Box R^\mu_\nu
    -
    \frac{1}{2}\nabla^\mu\nabla_\nu R
    -
    \frac{1}{2}\delta^\mu_\nu\Box R
    +
    4R^{\mu\lambda}_{\nu\rho}R^\rho_\lambda
    -
    \frac{3}{2}RR^\mu_\nu- 
    \delta^\mu_\nu
    \left(
        R_{\lambda\rho}R^{\lambda\rho}
        -
        \frac{3}{8}R^2
    \right)\,.
    \label{eq:NMG-K-tensor-Ricci}
\end{align}
Equations~\eqref{eq:NMG-conventional-field-equation} and
\eqref{eq:NMG-K-tensor-Ricci} provide the conventional Ricci-tensor
description of the NMG dynamics. The relative coefficient $3/8$ is
essential: it removes the independent scalar excitation from the
generic linearized spectrum and causes all curvature-derivative terms
to cancel from the trace of the nonlinear equations.

The theory preserves parity because its action contains no
orientation-dependent gravitational Chern--Simons term. On global
AdS$_3$, parity can be represented by the orientation-reversing
isometry
\begin{equation}
    (t,r,\phi)
    \longmapsto
    (t,r,-\phi)\,.
\end{equation}
The angular coordinate changes under this transformation, but the
background metric is preserved because
\(d\phi^2\mapsto d\phi^2\). Parity exchanges the helicities
\(+2\) and \(-2\); since both the NMG action and the AdS$_3$
background preserve this symmetry, the two massive helicities have
equal masses at generic couplings.

The absence of the scalar excitation and the equality of the
helicity masses do not by themselves guarantee unitarity. The signs
of the kinetic terms, the stability of the selected vacuum, and the
boundary conditions must also be examined. Moreover, the parameter
\(\mu\) in Eq.~\eqref{INMG} is a coupling scale and should not, in
general, be identified directly with the physical Fierz--Pauli mass.
On a curved background, the mass of the linearized excitation also
depends on the background curvature~\cite{Bergshoeff:2009aq}.

It is also necessary to distinguish the bare cosmological parameter
\(\lambda\), which appears in the action, from the effective
cosmological constant \(\Lambda\) characterizing a particular
maximally symmetric vacuum. For an AdS vacuum,
\begin{equation}
    R_{\mu\nu}
    =
    2\Lambda g_{\mu\nu}\,,
    \qquad
    \Lambda
    =
    -\frac{1}{\ell^2}\,,
    \label{eq:NMG-AdS-curvature}
\end{equation}
where \(\ell\) is the effective AdS radius. The relation between
\(\lambda\), \(\Lambda\), and the higher-curvature couplings follows
from the NMG field equations and can admit more than one maximally
symmetric vacuum.

Whenever the couplings admit an AdS vacuum, every geometry locally
isometric to that vacuum satisfies the local NMG field equations.
This includes the corresponding BTZ family. Nevertheless, the
conserved mass, angular momentum, and entropy assigned to a BTZ
geometry depend on the gravitational action and are not generally
equal to their Einstein values. The relation between the geometric
parameters and the NMG charges is examined in
Section~\ref{sec:NMG-BTZ}.

The NMG equations also admit non-Einstein configurations whose Ricci
tensor is not proportional to the metric. Their existence does not
contradict the algebraic vanishing of the Weyl tensor in three
dimensions: the vanishing of the Weyl tensor determines the
algebraic decomposition of the Riemann tensor, but it does not impose
the Einstein equations on the Ricci tensor. Consequently, the
allowed asymptotic coefficients must be determined from the NMG
field equations rather than inferred solely from the
Fefferman--Graham coordinate choice.

\subsection{Schouten formulation}

The conventional Ricci-tensor form of the NMG equations is sufficient
for describing the bulk dynamics. The boundary analysis, however,
requires the total derivative generated by the metric variation to be
retained. For this purpose, it is convenient to express the action and
its variation in terms of the Schouten and Cotton tensors.

The special algebraic structure of curvature in three dimensions makes the Schouten and Cotton tensors a convenient basis for expressing the NMG action and field equations~\cite{Tekin:2015rha}. Since the Weyl tensor vanishes identically, the Riemann tensor is completely determined by the Schouten tensor,
\begin{equation}
    R^{\mu\nu}_{\lambda\rho}
    =
    4\delta^{[\mu}_{[\lambda}
    S^{\nu]}_{\rho]}\,,
    \qquad
    S^\mu_\nu
    =
    R^\mu_\nu
    -
    \frac{1}{4}\delta^\mu_\nu R\,.
    \label{Schouten}
\end{equation}
These relations are purely geometric identities and do not require the NMG field equations. The Schouten tensor organizes the algebraic curvature dependence of the theory, while its covariant derivatives give rise to the Cotton tensor appearing in the fourth-order part of the equations of motion.

The quadratic-curvature interaction in the NMG action can also be written directly in terms of \(S^\mu_\nu\). In terms of the generalized Kronecker delta, one obtains
\begin{align}
\delta^{\mu\nu}_{\lambda\rho}
S^\lambda_\mu S^\rho_\nu
&=
\left(
\delta^\mu_\lambda\delta^\nu_\rho
-
\delta^\mu_\rho\delta^\nu_\lambda
\right)
S^\lambda_\mu S^\rho_\nu
\nonumber\ \,, \\
&=
\left(S^\mu_\mu\right)^2
-
S^\nu_\mu S^\mu_\nu
\nonumber\ \,, \\
&=
S^2-S_{\mu\nu}S^{\mu\nu}\,.
\label{eq:decompositionfirstnmg}
\end{align}
The first term is \(S^2=R^2/16\). The second follows from Eq.~\eqref{Schouten} 
\begin{align}
S_{\mu\nu}S^{\mu\nu}
&=
\left(
R_{\mu\nu}
-
\frac{1}{4}g_{\mu\nu}R
\right)
\left(
R^{\mu\nu}
-
\frac{1}{4}g^{\mu\nu}R
\right)
\nonumber\ \,,\\
&=
R_{\mu\nu}R^{\mu\nu}
-
\frac{1}{2}R^2
+
\frac{1}{16}
g_{\mu\nu}g^{\mu\nu}R^2\,.
\end{align}
Since \(g_{\mu\nu}g^{\mu\nu}=3\), one finds
\begin{equation}
S_{\mu\nu}S^{\mu\nu}
=
R_{\mu\nu}R^{\mu\nu}
-
\frac{5}{16}R^2.
\end{equation}
Substitution into Eq.~\eqref{eq:decompositionfirstnmg} then gives
\begin{equation}
\delta^{\mu\nu}_{\lambda\rho}
S^\lambda_\mu S^\rho_\nu
=
-R_{\mu\nu}R^{\mu\nu}
+
\frac{3}{8}R^2.
\label{SSchouten}
\end{equation}
The quadratic-curvature combination in the NMG action, therefore, satisfies
\begin{equation}
R_{\mu\nu}R^{\mu\nu}
-
\frac{3}{8}R^2
=
-
\delta^{\mu\nu}_{\lambda\rho}
S^\lambda_\mu S^\rho_\nu.
\end{equation}
Using \(R=4S\), the NMG Lagrangian can consequently be written entirely in terms of the Schouten tensor as follows
\begin{equation}
\mathcal L_{\rm NMG}
=
\kappa
\left(
4\sigma_1S
-
2\lambda
+
\frac{\sigma_2}{\mu^2}
\delta^{\mu\nu}_{\lambda\rho}
S^\lambda_\mu S^\rho_\nu
\right).
\label{LagNMG}
\end{equation}
Further conventions and identities involving generalized Kronecker deltas are collected in Appendix~\ref{app:asymptotic-expansions}.

\subsection{Metric variation and field equations}

The boundary analysis requires the complete metric variation of the
action, including the total derivative that is usually discarded when
deriving the bulk equations. An arbitrary variation can be separated
as
\begin{equation}
    \delta I_{\rm NMG}
    =
    \int_{\mathcal M}
    d^3x\,\sqrt{|g|}\,
    \mathcal E_{\mu\nu}\delta g^{\mu\nu}
    +
    \int_{\mathcal M}
    d^3x\,\sqrt{|g|}\,
    \nabla_\mu\Theta^\mu(g,\delta g)\,.
    \label{deltaINMG}
\end{equation}
The tensor equation \(\mathcal E_{\mu\nu}=0\) reproduces the bulk equation~\eqref{eq:NMG-conventional-field-equation}. The vector \(\Theta^\mu(g,\delta g)\) is the surface current, also called the presymplectic potential current, generated by the integrations by parts. On shell, the
bulk contribution vanishes and the variation reduces to the flux of
\(\Theta^\mu\) through the boundary.

Since the NMG Lagrangian depends algebraically on the Riemann tensor,
both contributions can be organized by
\begin{equation}
    E^{\mu\nu}_{\lambda\rho}
    :=
    \frac{\partial\mathcal L_{\rm NMG}}
         {\partial R^{\lambda\rho}_{\mu\nu}}\,.
    \label{eq:NMG-E-definition}
\end{equation}
The tensor \(E^{\mu\nu}_{\lambda\rho}\) inherits the algebraic symmetries of the Riemann tensor. In terms of this quantity, the bulk equations and the surface current take the general form~\cite{Padmanabhan:2011ex}
\begin{align}
    \mathcal E_{\mu\nu}
    &=
    E_\mu{}^{\lambda\rho\sigma}
    R_{\nu\lambda\rho\sigma}
    -
    \frac{1}{2}g_{\mu\nu}\mathcal L_{\rm NMG}
    -
    2\nabla^\lambda\nabla^\rho
    E_{\mu\lambda\rho\nu}\,,
    \label{Emunu}
    \\
    \Theta^\mu
    &=
    2\delta\Gamma^\lambda_{\nu\rho}
    E_\lambda{}^{\rho\mu\nu}
    -
    2\delta g_{\nu\sigma}
    \nabla_\rho E^{\nu\mu\rho\sigma}\,.
    \label{Thetamu}
\end{align}
These expressions follow from the Palatini identity and two integrations by parts.\footnote{The Palatini identity is
\(\delta R^\mu{}_{\nu\lambda\rho}
=2\nabla_{[\lambda}\delta\Gamma^\mu_{\rho]\nu}\).
Since the NMG Lagrangian contains no covariant derivatives of the curvature, \(E^{\mu\nu}_{\lambda\rho}\) is obtained by an ordinary derivative with respect to the Riemann tensor.}

For the Schouten form of the NMG Lagrangian in Eq.~\eqref{LagNMG}, direct differentiation gives
\begin{equation}
    E^{\mu\nu}_{\lambda\rho}
    =
    \frac{\kappa\sigma_1}{2}
    \delta^{\mu\nu}_{\lambda\rho}
    -
    \frac{2\kappa\sigma_2}{\mu^2}
    \left(
        \delta^{[\mu}_{[\lambda}
        S^{\nu]}_{\rho]}
        -
        \frac{1}{4}
        \delta^{\mu\nu}_{\lambda\rho}S
    \right)\,.
    \label{Etensor}
\end{equation}
The first term is the Einstein--Hilbert contribution. The second
contains the Schouten-tensor dependence generated by the
quadratic-curvature interaction.

To evaluate the divergence of Eq.~\eqref{Etensor}, recall that the contracted Bianchi identity implies
\(\nabla_\mu S^\mu_\nu=\nabla_\nu S\). Introducing the three-index Cotton tensor as
\(C_{\lambda\mu\nu}:=2\nabla_{[\nu}S_{\mu]\lambda}\), the divergence reduces to
\begin{equation}
    \nabla^\rho E^{\mu\nu}_{\lambda\rho}
    =
    -\frac{\kappa\sigma_2}{2\mu^2}
    C_\lambda{}^{\mu\nu}\,.
    \label{nablaEtensor}
\end{equation}
This relation isolates the derivative part of the higher-curvature contribution in a geometrically useful form.\footnote{In three dimensions, the Cotton tensor replaces the Weyl tensor as a local measure of conformal curvature. Its vanishing is the local condition for conformal flatness. The two-index dual
\(C_{\mu\nu}=\epsilon_\mu{}^{\alpha\beta}\nabla_\alpha S_{\beta\nu}\)
is symmetric, traceless, and covariantly conserved.}

Substitution of Eqs.~\eqref{Etensor} and
\eqref{nablaEtensor} into Eq.~\eqref{Emunu} gives the same field
equations in Schouten form
\begin{equation}
    \kappa^{-1}\mathcal E^\mu_\nu
    =
    \sigma_1G^\mu_\nu
    +
    \lambda\delta^\mu_\nu
    +
    \frac{\sigma_2}{2\mu^2}
    \left(
        \delta^{\mu\alpha\beta}_{\nu\lambda\rho}
        S^\lambda_\alpha S^\rho_\beta
        +
        2\nabla_\lambda C_\nu{}^{\lambda\mu}
    \right)\,.
    \label{eom2}
\end{equation}
The equivalence with
Eq.~\eqref{eq:NMG-conventional-field-equation} follows from the
three-dimensional curvature decomposition and the identity
\begin{equation}
    \mathcal K^\mu_\nu
    =
    -
    \left(
        \delta^{\mu\alpha\beta}_{\nu\lambda\rho}
        S^\lambda_\alpha S^\rho_\beta
        +
        2\nabla_\lambda C_\nu{}^{\lambda\mu}
    \right).
    \label{eq:NMG-K-tensor}
\end{equation}
The generalized-delta contraction is algebraic in the curvature,
whereas the divergence of the Cotton tensor contains fourth
derivatives of the metric. Expressing both terms in terms of the
Ricci tensor reproduces Eq.~\eqref{eq:NMG-K-tensor-Ricci}.

Taking the trace provides a useful consistency check. The Cotton
contribution has vanishing trace, and therefore no derivatives of the
curvature remain in the resulting equation. Using
\(G^\mu_\mu=-R/2=-2S\), one obtains
\begin{equation}
    2\sigma_1S
    =
    \frac{\sigma_2}{2\mu^2}
    \delta^{\mu\nu}_{\lambda\rho}
    S^\lambda_\mu S^\rho_\nu
    +
    3\lambda.
    \label{treom2}
\end{equation}
Although the complete field equations are fourth order, their trace
contains only algebraic contractions of the curvature. In terms of
the Ricci tensor, it is equivalently written as
\begin{equation}
    -\frac{\sigma_1}{2}R
    +
    3\lambda
    -
    \frac{\sigma_2}{2\mu^2}
    \left(
        R_{\mu\nu}R^{\mu\nu}
        -
        \frac{3}{8}R^2
    \right)
    =
    0.
\end{equation}

The surface current follows from the same metric variation. Using
\(
(g^{-1}\delta g)^\lambda_\nu
=
g^{\lambda\sigma}\delta g_{\sigma\nu}
\)
for the mixed-index metric variation, Eqs.~\eqref{Etensor} and
\eqref{nablaEtensor} reduce Eq.~\eqref{Thetamu} to
\begin{align}
    \Theta^\mu
    ={}&
    2\delta\Gamma^\sigma_{\nu\rho}g^{\rho\alpha}
    \left[
        \frac{\kappa\sigma_1}{2}
        \delta^{\mu\nu}_{\sigma\alpha}
        -
        \frac{2\kappa\sigma_2}{\mu^2}
        \left(
            \delta^{[\mu}_{[\sigma}
            S^{\nu]}_{\alpha]}
            -
            \frac{1}{4}
            \delta^{\mu\nu}_{\sigma\alpha}S
        \right)
    \right] -
    \frac{\kappa\sigma_2}{\mu^2}
    (g^{-1}\delta g)^\lambda_\nu
    C_\lambda{}^{\nu\mu}.
    \label{Thetamu2}
\end{align}
On shell, the variation of the bulk action is entirely determined by
the flux of this current through the boundary. Its radial expansion
therefore provides the starting point for identifying the boundary
terms required for a finite and well-posed variational principle.

\subsection{Constant-curvature spaces}
\label{sec:constant-curvature-spaces}

The simplest solutions of the NMG field equations are maximally symmetric spacetimes. Their curvature is characterized by a constant \(\Lambda\) through
\begin{equation}
    R^{\mu\nu}_{\lambda\rho}
    =
    \Lambda\,
    \delta^{\mu\nu}_{\lambda\rho}\,.
    \label{eq:NMG-constant-curvature}
\end{equation}
This condition fixes all contractions of the curvature tensor. In three dimensions, the Ricci tensor, Ricci scalar, Einstein tensor, and Schouten tensor reduce to
\begin{equation}
    R^\mu{}_\nu
    =
    2\Lambda\delta^\mu_\nu\,,
    \qquad
    R=6\Lambda\,,
    \qquad
    G^\mu{}_\nu
    =
    -\Lambda\delta^\mu_\nu\,,
    \qquad
    S^\mu{}_\nu
    =
    \frac{\Lambda}{2}\delta^\mu_\nu\,.
    \label{eq:NMG-constant-curvature-contractions}
\end{equation}
Since the Schouten tensor is covariantly constant on this background, the Cotton tensor vanishes:
\begin{equation}
    C_{\lambda\mu\nu}=0\,.
\end{equation}
Thus, the higher derivative contribution to the NMG equations disappears, and the vacuum condition becomes purely algebraic.
To evaluate the remaining curvature-squared term in Eq.~\eqref{eom2}, substitute \(S^\mu_\nu=\Lambda\delta^\mu_\nu/2\). The generalized-delta contraction becomes
\begin{align}
    \delta^{\mu\alpha\beta}_{\nu\lambda\rho}
    S^\lambda_\alpha S^\rho{}_\beta
    &=
    \frac{\Lambda^2}{4}
    \delta^{\mu\alpha\beta}_{\nu\alpha\beta} =
    \frac{\Lambda^2}{2}\delta^\mu_\nu\,.
    \label{eq:NMG-constant-Schouten-contraction}
\end{align}
In the second line, we used
\(\delta^{\mu\alpha\beta}_{\nu\alpha\beta}=2\delta^\mu_\nu\)
in three dimensions. The field equations consequently reduce to
\begin{equation}
    \mathcal E^\mu_\nu
    =
    \kappa
    \left(
        \lambda
        -
        \sigma_1\Lambda
        +
        \frac{\sigma_2\Lambda^2}{4\mu^2}
    \right)
    \delta^\mu_\nu\,.
\end{equation}
A maximally symmetric geometry is therefore an NMG vacuum if its effective cosmological constant satisfies
\begin{equation}
    \mathcal E_0(\Lambda)
    :=
    \lambda
    -
    \sigma_1\Lambda
    +
    \frac{\sigma_2\Lambda^2}{4\mu^2}
    =
    0\,.
    \label{Lambdaeffconstraint}
\end{equation}
The notation \(\mathcal E_0(\Lambda)\) emphasizes that this is the field equation evaluated on the constant-curvature ansatz. It also makes explicit the distinction between the bare cosmological parameter \(\lambda\), which is a coupling in the action, and the effective cosmological constant \(\Lambda\), which determines the curvature of a particular vacuum.

Multiplying Eq.~\eqref{Lambdaeffconstraint} by \(4\mu^2\) gives the quadratic equation
\begin{equation}
    \sigma_2\Lambda^2
    -
    4\sigma_1\mu^2\Lambda
    +
    4\mu^2\lambda
    =
    0\,.
\end{equation}
Using \(\sigma_1^2=\sigma_2^2=1\), its two roots can be written as
\begin{equation}
    \Lambda_\pm
    =
    2\sigma_2\mu^2
    \left(
        \sigma_1
        \pm
        \sqrt{\Delta}
    \right),
    \qquad
    \Delta
    :=
    1-\frac{\sigma_2\lambda}{\mu^2}\,.
    \label{Lambdapm}
\end{equation}
The discriminant \(\Delta\) determines the number of real maximally symmetric vacua. For \(\Delta>0\), the theory admits two distinct real vacua. At \(\Delta=0\), the two roots coincide, defining the degenerate point discussed in the following subsection. If \(\Delta<0\), the theory has no real maximally symmetric vacuum.

The present analysis is restricted to a negative root of Eq.~\eqref{Lambdapm}. The corresponding effective AdS radius is defined by $\Lambda=-\ell^{-2}\,.$ Once an AdS branch has been selected, two combinations of the couplings are particularly useful
\begin{equation}
    \varpi_\pm
    :=
    \sigma_1
    \pm
    \frac{\sigma_2}{2\mu^2\ell^2}\,.
    \label{eq:NMG-varpi}
\end{equation}
The combinations \(\varpi_+\) and \(\varpi_-\) have different physical meanings. The first controls the degeneracy of the vacuum polynomial, while the second determines the effective coupling of the Einstein sector. The role of \(\varpi_-\) can already be identified by evaluating the \(E\)-tensor~\eqref{Etensor} on the AdS background. For \(S^\mu_\nu=\Lambda\delta^\mu_\nu/2\), the Schouten-dependent expression in Eq.~\eqref{Etensor} satisfies
\begin{equation}
    \delta^{[\mu}_{[\lambda}
    S^{\nu]}_{\rho]}
    -
    \frac{1}{4}
    \delta^{\mu\nu}_{\lambda\rho}S
    =
    -\frac{\Lambda}{8}
    \delta^{\mu\nu}_{\lambda\rho}\,.
    \label{eq:NMG-E-AdS-intermediate}
\end{equation}
The Einstein and curvature-squared contributions therefore combine according to
\begin{align}
    \left.
    E^{\mu\nu}_{\lambda\rho}
    \right|_{\rm AdS}
    &=
    \left(
        \frac{\kappa\sigma_1}{2}
        +
        \frac{\kappa\sigma_2\Lambda}{4\mu^2}
    \right)
    \delta^{\mu\nu}_{\lambda\rho}
    \nonumber\\
    &=
    \frac{\kappa}{2}
    \left(
        \sigma_1
        +
        \frac{\sigma_2\Lambda}{2\mu^2}
    \right)
    \delta^{\mu\nu}_{\lambda\rho}
    \nonumber\\
    &=
    \frac{\kappa\varpi_-}{2}
    \delta^{\mu\nu}_{\lambda\rho}\,.
    \label{eq:NMG-E-AdS}
\end{align}
Thus, on a constant-curvature AdS background, the NMG \(E\)-tensor has the same tensorial structure as in Einstein gravity, multiplied by the effective factor \(\varpi_-\). This factor consequently appears in the boundary response and in the Noether--Wald charges of locally AdS solutions.\footnote{The terminology effective coupling refers to the overall factor multiplying the Einstein-sector quantities. Its sign is relevant for the kinetic normalization and the boundary central charge, but it does not by itself establish the unitarity of the complete theory.}

The same factor appears in the on-shell value of the NMG Lagrangian. From Eq.~\eqref{eq:NMG-constant-curvature-contractions},
\begin{equation}
    R_{\mu\nu}R^{\mu\nu}
    =
    12\Lambda^2,
    \qquad
    R^2=36\Lambda^2\,.
\end{equation}
The quadratic-curvature combination consequently reduces to
\begin{equation}
    R_{\mu\nu}R^{\mu\nu}
    -
    \frac{3}{8}R^2
    =
    12\Lambda^2
    -
    \frac{3}{8}
    \left(36\Lambda^2\right)
    =
    -\frac{3}{2}\Lambda^2\,.
    \label{eq:NMG-quadratic-AdS}
\end{equation}
Substitution into the Lagrangian gives
\begin{equation}
    \left.
    \mathcal L_{\rm NMG}
    \right|_{\rm AdS}
    =
    \kappa
    \left(
        6\sigma_1\Lambda
        -
        2\lambda
        +
        \frac{3\sigma_2\Lambda^2}{2\mu^2}
    \right)\,.
    \label{eq:NMG-AdS-Lagrangian-before-vacuum}
\end{equation}
The vacuum equation~\eqref{Lambdaeffconstraint} may be solved for the bare parameter,
\begin{equation}
    \lambda
    =
    \sigma_1\Lambda
    -
    \frac{\sigma_2\Lambda^2}{4\mu^2}\,.
\end{equation}
Using this relation in Eq.~\eqref{eq:NMG-AdS-Lagrangian-before-vacuum}, the on-shell Lagrangian becomes
\begin{align}
    \left.
    \mathcal L_{\rm NMG}
    \right|_{\rm AdS}
    &=
    \kappa
    \left(
        4\sigma_1\Lambda
        +
        \frac{2\sigma_2\Lambda^2}{\mu^2}
    \right)
    \nonumber \,,\\
    &=
    4\kappa\Lambda
    \left(
        \sigma_1
        +
        \frac{\sigma_2\Lambda}{2\mu^2}
    \right) \,, 
    \nonumber\\
    &=
    4\kappa\Lambda\varpi_-\,.
    \label{eq:NMG-on-shell-AdS-Lagrangian}
\end{align}
The occurrence of the same factor in Eqs.~\eqref{eq:NMG-E-AdS} and \eqref{eq:NMG-on-shell-AdS-Lagrangian} provides the constant-curvature basis for the effective rescaling of Einstein-sector boundary quantities in NMG.

\subsection{Degenerate point}

The condition $\Delta=0$ defines the degenerate point of NMG. It fixes
the bare cosmological parameter and causes the two effective
cosmological constants to coincide:
\begin{equation}
    \lambda=\sigma_2\mu^2\,,
    \qquad
    \Lambda_+=\Lambda_-
    =
    2\sigma_1\sigma_2\mu^2\,.
    \label{eq:NMG-degenerate-vacuum}
\end{equation}
It is convenient to introduce the dimensionless combinations
\begin{equation}
    \varpi_{\pm}
    :=
    \sigma_1
    \pm
    \frac{\sigma_2}{2\mu^2\ell^2}\,.
    \label{eq:varpi-minus-degenerate}
\end{equation}
For an AdS vacuum, the degeneracy condition is equivalently expressed
as
\begin{equation}
    \varpi_+=0\,.
    \label{degcondition}
\end{equation}

The combination $\varpi_+$ multiplies the first subleading terms in
the asymptotic field equations that, at generic couplings, impose
$g_{(1)ij}=0$. At the degenerate point, the rank of these equations is
reduced and a nonvanishing $g_{(1)ij}$ may therefore appear in the
Fefferman--Graham expansion
\cite{Alishahiha:2010bw,Giribet:2010ed,Kwon:2011jz,Cunliff:2013en}.
This coefficient produces a weaker fall-off than the standard
Brown--Henneaux behavior and defines an additional holographic source
at the conformal boundary. The same degenerate point admits the static
and rotating hairy black holes used below as the principal examples
of non-Einstein configurations
\cite{Oliva:2009ip,Giribet:2009qz}.

\subsection{Critical point}

A different special point of the parameter space is defined by
\begin{equation}
    \varpi_-=0\,.
    \label{eq:NMG-critical-condition}
\end{equation}
This effective coupling multiplies the conserved charges and the
central charge of the asymptotic symmetry algebra. Under
Brown--Henneaux boundary conditions, the left- and right-moving
Virasoro algebras have equal central charges
\cite{Brown:1986nw,Liu:2009pha,Liu:2009bk,Kraus:2006wn},
\begin{equation}
    c_L=c_R
    =
    \frac{3\ell}{2G}
    \left(
        \sigma_1
        -
        \frac{\sigma_2}{2\mu^2\ell^2}
    \right)
    =
    \frac{3\ell}{2G}\varpi_-\,.
    \label{eq:NMG-central-charge}
\end{equation}
Both central charges therefore vanish at the critical point. The
Wald charges of the BTZ black hole vanish there as well
\cite{Clement:2009gq,Wald:1993nt,Iyer:1994ys,Wald:1999wa}.

Nonvanishing charges can nevertheless arise when logarithmic terms
are admitted through relaxed boundary conditions
\cite{Grumiller:2008es,Grumiller:2008qz,Henneaux:2009pw,
Maloney:2009ck,Liu:2009kc}. In particular, NMG admits a black hole
solution with logarithmic fall-off at the critical point
\cite{Clement:2009ka,Hohm:2010jc,Giribet:2010ed}. 

\section{Asymptotic analysis and renormalization}
\label{sec:asymptotics}

The construction of boundary terms must be compatible with the
asymptotic solutions of the NMG field equations. In particular, the
radial dependence of the metric cannot be inferred solely from the
Fefferman--Graham coordinate choice. The field equations determine
which coefficients are allowed, which are constrained by the
boundary metric, and which represent independent asymptotic data.

The Gauss-normal decomposition and the Fefferman--Graham metric were
introduced in Eqs.~\eqref{eq:radial-metric-decomposition} and
\eqref{eq:fg-metric-background}, respectively. For the power-law
sector considered in this chapter, we use the asymptotic expansion
\begin{equation}
    \bar g_{ij}(z,x)
    =
    g_{(0)ij}(x)
    +
    \frac{z}{\ell}g_{(1)ij}(x)
    +
    \frac{z^2}{\ell^2}g_{(2)ij}(x)
    +
    \mathcal O(z^3).
    \label{eq:NMG-integer-expansion}
\end{equation}
Unlike the Einstein expansion~\eqref{eq:Einstein-FG-expansion},
this ansatz allows a term linear in $z$. Whether $g_{(1)ij}$ vanishes
is determined by the NMG field equations rather than by the
Fefferman--Graham gauge. Logarithmic terms, which may arise at the
critical point under relaxed boundary conditions, are excluded from
the present analysis.

The field equations are now decomposed into radial and tangential
components and evaluated using
Eq.~\eqref{eq:NMG-integer-expansion}. The resulting constraints
provide the input for the subsequent expansion of the on-shell action
and its boundary variation.

\subsection{Field equations}

The NMG equations~\eqref{eom2} separate naturally according to the radial foliation. The \(zz\) component acts as a radial or Hamiltonian constraint, the mixed \(zi\) component acts as a momentum constraint, and the tangential components govern the radial evolution of the induced metric.\footnote{The terms Hamiltonian and momentum constraint refer to the structure of the radial decomposition. The radial coordinate is spacelike and should not be interpreted as physical time.}

In Gauss-normal coordinates of Eq.~\eqref{eq:radial-metric-decomposition}, the independent components of the equations of motion can be organized as
\begin{subequations}
\label{eq:NMG-Gaussian-components}
\begin{align}
    \kappa^{-1}\mathcal E^z_z
    ={}&
    \sigma_1G^z_z
    +
    \lambda
    +
    \frac{\sigma_2}{\mu^2}
    \left[
        \frac{1}{2}
        \left(
            S^i_iS^j_j
            -
            S^i_jS^j_i
        \right)
        +
        \nabla_i C_z{}^{iz}
    \right]\,,
    \label{eq:NMG-Gaussian-zz}
    \\
    \kappa^{-1}\mathcal E^i_j
    ={}&
    \sigma_1G^i_j
    +
    \lambda\delta^i_j
    +
    \frac{\sigma_2}{2\mu^2}
    \Big[
        2S^i_kS^k_j
        +
        2S^i_zS^z_j
        -
        2S^i_j
        \left(
            S^k_k+S^z_z
        \right)
        +
        2\nabla_zC_j{}^{zi}
        +
        2\nabla_kC_j{}^{ki}
        \nonumber\\
    &\hspace{1.7cm}
        +
        \delta^i_j
        \left(
            S^k_k
            \left[
                S^m_m+2S^z_z
            \right]
            -
            2S^k_zS^z_k
            -
            S^k_mS^m_k
        \right)
    \Big]\,,
    \label{eq:NMG-Gaussian-ij}
    \\
    \kappa^{-1}\mathcal E^i_z
    ={}&
    \sigma_1G^i_z
    +
    \frac{\sigma_2}{\mu^2}
    \left[
        S^i_jS^j_z
        -
        S^k_kS^i_z
        +
        \nabla_zC_z{}^{zi}
        +
        \nabla_kC_z{}^{ki}
    \right]\,.
    \label{eq:NMG-Gaussian-iz}
\end{align}
\end{subequations}
These equations are exact and have not yet been expanded near the conformal boundary. The Gauss--Codazzi relations express their curvature tensors in terms of the intrinsic and extrinsic geometry of the constant-\(z\) hypersurfaces. The required component expansions are collected in Appendix~\ref{sec:GN-FG}. In particular, the Einstein tensor may be recovered from the Schouten tensor through
\begin{equation}
    G^\mu_\nu
    =
    S^\mu_\nu
    -
    S\delta^\mu_\nu\,.
\end{equation}

We now insert the power-law metric expansion given in Eq.~\eqref{eq:NMG-integer-expansion}. All boundary indices are raised and lowered with \(g_{(0)ij}\), while \(\mathcal D_{(0)i}\) and \(\mathcal R_{(0)}\) denote the covariant derivative and Ricci scalar constructed from this boundary metric. We also use the compact contractions
\[
    \text{tr }g_{(1)}
    :=
    g_{(0)}^{ij}g_{(1)ij}\,,
    \qquad
    \text{tr }g_{(1)}^2
    :=
    g^i_{(1)j}g^j_{(1)i}\,,
\]
and similarly for \(g_{(2)ij}\).

After substituting the curvature expansions into Eqs.~\eqref{eq:NMG-Gaussian-components}, the radial constraint becomes
\begin{align}
    \kappa^{-1}\mathcal E^z_z
    ={}&
    \mathcal E_0
    -
    \frac{z}{2\ell^3}
    \varpi_+\text{tr }g_{(1)}
    -
    \frac{z^2}{8\ell^4}
    \Bigg\{
        \varpi_-
        \left[
            \left(\text{tr }g_{(1)}\right)^2
            -
            \text{tr }g_{(1)}^2
        \right]
        \nonumber\\
    &\hspace{2.0cm}
        -
        2\varpi_+
        \left[
            \text{tr }g_{(1)}^2
            +
            \left(\text{tr }g_{(1)}\right)^2
            -
            4\text{tr }g_{(2)}
            -
            2\ell^2\mathcal R_{(0)}
        \right]
    \Bigg\}
    +
    \mathcal O(z^3)\,.
    \label{eomzz}
\end{align}
The tangential equations have the corresponding expansion
\begin{align}
    \kappa^{-1}\mathcal E^i_j
    ={}&
    \mathcal E_0\delta^i_j
    +
    \frac{\varpi_+}{2\ell^3}
    \Bigg[
        \left(
            g^i_{(1)j}
            -
            \delta^i_j\text{tr }g_{(1)}
        \right)z
        -
        \frac{z^2}{2\ell}
        \left(
            g^i_{(1)k}g^k_{(1)j}
            -
            \delta^i_j\text{tr }g_{(1)}^2
        \right)
    \Bigg]
    +
    \mathcal O(z^3)\,,
    \label{eomij}
\end{align}
while the mixed equation begins at quadratic order:
\begin{equation}
    \kappa^{-1}\mathcal E^i_z
    =
    \frac{\varpi_+z^2}{2\ell^3}
    \left(
        \mathcal D_{(0)j}g_{(1)}^{ji}
        -
        \mathcal D_{(0)}^i\text{tr }g_{(1)}
    \right)
    +
    \mathcal O(z^3)\,.
    \label{eomzi}
\end{equation}
The quantities \(\mathcal E_0\) and \(\varpi_\pm\) are those defined in Eqs.~\eqref{Lambdaeffconstraint} and \eqref{eq:NMG-varpi}. Their appearance at different radial orders separates the condition selecting the leading AdS vacuum from the conditions controlling the subleading metric coefficients.

At leading order, Eqs.~\eqref{eomzz} and \eqref{eomij} require
\begin{equation}
    \mathcal E_0=0.
    \label{eq:NMG-leading-vacuum-condition}
\end{equation}
Thus, the same polynomial that determines the maximally symmetric vacua also fixes the leading curvature radius of an asymptotically AdS solution. In particular, the radius \(\ell\) appearing in the Fefferman--Graham metric must correspond to one of the AdS roots of Eq.~\eqref{Lambdapm}; it is not an independent parameter.

Away from the degenerate point, \(\varpi_+\neq0\), the coefficient of \(z\) in the radial equation~\eqref{eomzz} then imposes
\begin{equation}
    \text{tr }g_{(1)}=0\,.
    \label{eq:NMG-g1-trace-zero}
\end{equation}
Using this result in the order-\(z\) part of Eq.~\eqref{eomij} gives
\begin{equation}
    g_{(1)ij}=0\,.
    \label{eq:NMG-g1-zero}
\end{equation}
The mixed equation~\eqref{eomzi} is then automatically satisfied at the displayed order.

Once \(g_{(1)ij}=0\) is imposed, the coefficient of \(z^2\) in Eq.~\eqref{eomzz} reduces to
\begin{equation}
    -\frac{\varpi_+z^2}{\ell^4}
    \left(
        \text{tr }g_{(2)}
        +
        \frac{\ell^2}{2}\mathcal R_{(0)}
    \right)
    =
    0\,.
\end{equation}
For \(\varpi_+\neq0\), this gives
\begin{equation}
    \text{tr }g_{(2)}
    =
    -\frac{\ell^2}{2}\mathcal R_{(0)}\,.
    \label{eq:NMG-g2-trace}
\end{equation}
The result has the same form as the Einstein relation~\eqref{trg2condition}, but \(\ell\) is now the effective AdS radius selected by the NMG vacuum equation.

The generic power-law sector is therefore characterized by
\begin{equation}
    \mathcal E_0=0\,,
    \qquad
    g_{(1)ij}=0\,,
    \qquad
    \text{tr }g_{(2)}
    =
    -\frac{\ell^2}{2}\mathcal R_{(0)}\,.
    \label{g-solNMG}
\end{equation}
These relations are compatible with the standard Brown--Henneaux
power-law fall-offs after the corresponding gauge conditions are
imposed~\cite{Ciambelli:2020ftk,Ciambelli:2022vot}. They do not
determine the complete coefficient $g_{(2)ij}$: its trace is fixed
by $g_{(0)ij}$, while its remaining part distinguishes different
bulk solutions and determines the holographic stress tensor. We
refer to this as the Brown--Henneaux, or equivalently
$g_{(1)ij}=0$, sector
\cite{Ayon-Beato:2009cgh,Giribet:2010ed,Alishahiha:2010bw,
Kwon:2011jz,Cunliff:2013en}.

The conclusion \(g_{(1)ij}=0\) changes at the degenerate point. When \(\varpi_+=0\), all the terms linear in \(z\) displayed in Eqs.~\eqref{eomzz} and \eqref{eomij} vanish identically. The field equations at this order therefore lose the conditions that removed \(g_{(1)ij}\) in the generic theory. The coefficient \(g_{(1)ij}\) becomes additional asymptotic data, although it remains subject to constraints at higher radial orders.

At the degenerate point, Eq.~\eqref{degcondition} gives \(\varpi_-=2\sigma_1\neq0\). The order-\(z^2\) term in Eq.~\eqref{eomzz} consequently imposes
\begin{equation}
    \text{tr }g_{(1)}^2
    =
    \left(\text{tr }g_{(1)}\right)^2\,.
    \label{trg1}
\end{equation}
This condition has a simple interpretation in terms of the two-dimensional mixed tensor
\(g^i_{(1)j}\). The Cayley--Hamilton identity for a \(2\times2\) matrix gives
\begin{equation}
    g_{(1)}^2
    -
    \left(\text{tr }g_{(1)}\right)g_{(1)}
    +
    \det\left(g^i_{(1)j}\right) \mathbb{I}
    =
    0\,,
\end{equation}
together with
\begin{equation}
    2\det\left(g^i_{(1)j}\right)
    =
    \left(\text{tr }g_{(1)}\right)^2
    -
    \text{tr }g_{(1)}^2\,.
\end{equation}
Equation~\eqref{trg1} therefore implies
\begin{equation}
    \det\left(g^i_{(1)j}\right)=0\,,
\end{equation}
and the Cayley--Hamilton identity reduces to
\begin{equation}
    g^k_{(1)i}g_{(1)jk}
    =
    \left(\text{tr }g_{(1)}\right)
    g_{(1)ij}\,.
    \label{eq:NMG-g1-Cayley-Hamilton}
\end{equation}
The field equations thus permit a nonzero \(g_{(1)ij}\), but do not leave it completely unconstrained.

The asymptotic solutions relevant to the renormalization problem separate into three sectors. The first is the constant-curvature Einstein sector described in Section~\ref{sec:constant-curvature-spaces}. It contains AdS$_3$ and locally AdS quotients such as the BTZ black hole~\cite{Banados:1992wn,Banados:1992gq}.

The second sector contains non-Einstein geometries with the standard power-law behavior \(g_{(1)ij}=0\). Their Schouten tensor is not covariantly constant, but their leading radial fall-off agrees with the generic sector described by Eq.~\eqref{g-solNMG}.

The third sector consists of non-Einstein solutions at the degenerate point for which \(g_{(1)ij}\neq0\), subject to Eq.~\eqref{trg1}. This sector includes the static and rotating hairy black holes~\cite{Oliva:2009ip,Giribet:2009qz,Bergshoeff:2009aq}. Some of these geometries are conformally flat and therefore have a vanishing Cotton tensor, despite their non-Einstein nature. \footnote{In three dimensions, \(C_{\lambda\mu\nu}=0\) implies local conformal flatness, not constant curvature. A conformally flat metric may therefore be non-Einstein.}

The presence of \(g_{(1)ij}\) changes the radial orders appearing separately in the bulk and boundary contributions to the action. It must therefore be retained throughout the renormalization analysis. No conclusion about an additional divergence follows from \(g_{(1)ij}\neq0\) alone: the field-equation constraints must first be imposed and all contributions at the same radial order must be combined.

\subsection{Asymptotic expansion of the action and its variation}
\label{subsec:NMG-action-expansion}

The asymptotic field equations determine which coefficients of the
Fefferman--Graham expansion are allowed in each sector of NMG. We now use those results to identify the divergent terms in the
on-shell action at a finite radial cutoff and to determine whether they can be removed by
boundary terms compatible with the prescribed asymptotic data.

To keep the normalization of the action explicit, we expand the scalar
Lagrangian with the overall factor $\kappa$ removed:
\begin{equation}
    \kappa^{-1}\mathcal L_{\rm NMG}
    =
    \mathcal L_{(0)}
    +z\mathcal L_{(1)}
    +z^{2}\mathcal L_{(2)}
    +\mathcal O(z^{3})\,.
    \label{LagFG}
\end{equation}
Substitution of the Fefferman--Graham expansion into
Eq.~\eqref{LagNMG} gives
\begin{subequations} \label{Lag-coeff}
    \begin{align}
  \mathcal L_{(0)}	&= -2\mathcal{E}_0 -\frac{4\varpi_{-}}{\ell^2}\,, \qquad \mathcal L_{(1)}	=\frac{2\varpi_{-}\text{tr }  g_{(1)}}{\ell^{3}}\,, \label{L0L1} \\
    \mathcal L_{(2)}	&=\frac{\varpi_{+}}{4\ell^{4}}\left([\text{tr }  g_{(1)}]^{2}-\text{tr }  g_{(1)}^{2}\right)
	-\frac{\varpi_{-}}{2\ell^{4}}\left([\text{tr } g_{(1)}]^{2}+2\text{tr } g_{(1)}^{2}-2\left(2\text{tr } g_{(2)}+\ell^{2}\mathcal{R}_{(0)}\right)\right) \label{L2} \,. 
    \end{align}
\end{subequations}
These expressions are obtained directly from the curvature expansion.
The conditions following from the field equations have not yet been
imposed.

The volume element contributes additional powers of the radial
coordinate. Its expansion is
\begin{align}
    \sqrt{|g|}
    ={}&
    \frac{\ell^{3}}{z^{3}}
    \sqrt{|g_{(0)}|}
    \Bigg[
        1
        +\frac{z}{2\ell}\text{tr }g_{(1)}
        +\frac{z^{2}}{2\ell^{2}}
        \left(
            \text{tr }g_{(2)}
            +\frac{1}{4}
            \big(\text{tr }g_{(1)}\big)^{2}
            -\frac{1}{2}\text{tr }g_{(1)}^{2}
        \right)
        +\mathcal O(z^{3})
    \Bigg]\,.
    \label{sqrtdetgexp}
\end{align}
This result follows from the matrix identity
\begin{equation}
    \sqrt{\det\!\left(
        \mathbf 1+\epsilon A+\epsilon^{2}B
    \right)}
    =
    1+\frac{\epsilon}{2}\text{tr }A
    +\epsilon^{2}
    \left[
        \frac{1}{2}\text{tr }B
        +\frac{1}{8}\big(\text{tr }A\big)^{2}
        -\frac{1}{4}\text{tr }A^{2}
    \right]
    +\mathcal O(\epsilon^{3})\,.
\end{equation}

Multiplying Eqs.~\eqref{LagFG} and \eqref{sqrtdetgexp}, the bulk
action evaluated on $\mathcal M_\epsilon$ takes the form
\begin{equation}
    I_{\rm NMG}
    =
    \kappa
    \int_{\mathcal M_\epsilon}d^{3}x\,
    \sqrt{|g_{(0)}|}
    \left[
        \frac{\mathcal L_{(-3)}}{z^{3}}
        +\frac{\mathcal L_{(-2)}}{z^{2}}
        +\frac{\mathcal L_{(-1)}}{z}
        +\mathcal O(1)
    \right]\,,
    \label{eq:NMG-bulk-action-expansion}
\end{equation}
where
\begin{subequations}
\label{eq:NMG-integrand-coefficients}
    \begin{align}
        \mathcal{L}_{(-3)} &= \ell^3 \mathcal{L}_{(0)}\,, \qquad \mathcal{L}_{(-2)}= \frac{\ell^2}{2} (2\ell \mathcal{L}_{(1)} + \text{tr } g_{(1)} \mathcal{L}_{(0)} )\,, \\
        \mathcal{L}_{(-1)}	&=\frac{\ell}{8}\left(\left[(\text{tr } g_{(1)})^{2}-2\text{tr } g_{(1)}^{2}+4\text{tr } g_{(2)}\right]\mathcal{L}_{(0)}+4\ell\,\text{tr } \,g_{(1)}\mathcal{L}_{(1)}+8\ell^{2}\mathcal{L}_{(2)}\right) \,,
    \end{align}
\end{subequations}
These coefficients distinguish the different types of cutoff
dependence. Radial integration from $z=\epsilon$ produces an
$\epsilon^{-2}$ divergence from the $z^{-3}$ term, an
$\epsilon^{-1}$ divergence from the $z^{-2}$ term, and a logarithmic
term from the $z^{-1}$ contribution.

For the nondegenerate branch, we find the relations given in Eq.~\eqref{g-solNMG}. Using them, the coefficients in Eq.~\eqref{eq:NMG-integrand-coefficients} reduce to
\begin{equation}
    \mathcal L_{(-3)}
    =
    -4\ell\varpi_{-},
    \qquad
    \mathcal L_{(-2)}
    =
    0,
    \qquad
    \mathcal L_{(-1)}
    =
    \ell\varpi_{-}\mathcal R_{(0)}.
    \label{eq:NMG-generic-divergent-coefficients}
\end{equation}
Thus, before the boundary terms are added, the on-shell bulk
action contains the leading volume divergence and, for a curved
boundary representative, a logarithmic divergence proportional
to the two-dimensional Euler density. On a closed boundary, the
integral of this density is topological, so its variation does not
modify the local stress tensor. For the flat boundary metrics used
in the explicit solutions, $\mathcal R_{(0)}=0$ and this logarithmic
term is absent.

The cancellation of these divergences is only one requirement on the
boundary completion. The on-shell variation must also be finite and
must depend exclusively on variations of the prescribed boundary
sources. For the sectors satisfying $g_{(1)ij}=0$, the only independent
source is $g_{(0)ij}$. At the degenerate point, the field equations
allow $g_{(1)ij}\neq0$, and the generalized Dirichlet problem fixes
both $g_{(0)ij}$ and $g_{(1)ij}$, subject to their asymptotic
constraints. These sectors are considered separately below.

\subsubsection{Constant-curvature spaces}

For the constant-curvature AdS sector described by
Eqs.~\eqref{eq:NMG-constant-curvature} and
\eqref{eq:NMG-constant-curvature-contractions}, the Schouten tensor is
covariantly constant and the Cotton tensor vanishes. The on-shell NMG
Lagrangian is therefore given by
Eq.~\eqref{eq:NMG-on-shell-AdS-Lagrangian} and is proportional to
$\varpi_-$. Combining this result with the expansion of the volume
element gives
\begin{align}
    I_{\rm NMG}\big|_{\rm E}
    ={}&
    -\frac{4\kappa\varpi_-}{\ell^{2}}
    \int_{\mathcal M_\epsilon}d^{3}x\,
    \sqrt{|g_{(0)}|}
    \left[
        \frac{\ell^{3}}{z^{3}}
        +\frac{\ell}{2z}\text{tr }g_{(2)}
        +\mathcal O(z)
    \right].
    \label{INMGE}
\end{align}
The first term produces the leading volume divergence. Since a
constant-curvature geometry also satisfies the Einstein equations,
the trace relation~\eqref{trg2condition} converts the second term into
the logarithmic contribution proportional to the two-dimensional
Euler density, in agreement with
Eq.~\eqref{eq:NMG-generic-divergent-coefficients}.

The power-law divergence and the corresponding divergent part of the
variation are removed by the extrinsic boundary term
\begin{equation}
    I_K
    =
    \kappa\varpi_-
    \int_{\partial \Sigma_\epsilon}
    d^2x\,\sqrt{|h|}\,K.
    \label{KT-NMG-E}
\end{equation}
In the limit $\mu\rightarrow\infty$, its coefficient approaches
$\kappa\sigma_{1}$, reproducing the one-half
Gibbons--Hawking--York prescription with the normal and
extrinsic-curvature conventions adopted in Chapter~\ref{Chapter4}.

On a constant-curvature background, the NMG $E$-tensor is the
Einstein $E$-tensor multiplied by $\varpi_-$. The bulk surface
term and the variation of Eq.~\eqref{KT-NMG-E} therefore combine
into $\varpi_-$ times the one-half Gibbons--Hawking--York
variation discussed around Eq.~\eqref{GHY/2}. Using the same
normal $n_\mu=N\delta_\mu^z$ and the boundary orientation fixed
in Chapter~\ref{Chapter4}, no additional normal convention is
required.

After inserting the asymptotic expansion and imposing the field
equations, the finite variation is
\begin{equation}
    \delta I_{\rm ren}\big|_{\rm EOM}
    =
    \frac{1}{2}
    \int_{\partial\mathcal M}d^{2}x\,
    \sqrt{|g_{(0)}|}
    \,\tau^{ij}\delta g_{(0)ij},
\end{equation}
with
\begin{equation}
    \tau^{ij}
    =
    \frac{2\kappa\varpi_{-}}{\ell}
    \left(
        g_{(2)}^{ij}
        -g_{(0)}^{ij}\text{tr }g_{(2)}
    \right).
    \label{tauijNMG}
\end{equation}
The variation of the integrated two-dimensional Euler density has been
omitted from this local response. Equation~\eqref{tauijNMG} is finite
and depends only on the boundary source $g_{(0)ij}$. It therefore
vanishes under Dirichlet boundary conditions,
$\delta g_{(0)ij}=0$. At the critical point
$\varpi_{-}=0$, the stress tensor vanishes for constant-curvature
solutions satisfying Brown--Henneaux boundary conditions.

\subsubsection{Nonconstant-curvature spaces with
\texorpdfstring{$g_{(1)ij}=0$}{g1 equal 0}}

For a nonconstant-curvature solution, the Schouten tensor is not
proportional to the identity, and the Cotton tensor need not vanish.
The constant-curvature reduction used above is therefore
unavailable, and the complete presymplectic potential must be
retained.

In Gauss normal coordinates, its contraction with the normal is
\begin{align}
    n_{\mu}\Theta^{\mu}
    ={}&
    2N\delta\Gamma^{k}_{ij}
    h^{j\ell}E^{zi}_{k\ell}
    +4\delta K^{i}_{j}E^{zj}_{zi}+
    \left(
        2K^{j}_{k}E^{zk}_{zi}
        -\frac{\kappa\sigma_{2}}{\mu^{2}}NC_{i}{}^{zj}
    \right)
    (h^{-1}\delta h)^{i}_{j}\,.
    \label{bdytermnonconstant}
\end{align}
The components of the $E$-tensor entering this expression follow from
Eq.~\eqref{Etensor}:
\begin{align}
    E^{zi}_{zj}
    &=
    \kappa
    \left[
        \frac{\sigma_{1}}{2}\delta^{i}_{j}
        -\frac{\sigma_{2}}{2\mu^{2}}
        \left(
            S^{i}_{j}
            +\delta^{i}_{j}S^{z}_{z}
            -\delta^{i}_{j}S
        \right)
    \right]\,,
    \nonumber\\
    E^{z\ell}_{kj}
    &=
    \frac{\kappa\sigma_{2}}{\mu^{2}}
    \delta^{\ell}_{[k}S^{z}_{j]}\,.
    \label{eq:NMG-relevant-E-components}
\end{align}

It is useful to express the result in terms of the traceless part of
the extrinsic curvature and the mixed Gauss--Codazzi combination,
\begin{equation}
    \overline K^{i}_{j}
    =
    K^{i}_{j}
    -\frac{1}{2}\delta^{i}_{j}K,
    \qquad
    \mathcal U_{i}
    =
    \mathcal D_{j}K^{j}_{i}
    -\mathcal D_{i}K.
    \label{eq:NMG-U-Kbar}
\end{equation}
Using the Gauss--Codazzi identities collected in
Appendix~\ref{app:asymptotic-expansions}, the variation of
$I_{\rm NMG}+I_{K}$ can be organized as
\begin{align}
   \delta\left(I_{\rm NMG}+I_K\right)
    ={}&
    \kappa
    \int_{\partial\mathcal M}d^{2}x\,
    \sqrt{|h|}
    \Bigg\{
        \sigma_{1}
        \left[
            \delta^{i}_{j}\delta K^{j}_{i}
            +\overline K^{i}_{j}
            (h^{-1}\delta h)^{j}_{i}
        \right]
        +\frac{\sigma_{2}}{\mu^{2}}
        \Bigg[
            \left(
                \left[
                    \frac{1}{2\ell^{2}}
                    +2S^{k}{}_{k}
                \right]\delta^{i}_{j}
                -2S^{i}{}_{j}
            \right)\delta K^{j}_{i}
        \nonumber\\
    &+
        \Bigg(
            \frac{1}{N}\partial_{z}S^{i}{}_{j}
            -2K^{i}_{k}S^{k}{}_{j}
            +SK^{i}_{j}
            +2\mathcal D^{i}\mathcal U_{j}
            +
            \left[
                \frac{K}{2\ell^{2}}
                -\mathcal D_{k}\mathcal U^{k}
            \right]\delta^{i}_{j}
        \Bigg)
        (h^{-1}\delta h)^{j}_{i}
        \Bigg]
    \Bigg\}\,.
    \label{eq:NMG-nonconstant-variation}
\end{align}
This form separates the coefficients of $\delta K^{i}_{j}$ from those
of $(h^{-1}\delta h)^{i}_{j}$ and is therefore suited to the
asymptotic expansion.

For $g_{(1)ij}=0$, the deviations of $K^{i}_{j}$ and
$S^{i}{}_{j}$ from their AdS values begin at order $z^{2}$. The same
is true for $\delta K^{i}_{j}$ under variations of the asymptotic
data. These factors compensate the behavior
$\sqrt{|h|}\sim z^{-2}$. Substitution of the expansions into
Eq.~\eqref{eq:NMG-nonconstant-variation}, followed by use of
Eq.~\eqref{g-solNMG}, cancels all negative powers of $z$ and leaves
\begin{equation}
    \delta I_{\rm ren}\big|_{\rm EOM}
    =
    \frac{1}{2}
    \int_{\partial\mathcal M}d^{2}x\,
    \sqrt{|g_{(0)}|}
    \,\tau^{ij}\delta g_{(0)ij}\,,
\end{equation}
where $\tau^{ij}$ is again given by Eq.~\eqref{tauijNMG}. Thus, the
same extrinsic boundary term provides a finite Dirichlet variation
throughout the  branch with $g_{(1)ij}=0$, including
solutions that are not locally Einstein.

\subsubsection{Degenerate case
\texorpdfstring{$g_{(1)ij}\neq0$}{g1 not equal 0}}

At the degenerate point, $\varpi_{+}=0$, while
$\varpi_{-}=2\sigma_{1}$ and $\mathcal E_{0}=0$. The asymptotic field
equations no longer require $g_{(1)ij}$ to vanish. Equations
\eqref{Lag-coeff} reduce to
\begin{subequations}
\label{eq:NMG-degenerate-Lagrangian-coefficients}
\begin{align}
    \mathcal L_{(0)}
    &=
    -\frac{8\sigma_{1}}{\ell^{2}}\,,
    \qquad \mathcal L_{(1)}
    =
    \frac{4\sigma_{1}}{\ell^{3}}
    \text{tr }g_{(1)}\,,
    \\
    \mathcal L_{(2)}
    &=
    \frac{2\sigma_{1}}{\ell^{4}}
    \left[
        2\text{tr }g_{(2)}
        +\ell^{2}\mathcal R_{(0)}
        -\frac{3}{2}\text{tr }g_{(1)}^{2}
    \right]\,.
\end{align}
\end{subequations}

The possible $z^{-2}$ term in the bulk integrand cancels directly:
\begin{align}
    \mathcal L_{(-2)}
    &=
    \frac{\ell^{2}}{2}
    \left[
        2\ell
        \left(
            \frac{4\sigma_{1}}{\ell^{3}}
            \text{tr }g_{(1)}
        \right)
        -
        \frac{8\sigma_{1}}{\ell^{2}}
        \text{tr }g_{(1)}
    \right]
    =0.
    \label{eq:NMG-degenerate-Lminus2}
\end{align}
At the next order,
\begin{equation}
    \mathcal L_{(-1)}
    =
    \frac{\sigma_{1}}{\ell}
    \left[
        \big(\text{tr }g_{(1)}\big)^{2}
        -\text{tr }g_{(1)}^{2}
        +2\ell^{2}\mathcal R_{(0)}
    \right].
\end{equation}
We find that the degenerate sector imposes Eq.~\eqref{trg1}. Consequently
\begin{equation}
    \mathcal L_{(-1)}
    =
    2\sigma_{1}\ell\mathcal R_{(0)}\,.
\end{equation}
The divergent part of the bulk action is therefore
\begin{equation}
    I_{\rm NMG}\big|_{\rm deg}
    =
    \kappa\sigma_{1}
    \int_{\mathcal M_\epsilon}d^{3}x\,
    \sqrt{|g_{(0)}|}
    \left[
        -\frac{8\ell}{z^{3}}
        +\frac{2\ell}{z}\mathcal R_{(0)}
        +\mathcal O(1)
    \right]\,.
    \label{eq:NMG-degenerate-bulk-action}
\end{equation}
Hence, the nonvanishing coefficient $g_{(1)ij}$ does not introduce an
additional power-law divergence in the bulk action. The $z^{-1}$ term
is again the logarithmic Euler contribution.

The situation is different for the boundary term. Using the
determinant and extrinsic-curvature expansions gives
\begin{equation}
    I_{K}\big|_{\rm deg}
    =
    \kappa\varpi_{-}
    \int_{\partial \Sigma_\epsilon}d^{2}x\,
    \sqrt{|g_{(0)}|}
    \left[
        \frac{2\ell}{z^{2}}
        +\frac{\text{tr }g_{(1)}}{2z}
        +\mathcal O(1)
    \right]\,.
    \label{Ikdegenereque}
\end{equation}
The first term is the leading boundary divergence. Indeed, radial
integration of the $z^{-3}$ term in
Eq.~\eqref{eq:NMG-degenerate-bulk-action} produces
\begin{equation}
    -\frac{4\kappa\sigma_{1}\ell}{\epsilon^{2}}
    =
    -\frac{2\kappa\varpi_{-}\ell}{\epsilon^{2}}\,,
\end{equation}
whereas Eq.~\eqref{Ikdegenereque} contributes
$2\kappa\varpi_{-}\ell/\epsilon^{2}$. These two terms cancel. The
remaining $\epsilon^{-1}$ term is a subleading inverse-cutoff divergence
proportional to $\text{tr }g_{(1)}$.

This contribution is removed by the quadratic extrinsic counterterm
\begin{equation}
    I_{\rm ct}
    =
    -\kappa\varpi_{-}
    \int_{\partial \Sigma_\epsilon}d^{2}x\,
    \sqrt{|h|}
    \left(
        \frac{1}{\ell}
        -\frac{\ell}{2}
        \delta^{ik}_{jl}K^{j}_{i}K^{l}_{k}
    \right)\,.
    \label{GBdyaction}
\end{equation}
In two boundary dimensions, the generalized-delta contraction is
\begin{equation}
    \delta^{ik}_{jl}K^{j}_{i}K^{l}_{k}
    =
    K^{2}-K^{i}_{j}K^{j}_{i}.
\end{equation}
Its asymptotic expansion gives
\begin{equation}
    \sqrt{|h|}
    \left(
        \frac{1}{\ell}
        -\frac{\ell}{2}
        \delta^{ik}_{jl}K^{j}_{i}K^{l}_{k}
    \right)
    =
    \sqrt{|g_{(0)}|}
    \left[
        \frac{\text{tr }g_{(1)}}{2z}
        +\mathcal O(1)
    \right]\,.
    \label{eq:NMG-degenerate-counterterm-expansion}
\end{equation}
The minus sign in Eq.~\eqref{GBdyaction} therefore cancels the
$z^{-1}$ contribution in Eq.~\eqref{Ikdegenereque}. Since the
combination inside parentheses has no constant leading term, this
counterterm does not modify the cancellation of the
$\epsilon^{-2}$ volume divergence.

Substituting the Fefferman--Graham expansions into the variation of
$I_{\rm NMG}+I_{K}+I_{\rm ct}$ and imposing the asymptotic field
equations yields
\begin{align}
    \delta
   & (I_{\rm NMG}+I_{K}+I_{\rm ct})
    \big|_{\rm EOM}
    ={}
    -\frac{\kappa\varpi_{-}}{4\ell}
    \int_{\partial\mathcal M}d^{2}x\,
    \sqrt{|g_{(0)}|}
    \Bigg\{
        \Bigg[
            \text{tr }g_{(1)}g_{(1)}^{ij}
            -4g_{(2)}^{ij}
        \nonumber\\
    &\quad
        +
        \left(
            2\text{tr }g_{(2)}
            -\ell^{2}\mathcal R_{(0)}
            -\frac{1}{2}\text{tr }g_{(1)}^{2}
        \right)g_{(0)}^{ij}
        \Bigg]\delta g_{(0)ij}
        -
        \left(
            g_{(1)}^{ij}
            -\text{tr }g_{(1)}g_{(0)}^{ij}
        \right)\delta g_{(1)ij}
    \Bigg\}\,.
    \label{eq:NMG-degenerate-finite-variation}
\end{align}
The two independent responses are defined by writing this result as
\begin{equation}
    \delta I_{\rm ren}\big|_{\rm EOM}
    =
    \frac{1}{2}
    \int_{\partial\mathcal M}d^{2}x\,
    \sqrt{|g_{(0)}|}
    \left(
        T^{ij}\delta g_{(0)ij}
        +P^{ij}\delta g_{(1)ij}
    \right)\,.
    \label{eq:NMG-degenerate-response-definition}
\end{equation}
Comparison with Eq.~\eqref{eq:NMG-degenerate-finite-variation} gives
\begin{subequations}
\label{eq:NMG-degenerate-responses}
\begin{align}
    T^{ij}
    &=
    \frac{\kappa\varpi_{-}}{2\ell}
    \Bigg[
        4g_{(2)}^{ij}
        -\text{tr }g_{(1)}g_{(1)}^{ij}
        -
        \left(
            2\text{tr }g_{(2)}
            -\ell^{2}\mathcal R_{(0)}
            -\frac{1}{2}\text{tr }g_{(1)}^{2}
        \right)g_{(0)}^{ij}
    \Bigg]\,,
    \\
    P^{ij}
    &=
    \frac{\kappa\varpi_{-}}{2\ell}
    \left(
        g_{(1)}^{ij}
        -\text{tr }g_{(1)}g_{(0)}^{ij}
    \right)\,.
\end{align}
\end{subequations}
The generalized Dirichlet conditions are $\delta g_{(0)ij}=0$ and $\delta g_{(1)ij}=0$; they therefore make the finite on-shell variation vanish. Restricting the
phase space to $g_{(1)ij}=0$ consistently removes the second source
and its conjugate response.

Boundary diffeomorphism invariance provides a consistency condition
relating these responses. Under a boundary vector field
$\xi=\xi^{i}\partial_{i}$, the sources transform as
\begin{equation}
    \delta_{\xi}g_{(0)ij}
    =
    \mathcal L_{\xi}g_{(0)ij}\,,
    \qquad
    \delta_{\xi}g_{(1)ij}
    =
    \mathcal L_{\xi}g_{(1)ij}\,.
\end{equation}
Inserting these variations into
Eq.~\eqref{eq:NMG-degenerate-response-definition} and integrating by
parts gives the Ward identity
\begin{equation}
    \mathcal D^{(0)}_{j}
    \left(
        T^{ij}
        +P^{jk}g^{i}_{(1)k}
    \right)
    =
    \frac{1}{2}
    P^{jk}\mathcal D^{i}_{(0)}g_{(1)jk}\,.
    \label{eq:NMG-degenerate-Ward-identity}
\end{equation}
For an asymptotic Killing vector that preserves both sources, the
corresponding charge is
\begin{equation}
    Q[\xi]
    =
    \int_{\Sigma}dx\,
    \sqrt{|\gamma|}\,
    u^{i}
    \left(
        T_{ij}
        +P^{k}_{i}g_{(1)jk}
    \right)\xi^{j}\,.
    \label{chargedegTijPij}
\end{equation}
Thus, the relaxed degenerate branch contains an additional boundary
response associated with $g_{(1)ij}$, and this response contributes to
the conserved charges whenever the additional source is nonzero.

\subsection{Comparison with the Hohm--Tonni stress tensor}
\label{subsec:Hohm-Tonni-comparison}

The auxiliary-field formulation introduced in
Ref.~\cite{Hohm:2010jc} provides an independent construction of the
holographic stress tensor of NMG. Instead of working directly with
the fourth-order metric equations, this formulation introduces a
symmetric tensor $f_{\mu\nu}$ as an independent field. The resulting
action is second order in derivatives of the metric, while the
equation for $f_{\mu\nu}$ is algebraic.

In the conventions adopted here, the auxiliary-field action is
\begin{equation}
    I_{\rm aux}
    =
    \kappa
    \int_{\mathcal M}d^{3}x\,\sqrt{|g|}
    \left[
        \sigma_{1}R
        -2\lambda
        +f^{\mu\nu}G_{\mu\nu}
        +\frac{\sigma_{2}\mu^{2}}{4}
        \left(
            f_{\mu\nu}f^{\mu\nu}-f^{2}
        \right)
    \right]\,,
    \label{eq:NMG-auxiliary-action}
\end{equation}
where $f=g^{\mu\nu}f_{\mu\nu}$. Varying the action with respect to
$f^{\mu\nu}$ gives
\begin{equation}
    G_{\mu\nu}
    +\frac{\sigma_{2}\mu^{2}}{2}
    \left(
        f_{\mu\nu}-fg_{\mu\nu}
    \right)
    =0\,.
    \label{eq:NMG-auxiliary-equation}
\end{equation}
This equation can be solved without using the metric field equations.
Taking its trace in three dimensions gives $  -\tfrac{R}{2}
    -\sigma_{2}\mu^{2}f
    =0,$
and hence
\begin{equation}
    f=-\frac{\sigma_{2}}{2\mu^{2}}R\,.
\end{equation}
Substituting this quantity into Eq.~\eqref{eq:NMG-auxiliary-equation} then yields
\begin{equation}
    f_{\mu\nu}
    =
    -\frac{2\sigma_{2}}{\mu^{2}}
    \left(
        R_{\mu\nu}
        -\frac{1}{4}g_{\mu\nu}R
    \right)
    =
    -\frac{2\sigma_{2}}{\mu^{2}}S_{\mu\nu}\,.
    \label{auxiliarySchouten}
\end{equation}

The equivalence with the original NMG action follows directly from
this relation. In three dimensions,
\begin{equation}
    S^{\mu\nu}G_{\mu\nu}
    =
    R_{\mu\nu}R^{\mu\nu}
    -\frac{3}{8}R^{2}\,, \qquad S_{\mu\nu}S^{\mu\nu}-S^{2}
    =
    R_{\mu\nu}R^{\mu\nu}
    -\frac{3}{8}R^{2}\,.
\end{equation}
Consequently,
\begin{align}
    f^{\mu\nu}G_{\mu\nu}
    &=
    -\frac{2\sigma_{2}}{\mu^{2}}
    \left(
        R_{\mu\nu}R^{\mu\nu}
        -\frac{3}{8}R^{2}
    \right)\,,\qquad 
    \frac{\sigma_{2}\mu^{2}}{4}
    \left(
        f_{\mu\nu}f^{\mu\nu}-f^{2}
    \right)
    =
    \frac{\sigma_{2}}{\mu^{2}}
    \left(
        R_{\mu\nu}R^{\mu\nu}
        -\frac{3}{8}R^{2}
    \right)\,.
\end{align}
Their sum reproduces the curvature-squared term in
Eq.~\eqref{INMG}:
\begin{equation}
    f^{\mu\nu}G_{\mu\nu}
    +\frac{\sigma_{2}\mu^{2}}{4}
    \left(
        f_{\mu\nu}f^{\mu\nu}-f^{2}
    \right)
    =
    -\frac{\sigma_{2}}{\mu^{2}}
    \left(
        R_{\mu\nu}R^{\mu\nu}
        -\frac{3}{8}R^{2}
    \right)\,.
\end{equation}
The auxiliary formulation is therefore dynamically equivalent to
the metric formulation after $f_{\mu\nu}$ is eliminated.

To formulate the boundary problem, the auxiliary tensor is
decomposed into components tangent and normal to the constant-$z$
hypersurfaces. Using the tangential projector introduced in Chapter~\ref{Chapter4}, these projections are
\begin{equation}
    \widehat f^{ij}
    =
    f^{\mu\nu}h^{i}_{\mu}h^{j}_{\nu}\,,
    \qquad
    \widehat h^{i}
    =
    f^{\mu\nu}h^{i}_{\mu}n_{\nu}\,,
    \qquad
    \widehat s
    =
    f^{\mu\nu}n_{\mu}n_{\nu}\,.
    \label{auxiliaryProjections}
\end{equation}
The tangential trace is
\begin{equation}
    \widehat f=h_{ij}\widehat f^{ij}.
\end{equation}
Indices intrinsic to the cutoff hypersurface are raised and lowered
with $h_{ij}$, and $\mathcal D_i$ denotes the corresponding compatible
covariant derivative.

The generalized Gibbons--Hawking term is chosen so that the normal
derivatives of the metric variation are removed when the induced
metric and the appropriate auxiliary-field data are fixed at the
boundary. Including the intrinsic volume counterterm, the completed
action is
\begin{align}
I_{\rm HT}
=
I_{\rm NMG}
-
\kappa
\int_{\partial\mathcal M}d^2x\,\sqrt{|h|}
\left(
    2\sigma_1K
    +\widehat f^{ij}K_{ij}
    -\widehat f K
\right)
-
\frac{2\kappa\varpi_-}{\ell}
\int_{\partial\mathcal M}d^2x\,\sqrt{|h|}.
    \label{ITonniHohm}
\end{align}
The term proportional to $\sigma_{1}$ is the Gibbons--Hawking contribution from the Einstein--Hilbert sector. The
terms containing $\widehat f^{ij}$ extend the boundary completion to
the curvature-squared sector, while the final term subtracts the
leading asymptotic volume divergence.

When the auxiliary boundary data are held fixed, the metric variation
defines the cutoff stress tensor through
\begin{equation}
    \delta I_{\rm HT}
    =
    \frac{1}{2}
    \int_{\partial\mathcal M}d^{2}x\,\sqrt{|h|}
    \,T_{\rm ren}^{ij}\delta h_{ij}\,.
    \label{variationHT}
\end{equation}
Before including the intrinsic counterterm, the contribution from the
bulk action and the generalized Gibbons--Hawking term is
\begin{align}
    T_{\rm HT}^{ij}
    =
    2\kappa\bigg[&
        -\sigma_{1}
        \left(
            K^{ij}-Kh^{ij}
        \right)
        +\frac{1}{2}\widehat f K^{ij}
        +\mathcal D^{(i}\widehat h^{j)}
        -\frac{1}{2N}\partial_{z}\widehat f^{ij}
        +K_{k}^{(i}\widehat f^{j)k}
        -\frac{1}{2}\widehat s K^{ij}
        \nonumber\\
    &+
        h^{ij}
        \left(
            -\mathcal D_{k}\widehat h^{k}
            +\frac{1}{2}\widehat s K
            -\frac{1}{2}\widehat f K
            +\frac{1}{2N}\partial_{z}\widehat f
        \right)
    \bigg].
    \label{stressHT}
\end{align}
The variation of the intrinsic volume counterterm shifts this tensor
by a term proportional to the induced metric:
\begin{equation}
    T_{\rm ren}^{ij}
    =
    T_{\rm HT}^{ij}
    -\frac{2\kappa\varpi_{-}}{\ell}h^{ij}\,.
    \label{Tijren}
\end{equation}

The cancellation of the leading term can be checked directly on the
AdS background. Equation~\eqref{auxiliarySchouten} gives
\begin{equation}
    \widehat f^{ij}
    =
    \frac{\sigma_{2}}{\mu^{2}\ell^{2}}h^{ij}\,,
    \qquad
    \widehat h^{i}=0\,,
    \qquad
    \widehat s
    =
    \frac{\sigma_{2}}{\mu^{2}\ell^{2}}\,.
\end{equation}
Substituting these expressions into Eq.~\eqref{stressHT}, including
the radial derivative of $\widehat f^{ij}$, gives
\begin{equation}
    T_{\rm HT}^{ij}
    =
    \frac{2\kappa}{\ell}
    \left(
        \sigma_{1}
        -\frac{\sigma_{2}}{2\mu^{2}\ell^{2}}
    \right)h^{ij}
    +\text{subleading terms}
    =
    \frac{2\kappa\varpi_{-}}{\ell}h^{ij}
    +\text{subleading terms}\,.
\end{equation}
The intrinsic counterterm in Eq.~\eqref{Tijren} cancels this leading
contribution.

For the complete nondegenerate branch, Eq.~\eqref{auxiliarySchouten}
is inserted into Eq.~\eqref{stressHT} together with the
Fefferman--Graham expansions and the asymptotic relations
\eqref{g-solNMG}. After the leading term has been removed, the
remaining cutoff stress tensor behaves as
\begin{equation}
    T_{\rm ren}^{ij}
    =
    \frac{2\kappa\varpi_{-}z^{4}}{\ell^{5}}
    \left(
        g_{(2)}^{ij}
        -g_{(0)}^{ij}\text{tr }g_{(2)}
    \right)
    +\mathcal O(z^{6})\,.
    \label{eq:HT-stress-FG-expansion}
\end{equation}
Although this tensor vanishes as $z^{4}$, the variation
\eqref{variationHT} has a finite limit because
\begin{equation}
    \sqrt{|h|}
    =
    \frac{\ell^{2}}{z^{2}}
    \sqrt{|g_{(0)}|}
    +\mathcal O(1)\,,
    \qquad
    \delta h_{ij}
    =
    \frac{\ell^{2}}{z^{2}}
    \delta g_{(0)ij}
    +\mathcal O(1)\,.
\end{equation}
The factors of $z$ therefore cancel in
$\sqrt{|h|}T_{\rm ren}^{ij}\delta h_{ij}$. Taking the limit
$z\rightarrow0$ gives
\begin{equation}
    \delta I_{\rm HT}\big|_{\rm EOM}
    =
    \frac{\kappa\varpi_{-}}{\ell}
    \int_{\partial\mathcal M}d^{2}x\,
    \sqrt{|g_{(0)}|}
    \left(
        g_{(2)}^{ij}
        -g_{(0)}^{ij}\text{tr }g_{(2)}
    \right)
    \delta g_{(0)ij}\,.
    \label{variationHTFG}
\end{equation}
Comparison with the definition
\begin{equation}
    \delta I_{\rm HT}\big|_{\rm EOM}
    =
    \frac{1}{2}
    \int_{\partial\mathcal M}d^{2}x\,
    \sqrt{|g_{(0)}|}
    \,\tau_{\rm HT}^{ij}\delta g_{(0)ij} \,,
\end{equation}
therefore gives
\begin{equation}
    \tau_{\rm HT}^{ij}
    =
    \frac{2\kappa\varpi_{-}}{\ell}
    \left(
        g_{(2)}^{ij}
        -g_{(0)}^{ij}\text{tr }g_{(2)}
    \right)\,.
\end{equation}
This is precisely the stress tensor obtained from the extrinsic
prescription in Eq.~\eqref{tauijNMG}. The two boundary constructions
therefore give the same finite response for the nondegenerate
power-law sector satisfying Eq.~\eqref{g-solNMG}, including both
constant-curvature solutions and nonconstant-curvature solutions with
$g_{(1)ij}=0$.

\section{Noether--Wald charges}
\label{sec:NW}

The boundary terms obtained in the previous section cancel the
power-law divergences and render the on-shell variation finite under
the asymptotic conditions considered here. For a curved boundary
representative, the logarithmic Euler contribution identified in
Eq.~\eqref{eq:NMG-generic-divergent-coefficients} must be treated
separately. The boundary completion also contributes to the conserved
charges. In this section, the Noether--Wald prescription is applied
to NMG in the constant-curvature, nonconstant-curvature, and
degenerate sectors.

For a vector field $\xi=\xi^{\mu}\partial_{\mu}$, diffeomorphism invariance defines the Noether current
\begin{equation}
    J_{\xi}^{\mu}
    =
    \Theta^{\mu}
    \left(
        g,\mathcal L_{\xi}g
    \right)
    -
    \xi^{\mu}\mathcal L_{\rm NMG}\,.
    \label{eq:NMG-Noether-current}
\end{equation}
When the metric satisfies the field equations, this current is conserved and can be written locally as the divergence of an antisymmetric prepotential:
\begin{equation}
    \nabla_{\mu}J_{\xi}^{\mu}=0\,,
    \qquad
    J_{\xi}^{\mu}
    =
    \nabla_{\nu}q_{\xi}^{\mu\nu}\,,
    \qquad
    q_{\xi}^{\mu\nu}
    =
    -q_{\xi}^{\nu\mu}\,.
\end{equation}
For a Lagrangian depending algebraically on the Riemann tensor, the corresponding prepotential is~\cite{Wald:1993nt,Iyer:1994ys}
\begin{equation}
    q_{\xi}^{\mu\nu}
    =
    -2
    \left(
        E^{\mu\nu}_{\lambda\rho}
        \nabla^{\lambda}\xi^{\rho}
        +
        2\xi^{\lambda}
        \nabla^{\rho}
        E^{\mu\nu}_{\lambda\rho}
    \right)\,.
    \label{Noetherprep}
\end{equation}
The second contribution vanishes when the $E$-tensor is covariantly constant, but it must be retained for a generic NMG solution.

The charge also receives a contribution from the boundary completion of the action. Denoting the corresponding boundary scalar by $\mathcal B$, the integrable charge associated with $\xi$ takes the form
\begin{equation}
    \mathcal Q[\xi]
    =
    \int_{\Sigma_{\infty}}
    \left(
        q_{\xi}^{\mu\nu}
        -
        2\xi^{[\mu}n^{\nu]}\mathcal B
    \right)
    d\Sigma_{\mu\nu}\,.
    \label{QNW}
\end{equation}
Here, $\Sigma_{\infty}$ is a codimension-two spatial section of the asymptotic boundary, and its oriented surface element is defined by
\begin{equation}
    d\Sigma_{\mu\nu}
    =
    \frac{1}{(D-2)!}
    \mathcal{E}_{\mu\nu\lambda_{3}\cdots\lambda_{D}}\,
    dx^{\lambda_{3}}\wedge\cdots\wedge dx^{\lambda_{D}}\,.
    \label{eq:codimension-two-surface-element}
\end{equation}
The Levi--Civita tensor is related to the completely antisymmetric symbol by
\begin{equation}
    \mathcal{E}_{\mu_{1}\cdots\mu_{D}}
    =
    \sqrt{|g|}\,
\varepsilon_{\mu_{1}\cdots\mu_{D}}\,,
\end{equation}
where $\varepsilon_{\mu_{1}\cdots\mu_{D}}$ is the Levi-Civita symbol and takes the values $0$ and $\pm1$, with its sign fixed by the orientation of the spacetime.

In three bulk dimensions, $\Sigma_{\infty}$ is a spatial curve contained in the conformal boundary, and the surface element can equivalently be expressed as
\begin{equation}
    d\Sigma_{\mu\nu}
    =
    2n_{[\mu}u_{\nu]}
    \sqrt{|\gamma|}\,dx\,.
\end{equation}
In this expression, $u_\mu$ and $\gamma$ are defined using the
metric induced on the finite-cutoff surface. In Fefferman--Graham
coordinates,
$n_z=\mathcal O(z^{-1})$, $u_i=\mathcal O(z^{-1})$, and
$\sqrt{|\gamma|}=\mathcal O(z^{-1})$, so that
$d\Sigma_{zi}=\mathcal O(z^{-3})$. After the finite boundary
limit is taken, the symbols $u_i$ and $\gamma$ in the charge
formulas denote the corresponding quantities constructed from
the boundary representative $g_{(0)ij}$. The sign is fixed by
requiring this expression to agree with the orientation chosen
in Eq.~\eqref{eq:codimension-two-surface-element}.

For the $g_{(1)ij}=0$ sector, the required boundary density is determined by the extrinsic term in Eq.~\eqref{KT-NMG-E}. At the degenerate point, the quadratic contribution in Eq.~\eqref{GBdyaction} must also be included. Thus, the boundary terms that cancel the divergences of the action and its variation also contribute to the finite conserved charges.

Substituting the NMG $E$-tensor~\eqref{Etensor} into Eq.~\eqref{Noetherprep} gives
\begin{align}
    q_{\xi}^{\mu\nu}
    ={}&
    -2\kappa
    \left[
        \frac{1}{2}
        \left(
            \sigma_{1}
            +
            \frac{\sigma_{2}}{\mu^{2}}S
        \right)
        \delta^{\mu\nu}_{\lambda\rho}
        -
        \frac{2\sigma_{2}}{\mu^{2}}
        S^{[\mu}_{[\lambda}
        \delta^{\nu]}_{\rho]}
    \right]
    \nabla^{\lambda}\xi^{\rho}
    +
    \frac{2\kappa\sigma_{2}}{\mu^{2}}
    \xi^{\lambda}C_{\lambda}{}^{\mu\nu}\,.
    \label{eq:pnoether}
\end{align}
The Cotton term originates from the covariant derivative of the $E$-tensor and therefore vanishes for Einstein spaces, but not for a generic nonconstant-curvature solution.

\subsection{Constant-curvature sector}

For the constant-curvature sector defined by
Eqs.~\eqref{eq:NMG-constant-curvature} and
\eqref{eq:NMG-constant-curvature-contractions}, the Cotton tensor vanishes and the $E$-tensor reduces to the covariantly constant expression~\eqref{eq:NMG-E-AdS}. The prepotential consequently becomes
\begin{equation}
    q_{\xi}^{\mu\nu}\big|_{\rm E}
    =
    -2\kappa\varpi_{-}
    \nabla^{[\mu}\xi^{\nu]}\,.
    \label{eq:NMG-Einstein-prepotential}
\end{equation}
Both the bulk prepotential and the corresponding boundary contribution are therefore equal to their Einstein-gravity counterparts multiplied by $\varpi_{-}$. It follows that the charges of locally AdS solutions satisfying Brown--Henneaux boundary conditions vanish at the critical point $\varpi_{-}=0$.

A similar mechanism occurs in higher-dimensional critical gravities. In four-dimensional Critical Gravity, the massless and massive spin-two operators become degenerate, and the Noether--Wald energy and entropy of Einstein black holes vanish at the critical coupling~\cite{Lu:2011zk,Anastasiou:2017rjf}. In six dimensions, the Noether prepotential likewise vanishes when evaluated on the Einstein sector~\cite{Anastasiou:2021tlv}. These statements apply to Einstein solutions at the critical point; non-Einstein or logarithmic modes admitted by relaxed boundary conditions may carry nonvanishing charges.

\subsection{Nonconstant-curvature sector with
\texorpdfstring{$g_{(1)ij}=0$}{g1 equal 0}}

For a nonconstant-curvature solution, the Schouten tensor is not generally covariantly constant. The Cotton contribution and the derivative of the $E$-tensor in Eq.~\eqref{Noetherprep} must therefore be retained. Taking an asymptotic Killing vector tangent to the radial hypersurfaces, so that $\xi^{z}=0$, the component entering the charge is
\begin{align}
    q_{\xi}^{zi}
    ={}&
    -\kappa
    \left[
        \sigma_{1}
        +
        \frac{\sigma_{2}}{\mu^{2}}
        \left(
            S-S^{z}_{z}
        \right)
    \right]
    \nabla^{z}\xi^{i}
    +
    \frac{\kappa\sigma_{2}}{\mu^{2}}
    S^{i}_{j}\nabla^{z}\xi^{j}-
    \frac{2\kappa\sigma_{2}}{\mu^{2}}
    \left(
        S^{z}_{j}\nabla^{[i}\xi^{j]}
        -
        \xi^{j}C_{j}{}^{zi}
    \right)\,,
    \label{qzi1}
\end{align}
Before imposing the asymptotic field equations, the radial derivative of the symmetry generator has the expansion
\begin{align}
    \nabla^{[z}\xi^{i]}
    ={}&
    -\frac{z}{\ell^{2}}\xi^{i}
    +
    \frac{z^{2}}{2\ell^{3}}
    \left(
        g^{i}_{(1)j}\xi^{j}
        +
        \ell\,\partial_{z}\xi^{i}
    \right)+
    \frac{z^{3}}{2\ell^{4}}
    \left(
        2g^{i}_{(2)j}
        -
        g^{i}_{(1)k}g^{k}_{(1)j}
    \right)\xi^{j}
    +
    \mathcal O(z^{4})\,.
    \label{eq:NMG-Killing-derivative-expansion}
\end{align}
In the nondegenerate power-law branch, the asymptotic equations~\eqref{g-solNMG} set $g_{(1)ij}=0$. Substituting the resulting expansions into Eqs.~\eqref{qzi1} and~\eqref{QNW}, the divergent terms in the bulk prepotential are canceled by the contribution of the boundary term~\eqref{KT-NMG-E}. The finite part is
\begin{equation}
    \mathcal Q[\xi]
    =
    \frac{2\kappa\varpi_{-}}{\ell}
    \int_{\Sigma_{\infty}}
    dx\,
    \sqrt{|\gamma|}\,
    u_{i}\xi^{j}
    \left(
        g^{i}_{(2)j}
        +
        \frac{\ell^{2}}{2}
        \delta^{i}_{j}\mathcal R_{(0)}
    \right)\,.
    \label{eq:NMG-NW-charge-g1zero}
\end{equation}
Using the trace relation~\eqref{eq:NMG-g2-trace}, this expression agrees with the charge~\eqref{eq:Einstein-boundary-charge} constructed from the holographic stress tensor~\eqref{tauijNMG}. The Noether--Wald and holographic prescriptions therefore produce the same conserved charges throughout the power-law sector with $g_{(1)ij}=0$.

\subsection{Degenerate sector}

At the degenerate point, the asymptotic equations allow a nonvanishing coefficient $g_{(1)ij}$. The charge must then include the full boundary completion formed by the extrinsic term~\eqref{KT-NMG-E} and the quadratic counterterm~\eqref{GBdyaction}. After inserting the corresponding asymptotic expansions into Eq.~\eqref{QNW}, the divergent contributions cancel and the finite charge becomes
\begin{align}
    \mathcal Q[\xi]
    ={}&
    \frac{2\kappa\varpi_{-}}{\ell}
    \int_{\Sigma_{\infty}}
    dx\,
    \sqrt{|\gamma|}\,
    u_{i}\xi^{j}
    \Bigg[
        g^{i}_{(2)j}
        -
        \frac{1}{2}
        \text{tr }g_{(2)}
        \delta^{i}_{j}
        \nonumber\\
    &\hspace{2.5cm}
        -
        \frac{1}{4}
        \text{tr }g_{(1)}
        \left(
            g^{i}_{(1)j}
            -
            \frac{1}{2}
            \text{tr }g_{(1)}
            \delta^{i}_{j}
        \right)
        +
        \frac{\ell^{2}}{4}
        \delta^{i}_{j}\mathcal R_{(0)}
    \Bigg]\,.
    \label{Chargefromaction}
\end{align}
This result can be compared with the charge obtained from the holographic responses in Eq.~\eqref{chargedegTijPij}. In two boundary dimensions, the constraint~\eqref{trg1} implies that the determinant of the mixed tensor $g^{i}_{(1)j}$ vanishes. Using~\eqref{eq:NMG-g1-Cayley-Hamilton} in the combination of responses appearing in Eq.~\eqref{chargedegTijPij} reproduces the integrand of Eq.~\eqref{Chargefromaction}. The Noether--Wald charge therefore agrees with the charge following from the holographic Ward identity. The terms involving $g_{(1)ij}$ account for the contribution of the additional boundary source and its conjugate response in the degenerate sector.

\section{Black holes and AdS waves: Conserved charges}
\label{Sec.BHandConservedCharges}

The conserved charges derived in the previous section can be evaluated on representative solutions of the different asymptotic sectors of NMG. Constant curvature black holes provide a direct test of the effective coupling \(\varpi_-\), whereas nonconstant-curvature geometries probe contributions from the Schouten and Cotton tensors. The examples considered below are the BTZ black hole, an AdS-wave solution satisfying Brown--Henneaux fall-offs for a suitable range of parameters, and the rotating hairy black hole at the degenerate point. The last solution possesses a nonvanishing coefficient \(g_{(1)ij}\) and therefore also tests the quadratic extrinsic counterterm introduced in Eq.~\eqref{GBdyaction}.

\subsection{BTZ black hole}
\label{sec:NMG-BTZ}

The discovery of the BTZ black hole showed that the absence of local gravitational waves does not preclude black-hole solutions in three-dimensional Einstein gravity~\cite{Banados:1992wn,Banados:1992gq}. In stationary and axisymmetric coordinates, its metric is
\begin{equation}
    ds^{2}
    =
    -N^{2}(r)dt^{2}
    +
    \frac{dr^{2}}{N^{2}(r)}
    +
    r^{2}
    \left(
        d\phi+N^{\phi}(r)dt
    \right)^{2},
    \qquad
    \phi\sim\phi+2\pi,
    \label{eq:btz-metric}
\end{equation}
with metric functions
\begin{equation}
    N^{2}(r)
    =
    -8GM
    +
    \frac{r^{2}}{\ell^{2}}
    +
    \frac{16G^{2}J^{2}}{r^{2}},
    \qquad
    N^{\phi}(r)
    =
    -\frac{4GJ}{r^{2}}.
    \label{eq:btz-lapse-shift}
\end{equation}
In Einstein gravity, the parameters \(M\) and \(J\) are the conserved mass and angular momentum. The zeros of \(N^{2}(r)\) occur at
\begin{equation}
    r_{\pm}^{2}
    =
    4G\ell^{2}
    \left(
        M
        \pm
        \sqrt{
            M^{2}
            -
            \frac{J^{2}}{\ell^{2}}
        }
    \right),
    \label{eq:btz-horizon-radii}
\end{equation}
so the lapse function can be factorized as
\begin{equation}
    N^{2}(r)
    =
    \frac{
        (r^{2}-r_{+}^{2})
        (r^{2}-r_{-}^{2})
    }{
        \ell^{2}r^{2}
    }.
    \label{eq:btz-factorized-lapse}
\end{equation}
A black hole with a nonzero outer-horizon radius requires \(M>0\) and \(\lvert J\rvert\leq M\ell\). For a nonextremal rotating solution, \(r_{+}\) is the outer event horizon and \(r_{-}\) is the inner Cauchy horizon. In the nonrotating case, \(r_{-}=0\), so there is no separate inner horizon at positive radius. Extremality occurs when \(\lvert J\rvert=M\ell\), for which the two horizons coincide. Conversely, the metric parameters can be written as
\begin{equation}
    M
    =
    \frac{r_{+}^{2}+r_{-}^{2}}{8G\ell^{2}},
    \qquad
    \lvert J\rvert
    =
    \frac{r_{+}r_{-}}{4G\ell},
    \label{eq:btz-parameters-horizons}
\end{equation}
where \(r_{\pm}\) are taken to be nonnegative. The sign of \(J\) specifies the orientation of rotation.

Despite the presence of horizons, the BTZ spacetime has constant negative curvature,
\begin{equation}
    R^{\mu\nu}_{\lambda\rho}
    =
    -\frac{1}{\ell^{2}}
    \delta^{\mu\nu}_{\lambda\rho}.
    \label{eq:btz-constant-curvature}
\end{equation}
Its horizons are therefore not curvature singularities. The distinction between the BTZ black hole and global \(\mathrm{AdS}_{3}\) is instead global: the black hole is obtained as a quotient of an appropriate region of the universal covering space of \(\mathrm{AdS}_{3}\) by a discrete subgroup of its isometry group~\cite{Banados:1992gq}. Its causal structure and the behavior at \(r=0\) are consequently determined by this quotient and are not detected by divergent local curvature invariants.

Global \(\mathrm{AdS}_{3}\) must also be distinguished from the massless BTZ geometry. With the normalization used in Eq.~\eqref{eq:btz-lapse-shift}, global \(\mathrm{AdS}_{3}\) corresponds to
\begin{equation}
    M_{\mathrm{AdS}}
    =
    -\frac{1}{8G},
    \qquad
    J_{\mathrm{AdS}}=0,
    \label{eq:global-AdS-BTZ-parameters}
\end{equation}
whereas the massless BTZ solution has \(M=J=0\). These geometries are locally AdS but are not globally equivalent.

Because the BTZ geometry satisfies Eq.~\eqref{eq:btz-constant-curvature}, it also solves the NMG field equations whenever the AdS radius \(\ell\) satisfies the vacuum equation~\eqref{Lambdaeffconstraint}. It therefore belongs to the constant-curvature sector analyzed above. In particular, the Cotton tensor vanishes and the \(E\)-tensor takes the constant-curvature value given in Eq.~\eqref{eq:NMG-E-AdS}. The NMG Noether prepotential consequently reduces to the Einstein prepotential multiplied by the effective coupling \(\varpi_{-}\),
\begin{equation}
    q_{\xi}^{\mu\nu}\big|_{\rm BTZ}
    =
    \varpi_{-}
    q_{\xi,{\rm E}}^{\mu\nu}.
    \label{eq:BTZ-prepotential-scaling}
\end{equation}
The boundary contribution entering the complete charge~\eqref{QNW} carries the same overall factor. Hence,
\begin{equation}
    \mathcal Q_{\rm NMG}[\xi]
    =
    \varpi_{-}
    \mathcal Q_{\rm E}[\xi].
    \label{eq:BTZ-charge-scaling}
\end{equation}
No Cotton contribution is present, and the coefficient \(g_{(1)ij}\) vanishes in this constant-curvature sector.

Applying Eq.~\eqref{eq:BTZ-charge-scaling} to the stationary and rotational Killing vectors gives
\begin{subequations}
\label{NWBTZ}
\begin{align}
    \mathcal Q_{\rm NMG}[\partial_t]
    &:=
    M_{\rm NMG}
    =
    \varpi_{-}M
    =
    M
    \left(
        \sigma_{1}
        -
        \frac{\sigma_{2}}{2\mu^{2}\ell^{2}}
    \right),
    \\
    \mathcal Q_{\rm NMG}[\partial_\phi]
    &:=
    J_{\rm NMG}
    =
    \varpi_{-}J
    =
    J
    \left(
        \sigma_{1}
        -
        \frac{\sigma_{2}}{2\mu^{2}\ell^{2}}
    \right).
\end{align}
\end{subequations}
Thus, \(M\) and \(J\) continue to determine the BTZ metric and its horizon radii, whereas \(M_{\rm NMG}\) and \(J_{\rm NMG}\) are the conserved charges assigned to the same geometry by the NMG action.\footnote{The location of the horizons depends only on the metric parameters \(M\) and \(J\). The conserved charges, however, depend on the gravitational theory used to evaluate them. Consequently, the same locally AdS geometry can be assigned different energies and angular momenta by theories with different effective couplings.}

Care is required when comparing Eq.~\eqref{NWBTZ} with results obtained using different signs for the curvature-squared interaction in Eq.~\eqref{INMG}. The conventions of Hohm and Tonni~\cite{Hohm:2010jc} correspond to $(\sigma_1,\sigma_2)=(1,-1)$ and therefore give $\varpi_-=1+1/(2\mu^2\ell^2)$, whereas those of Ref.~\cite{Clement:2009gq} correspond to $(\sigma_1,\sigma_2)=(1,1)$ and give $\varpi_-=1-1/(2\mu^2\ell^2)$. After accounting for this relative sign, both results agree with the general charges $M_{\rm NMG}=\varpi_-M$ and $J_{\rm NMG}=\varpi_-J$ in Eq.~\eqref{NWBTZ}.

At the critical point, defined by \(\varpi_{-}=0\), Eq.~\eqref{NWBTZ} yields
\begin{equation}
    M_{\rm NMG}=0,
    \qquad
    J_{\rm NMG}=0.
\end{equation}
The vanishing of these charges does not make the BTZ geometry trivial: its horizons, quotient structure, and local curvature remain unchanged. Instead, the effective normalization assigned by the NMG action to the locally AdS sector vanishes at this coupling. Away from the critical point, the sign of \(\varpi_{-}\) determines whether the NMG charges have the same sign as their Einstein counterparts.

For completeness, the Wald entropy acquires the same multiplicative correction~\cite{Clement:2009gq},
\begin{equation}
    S_{\rm BTZ}^{\rm NMG}
    =
    \frac{\varpi_{-}A_{+}}{4G},
    \qquad
    A_{+}=2\pi r_{+},
    \label{eq:BTZ-Wald-entropy}
\end{equation}
where \(A_{+}\) is the length of the outer horizon. This scaling follows directly from the constant-curvature value of the \(E\)-tensor, since the Wald entropy is linear in its contraction with the horizon binormal.

For fixed NMG couplings, the charges in Eq.~\eqref{NWBTZ} and the Wald entropy satisfy the first law of black hole thermodynamics.

\subsection{Nonconstant-curvature AdS waves}

AdS waves form a class of exact solutions describing gravitational waves propagating on an AdS background. In dimensions \(D\geq4\), solutions of this type in general relativity with a negative cosmological constant were obtained in Refs.~\cite{Garcia:1981ads,Salazar:1983ads,Ozsvath:1985qn}. In three-dimensional vacuum Einstein gravity, the corresponding geometries remain locally AdS and are therefore locally related to the AdS vacuum by a diffeomorphism~\cite{Ayon-Beato:2005gdo}. This conclusion changes when additional gravitational or matter degrees of freedom are present. Nonconstant-curvature AdS waves have been obtained in theories with conformally coupled scalar fields, in Topologically Massive Gravity, and in NMG~\cite{Ayon-Beato:2005gdo,Ayon-Beato:2009cgh}.

The NMG solution can be written in Kerr--Schild form as
\begin{equation}
    ds^2
    =
    \left(
        g_{\mu\nu}^{\rm AdS}
        -
        F k_\mu k_\nu
    \right)
    dx^\mu dx^\nu
    =
    \frac{\ell^2}{z^2}
    \left[
        dz^2
        -
        2\,du\,dv
        -
        F(u,z)\,du^2
    \right]\,.
    \label{eq:AdS-wave-metric}
\end{equation}
The null Kerr--Schild one-form is
\begin{equation}
    k_\mu dx^\mu
    =
    \frac{\ell}{z}\,du\,.
    \label{eq:AdS-wave-null-form}
\end{equation}
It is null with respect to both the AdS background and the complete metric, and its metric-dual vector is geodesic.\footnote{With the background metric used in Eq.~\eqref{eq:AdS-wave-metric}, the metric-dual vector is proportional to \(\partial_v\), rather than to \(\partial_u\). It is therefore important to distinguish the null one-form \(k_\mu dx^\mu\) from its associated vector \(k^\mu\partial_\mu\).}

Substitution of Eq.~\eqref{eq:AdS-wave-metric} into the NMG field equations~\eqref{eom2} reduces the problem to a radial differential equation for \(F(u,z)\). Since no derivatives with respect to \(u\) constrain the profile, the integration coefficients remain arbitrary functions of \(u\). Taking a power-law ansatz
\begin{equation}
    F(u,z)=F_\Delta(u)z^\Delta \,,
\end{equation}
gives the radial indicial equation
\begin{equation}
    \Delta(\Delta-2)
    \left[
        (\Delta-1)^2
        -
        \ell^2\mu^2\sigma_2\varpi_+
    \right]
    =
    0\,.
    \label{eq:AdS-wave-indicial-equation}
\end{equation}
The four roots are consequently
\begin{equation}
    \Delta=0\,,
    \qquad
    \Delta=2\,,
    \qquad
    \Delta_\pm
    =
    1
    \pm
    \ell\mu\sqrt{\sigma_2\varpi_+}\,.
\end{equation}

The branches with \(\Delta=0\) and \(\Delta=2\) solve the three-dimensional Einstein equations and correspond to locally AdS contributions. They can be removed locally by the coordinate transformations described in Ref.~\cite{Ayon-Beato:2005gdo}. After suppressing these two branches, the non-Einstein profile is
\begin{equation}
    F(u,z)
    =
    F_+(u)
    z^{1+\ell\mu\sqrt{\sigma_2\varpi_+}}
    +
    F_-(u)
    z^{1-\ell\mu\sqrt{\sigma_2\varpi_+}}\,,
    \label{AdSwave}
\end{equation}
where \(F_\pm(u)\) are arbitrary functions of the null coordinate \(u\).\footnote{If dimensional powers of \(z\) are to be displayed explicitly, one may replace \(z^\Delta\) by \((z/\ell)^\Delta\) and absorb the corresponding powers of \(\ell\) into \(F_\pm(u)\).}

Reality of the characteristic exponents requires
\begin{equation}
    \sigma_2\varpi_+\geq0\,.
\end{equation}
For the choice \(\sigma_2=1\) and \(\varpi_+>0\), the branch proportional to \(F_-(u)\) has the slower radial decay. Setting
\begin{equation}
    F_-(u)=0
\end{equation}
selects the faster-decaying solution discussed in Ref.~\cite{Ayon-Beato:2009cgh}. More precisely, the rescaled metric component is
\begin{equation}
    \bar g_{uu}
    =
    -F_+(u)
    z^{1+\ell\mu\sqrt{\sigma_2\varpi_+}}\,.
\end{equation}
Strict Brown--Henneaux fall-offs require this correction to be of order \(z^2\) or faster. Therefore, in addition to \(F_-(u)=0\), one requires
\begin{equation}
    1+\ell\mu\sqrt{\sigma_2\varpi_+}
    \geq2\,.
    \label{eq:AdS-wave-BH-condition}
\end{equation}
The weaker conditions \(\sigma_2=1\) and \(\varpi_+>0\) ensure a real decaying \(F_+\) branch, but they do not by themselves guarantee the standard Brown--Henneaux rate when the exponent lies between \(1\) and \(2\).\footnote{When two indicial roots coincide, the radial equation may also admit a logarithmic solution. Such logarithmic branches require separate boundary conditions and are not included in the power-law sector considered here.}

The metric~\eqref{eq:AdS-wave-metric} admits the null Killing vector
\begin{equation}
    \xi=\partial_v\,.
\end{equation}
The charge associated with this generator can be evaluated either from Eq.~\eqref{QNW} or from the holographic stress tensor. The wave profile modifies only the \(uu\) component of the metric, whereas the generator has support along the \(v\) direction. Their contraction in the boundary charge therefore vanishes. Explicitly, the finite charge density contains the contraction
\begin{equation}
    u_i\xi^j
    \left(
        g^i_{(2)j}
        +
        \frac{\ell^2}{2}
        \delta^i_j\mathcal R_{(0)}
    \right)\,.
\end{equation}
For \(\xi^j=\delta^j_v\), the wave-dependent coefficient has no lower \(v\) index, and the contribution of the profile vanishes. With the AdS vacuum charge chosen as the reference value, one obtains
\begin{equation}
    \mathcal Q[\partial_v]=0\,.
    \label{eq:AdS-wave-null-charge}
\end{equation}
The charge is therefore finite but equal to zero. This behavior agrees with the Abbott--Deser--Tekin construction~\cite{Abbott:1981ff,Deser:2002rt,Deser:2002jk} and with topologically renormalized Noether--Wald charges for the corresponding four-dimensional AdS-wave solutions in quadratic gravity~\cite{Giribet:2018hck}.

\subsection{The rotating hairy black hole}

At the degenerate point $ \varpi_+=0,$ NMG admits a nonconstant-curvature black hole with relaxed AdS boundary conditions~\cite{Oliva:2009ip,Giribet:2009qz}. The rotating solution is
\begin{equation}
    ds^2
    =
    -N(r)f(r)\,dt^2
    +
    \frac{dr^2}{f(r)}
    +
    r^2
    \left[
        d\varphi
        +
        N^\varphi(r)\,dt
    \right]^2,
    \label{eq:rotating-hairy-metric}
\end{equation}
where the metric functions are
\begin{align}
    N(r)
    &=
    \left[
        1
        +
        \frac{b\ell^2}{4H(r)}
        (1-\eta)
    \right]^2\,,
    \nonumber\\
    N^\varphi(r)
    &=
    -\frac{a}{2r^2}
    \left[
        4Gm-bH(r)
    \right]\,,
    \label{eq:hairy-lapse-shift}
    \\
    f(r)
    &=
    \frac{H^2(r)}{r^2}
    \left[
        \frac{H^2(r)}{\ell^2}
        +
        \frac{b}{2}(1+\eta)H(r)
        +
        \frac{b^2\ell^2}{16}(1-\eta)^2
        -
        4Gm\eta
    \right]\,,
    \nonumber\\
    H(r)
    &:=
    \left[
        r^2
        -
        2Gm\ell^2(1-\eta)
        -
        \frac{b^2\ell^4}{16}(1-\eta)^2
    \right]^{1/2}\,,
    \qquad
    \eta
    =
    \sqrt{1-\frac{a^2}{\ell^2}}\,.
    \label{eq:hairy-radial-functions}
\end{align}
The constants $m$, $a$, and $b$ are integration parameters. The
parameter $a$ controls rotation and must satisfy $|a|\leq\ell$ for
$\eta$ to be real, while $b$ determines the additional gravitational
hair. When $b=0$, the weakened asymptotic mode is absent and the
geometry reduces to the corresponding BTZ family. 

The role of \(b\) in the asymptotic expansion can be seen directly at large \(r\). Expanding \(H(r)\) gives
\begin{align}
    H(r)
    =
    r
    -
    \frac{1}{r}
    \left[
        Gm\ell^2(1-\eta)
        +
        \frac{b^2\ell^4}{32}(1-\eta)^2
    \right]
    +
    \mathcal O(r^{-3})\,.
    \label{eq:hairy-H-expansion}
\end{align}
The remaining functions consequently behave as
\begin{align}
    N(r)
    &=
    1
    +
    \frac{b\ell^2}{2r}(1-\eta)
    +
    \mathcal O(r^{-2})\,,
    \\
    N^\varphi(r)
    &=
    \frac{ab}{2r}
    -
    \frac{2aGm}{r^2}
    +
    \mathcal O(r^{-3})\,,
    \\
    f(r)
    &=
    \frac{r^2}{\ell^2}
    +
    \frac{b}{2}(1+\eta)r
    +
    \mathcal O(1)\,.
    \label{eq:hairy-function-expansions}
\end{align}
Combining the first and third expressions gives
\begin{equation}
    N(r)f(r)
    =
    \frac{r^2}{\ell^2}
    +
    br
    +
    \mathcal O(1)\,,
    \label{eq:hairy-Nf-expansion}
\end{equation}
while the off-diagonal part behaves as
\begin{equation}
    r^2N^\varphi(r)
    =
    \frac{ab}{2}r
    -
    2aGm
    +
    \mathcal O(r^{-1})\,.
    \label{eq:hairy-shift-expansion}
\end{equation}

Near the conformal boundary, the leading radial transformation is
\begin{equation}
    r\sim\frac{\ell^2}{z}\,.
\end{equation}
The terms proportional to \(r\) in Eqs.~\eqref{eq:hairy-Nf-expansion} and \eqref{eq:hairy-shift-expansion} therefore become terms proportional to \(z^{-1}\) in the physical metric. After extracting the overall Fefferman--Graham factor \(\ell^2/z^2\), these terms are linear in \(z\) in the rescaled metric \(\bar g_{ij}\). The parameter \(b\) therefore activates \(g_{(1)ij}\), producing the weakened AdS fall-off allowed by the degenerate field equations.

Since this solution belongs to the \(g_{(1)ij}\neq0\) sector, its conserved charges must be evaluated using Eq.~\eqref{Chargefromaction}, including the contribution of the quadratic extrinsic counterterm~\eqref{GBdyaction}. The Killing vectors \(\partial_t\) and \(\partial_\varphi\) give
\begin{subequations}
\label{chargesOTT}
\begin{align}
    \mathcal Q[\partial_t]
    &:=
    M_{\rm OTT}
    =
    m
    +
    \frac{b^2\ell^2}{16G}\,,
    \\
    \mathcal Q[\partial_\varphi]
    &:=
    J_{\rm OTT}
    =
    a M_{\rm OTT}\,.
\end{align}
\end{subequations}
Here the normalization of the parameters follows Refs.~\cite{Oliva:2009ip,Giribet:2009qz}.\footnote{The notation OTT commonly refers to the static black hole introduced by Oliva, Tempo, and Troncoso. Its rotating extension and the corresponding conserved charges were analyzed in Ref.~\cite{Giribet:2009qz}.}

Although \(b\) first appears linearly in the asymptotic coefficient \(g_{(1)ij}\), its contribution to the conserved charges is quadratic. This behavior follows directly from the \(g_{(1)}^2\) terms in the degenerate charge~\eqref{Chargefromaction}. In particular, $g_{(1)}\neq 0$ shifts the mass. 

The charges in Eq.~\eqref{chargesOTT} agree with the Abbott--Deser--Tekin results obtained in Refs.~\cite{Oliva:2009ip,Giribet:2009qz}. They also agree with the Hohm--Tonni quasilocal stress tensor supplemented by the counterterms required by the weakened asymptotic behavior~\cite{Giribet:2010ed}. The comparison shows that the extrinsic boundary completion reproduces the known charges in both the nondegenerate and degenerate sectors. Together with the BTZ and AdS-wave calculations, these examples test the prescription on constant-curvature geometries, nonconstant-curvature solutions with \(g_{(1)ij}=0\), and relaxed solutions with \(g_{(1)ij}\neq0\).

\chapter{Conclusion and Future Directions} \label{chapter6}

This thesis has examined two independent problems in which local field equations must be supplemented by global or boundary information to determine the relevant observables. The first concerned the axial anomaly of charged Dirac fermions on
Einstein--ModMax instantons. The second concerned the construction of finite actions, boundary responses, and conserved charges in asymptotically AdS New Massive Gravity. Although the two parts of this thesis involve different theories and methods, both show why the geometric setting and the admissible boundary data are essential parts of a physical calculation.

In the first work of this thesis, we have shown that gravitational instantons sourced by nonlinear conformal electrodynamics induce the axial anomaly when coupled to Einstein gravity. To this end, we first reviewed the Taub-NUT solution of ModMax electrodynamics found in Refs.~\cite{BallonBordo:2020jtw,Flores-Alfonso:2020nnd}. In the asymptotic region, we verified that the corresponding ModMax gauge field approaches its Maxwell counterpart in this background. Moreover, this solution is continuously connected to Einstein-Maxwell one in the limit $\gamma\to0$, where $\gamma$  is the parameter that controls the nonlinearity of the U(1) sector. We then constructed a new Eguchi-Hanson instanton in Einstein-ModMax theory. Similar to the Taub-NUT case, we found that the $\gamma-$ parameter effectively screens the electric and magnetic charges by the nonlinear $U(1)$ field. By additionally demanding (anti-)self duality together with the absence of conical singularities, we obtained a nonlinearly charged, geodesically complete lens space that is asymptotically locally Euclidean. 

We then used the Atiyah-Patodi-Singer theorem to compute the index of the Dirac operator for charged spinors propagating on these backgrounds. In the Eguchi-Hanson instanton, the ModMax field does not modify the $\eta_{D}-$invariant, but the nontrivial topology of the bolt gives rise to a nonvanishing $U(1)$ Chern-Pontryagin index, which contributes to the anomaly and captures the effect of the nonlinear theory at the level of the topological charge. In this sense, the nonlinear electrodynamics produce an effect analogous to charge screening. By contrast, in the (anti-) self-dual Taub-NUT instanton, the gauge potential contributes nontrivially to the $\eta_{D}-$ invariant, while the $U(1)$ Chern-Pontryagin index remains independent of the nonlinear parameter. We computed the $\eta_{D}-$ invariant by analytically continuing the meromorphic function in Eq.~\eqref{etasalpha} to $s=0$, which amounts to evaluating the spectral asymmetry of the boundary Dirac operator. As a result, we found that both instantons induce an excess of positive chirality zero modes over negative ones [cf.~\eqref{DiffEH} and~\eqref{DiffTN}].

Several interesting questions remain open in this part of the thesis. First, it would be worthwhile to study harmonic forms and spinors in the nonlinearly charged Taub-Bolt background, along the lines of the analysis performed in Ref.~\cite{Franchetti:2018woq}. Second, although the corresponding computation in the presence of a negative cosmological constant is considerably more involved, it may provide useful insight into the role of nonlinear $U(1)$ fields in the axial anomaly from the viewpoint of conformal field theory and holography~\cite{Maldacena:1997re,Gubser:1998bc,Witten:1998qj}. Some progress in this direction has already been made in Ref.~\cite{Franchetti:2022tga}, where the space of $SU(2)$-invariant harmonic $2-$ forms was determined for the Taub-NUT-AdS solution in Einstein-Maxwell theory. Finally, it would be very interesting to understand whether quantum effects spoil the conformal symmetry of ModMax electrodynamics. 

In the second work, an extrinsic renormalization prescription for asymptotically AdS solutions of NMG was developed. The construction is particularly relevant at the degenerate point, where the two maximally symmetric vacua coincide and the Fefferman--Graham expansion may contain a nonvanishing coefficient \(g_{(1)ij}\). This coefficient weakens the standard AdS fall-off and acts as an additional boundary source.

The construction was motivated by three-dimensional Einstein-AdS gravity. Under the asymptotic conditions considered here, the usual Gibbons--Hawking--York term and intrinsic volume counterterm can be reorganized as a prescription involving one-half of the Gibbons--Hawking--York term~\cite{Miskovic:2006tm}. The difference is a finite contribution proportional to the two-dimensional Euler density. This relation provides the natural starting point for extending the extrinsic prescription to NMG.

For the sector with \(g_{(1)ij}=0\), a boundary term linear in the extrinsic curvature renders both the on-shell action and its variation finite. This result applies not only to locally AdS geometries but also to nonconstant-curvature solutions satisfying Brown--Henneaux fall-offs. The resulting holographic stress tensor has the Einstein form multiplied by the effective coupling \(\varpi_-\), explaining the vanishing of the Einstein-sector stress tensor and conserved charges at the critical point.

The stress tensor obtained from this prescription agrees with the auxiliary-field result of Hohm and Tonni~\cite{Hohm:2010jc}. Once the auxiliary field is replaced on shell by the Schouten tensor, its asymptotic expansion reproduces the same response to \(g_{(0)ij}\). Thus, throughout the \(g_{(1)ij}=0\) sector, both renormalization schemes encode the same holographic data, while the extrinsic formulation avoids introducing auxiliary variables.

At the degenerate point \(\varpi_+=0\), the asymptotic equations no longer require \(g_{(1)ij}=0\). Although this mode does not introduce a new divergence in the bulk action, it produces a subleading \(z^{-1}\) contribution in the boundary term linear in the extrinsic curvature. The quadratic counterterm in Eq.~\eqref{GBdyaction} removes this contribution. The finite variation then contains two responses: \(T^{ij}\), conjugate to the boundary metric \(g_{(0)ij}\), and \(P^{ij}\), conjugate to the additional source \(g_{(1)ij}\).

The Noether--Wald charges provide an independent verification of the construction. For constant-curvature solutions, the prepotential reduces to the Einstein expression multiplied by \(\varpi_-\), reproducing the standard NMG charges of the BTZ black hole. For nonconstant-curvature AdS waves satisfying Brown--Henneaux conditions, the charge associated with the null Killing vector vanishes. At the degenerate point, the prescription gives the rotating hairy-black-hole charges in Eq.~\eqref{chargesOTT}, in agreement with results obtained using other methods~\cite{Oliva:2009ip,Giribet:2009qz,Giribet:2010ed}.

The present analysis is restricted to power-law asymptotic expansions and excludes the logarithmic modes admitted at the critical point~\cite{Clement:2009ka,Hohm:2010jc,Giribet:2010ed}. Extending the extrinsic prescription to logarithmic boundary conditions is therefore a natural next step. More generally, it would be useful to investigate whether the relation between vacuum degeneracy, additional holographic sources, and extrinsic boundary terms persists in other higher-curvature theories with weakened AdS fall-offs.

\appendix
\appendixpage
\addappheadtotoc
\addtocontents{toc}{\protect\setcounter{tocdepth}{0}}

\chapter{Path integral formalism} \label{pathintegralformalism}

\begin{figure}[ht]
    \centering
    \includegraphics[width=0.8\linewidth]{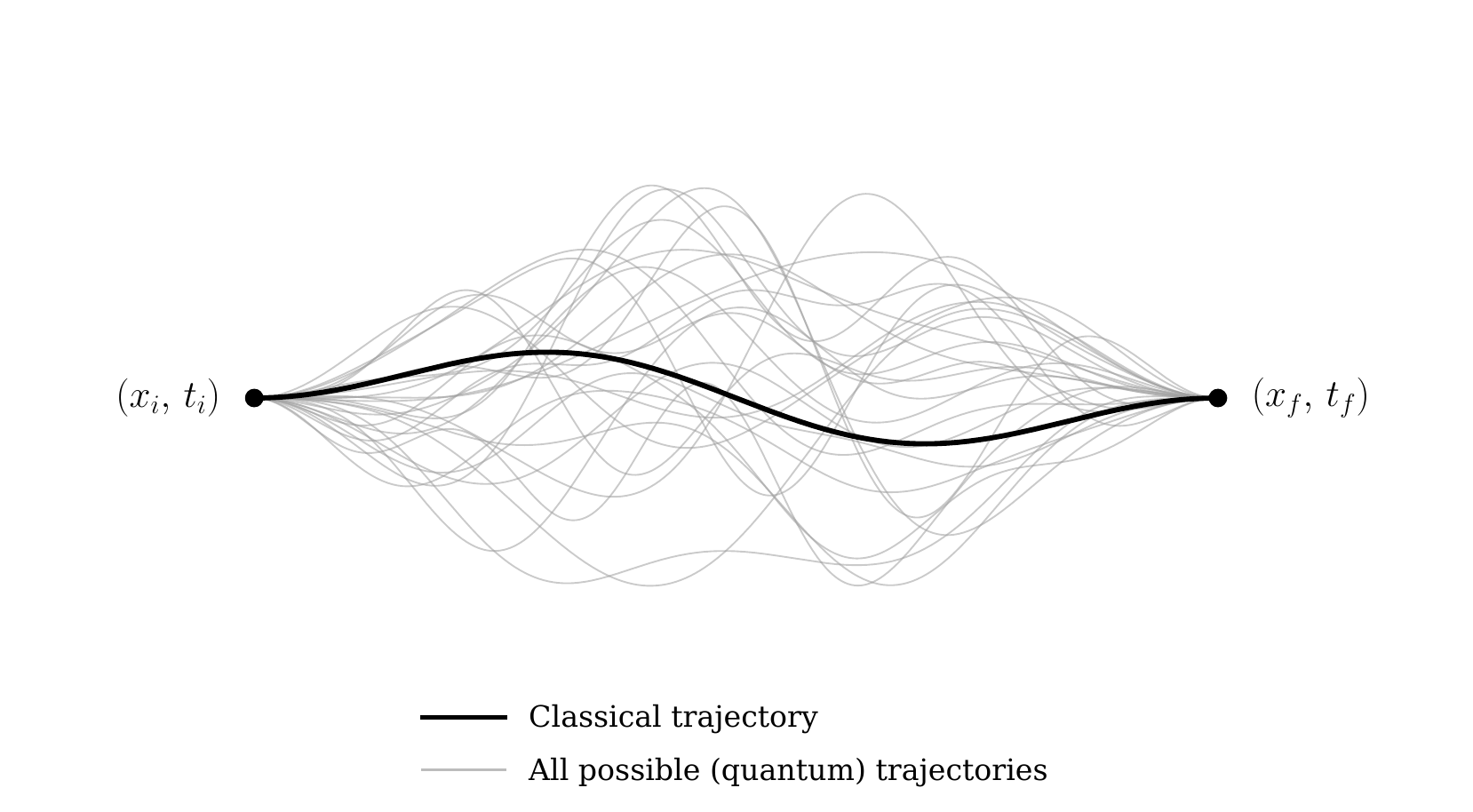}
    \caption{Representation of all possible trajectories between two points.}
    \label{fig:path_integral}
\end{figure}

We now introduce the path integral formalism. In quantum mechanics, Feynman proposed that a particle propagating from an initial point to a final point does not follow a single trajectory (See figure~\ref{fig:path_integral}). Instead, the transition amplitude receives contributions from all possible paths connecting the two endpoints. Each path is weighted by a complex phase factor $e^{iI/\hbar}$, where $I$ is the classical action evaluated along the path under consideration. The total transition amplitude is then obtained by summing over all possible histories. This formulation is known as the \textbf{sum over histories}. In what follows, we work in natural units, i.e. $\hbar=1$. 

\section{Quantum Mechanics}

Let us consider the transition amplitude between an initial position eigenstate $\ket{x_i}$ at time $t_i$ and a final position eigenstate $\ket{x_f}$ at time $t_f$, with $t_f>t_i$. For a time-independent Hamiltonian, this amplitude is the position-space matrix element of the evolution operator,
\begin{equation} \label{amplitude1}
    \mathcal{Z}\left(x_{f},t_{f};x_{i},t_{i}\right)=\bra{x_{f}}e^{-i\hat{H}\left(t_{f}-t_{i}\right)}\ket{x_{i}}\,,
\end{equation}
where in one-dimensional non-relativistic quantum mechanics, the Hamiltonian is 
\begin{equation} \label{HamiltonianNormal}
    \hat{H}=\frac{\hat{p}^{2}}{2m}+V(\hat{x})\,,
\end{equation}
where $\hat p$ is the momentum operator and $V(\hat x)$ acts by multiplication in the position representation.

 Although the evolution operator is known in this case, the matrix element is not straightforward to evaluate because the kinetic and potential operator generally do not commute, $[\hat{p}^{2}/2m,V(\hat{x})]\neq0$. Therefore, the exponential cannot be factorized exactly into the kinetic and potential contributions. To circumvent this difficulty, one divides the time interval into $N$ infinitesimal steps of length $\Delta t=(t_{f}-t_{i})/N$. This allows one to use the Suzuki-Trotter decomposition \footnote{The Suzuki-Trotter decomposition, also known as a splitting method, allows one to approximate the exponential of a sum of non-commuting operators by a product of exponentials. More precisely, 
\begin{equation*}
    e^{\left(A+B\right)t}=\lim_{N\to\infty}\left(e^{\frac{t}{N}A}e^{\frac{t}{N}B}\right)^{N}\,.
\end{equation*}
For a single infinitesimal step $h$, this implies the short-time approximation 
\begin{equation*}
    e^{h(A+B)}=e^{hA}e^{hB}+\mathcal{O}(h^{2})\,.
\end{equation*}
This allows one to separate the kinetic and potential contributions at each time slice, with the approximation becoming exact in the continuum limit $N\to\infty$, $\Delta t\to0$.}, which factorize the kinetic and potential contributions at the level of the short-time evolution operator. Thus, the transition amplitude can be written as 
\begin{equation} \label{Amplitude2}
    \mathcal{Z}\left(x_{f},t_{f};x_{i},t_{i}\right)=\bra{x_{f}}\Big(e^{-i\hat{H}\Delta t}\Big)^{N}\ket{x_{i}}\,.
\end{equation}
For a time-dependent potential, the Hamiltonian and evolution operator are instead
\begin{align}
    \hat H(t) &= \frac{\hat p^2}{2m}+V(\hat x,t)\,, \label{TimeDependentH} \\
    \hat U(t_f,t_i) &= \mathcal T \exp\left[  -i\int_{t_i}^{t_f}dt\,\hat H(t) \right]\,.
\end{align}
The symbol $\mathcal T$ orders later times to the left. For sufficiently regular time dependence, the evolution operator can be obtained as the limit of a product in which the Hamiltonian is evaluated at the initial time of each interval, namely 
\begin{equation}
    \hat U(t_f,t_i)  =  \lim_{N\to\infty} e^{-i\Delta t\hat H(t_{N-1})}  \cdots e^{-i\Delta t\hat H(t_1)} e^{-i\Delta t\hat H(t_0)}\,, \qquad
    t_n=t_i+n\Delta t\,.
\end{equation}
At finite $N$, this product approximates the time-ordered evolution. Its ordering ensures that the earliest evolution factor acts first on the initial state. The following construction therefore applies to both time-independent and time-dependent potentials.

To express the operator product as an integral,  we insert position resolutions of the identity between successive evolution factors 
\begin{equation}
    \int_{-\infty}^{\infty}dx_n\, \ket{x_n}\bra{x_n}  = \mathbb I, \qquad \langle x|x'\rangle=\delta(x-x').
    \label{CompletenessRelation}
\end{equation}
Each insertion introduces an intermediate position $x_n$ and splits the matrix element into amplitudes for adjacent time intervals. In order to illustrate this, we consider an example of two factors that are joined through
\begin{align}
    \bra{x_2} e^{-i\Delta t\hat H(t_1)} e^{-i\Delta t\hat H(t_0)} \ket{x_0} = \int_{-\infty}^{\infty}dx_1\, \bra{x_2}
    e^{-i\Delta t\hat H(t_1)}
    \ket{x_1}
    \bra{x_1}
    e^{-i\Delta t\hat H(t_0)}
    \ket{x_0}\,.
\end{align}
Repeating this operation $N-1$ times gives
\begin{align}
        \mathcal{Z}\left(x_{f},t_{f};x_{i},t_{i}\right)&= \lim_{N\to\infty} \int dx_{1}....dx_{N-1}\langle x_{f}|e^{-i\Delta t\hat{H}(t_{N-1})}|x_{N-1}\rangle... \nonumber\\
        &\times...\langle x_{n+1}|e^{-i\Delta t\hat{H}(t_{n})}|x_{n}\rangle...\langle x_{1}|e^{-i\Delta t\hat{H}(t_{i})}|x_{i}\rangle\,.\label{AmplitudeGral}
\end{align}
The endpoint positions in this expression are fixed, whereas the intermediate positions are integrated over. The limit is needed for the time-dependent approximation and for the subsequent splitting of the kinetic and potential operators.

We now focus on a single short-time transition amplitude, 
\begin{equation} \label{Amplitude3}
    \mathcal{Z}_{n}	=\langle x_{n+1}|e^{-i\Delta t\hat{H}(t_{n})}|x_{n}\rangle=\langle x_{n+1}|e^{-i\Delta t\Big(\frac{\hat{p}^{2}}{2m}+V(\hat{x},t_{n})\Big)}|x_{n}\rangle\,.
\end{equation}
For a time-dependent Hamiltonian, $\mathcal Z_n$ uses the Hamiltonian frozen at $t_n$. Applying the short-time splitting gives, in the expansion described above,
\begin{equation} \label{SusukiTrotterDecom}
    e^{-i\Delta t\Big(\frac{\hat{p}^{2}}{2m}+V(\hat{x},t_{n})\Big)}=e^{-i\Delta t\frac{\hat{p}^{2}}{2m}}e^{-i\Delta tV(\hat{x},t_{n})}+\mathcal{O}(\Delta t^{2})\,.
\end{equation}
The chosen ordering places the potential exponential next to the initial position eigenstate. Since a function of $\hat x$ acts diagonally in the position basis, its action is
\begin{equation} \label{OperatorToEigenvalues}
    e^{-i\hat{V}(\hat{x},t_{n})\Delta t}|x_{n}\rangle=e^{-iV(x_{n},t_{n})\Delta t}|x_{n}\rangle\,.
\end{equation}
Therefore, the single short-time transition amplitude can be written as
\begin{equation} \label{MiniPropagator}
    \mathcal{Z}_{n}=e^{-iV(x_{n},t_{n})\Delta t}\langle x_{n+1}|e^{-i\frac{\hat{p}^{2}}{2m}\Delta t}|x_{n}\rangle\,.
\end{equation}
The appearance of $V(x_n,t_n)$ is therefore a consequence of the ordering in \eqref{SusukiTrotterDecom}. It corresponds to evaluating the potential at the initial point of each time interval.
The kinetic operator is diagonal in the momentum basis. We choose the normalization $\langle p|p'\rangle=2\pi\delta(p-p')$, which fixes the momentum completeness relation and plane waves overlaps as
\begin{align}
    \int_{-\infty}^{\infty}\frac{dp}{2\pi} \ket p\bra p  &= \mathbb I\,, \label{MomentumCompleteness} \\ 
    \langle p|x_n\rangle  =  e^{-ipx_n}\,,  \qquad & \langle x_n|p\rangle  = e^{ipx_n}\,.
    \label{Conventionspxn}
\end{align}
These conventions reproduce the position normalization through the Fourier identity $\int dp\,e^{ip(x-x')}=2\pi\,\delta(x-x')$. Inserting the identity on a position eigenstate thus expresses it as a superposition of momentum eigenstates
\begin{equation}
    \ket{x_n} = \int_{-\infty}^{\infty}\frac{dp}{2\pi} \ket p\,\langle p|x_n\rangle  = \int_{-\infty}^{\infty}\frac{dp}{2\pi} \ket p\,e^{-ipx_n}\,.
    \label{StatexnRelation}
\end{equation}
Each momentum eigenstate acquires the kinetic phase $e^{-ip^2\Delta t/(2m)}$. Projecting onto the final position eigenstate supplies the remaining plane-wave factor
\begin{align}
    \bra{x_{n+1}} e^{-i\frac{\hat p^2}{2m}\Delta t}  \ket{x_n} &= \int_{\infty}^{-\infty}\frac{dp}{2\pi}  \langle x_{n+1}|p\rangle  e^{ -i\frac{p^2}{2m}\Delta t} \langle p|x_n\rangle \,, \nonumber\\
    &= \int_{-\infty}^{\infty}\frac{dp}{2\pi} \exp\left[  -i\frac{\Delta t}{2m}p^2+ip(x_{n+1}-x_n) \right]\,.
\end{align}
The position dependence appears only through the difference between adjacent positions. Defining $\Delta x_n:=x_{n+1}-x_n$, the short-time amplitude becomes
\begin{equation}
    \mathcal Z_n \simeq e^{-iV(x_n,t_n)\Delta t}  \int_{-\infty}^{\infty}\frac{dp}{2\pi} \exp\left[ -i\frac{\Delta t}{2m}p^2 +ip\Delta x_n \right].
\end{equation}
The momentum integral is Gaussian.\footnote{ For $\operatorname{Re}a>0$, we use
\begin{equation*}
    \int_{-\infty}^{\infty}dp\,
    \exp\left[-\frac a2p^2+bp\right]
    =
    \sqrt{\frac{2\pi}{a}}\,
    \exp\left[\frac{b^2}{2a}\right]\,.
\end{equation*}
In the present case, the formal identifications are
\begin{equation*}
    a=\frac{i\Delta t}{m}\,,
    \qquad
    b=i(x_{n+1}-x_n)\,.
\end{equation*}
Since $a$ is imaginary, the integral is defined by replacing $a$ with $\epsilon+i\Delta t/m$ and taking $\epsilon\to0^+$ after integration.
}
To make its evaluation explicit, we complete the square as follows
\begin{align}
    -i\frac{\Delta t}{2m}p^2+ip\Delta x_n &= -i\frac{\Delta t}{2m} \left[ p^2-\frac{2m\Delta x_n}{\Delta t}p \right] \,, \nonumber\\
    &= -i\frac{\Delta t}{2m} \left( p-\frac{m\Delta x_n}{\Delta t} \right)^2 + i\frac{m(\Delta x_n)^2}{2\Delta t}\,.
\end{align}
This separates the position-dependent phase from the Gaussian normalization. A shift of the integration variable to $q=p-m\Delta x_n/\Delta t$, understood with the prescription stated in the previous footnote, then gives
\begin{align}
    \mathcal Z_n
    &\simeq
    \exp\left[
        i\frac{m(\Delta x_n)^2}{2\Delta t}
        -iV(x_n,t_n)\Delta t
    \right]
    \int_{-\infty}^{\infty}\frac{dq}{2\pi}
    e^{-i\frac{\Delta t}{2m}q^2}\,,
    \\
    \int_{-\infty}^{\infty}\frac{dq}{2\pi}
    e^{-i\frac{\Delta t}{2m}q^2}
    &=
    \lim_{\epsilon\to0^+}
    \frac{1}{\sqrt{2\pi}}
    \left(
        \epsilon+\frac{i\Delta t}{m}
    \right)^{-1/2}
    =
    \left(\frac{m}{2\pi i\Delta t}\right)^{1/2}\,.
\end{align}
For $\Delta t>0$, the prescription selects $(1/i)^{1/2}=e^{- i\pi/4}$. The short-time amplitude is therefore
\begin{equation}
    \mathcal Z_n
    \simeq
    \left(-\frac{im}{2\pi\Delta t}\right)^{1/2}
    e^{-iV(x_n,t_n)\Delta t}
    e^{i\frac{m}{2}
        \frac{(x_{n+1}-x_n)^2}{\Delta t}}.
\end{equation}
The kinetic phase depends quadratically on the position difference, while the potential contributes a phase proportional to the duration of the interval. Combining them makes the connection with the classical Lagrangian more explicit
\begin{equation}
    \mathcal Z_n
    \simeq
    \mathcal N
    \exp\left[
        i\Delta t
        \left(
            \frac m2
            \left(
                \frac{x_{n+1}-x_n}{\Delta t}
            \right)^2
            -
            V(x_n,t_n)
        \right)
    \right]\,.
\end{equation}
Substituting this expression into~\eqref{AmplitudeGral}, each of the $N$ short-time kernels contributes one normalization factor. Their phases add because the kernels are ordinary complex valued functions after the intermediate states have been inserted. The transition amplitude then takes the following form 
\begin{equation} \label{Zfinal}
    \mathcal{Z}=\lim_{N\to\infty}\mathcal{N}^{N}\int\left(\prod_{n=1}^{N-1}dx_n\right)\exp\left[i\sum_{n=0}^{N-1}\Delta t\left(\frac{m}{2}\left(\frac{x_{n+1}-x_{n}}{\Delta t}\right)^{2}-V(x_{n},t_{n})\right)\right]\,,
\end{equation}
where $\mathcal N$ is the normalization factor arising from each momentum integral,
\begin{equation}
    \mathcal N
    \equiv
    \left(\frac{m}{2\pi i\Delta t}\right)^{1/2}
    =
    e^{-i\pi/4}
    \sqrt{\frac{m}{2\pi\Delta t}}\,,
    \qquad
    \Delta t=\frac{t_f-t_i}{N}\,.
\end{equation}
Its dependence on $\Delta t$, and hence on $N$, is implicit in the notation. Although there are $N$ time intervals, only $N-1$ positions are integrated over because the endpoints are fixed.\footnote{Notice that we define here $x_{0}=x_{i}$ and $x_{N}=x_{f}$.}

The sum appearing in \eqref{Zfinal} is the discretized action,
\begin{equation}
    I_N[x_0,\ldots,x_N]
    :=
    \sum_{n=0}^{N-1}\Delta t
    \left[
        \frac m2
        \left(
            \frac{x_{n+1}-x_n}{\Delta t}
        \right)^2
        -
        V(x_n,t_n)
    \right]\,.
\end{equation}
In the formal continuum limit, the finite difference expression is identified with the velocity $\dot{x}(t)$, while the discrete sum approaches the time integral of the classical Lagrangian. This motivates the continuum notation
\begin{align}
    \mathcal Z(x_f,t_f;x_i,t_i)
    &=
    \int_{x(t_i)=x_i}^{x(t_f)=x_f}
    \mathcal D x(t)\,e^{iI[x]}\,,
    \\
    I[x]
    &=
    \int_{t_i}^{t_f}dt\,
    \left[
        \frac m2\dot x^{\,2}-V(x,t)
    \right]\,.
\end{align}
Here, $\mathcal D x(t)$ is not an ordinary finite-dimensional measure.
Within the time-slicing prescription introduced above, it is defined by
\begin{equation}
    \int_{x(t_i)=x_i}^{x(t_f)=x_f}
    \mathcal D x(t)\,F[x]
    :=
    \lim_{N\rightarrow\infty}
    \mathcal N^{N}
    \prod_{n=1}^{N-1}
    \int_{-\infty}^{\infty}dx_n\,
    F[x_0,x_1,\ldots,x_N]\,,
    \label{FunctionalMeasureDefinition}
\end{equation}
where
\begin{equation}
    \Delta t=\frac{t_f-t_i}{N}\,,
    \qquad
    x_0=x_i\,,
    \qquad
    x_N=x_f\,,
    \qquad
    \mathcal N=
    \left(
        \frac{m}{2\pi i\Delta t}
    \right)^{1/2}\,,
\end{equation}
and $F[x]$ denotes an arbitrary functional of the path. Thus,
$\mathcal D x(t)$ is a compact notation for the regulated limit of the
product of integrations over all intermediate positions, together with
the normalization factors generated by the short-time propagators. The
endpoint values are held fixed and are therefore not integrated over.

This expression defines the configuration space \textbf{path integral}. Although
its continuum form is compact, its meaning is inherited from a chosen
regularization, such as the time-slicing prescription above. In real
time, the resulting integral is oscillatory and is understood with an
appropriate convergence prescription, or equivalently by continuation
from Euclidean time. The dependence of the functional measure on its
regularization becomes particularly important in quantum field theory,
where a classical symmetry may fail to preserve the regulated measure.
This failure plays a central role in the path-integral description of
quantum anomalies.
\section{Path Integral Formulation of QFT}

This construction can be extended to quantum field theory. The derivation is quite similar to the quantum mechanical case, with the position and momentum variables replaced by a field and its conjugate momentum. In the Schrödinger picture, they are given by 
\begin{align}
    \hat{\phi}(\vec{x})	&=\int\frac{d^{3}p}{(2\pi)^{3}}\frac{1}{\sqrt{2\omega_{p}}}\left(a_{p}e^{i\vec{p}\cdot\vec{x}}+a_{p}^{\dagger}e^{-i\vec{p}\cdot\vec{x}}\right)\,, \\
\hat{\pi}(\vec{x})	&=-i\int\frac{d^{3}p}{(2\pi)^{3}}\sqrt{\frac{\omega_{p}}{2}}\left(a_{p}e^{i\vec{p}\cdot\vec{x}}-a_{p}^{\dagger}e^{-i\vec{p}\cdot\vec{x}}\right)\,,
\end{align}
respectively. These canonical operators satisfy
\begin{equation}
    \left[\hat{\phi}(\vec{x}),\hat{\pi}(\vec{y})\right]=i\delta^{(3)}\left(\vec{x}-\vec{y}\right)\,.
\end{equation}
We introduce field eigenstates $|\Phi\rangle$  and conjugate-momentum eigenstates $|\Pi\rangle$, defined by 
\begin{equation}
    \hat{\phi}(\vec{x})|\Phi\rangle=\Phi(\vec{x})|\Phi\rangle\,,\qquad\text{and}\qquad\hat{\pi}(\vec{x})|\Pi\rangle=\Pi(\vec{x})|\Pi\rangle\,.
\end{equation}
They obey the following relation 
\begin{equation}
    \langle\Pi|\Phi\rangle=e^{-i\int d^{3}x\Pi(\vec{x})\Phi(\vec{x})}\,.
\end{equation}
For simplicity, let us consider a real scalar field. Its Lagrangian density is 
\begin{equation}
    \mathcal{L}=\frac{1}{2}\dot{\phi}^{2}-\frac{1}{2}\left(\nabla\phi\right)^{2}-V(\phi)\,,
\end{equation}
and the corresponding Hamiltonian is given by 
\begin{equation}
    H=\int d^{3}x\left(\frac{1}{2}\left[\pi^{2}+\left(\nabla\phi\right)^{2}\right]+V(\phi)\right)\,,
\end{equation}
where the canonical momentum is $\pi=\dot{\phi}$. Promoting the fields to operators the Hamiltonian becomes 
\begin{equation}
    \hat{H}=\int d^{3}x\left(\frac{1}{2}\hat{\pi}^{2}+\mathcal{V}(\hat{\phi})\right)\,,
\end{equation}
where $\mathcal{V}(\hat{\phi})=\tfrac{1}{2}\left(\nabla\phi\right)^{2}+V(\phi)\,.$

Consider now the transition amplitude between eigenstates of the Schrödinger-picture field operator $\hat{\phi}(\mathbf{x})$. More precisely, $\ket{\Phi_i}$ and $\ket{\Phi_f}$ denote simultaneous generalized eigenstates satisfying
\begin{equation}
\hat{\phi}(\vec{x})\ket{\Phi_{i,f}}
=
\Phi_{i,f}(\vec{x})\ket{\Phi_{i,f}},,
\end{equation}
where each eigenvalue $\Phi_{i,f}(\vec{x})$ represents an entire field configuration on the corresponding spatial hypersurface. The transition amplitude between these configurations is
\begin{equation}
\mathcal{Z}\left(\Phi_f,t_f;\Phi_i,t_i\right)
=
\langle\Phi_f|
e^{-i\hat{H}(t_f-t_i)}
|\Phi_i\rangle,.
\end{equation}
Similarly as in quantum mechanics, we divide the time interval into $N$ infinitesimal steps and insert complete sets of field eigenstates, 
\begin{equation}
    \int\mathcal{D}\Phi_{n}|\Phi_{n}\rangle\langle\Phi_{n}|=\mathbb{I}\,.
\end{equation}
This gives, 
\begin{align}
    \mathcal{Z}\left(\Phi_{f},t_{f};\Phi_{i},t_{i}\right)&=\int\mathcal{D}\Phi_{1}(x)...\mathcal{D}\Phi_{N-1}(x)\langle\Phi_{f}|e^{-i\Delta t\hat{H}(t_{N-1})}|\Phi_{N-1}\rangle...\times \nonumber \\
    &...\times\langle\Phi_{n+1}|e^{-i\Delta t\hat{H}(t_{n})}|\Phi_{n}\rangle...\langle\Phi_{1}|e^{-i\Delta t\hat{H}(t_{0})}|\Phi_{i}\rangle\,.
\end{align}
We now analyze a single short-time amplitude, 
\begin{equation}
    \mathcal{Z}_{n}=\langle\Phi_{n+1}|e^{-i\Delta t\hat{H}(t_{n})}|\Phi_{n}\rangle\,.
\end{equation}
Using the Suzuki-Trotter decomposition~\cite{Trotter:1959ytf,Suzuki:1976be}, we can separate the momentum term from the potential 
\begin{equation}
    e^{-i\Delta t\hat{H}(t_{n})}=\exp\left[-\frac{i}{2}\Delta t\int d^{3}x\hat{\pi}^{2}\right]\exp\left[-i\Delta t\int d^{3}x\mathcal{V}(\hat{\phi})\right]+\mathcal{O}\left(\Delta t^{2}\right)\,.
\end{equation}
Since $|\Phi_{n}\rangle$  is an eigenstate of $\hat{\Phi}(\vec{x})$, the potential term acts diagonally as follows 
\begin{equation}
    e^{-i\Delta t\int d^{3}x\mathcal{V}(\hat{\phi})}|\Phi_{n}\rangle=e^{-i\Delta t\int d^{3}x\mathcal{V}(\Phi_{n})}|\Phi_{n}\rangle\,.
\end{equation}
Thus, 
\begin{equation}
    \mathcal{Z}_{n}=e^{-i\Delta t\int d^{3}x\mathcal{V}(\Phi_{n})}\langle\Phi_{n+1}|e^{-\frac{i}{2}\Delta t\int d^{3}x\hat{\pi}^{2}}|\Phi_{n}\rangle\,.
\end{equation}
The remaining matrix element involves the conjugate momentum operator. Since the field eigenstate is not an eigenstate of $\hat{\pi}$, we insert a complete set of momentum eigenstates 
\begin{equation}
    \int\mathcal{D}\Pi_{n}|\Pi_{n}\rangle\langle\Pi_{n}|=\mathbb{I}\,.
\end{equation}
Using the fact that 
\begin{equation}
    |\Phi_{n}\rangle	=\int\mathcal{D}\Pi_{n}|\Pi_{n}\rangle e^{-i\int d^{3}x\Pi_{n}(\vec{x})\Phi_{n}(\vec{x})}\,,
\end{equation}
we obtain
\begin{align} \label{ShortTimeAmpli}
    \mathcal{Z}_{n}	&=e^{-i\Delta t\int d^{3}x\mathcal{V}(\Phi_{n})}\int\mathcal{D}\Pi_{n}\langle\Phi_{n+1}|\exp\left[-\frac{i}{2}\Delta t\int d^{3}x\hat{\pi}^{2}\right]|\Pi_{n}\rangle\exp\left[-i\int d^{3}x\Pi_{n}\Phi_{n}\right]\,, \nonumber \\
	&=e^{-i\Delta t\int d^{3}x\mathcal{V}(\Phi_{n})}\int\mathcal{D}\Pi_{n}\langle\Phi_{n+1}|\Pi_{n}\rangle\exp\left[-\frac{i}{2}\Delta t\int d^{3}x\Pi_{n}^{2}\right]\exp\left[-i\int d^{3}x\Pi_{n}\Phi_{n}\right]\,, \nonumber \\
	&=e^{-i\Delta t\int d^{3}x\mathcal{V}(\Phi_{n})}\int\mathcal{D}\Pi_{n}\exp\left[i\int d^{3}x\left(\Pi_{n}\left[\Phi_{n+1}-\Phi_{n}\right]-\frac{1}{2}\Delta t\Pi_{n}^{2}\right)\right]\,.
\end{align}
The integral over $\Pi_{n}$ is Gaussian. To evaluate it \footnote{The generalization of the Gauss integral for multi-dimensional integrals is given by 
\begin{equation*}
    \int d^{N}x\exp\left[-\frac{1}{2}\boldsymbol{x}^{T}\boldsymbol{A}\boldsymbol{x}+\boldsymbol{b}^{T}\boldsymbol{x}\right]=\left(\frac{(2\pi)^{N}}{\det\boldsymbol{A}}\right)^{1/2}\exp\left[\frac{1}{2}\boldsymbol{b}^{T}\boldsymbol{A}^{-1}\boldsymbol{b}\right]\,,
\end{equation*}
where $\boldsymbol{x}^{T}$ is the transpose of $\boldsymbol{x}$, and $\boldsymbol{A}$ is an invertible matrix, usually assumed symmetric with positive real part. Its functional form is 
\begin{equation*}
    \int\mathcal{D}[f(x)]\exp\left[-\frac{1}{2}\int dxdyf(x)A(x,y)f(y)+\int dxb(x)f(x)\right]=\left(\frac{(2\pi)^{N}}{\det A(x,y)}\right)^{1/2}\exp\left[\frac{1}{2}\int dxdyb(x)A^{-1}(x,y)b(y)\right]\,.
\end{equation*}}, we write the exponent as
\begin{equation}
    \int \mathcal{D}\Pi_n \exp \left[i\int d^3x \left( \Pi_n \Delta \Phi_n \right)-\frac{\Delta t}{2}\Pi_{n}^{2}\right)=\mathcal{N}\exp\left[\frac{i}{2}\Delta t\int d^{3}x\left(\frac{\Phi_{n+1}-\Phi_{n}}{\Delta t}\right)^{2}\right]\,,
\end{equation}
with $\mathcal{N}$ as the normalization factor. Thus, the short-time amplitude~\eqref{ShortTimeAmpli} becomes 
\begin{equation}
    \mathcal{Z}_{n}	=\mathcal{N}\exp\left[i\Delta t\int d^{3}x\left(\frac{1}{2}\left(\frac{\Phi_{n+1}-\Phi_{n}}{\Delta t}\right)^{2}-\mathcal{V}(\Phi_{n})\right)\right]\,.
\end{equation}
Notice that, in the continuum limit $\Delta t\to0$ and $N\to\infty$, 
\begin{equation}
    \frac{\Phi_{n+1}-\Phi_{n}}{\Delta t}\to\partial_{t}\Phi\,,
\end{equation}
and therefore the product of all short-time amplitudes gives 
\begin{equation} \label{PathIntegralField}
    \mathcal{Z}\left(\Phi_{f},t_{f};\Phi_{i},t_{i}\right)=\int_{\Phi(t_{i})=\Phi_{i}}^{\Phi(t_{f})=\Phi_{f}}\mathcal{D}\Phi e^{iI[\Phi]}\,,
\end{equation}
where $I[\phi]=\int d^{4}x\mathcal{L}[\Phi]$, with 
\begin{equation} \label{LagrangianField}
    \mathcal{L}[\Phi]=\frac{1}{2}\left(\partial_{t}\Phi\right)^{2}-\mathcal{V}\left(\Phi\right)\,.
\end{equation}
This corresponds to the path integral in quantum field theory. It tells us to integrate over all field configurations satisfying the prescribed boundary conditions. Another way to write~\eqref{PathIntegralField}, and it is more common to write
\begin{equation}
    \mathcal{Z}=\mathcal{N}\int\mathcal{D}\Phi e^{iI[\Phi]}\,.
\end{equation}

An important application of the path integral formalism in quantum field theory is the computation of correlation functions. To do so, we introduce the notion of external sources. Physically, a source can be understood as an external classical influence coupled to the quantum field. For instance, if we consider a scalar field $\phi(x)$, then the source $J(x)$ is an external spacetime dependent function that probes the field. This object is useful because it provides a systematic way to generate correlation functions. 

For a scalar field, we define the generating functional as 
\begin{equation}
    \mathcal{Z}\left[J\right]=\int\mathcal{D}\phi\exp\left[iI[\phi]+i\int d^{4}xJ(x)\phi(x)\right]\,,
\end{equation}
where $J(x)$ denotes the external source. Setting $J=0$, one obtains the vacuum-to-vacuum amplitude without sources, 
\begin{equation} \label{VacuumZ0}
    \mathcal{Z}\left[0\right]=\int\mathcal{D}\phi e^{iI[\phi]}\,.
\end{equation}
This quantity is used to normalize correlation functions. Indeed, the overall normalization of the functional measure is arbitrary, and the expectation value of the identity operator must satisfy the relation $\langle\Omega|\Omega\rangle=1$. Therefore, path integrals with operator insertions must be divided by Eq.~\eqref{VacuumZ0}. 

As we mentioned before, the source term allows us to insert fields into the path integral by differentiating with respect to $J(x)$. Since $J(x)$ is coupled linearly to $\phi(x)$, a functional derivative with respect to the source brings down a factor of the field. This is, 
\begin{equation}
    \frac{\delta}{\delta J(x_{1})}\left[i\int d^{4}xJ(x)\phi(x)\right]=i\phi(x_{1})\,.
\end{equation}
Therefore, 
\begin{equation}
    -i\frac{\delta \mathcal{Z}}{\delta J(x_{1})}=\int\mathcal{D}\phi\exp\left[iI[\phi]+i\int d^{4}xJ(x)\phi(x)\right]\phi(x_{1})\,.
\end{equation}
Setting $J=0$, one obtains 
\begin{equation}
    -i\frac{\partial\mathcal{Z}}{\partial J(x_{1})}\Big|_{J=0}=\int\mathcal{D}\phi e^{iI[\phi]}\phi(x_{1})\,.
\end{equation}
Thus, the normalized vacuum expectation value of one field operator is 
\begin{equation}
    \langle\Omega|\hat{\phi}(x_{1})|\Omega\rangle=-i\frac{1}{\mathcal{Z}[0]}\frac{\delta\mathcal{Z}}{\delta J(x_{1})}\Big|_{J=0}=\frac{\int\mathcal{D}\phi e^{iI[\phi]}\phi(x_{1})}{\int\mathcal{D}\phi e^{iI[\phi]}}\,.
\end{equation}
This object is known as the one-point correlation function, or one-point function, of the scalar field. More generally, by taking several functional derivatives with respect to the source, one obtains the $n-$point correlation functions, 
\begin{equation}
    \langle\Omega|T\{\hat{\phi}(x_{1})...\hat{\phi}(x_{n})\}|\Omega\rangle=\left(-i\right)^{n}\frac{1}{\mathcal{Z}[0]}\frac{\delta ^{n}\mathcal{Z}}{\delta J(x_{1})...\delta J(x_{n})}\Big|_{J=0}\,.
\end{equation}
Here $T$ denotes the time-ordering operator. It orders the field operators according to their time coordinates, placing later times to the left and earlier times to the right. This operator is a consequence of the construction of the path integral from time-evolution operators. It is worth mentioning that the correlation functions generated by $\mathcal{Z}[J]$ are time-ordered vacuum correlation functions, also known as Green functions.

\chapter{Algebraic Decomposition of the Riemann Tensor} \label{AlgebraicDecomposition}

This appendix derives the algebraic decomposition of the Riemann tensor using the mixed-index notation $R_{\lambda\rho}^{\mu\nu}$. The purpose is to establish explicitly that, in three spacetime dimensions, the Riemann tensor is completely determined by the Ricci tensor and its trace. This algebraic identity must be distinguished from the dynamical statement that three dimensional Einstein gravity has no local propagating degrees of freedom. 

Throughout this appendix, $g_{\mu\nu}$ is a nondegenerate metric and $\nabla$  denotes its Levi-Civita connection. The curvature convention is 
\begin{equation}
    [\nabla_{\lambda},\nabla_{\rho}]V^{\mu}=R^{\mu}{}_{\sigma\lambda\rho}V^{\sigma},\qquad R^{\mu\nu}_{\lambda\rho}:=g^{\nu\sigma}R^{\mu}{}_{\sigma\lambda\rho}\,.
\end{equation}
The Ricci tensor and scalar curvature are obtained through contractions 
\begin{equation}
    R^{\mu}{}_{\lambda}:=R^{\mu\nu}_{\lambda\nu},\qquad R:=R^{\mu}_{\mu}=R^{\mu\nu}_{\mu\nu}\,.
\end{equation}
The decompositions below assume $d\geq3$. We use normalized antisymmetrization, 
\begin{equation}
    A_{[\mu\nu]}:=\frac{1}{2}\left(A_{\mu\nu}-A_{\nu\mu}\right)\,,
\end{equation}
and define the generalized Kronecker delta by 
\begin{equation}
    \delta_{\lambda\rho}^{\mu\nu}:=\delta_{\lambda}^{\mu}\delta_{\rho}^{\nu}-\delta_{\rho}^{\mu}\delta_{\lambda}^{\nu}\,.
\end{equation}
Its relevant contractions are 
\begin{equation}
    \delta_{\lambda\nu}^{\mu\nu}=(d-1)\delta_{\lambda}^{\mu},\qquad\delta_{\mu\nu}^{\mu\nu}=d(d-1)\,.
\end{equation}

\section{Independent Components of the Curvature Tensor}

The Riemann tensor is antisymmetric in each index pair 
\begin{equation}
    R^{\mu\nu}_{\lambda\rho}=-R^{\nu\mu}_{\lambda\rho}=-R^{\mu\nu}_{\rho\lambda}\,.
\end{equation}
The pair-exchange symmetry and the algebraic Bianchi identity are most directly expressed after lowering the first pair, namely, $R_{\mu\nu\lambda\rho}:=g_{\mu\alpha}g_{\nu\beta}R^{\alpha\beta}_{\lambda\rho}\,.$ Thus, 
\begin{equation}
    R_{\mu\nu\lambda\rho}=R_{\lambda\rho\mu\nu},\qquad R_{\mu[\nu\lambda\rho]}=0\,.
\end{equation}
Notice that these properties do not depend on the signature of the metric. An antisymmetric pair of indices has 
\begin{equation}
    n_{A}=\frac{d(d-1)}{2}\,,
\end{equation}
independent values. Before imposing the Bianchi identity, the pair-exchange symmetry allows the curvature tensor to be regarded as a symmetric bilinear form on the space of two-forms. It therefore has $n_{A}(n_{A}+1)/2$ independent components. The Bianchi identity removes the totally antisymmetric four-index part, imposing $\binom{d}{4}$ independent conditions. Consequently,
\begin{equation}
    N_{R}(d)=\frac{n_{A}(n_{A}+1)}{2}-\binom{d}{4}=\frac{d^{2}(d^{2}-1)}{12}\,.
\end{equation}
Here, $\binom{d}{4}$ for $d<4$ is understood to vanish. The Ricci tensor with both indices lowered is symmetric, 
\begin{equation}
    R_{\mu\nu}:=g_{\mu\alpha}R^{\alpha}_{\nu}=R_{\nu\mu}\,.
\end{equation}
It therefore possesses 
\begin{equation}
    N_{{\rm Ricci}}(d)=\frac{d(d+1)}{2}\,,
\end{equation}
independent components. The curvature information not contained in the Ricci tensor is encoded in the Weyl tensor, as established by the following decomposition. 

\section{Weyl and Ricci Decomposition}

We first separate the Ricci tensor into its trace and traceless parts 
\begin{equation}
    R_{\nu}^{\mu}=H_{\nu}^{\mu}+\frac{R}{d}\delta_{\nu}^{\mu}\,,\qquad H_{\mu}^{\mu}=0\,.
\end{equation}
The Riemann tensor can then be decomposed as 
\begin{equation} \label{eq:riemann-general-ansatz}
    R_{\lambda\rho}^{\mu\nu}	=C_{\lambda\rho}^{\mu\nu}+\alpha\left(\delta_{\lambda}^{\mu}H_{\rho}^{\nu}-\delta_{\rho}^{\mu}H_{\lambda}^{\nu}-\delta_{\lambda}^{\nu}H_{\rho}^{\mu}+\delta_{\rho}^{\nu}H_{\lambda}^{\mu}\right)+\beta R\delta_{\lambda\rho}^{\mu\nu}\,,
\end{equation}
where $C_{\lambda\rho}^{\mu\nu}$ is the Weyl tensor. It has the algebraic symmetries of the Riemann tensor and is completely traceless. In particular, $C_{\lambda\nu}^{\mu\nu}=0$. 

To determine $\alpha$  and $\beta$, we contract the second upper index with the second lower index in Eq.~\eqref{eq:riemann-general-ansatz}. Using $H_{\mu}^{\mu}=0$, we obtain 
\begin{align}
    R_{\lambda}^{\mu}&=\alpha\left(\delta_{\lambda}^{\mu}H_{\nu}^{\nu}-\delta_{\nu}^{\mu}H_{\lambda}^{\nu}-\delta_{\lambda}^{\nu}H_{\nu}^{\mu}+\delta_{\nu}^{\nu}H_{\lambda}^{\mu}\right)+\beta R\delta_{\lambda\nu}^{\mu\nu}\,, \nonumber \\
	&=\alpha(d-2)H_{\lambda}^{\mu}+\beta(d-1)R\delta_{\lambda}^{\mu}\,.
\end{align}
Comparison with 
\begin{equation}
    R_{\lambda}^{\mu}=H_{\lambda}^{\mu}+\frac{R}{d}\delta_{\lambda}^{\mu}\,,
\end{equation}
fixes the coefficients 
\begin{equation}
    \alpha=\frac{1}{d-2},\qquad\beta=\frac{1}{d(d-1)}\,.
\end{equation}
Substituting these expressions and eliminating the traceless Ricci tensor gives 
\begin{equation} \label{eq:general-riemann-decomposition}
    R_{\lambda\rho}^{\mu\nu}=C_{\lambda\rho}^{\mu\nu}+\frac{1}{d-2}\left(\delta_{\lambda}^{\mu}R_{\rho}^{\nu}-\delta_{\rho}^{\mu}R_{\lambda}^{\nu}-\delta_{\lambda}^{\nu}R_{\rho}^{\mu}+\delta_{\rho}^{\nu}R_{\lambda}^{\mu}\right)-\frac{R}{(d-1)(d-2)}\delta_{\lambda\rho}^{\mu\nu}\,.
\end{equation}
For any mixed tensor $X^{\mu}{}_{\nu}$, normalized antisymmetrization over the two independent index pairs gives 
\begin{equation}
    4\delta_{[\lambda}^{[\mu}X^{\nu]}{}_{\rho]}=\delta_{\lambda}^{\mu}X^{\nu}{}_{\rho}-\delta_{\rho}^{\mu}X^{\nu}{}_{\lambda}-\delta_{\lambda}^{\nu}X^{\mu}{}_{\rho}+\delta_{\rho}^{\nu}X^{\mu}{}_{\lambda}\,.
\end{equation}
The Riemann decomposition can therefore be written compactly as 
\begin{equation}
    R_{\lambda\rho}^{\mu\nu}=C_{\lambda\rho}^{\mu\nu}+\frac{4}{d-2}\delta_{[\lambda}^{[\mu}R_{\rho]}^{\nu]}-\frac{R}{(d-1)(d-2)}\delta_{\lambda\rho}^{\mu\nu}\,.
\end{equation}
This decomposition separates the independent Ricci data from the trace–free curvature. For $d\geq3$, the number of independent Weyl components is consequently 
\begin{equation}
    N_{C}(d)=N_{R}(d)-N_{{\rm Ricci}}(d)=\frac{d(d+1)(d+2)(d-3)}{12}\,.
\end{equation}
An equivalent expression follows by introducing the Schouten tensor, 
\begin{equation}
    S_{\nu}^{\mu}=\frac{1}{d-2}\left(R_{\nu}^{\mu}-\frac{R}{2(d-1)}\delta_{\nu}^{\mu}\right)\,.
\end{equation}
The Riemann tensor then takes the form 
\begin{equation}
    R_{\lambda\rho}^{\mu\nu}=C_{\lambda\rho}^{\mu\nu}+4\delta_{[\lambda}^{[\mu}S_{\rho]}^{\nu]}\,.
\end{equation}

\subsection{The Road to Three Dimensions}

In three dimensions, 
\begin{equation}
    N_{R}(3)=6,\qquad N_{{\rm Ricci}}(3)=6,\qquad N_{C}(3)=0.\,
\end{equation}
Consequently, the Weyl tensor vanishes identically 
\begin{equation}
    C_{\lambda\rho}^{\mu\nu}\equiv0\,.
\end{equation}
This is an algebraic identity and does not require the use of any gravitational field equation. Setting $d=3$ in Eq.~\eqref{eq:general-riemann-decomposition} gives 
\begin{equation}
    R_{\lambda\rho}^{\mu\nu}	=\delta_{\lambda}^{\mu}R_{\rho}^{\nu}-\delta_{\rho}^{\mu}R_{\lambda}^{\nu}-\delta_{\lambda}^{\nu}R_{\rho}^{\mu}+\delta_{\rho}^{\nu}R_{\lambda}^{\mu}-\frac{R}{2}\delta_{\lambda\rho}^{\mu\nu}\,.
\end{equation}
Equivalently, 
\begin{equation} \label{eq:riemann-mixed-three-dimensions}
    R_{\lambda\rho}^{\mu\nu}=4\delta_{[\lambda}^{[\mu}R_{\rho]}^{\nu]}-\frac{R}{2}\delta_{\lambda\rho}^{\mu\nu}\,.
\end{equation}
Thus, the three dimensional Riemann tensor contains no algebraically independent information beyond the Ricci tensor. 

The compatibility of this identity with the contraction convention used in this thesis can be checked directly. Contracting Eq.~\eqref{eq:riemann-mixed-three-dimensions} over $\nu$  and $\rho$ , and using the fact that $\delta_{\mu}^{\mu}=3$ yields 
\begin{equation}
    R_{\lambda\nu}^{\mu\nu}=R\delta_{\lambda}^{\mu}-R_{\lambda}^{\mu}-R_{\lambda}^{\mu}+3R_{\lambda}^{\mu}-R\delta_{\lambda}^{\mu}=R_{\lambda}^{\mu}\,.
\end{equation}
A further contraction gives $R_{\mu\nu}^{\mu\nu}=R$, as required. 

Another quantity that it is worth pointing out is the three-dimensional Schouten tensor. This is given by 
\begin{equation}
    S_{\nu}^{\mu}=R_{\nu}^{\mu}-\frac{1}{4}R\delta_{\nu}^{\mu}\,,
\end{equation}
so the curvature identity also becomes
\begin{equation}
    R_{\lambda\rho}^{\mu\nu}=4\delta_{[\lambda}^{[\mu}S_{\rho]}^{\nu]}\,.
\end{equation}
Although the Weyl tensor vanishes identically, a general three-dimensional metric need not be conformally flat. The local obstruction to conformal flatness is instead measured by the Cotton tensor, which is defined as follows 
\begin{equation}
    C_{\mu\nu\rho}:=2\nabla_{[\rho}S_{\nu]\mu}\qquad S_{\mu\nu}:=g_{\mu\alpha}S_{\nu}^{\alpha}\,.
\end{equation}
Its two-index dual can be written as
\begin{equation}
    \mathscr{C}_{\nu}^{\mu}=\frac{1}{2}\epsilon^{\mu\alpha\beta}C_{\nu\beta\alpha}=\epsilon^{\mu\alpha\beta}\nabla_{\alpha}S_{\beta\nu}\,,
\end{equation}
where $\epsilon^{\mu\alpha\beta}$ is the metric volume tensor for a chosen orientation. The tensor
\begin{equation}
    \mathscr{C}_{\mu\nu}=g_{\mu\alpha}\mathscr{C}_{\nu}^{\alpha}
\end{equation}
is symmetric. A three-dimensional metric is locally conformally flat if and only if its Cotton tensor vanishes. 

\subsection{Einstein Spaces and Constant Curvature}

In mixed index notation, the vacuum Einstein equations with a cosmological constant are
\begin{equation}
    R_{\nu}^{\mu}+\left(\Lambda-\frac{1}{2}R\right)\delta_{\nu}^{\mu}=0\,. 
\end{equation}
Taking the trace in $d \geq 3 $ dimensions gives
\begin{equation}
    R=\frac{2d}{d-2}\Lambda\,.
\end{equation}
Substituting this result into the field equations yields
\begin{equation}
    R_{\nu}^{\mu}=\frac{2\Lambda}{d-2}\delta_{\nu}^{\mu}\,.
\end{equation}
The traceless Ricci tensor, therefore, vanishes, and the general curvature decomposition reduces to
\begin{equation}
    R_{\lambda\rho}^{\mu\nu}=C_{\lambda\rho}^{\mu\nu}+\frac{2\Lambda}{(d-1)(d-2)}\delta_{\lambda\rho}^{\mu\nu}\,.
\end{equation}
For $d>3$, an Einstein metric need not have constant sectional curvature because its Weyl tensor can be nonzero.

In three dimensions, however, $C_{\lambda\rho}^{\mu\nu}\equiv0$, and hence 
\begin{equation}
    R_{\nu}^{\mu}=2\Lambda\delta_{\nu}^{\mu}\,,\qquad R=6\Lambda\,,
\end{equation}
together with
\begin{equation}
    R_{\lambda\rho}^{\mu\nu}=\Lambda\delta_{\lambda\rho}^{\mu\nu}\,.
\end{equation}
Every smooth vacuum solution of three dimensional Einstein gravity is therefore locally maximally symmetric.

On the other hand, the Schouten tensor is given by 
\begin{equation}
    S_{\nu}^{\mu}=\frac{\Lambda}{2}\delta_{\nu}^{\mu}\,.
\end{equation}
Because $\Lambda$  is constant and the connection is metric compatible, the Cotton tensor also vanishes. Thus, these vacuum geometries are locally conformally flat. 

For a negative cosmological constant, define the AdS radius $\ell$  by $\Lambda=-\tfrac{1}{\ell^{2}}$. The curvature tensors then satisfy
\begin{equation}
    R_{\lambda\rho}^{\mu\nu}=-\frac{1}{\ell^{2}}\delta_{\lambda\rho}^{\mu\nu},\qquad R_{\nu}^{\mu}=-\frac{2}{\ell^{2}}\delta_{\nu}^{\mu},\qquad R=-\frac{6}{\ell^{2}}\,.
\end{equation}
These relations establish local equivalence to $\mathrm{AdS}_{3}$, but do not imply global equivalence. Geometries with the same local curvature can differ in their global identifications, topology, causal structure, holonomies, and conserved charges. The BTZ black hole provides a clear example of this difference. 

Finally, Eq.~\eqref{eq:riemann-mixed-three-dimensions} is valid for every three-dimensional metric, independently of the gravitational dynamics. It therefore remains valid in higher-derivative theories such as New Massive Gravity. What changes in those theories is the field equation obeyed by the Ricci tensor, which need not be proportional to the identity in vacuum. The algebraic vanishing of the Weyl tensor is consequently compatible with the existence of local propagating gravitational modes.

\chapter{Asymptotic Expansion Quantities}
\label{app:asymptotic-expansions}

This appendix collects the algebraic identities, Gauss--normal
decompositions, and Fefferman--Graham expansions used in
Chapter~\ref{chapter5}. The conventions for the radial normal and
the extrinsic curvature are those introduced in
Section~\ref{sec:radial-foliations-fg}. The Gauss--normal relations
presented below are geometric identities and do not use the field
equations. Likewise, the asymptotic expansions are initially kept
off shell; the sector-dependent constraints derived from the NMG
field equations must be imposed separately.

\section{Generalized Kronecker delta}
\label{sec:gen-kdelta}

The generalized Kronecker delta of rank \(p\) is defined by
\begin{equation}
    \delta^{\mu_1\ldots\mu_p}_{\nu_1\ldots\nu_p}
    :=
    p!\,
    \delta^{[\mu_1}_{\nu_1}
    \cdots
    \delta^{\mu_p]}_{\nu_p}.
    \label{eq:app-generalized-delta}
\end{equation}
Equivalently, it can be written as the determinant
\begin{equation}
    \delta^{\mu_1\ldots\mu_p}_{\nu_1\ldots\nu_p}
    =
    \det
    \begin{pmatrix}
        \delta^{\mu_1}_{\nu_1}
        & \cdots
        & \delta^{\mu_1}_{\nu_p}
        \\
        \vdots
        & \ddots
        & \vdots
        \\
        \delta^{\mu_p}_{\nu_1}
        & \cdots
        & \delta^{\mu_p}_{\nu_p}
    \end{pmatrix}.
\end{equation}
This representation makes its total antisymmetry in both sets of
indices manifest. In \(D\geq p\geq k\) dimensions, contracting
\(k\) pairs of indices gives
\begin{equation}
    \delta^{\mu_1\ldots\mu_p}_{\nu_1\ldots\nu_p}
    \delta^{\nu_1}_{\mu_1}
    \cdots
    \delta^{\nu_k}_{\mu_k}
    =
    \frac{(D-p+k)!}{(D-p)!}\,
    \delta^{\mu_{k+1}\ldots\mu_p}
           _{\nu_{k+1}\ldots\nu_p}.
    \label{footnoteKdelta}
\end{equation}
The identity follows by performing one contraction at a time. Each
contraction produces a factor \(D-p+r\), where \(r=1,\ldots,k\).

\section{Gauss--normal foliation}
\label{sec:GN-FG}

Consider the Gauss--normal metric introduced in
Eq.~\eqref{eq:radial-metric-decomposition}, with lapse \(N=N(z)\) and vanishing shift. In
terms of the extrinsic curvature convention adopted below
Eq.~\eqref{eq:extrinsic-curvature-convention}, the nonvanishing Christoffel symbols are
\begin{subequations}
\label{ChristoffelGN}
\begin{align}
    \Gamma^z{}_{zz}
    &=
    \frac{\partial_zN}{N},
    &
    \Gamma^z{}_{ij}
    &=
    \frac{1}{N}K_{ij},
    \\
    \Gamma^i{}_{zj}
    &=
    -NK^i{}_j,
    &
    \Gamma^i{}_{jk}
    &=
    \varGamma^i{}_{jk}(h).
\end{align}
\end{subequations}
Here \(\varGamma^i{}_{jk}(h)\) denotes the Levi--Civita connection
of the induced metric \(h_{ij}\).

The independent components of the bulk Riemann tensor are obtained
by substituting Eqs.~\eqref{ChristoffelGN} into its definition. They
take the Gauss--Codazzi form
\begin{subequations}
\label{RiemannGN}
\begin{align}
    R^{ij}_{kl}
    &=
    \mathcal R^{ij}_{kl}
    -K^i_kK^j_l
    +K^i_lK^j_k,
    \label{eq:app-Gauss}
    \\
    R^{zi}_{zj}
    &=
    \frac{1}{N}\partial_zK^i_j
    -K^i_kK^k_j,
    \label{eq:app-Ricci-normal}
    \\
    R^{zi}_{jk}
    &=
    \frac{2}{N}\mathcal D_{[j}K^i_{k]}.
    \label{eq:app-Codazzi}
\end{align}
\end{subequations}
The quantities \(\mathcal R^{ij}{}_{kl}\) and \(\mathcal D_i\) are,
respectively, the intrinsic Riemann tensor and the covariant
derivative associated with \(h_{ij}\).

Contracting the Gauss--Codazzi relations gives
\begin{subequations}
\label{eq:app-GN-Ricci}
\begin{align}
    R^z_z
    &=
    -K^i_jK^j_i
    +\frac{1}{N}\partial_zK,
    \\
    R^z_i
    &=
   - \frac{1}{N}
    \mathcal D_j
    \left(
        K^j_i-\delta^j_iK
    \right),
    \\
    R^i_j
    &=
    \mathcal R^i_j
    -KK^i_j
    +\frac{1}{N}\partial_zK^i_j,
    \\
    R
    &=
    \mathcal R
    -K^2
    -K^i_jK^j_i
    +\frac{2}{N}\partial_zK.
\end{align}
\end{subequations}

Using the definition of the Schouten tensor in
Eq.~\eqref{Schouten}, these relations imply
\begin{subequations}
\label{Schouten-comp}
\begin{align}
    S^i_j
    &=
    \mathcal R^i{}_j
    -KK^i{}_j
    +\frac{1}{N}\partial_zK^i{}_j
    \nonumber\\
    &\quad
    +\frac{1}{4}\delta^i{}_j
    \left(
        K^k{}_lK^l{}_k
        +K^2
        -\frac{2}{N}\partial_zK
        -\mathcal R
    \right).
    \label{eq:app-GN-Schouten-tangential}
    \\
    S^z_z
    &=
    -\frac14
    \left(
        \mathcal R
        +3K^i_jK^j_i
        -K^2
        -\frac{2}{N}\partial_zK
    \right),
    \label{eq:app-GN-Schouten-radial}
    \\
    S^z_j
    &=
    -\frac{1}{N}
    \mathcal D_i
    \left(
        K^i_j-\delta^i_jK
    \right),
    \label{eq:app-GN-Schouten-mixed}
    \\
    S^i_i
    &=
    \frac12
    \left(
        \mathcal R
        -K^2
        +K^i_jK^j_i
    \right).
    \label{eq:app-GN-Schouten-trace}
\end{align}
\end{subequations}
These equations follow directly from the Gauss--normal decomposition
and are valid independently of the gravitational field equations.

\section{Fefferman--Graham expansions}

We now specialize the exact relations above to the power-law
Fefferman--Graham expansion~\eqref{eq:fg-metric-background}. All indices carried by the
coefficients \(g_{(n)ij}\) are raised and lowered with
\(g_{(0)ij}\). Its covariant derivative is denoted by \(D_i\), and
its Ricci scalar by \(\mathcal R_{(0)}\).

For compactness, we introduce
\begin{equation}
    a:=\text{tr }g_{(1)},
    \qquad
    b:=\text{tr }g_{(2)},
    \qquad
    c:=\text{tr }\!\left(g_{(1)}^2\right).
    \label{eq:app-trace-shorthand}
\end{equation}
Thus, \(a^2\) denotes the square of the trace, whereas \(c\) is the
trace of the matrix square. No relation between \(a^2\) and \(c\) is
assumed unless the degenerate-sector field equations are imposed.

\subsection{Metric and volume elements}

The inverse rescaled metric is
\begin{equation}
    \bar g^{ij}
    =
    g_{(0)}^{ij}
    -\frac{z}{\ell}g_{(1)}^{ij}
    +\frac{z^2}{\ell^2}
    \left(
        g_{(1)}{}^i{}_kg_{(1)}^{kj}
        -g_{(2)}^{ij}
    \right)
    +\mathcal O(z^3).
    \label{eq:app-inverse-metric}
\end{equation}
This follows by requiring
\(\bar g^{ik}\bar g_{kj}=\delta^i_j\) order by order in \(z\).

The bulk and induced volume elements are
\begin{align}
    \sqrt{|g|}
    &=
    \frac{\ell^3}{z^3}\sqrt{|g_{(0)}|}
    \left[
        1+\frac{z}{2\ell}a
        +\frac{z^2}{\ell^2}
        \left(
            \frac b2+\frac{a^2}{8}-\frac c4
        \right)
        +\mathcal O(z^3)
    \right],
    \label{eq:app-bulk-determinant}
    \\
    \sqrt{|h|}
    &=
    \frac{\ell^2}{z^2}\sqrt{|g_{(0)}|}
    \left[
        1+\frac{z}{2\ell}a
        +\frac{z^2}{\ell^2}
        \left(
            \frac b2+\frac{a^2}{8}-\frac c4
        \right)
        +\mathcal O(z^3)
    \right].
    \label{eq:app-induced-determinant}
\end{align}
The remainder is displayed inside the brackets because the overall
negative powers of \(z\) change its order in the complete volume
elements.

\subsection{Christoffel symbols}

Applying Eqs.~\eqref{ChristoffelGN} with \(N=\ell/z\) gives
\begin{align}
    \Gamma^z{}_{zz}
    &=
    -\frac1z\,, \qquad \Gamma^z{}_{zi}
    =
    0\,, \qquad \Gamma^i{}_{zz}
    =
    0\,,
    \\
    \Gamma^z{}_{ij}
    &=
    \frac{g_{(0)ij}}{z}
    +\frac{g_{(1)ij}}{2\ell}
    +\mathcal O(z^2),
    \label{eq:app-Gamma-zij}
    \\
    \Gamma^i{}_{zj}
    &=
    -\frac{\delta^i_j}{z}
    +\frac{1}{2\ell}g_{(1)}{}^i{}_j
    +\frac{z}{\ell^2}
    \left(
        g_{(2)}{}^i{}_j
        -\frac12
        g_{(1)}{}^i{}_kg_{(1)}{}^k{}_j
    \right)
    +\mathcal O(z^2).
    \label{eq:app-Gamma-izj}
\end{align}
In particular, the term proportional to
\(z\,g_{(2)ij}\) cancels from \(\Gamma^z{}_{ij}\).

The tangential connection is unchanged by the overall conformal
factor on each constant-\(z\) hypersurface:
\begin{equation}
    \Gamma^i{}_{jk}
    =
    \Gamma[\bar g]^i{}_{jk}
    =
    \Gamma[h]^i{}_{jk}.
\end{equation}
Its expansion is
\begin{equation}
    \Gamma^i{}_{jk}
    =
    \Gamma[g_{(0)}]^i{}_{jk}
    +\frac{z}{\ell}\gamma_{(1)}{}^i{}_{jk}
    +\frac{z^2}{\ell^2}\gamma_{(2)}{}^i{}_{jk}
    +\mathcal O(z^3),
    \label{eq:app-Gamma-tangential}
\end{equation}
where
\begin{align}
    \gamma_{(1)}{}^i{}_{jk}
    &=
    \frac12
    \left(
        D_jg_{(1)}{}^i{}_k
        +D_kg_{(1)}{}^i{}_j
        -D^ig_{(1)jk}
    \right),
    \\
    \gamma_{(2)}{}^i{}_{jk}
    &=
    \frac12
    \left(
        D_jg_{(2)}{}^i{}_k
        +D_kg_{(2)}{}^i{}_j
        -D^ig_{(2)jk}
    \right)
    -g_{(1)}{}^i{}_l
     \gamma_{(1)}{}^l{}_{jk}.
    \label{eq:app-Gamma-coefficients}
\end{align}

\subsection{Extrinsic curvature}

The mixed extrinsic curvature and its trace are
\begin{align}
    K^i_j
    &=
    \frac{1}{\ell}\delta^i_j
    -\frac{z}{2\ell^2}g_{(1)}^i{}_j
    +\frac{z^2}{\ell^3}
    \left(
        \frac{1}{2}
        g_{(1)}{}^i{}_k g_{(1)}{}^k{}_j
        -g_{(2)}{}^i{}_j
    \right)
    +\mathcal O(z^3),
    \label{eq:app-K-mixed}
    \\
    K
    &=
    \frac{2}{\ell}
    -\frac{z}{2\ell^2}a
    +\frac{z^2}{\ell^3}
    \left(
        \frac{c}{2}-b
    \right)
    +\mathcal O(z^3).
    \label{eq:app-K-trace}
\end{align}

The quadratic contractions required by the extrinsic boundary terms
are
\begin{align}
    K^i_jK^j_i
    &=
    \frac2{\ell^2}
    -\frac{z}{\ell^3}a
    +\frac{z^2}{\ell^4}
    \left(
        \frac54c-2b
    \right)
    +\mathcal O(z^3),
    \\
    K^2
    &=
    \frac4{\ell^2}
    -\frac{2z}{\ell^3}a
    +\frac{z^2}{\ell^4}
    \left(
        \frac{a^2}{4}-4b+2c
    \right)
    +\mathcal O(z^3),
    \\
    K^2-K^i_jK^j_i
    &=
    \frac2{\ell^2}
    -\frac{z}{\ell^3}a
    +\frac{z^2}{\ell^4}
    \left(
        \frac{a^2}{4}-2b+\frac{3c}{4}
    \right)
    +\mathcal O(z^3).
    \label{eq:app-K-quadratic}
\end{align}

Combining these expressions with
Eq.~\eqref{eq:app-induced-determinant} produces the boundary
densities
\begin{align}
    \sqrt{|h|}\,K
    &=
    \sqrt{|g_{(0)}|}
    \left[
        \frac{2\ell}{z^2}
        +\frac{a}{2z}
        +\mathcal O(z)
    \right],
    \label{eq:app-density-K}
    \\
    \sqrt{|h|}
    \left(
        K^2-K^i_jK^j_i
    \right)
    &=
    \sqrt{|g_{(0)}|}
    \left[
        \frac2{z^2}
        +\frac1{\ell^2}
        \left(
            -b+\frac c4
        \right)
        +\mathcal O(z)
    \right].
    \label{eq:app-density-Kquadratic}
\end{align}
The absence of a \(z^{-1}\) term in the second expression results
from a cancellation between the determinant and extrinsic-curvature
expansions.

\subsection{Riemann tensor}

Substitution of the extrinsic-curvature expansion into
Eqs.~\eqref{RiemannGN} gives
\begin{align}
    R^{zi}_{zj}
    &=
    -\frac1{\ell^2}\delta^i_j
    +\frac{z}{2\ell^3}g_{(1)}{}^i{}_j
    -\frac{z^2}{4\ell^4}
     g_{(1)}{}^i{}_kg_{(1)}{}^k{}_j
    +\mathcal O(z^3),
    \label{eq:app-Riemann-radial}
    \\
    R^{zi}_{jk}
    &=
    -\frac{z^2}{\ell^3}
     D_{[j}g_{(1)}{}^i{}_{k]}
    +\mathcal O(z^3),
    \label{eq:app-Riemann-mixed-one}
    \\
    R^{ij}_{zk}
    &=
    -\frac{z^2}{\ell^3}
     D^{[i}g_{(1)}{}^{j]}{}_k
    +\mathcal O(z^3).
    \label{eq:app-Riemann-mixed-two}
\end{align}

Because the constant-\(z\) hypersurfaces are two-dimensional, their
intrinsic Riemann tensor is completely determined by their Ricci
scalar. The tangential bulk component consequently takes the form
\begin{equation}
    R^{ij}_{kl}
    =
    \left[
        -\frac1{\ell^2}
        +\frac{z}{2\ell^3}a
        +\frac{z^2}{\ell^4}
        \left(
            b-\frac{3c}{8}-\frac{a^2}{8}
            +\frac{\ell^2}{2}\mathcal R_{(0)}
        \right)
    \right]
    \delta^{ij}_{kl}
    +\mathcal O(z^3).
    \label{eq:app-Riemann-tangential}
\end{equation}

When \(g_{(1)ij}=0\), the quadratic contributions displayed above
vanish and the mixed components begin at cubic order:
\begin{align}
    R^{zi}_{jk}
    &=
    -\frac{2z^3}{\ell^4}
     D_{[j}g_{(2)}{}^i{}_{k]}
    +\mathcal O(z^4),
    \\
    R^{ij}_{zk}
    &=
    -\frac{2z^3}{\ell^4}
     D^{[i}g_{(2)}{}^{j]}{}_k
    +\mathcal O(z^4),
    \\
    R^{zi}_{zj}
    &=
    -\frac1{\ell^2}\delta^i_j
    +\mathcal O(z^3).
    \label{eq:app-Riemann-g1-zero}
\end{align}

\subsection{Ricci and Schouten tensors}

The Ricci components obtained by contracting the preceding
expressions are
\begin{align}
    R^z_z
    &=
    -\frac2{\ell^2}
    +\frac{z}{2\ell^3}a
    -\frac{z^2}{4\ell^4}c
    +\mathcal O(z^3),
    \\
    R^z_i
    &=
    \frac{z^2}{2\ell^3}
    \left(
        D_jg_{(1)}{}^j{}_i-D_i a
    \right)
    +\mathcal O(z^3),
    \\
    R^i_j
    &=
    -\frac2{\ell^2}\delta^i_j
    +\frac{z}{2\ell^3}
    \left(
        g_{(1)}{}^i{}_j+a\delta^i_j
    \right)
    \nonumber\\
    &\quad
    +\frac{z^2}{\ell^4}
    \left[
        -\frac a4g_{(1)}{}^i{}_j
        +\left(
            b-\frac c2
            +\frac{\ell^2}{2}\mathcal R_{(0)}
        \right)\delta^i_j
    \right]
    +\mathcal O(z^3).
    \label{eq:app-Ricci-expansions}
\end{align}
Their trace gives the scalar curvature
\begin{equation}
    R
    =
    -\frac6{\ell^2}
    +\frac{2z}{\ell^3}a
    +\frac{z^2}{\ell^4}
    \left(
        2b-\frac54c-\frac14a^2
        +\ell^2\mathcal R_{(0)}
    \right)
    +\mathcal O(z^3).
    \label{eq:app-Ricci-scalar}
\end{equation}

Using \(S^\mu_\nu=R^\mu_\nu-\tfrac14
\delta^\mu_\nu R\), one finds
\begin{align}
    S^z_z
    &=
    -\frac1{2\ell^2}
    +\frac{z^2}{16\ell^4}
    \left(
        c+a^2-8b-4\ell^2\mathcal R_{(0)}
    \right)
    +\mathcal O(z^3),
    \label{eq:app-Schouten-radial}
    \\
    S^i_j
    &=
    -\frac1{2\ell^2}\delta^i_j
    +\frac{z}{2\ell^3}g_{(1)}{}^i{}_j
    \nonumber\\
    &\quad
    +\frac{z^2}{16\ell^4}
    \left[
        -4a\,g_{(1)}{}^i{}_j
        +\left(
            8b-3c+a^2
            +4\ell^2\mathcal R_{(0)}
        \right)\delta^i_j
    \right]
    +\mathcal O(z^3),
    \label{eq:app-Schouten-tangential}
    \\
    S^z_i
    &=
    \frac{z^2}{2\ell^3}
    \left(
        D_jg_{(1)}{}^j{}_i-D_i a
    \right)
    +\mathcal O(z^3),
    \label{eq:app-Schouten-mixed}
    \\
    S
    &=
    -\frac3{2\ell^2}
    +\frac{z}{2\ell^3}a
    +\frac{z^2}{16\ell^4}
    \left(
        8b-5c-a^2
        +4\ell^2\mathcal R_{(0)}
    \right)
    +\mathcal O(z^3).
    \label{eq:app-Schouten-trace}
\end{align}
These expressions are the Fefferman--Graham expansion of the exact
Gauss--normal components in Eqs.~\eqref{Schouten-comp}.

\subsection{Expansion of the \texorpdfstring{$E$}{E}-tensor}

Using the normalization of
\(E^{\mu\nu}_{\lambda\rho}\) adopted in
Chapter~\ref{chapter5}, its independent components are
\begin{align}
    E^{zi}_{zj}
    &=
    \frac{\kappa\varpi_-}{2}\delta^i_j
    +\frac{\kappa\sigma_2z}{4\mu^2\ell^3}
    \left(
        a\delta^i_j-g_{(1)}{}^i{}_j
    \right)
    \nonumber\\
    &\quad
    +\frac{\kappa\sigma_2z^2}{32\mu^2\ell^4}
    \left[
        4g_{(1)}{}^i{}_kg_{(1)}{}^k{}_j
        +\left(
            8b-5c-a^2
            +4\ell^2\mathcal R_{(0)}
        \right)\delta^i_j
    \right]
    +\mathcal O(z^3),
    \label{eq:app-E-radial}
    \\
    E^{ij}_{kl}
    &=
    \left[
        \frac{\kappa\varpi_-}{2}
        +\frac{\kappa\sigma_2z^2}{32\mu^2\ell^4}
        \left(
            c+a^2-8b
            -4\ell^2\mathcal R_{(0)}
        \right)
    \right]
    \delta^{ij}_{kl}
    +\mathcal O(z^3),
    \label{eq:app-E-tangential}
    \\
    E^{zi}_{jk}
    &=
    \frac{\kappa\sigma_2z^2}{2\mu^2\ell^3}
    D_{[j}g_{(1)}{}^i{}_{k]}
    +\mathcal O(z^3),
    \\
    E^{ij}_{zk}
    &=
    \frac{\kappa\sigma_2z^2}{2\mu^2\ell^3}
    D^{[i}g_{(1)}{}^{j]}{}_k
    +\mathcal O(z^3).
    \label{eq:app-E-mixed}
\end{align}

\subsection{Cotton tensor}
\label{app:Cotton-expansion}

For the smooth power-law expansion considered here, the components
of the Cotton tensor with two contravariant indices satisfy
\begin{equation}
    C_i{}^{zj}
    =
    \mathcal O(z^4),
    \qquad
    C_i{}^{jk}
    =
    \mathcal O(z^4),
    \qquad
    C_z{}^{zi}
    =
    \mathcal O(z^4),
    \qquad
    C_z{}^{ij}
    =
    \mathcal O(z^4).
    \label{eq:app-Cotton-orders}
\end{equation}
One way to understand these orders is to use the conformal
invariance of the three-index covariant Cotton tensor in three
dimensions. Its covariant components remain finite for the regular
rescaled metric introduced in
Section~\ref{sec:radial-foliations-fg}, whereas raising two indices
with the physical metric supplies an overall factor
\(z^4/\ell^4\).

A direct radial expansion appears initially to produce
\begin{equation}
    C_i{}^{zj}
    =
    \frac{z^3}{8\ell^6}
    \left[
        2a\,g_{(1)}{}^j{}_i
        -2g_{(1)}{}^j{}_kg_{(1)}{}^k{}_i
        +(c-a^2)\delta^j_i
    \right]
    +\mathcal O(z^4).
    \label{eq:app-Cotton-cubic}
\end{equation}
The coefficient of \(z^3\), however, vanishes identically. Indeed,
the Cayley--Hamilton identity for the two-dimensional matrix
\(g_{(1)}{}^i{}_j\) is
\begin{equation}
    g_{(1)}{}^i{}_kg_{(1)}{}^k{}_j
    -a\,g_{(1)}{}^i{}_j
    +\frac12
    \left(
        a^2-c
    \right)\delta^i_j
    =
    0.
    \label{eq:app-Cotton-cancellation}
\end{equation}
Multiplication by \(-2\) shows that the expression in square
brackets in Eq.~\eqref{eq:app-Cotton-cubic} is zero.

This cancellation is a two-dimensional algebraic identity. It does
not require the NMG field equations or the degenerate-sector
condition \(c=a^2\). The quartic coefficients are not determined by
the truncation displayed here. Consequently,
Eq.~\eqref{eq:app-Cotton-orders} specifies the asymptotic fall-off
and does not imply that the Cotton tensor vanishes identically.

\cleardoublepage
\addcontentsline{toc}{chapter}{Bibliography}
\printbibliography
\end{document}